%% file: main.tex
\pdfoutput=1
\documentclass[12pt]{report}
\usepackage[a4paper]{geometry}
\usepackage[myheadings]{fullpage}
\usepackage{fancyhdr}
\usepackage{lastpage}
\usepackage{graphicx, wrapfig, subcaption, setspace, booktabs}
\usepackage[T1]{fontenc}
\usepackage[font=small, labelfont=bf]{caption}
\usepackage{fourier}
\usepackage[protrusion=true, expansion=true]{microtype}
\usepackage[english]{babel}
\usepackage{sectsty}
\usepackage[]{subcaption}
\usepackage{color}
\usepackage{placeins}
\usepackage{multirow}
\usepackage{url, lipsum}
\usepackage[colorinlistoftodos]{todonotes}
\usepackage{multicol}
\usepackage{amsmath}
\usepackage[separate-uncertainty = true,multi-part-units=single]{siunitx}
\usepackage[inline]{enumitem}
\newcommand{\MM}{Micromegas}
\newcommand{\MF}{Module Frame}
\newcommand{\TPC}{TPC}
\usepackage{siunitx}
\usepackage{color}
\usepackage{tikz}
\usepackage{authblk}

\usepackage{xspace}
\usepackage{hyperref} 

\newcommand{\HRule}[1]{\rule{\linewidth}{#1}}
\begin{document}

\title{ \normalsize \textsc{}
		\\ [2.0cm]
		\HRule{0.5pt} \\
		\LARGE \textbf{\uppercase{T2K ND280 Upgrade \\ Technical Design Report}}
		\HRule{2pt} \\ [0.5cm]
		\normalsize \today \vspace*{5\baselineskip}}

\input{authors}

\date{}


 \maketitle
 
 \abstract{
 In this document we present the Technical Design Report of the Upgrade of the T2K Near Detector ND280. The goal of this upgrade is to improve the Near Detector performance to measure the neutrino interaction rate and to constrain the neutrino interaction cross-sections so that the uncertainty in the number of predicted events at Super-Kamiokande is reduced to about 4\%. This will allow to improve the physics reach of the T2K-II project.

This goal is achieved by modifying the upstream part of the detector, adding a new highly granular scintillator detector (Super-FGD), two new TPCs (High-Angle TPC) and six TOF planes.
Details about the detector concepts, design and construction methods are presented, as well as a first look at the test-beam data taken in Summer 2018. An update of the physics studies is also presented.}
\tableofcontents
\newpage

\sectionfont{\scshape}


\input{executive-summary.tex}
\input{acro.tex}
\input{introduction.tex}

\input{Target.tex}

\input{TPC.tex}
\input{TestBeam.tex}

\input{ToF.tex}
\input{Integration.tex}

\input{Physics.tex}


\bibliographystyle{ieeetr}
\bibliography{Introduction,Physics,TPC,Target,ToF}


\end{document}

%% file: authors.tex
\author[48]{K.~Abe}
\author[47,23]{H.~Aihara}
\author[19]{A.~Ajmi}
\author[46,27]{C.~Andreopoulos}
\author[21]{M.~Antonova}
\author[24]{S.~Aoki}
\author[61]{Y.~Asada}
\author[25]{Y.~Ashida}
\author[46]{A.~Atherton}
\author[15]{E.~Atkin}

% ...
\affil[15]{Imperial College London, Department of Physics, London, United Kingdom}
% ...
\affil[19]{INFN Sezione di Padova and Universit\`a di Padova, Dipartimento di Fisica, Padova, Italy}
% ...
\affil[21]{Institute for Nuclear Research of the Russian Academy of Sciences, Moscow, Russia}
% ...
\affil[23]{Kavli Institute for the Physics and Mathematics of the Universe (WPI), The University of Tokyo Institutes for Advanced Study, University of Tokyo, Kashiwa, Chiba, Japan}
\affil[24]{Kobe University, Kobe, Japan}
\affil[25]{Kyoto University, Department of Physics, Kyoto, Japan}
% ...
\affil[27]{University of Liverpool, Department of Physics, Liverpool, United Kingdom}
% ...
\affil[46]{STFC, Rutherford Appleton Laboratory, Harwell Oxford,  and  Daresbury Laboratory, Warrington, United Kingdom}
\affil[47]{University of Tokyo, Department of Physics, Tokyo, Japan}
\affil[48]{University of Tokyo, Institute for Cosmic Ray Research, Kamioka Observatory, Kamioka, Japan}
% ...
\affil[61]{Yokohama National University, Faculty of Engineering, Yokohama, Japan}
% ...

%% file: executive-summary.tex
\chapter*{Executive summary}
\addcontentsline{toc}{chapter}{Executive Summary}
We present in this document the technical design report of the upgrade of the T2K Near Detector ND280 in order to reach a systematic uncertainty at the 4\% level, matching the needs of the T2K-II phase. This phase of the T2K experiment can provide a 3~$\sigma$
exclusion of CP conservation for 36 \% of the $\delta_{CP}$ phase space, if the neutrino mass ordering is known.

We have developed a detector design that significantly improves the performance provided by ND280. In particular we achieve full polar angle coverage for the muons produced in Charged Current events, improve the tracking efficiency of pions and protons stopping inside the scintillator detector and improve the separation of electrons from converted gammas required for electron neutrino studies. The downstream part of ND280 is not altered and will continue to provide useful information on the neutrino flux and cross-sections, as well as a comparison point with respect to T2K phase I data.  

The new detector consists of the addition of a highly granular scintillator detector, the Super-FGD  (small scintillator cubes, with  1 cm side,  each read out with WLS fibers in the three orthogonal directions). 
This detector is sandwiched between two High-Angle TPC, read out by resistive Micromegas detectors, with a compact and light field cage. These detectors are surrounded by six large TOF planes to determine the track direction and improve the PID. 

The Super-FGD is an innovative device with excellent detector performance. We have observed in the first tests that in realistic conditions a MIP crossing a single cube will produce more than 30 photoelectrons per WLS fiber. The timing resolution per fiber is better than 1 ns. With these precise information we will be able to track over 4$\pi$ solid angle pions and protons stopping in this detector. Moreover its high granularity will allow to distinguish electrons produced by electron neutrino interactions from converted photons. Study are ongoing to evaluate the potential to detect neutrons in this detector.

The TPC will measure charge, momentum, track angles and dE/dx with excellent efficiencies and low systematics. Preliminary measurement in the test beam show that the space point resolution is at the 300 $\mu$m level, to be compared to 600 $\mu$m for the existing TPC. 

The TOF, consisting of cast plastic scintillator readout by MPPC, will reach a time resolution of 150 ps.

Detector prototypes of the TPC, the Super-FGD and the TOF have been successfully tested in Summer 2018 at CERN. The analysis of these data is in progress but we have already demonstrated the main features thereby validating the  detector technologies and their performance.

The study of the integration of these new detectors is ongoing and a detailed visit to the detector as built has revealed no show-stopper. A detailed 3D model is being developed as well as the plan for the installation sequence and the commissioning.

The construction of these detectors will provide new high quality neutrino beam interaction data useful to constrain the cross section models. We have checked the effectiveness of the new detectors with detailed simulations. Propagating the new information by the upgrade Near Detector all the way to the prediction at the T2K Far Detector, we obtain a significant improvement both with respect to a fixed neutrino interaction model, and with respect to the capability to discriminate between different models. On average, the post-fit uncertainty after taking into account the data provided by the upgrade detector will be 30\% lower. Furthermore the near to far extrapolation will be much less model dependent.

The detector construction for the ND280 Upgrade will be performed in 2019-2020, for an installation in Japan in 2021.  

%% file: acro.tex
\chapter*{Acronyms and abbreviations}
\addcontentsline{toc}{chapter}{Acronyms and abbreviations}
\begin{tabular}{ll}

1p1h & one-particle-one-hole\\
2p2h & two-particles-two-holes\\
ADC & Analog to Digital Converter \\
ASIC & Application Specific Integrated Circuit \\
CC & Charged Current \\
CCQE & Charged Current Quasi-Elastic \\
CF & Carbon Fiber \\
CFD & Constant Fraction Discrimination\\
CLK & Clock signal \\
CTM & Cosmic Trigger Module \\
DDR3 & Double Data Rate 3rd generation memory\\
DIF & Digital Interface board \\
DLC & Diamond-Like-Carbon\\
DSEcal & Downstream Electromagnetic Calorimeter \\
ECAL & Electromagnetic Calorimeter \\
FEA & Finite Element Analysis \\
FEB & Front End Board \\
FEC & Front End Card \\
FEE & Front End Electronics \\
FEM & Finite Element Method (also Front End Module) \\
FGD & Fine Grained Detector \\
FPGA & Field Programmable Gate Array \\
FSI & Final State Interaction \\
FV & Fiducial Volume \\
GDCC & Gigabyte Data Concentrator Card \\
GTRIG & Global Trigger\\
HA-TPC & High-Angle TPC \\
HG & High Gain \\
ILC & International Linear Collider \\
J-PARC & Japan Proton Accelerator Research Complex \\
\end{tabular}
\vfill 

\begin{tabular}{ll}
LFG & Local Fermi Gas \\
LG & Low Gain \\
LPI & Lines-per-Inch\\
LY & Light Yield \\
MA & axial mass \\
MC & Monte Carlo\\
MCM & Master Clock Module \\
MCB & Master Clock Board \\

MFC & Mass Flow Controller\\
MM0 & Micromegas 0 \\
MM1 & Micromegas 1\\
MPGD & Micro Pattern Gaseous Detector \\
MPPC & Multi Pixel Photon Counter \\
ND280 & Near Detector at 280 m \\
ND & Near Detector \\
NM & Neutrino Monitor \\
P0D & Pi-zero Detector \\
PCB & Printed Circuit Board \\
PDC & Power Distribution Card \\
PDE & Photo-Detection Efficiency \\
p.e. & photo-electron \\
PID & Particle Identification \\
POPOP & 1,4-bis(5-phenyloxazol-2-yl) benzene  (scintillator)\\
PRF & Pad Response Function \\
PS & Proto Synchrotron \\
PTP & paraterphenyl \\
QC & Quality Control \\
QE & Quasi Elastic\\
RES & Resonant\\
RFG & Relativistic Fermi Gas\\
SCA & Switched Capacitor Array\\
SCM & Slave Clock Module \\
SMRD & Side Muod Range Detector \\
SoM & System-On-Module \\
SPI & Serial Peripheral Interface\\
SYNC & Synchronization (signal) \\
T2K & Tokai to Kamioka \\

\end{tabular}
\vfill
\begin{tabular}{ll}
TDCM & Trigger and Data Concentrator Module\\
TDM &  Time Division Multiplexing \\
TDR & Technical Design Report \\
TOF & Time of Flight \\ 
ToT & Time over Threshold \\
TPC & Time Projection Chamber \\
TRX & Optical Transceiver \\
WLS & Wavelength Shifting Fiber \\

 
\end{tabular}
\vfill

%% file: introduction.tex
\chapter{Introduction}

The T2K neutrino experiment at J-PARC (see Ref.~\cite{Abe2011106} for a description of the experiment and its near detector complex) 
has submitted a proposal~\cite{t2kproposal} for an extension of the T2K running  accumulating $20 \times 10^{21}$ protons-on-target, that is 6 times the present exposure, which has received phase-I approval. This aims at initial observation of CP violation at the  3 $\sigma$ level or higher significance if the CP violation is maximal. A further increase by a factor 10 will come with the Hyper-Kamiokande detector, increasing the far detector mass from 22.5 kt to more than 200 kt~\cite{Abe:2011ts,Abe:2015zbg, HK-DR}. 

While the present configuration of ND280 leads to systematic errors of the order of 6\%, the goal is to bring this number down to $\sim4$\% for T2K-II~\cite{Blondel:2299599}, and to $\sim3$\% or below for Hyper-Kamiokande.    

The design described in this report has been developed by a dedicated team over a period of two years. First, a T2K task force studied the possible upgrade configurations while at the same time developing the software tools needed to provide a full simulation and detector response, as well as comparing the performances for each configuration.
This first period ended with the task force report \cite{TN303} in January 2016, endorsed by the T2K collaboration, which issued a public statement officially launching the upgrade project.

We then opened the project to the particle physicists community outside of T2K by launching a series of open workshops \cite{upgrade-workshops}, alternating between CERN and J-PARC (Japan). In the process, we prepared and submitted to CERN SPSC the Expression of Interest CERN-SPSC-EOI-15~\cite{EOI}, followed by  a proposal (P357) for the upgrade of the near detector ND280~\cite{Blondel:2299599}.
This TDR embodies the studies, discussions and suggestions generated during this process. 

\begin{figure} [htbp!] \centering 
\includegraphics[width=0.8\linewidth]{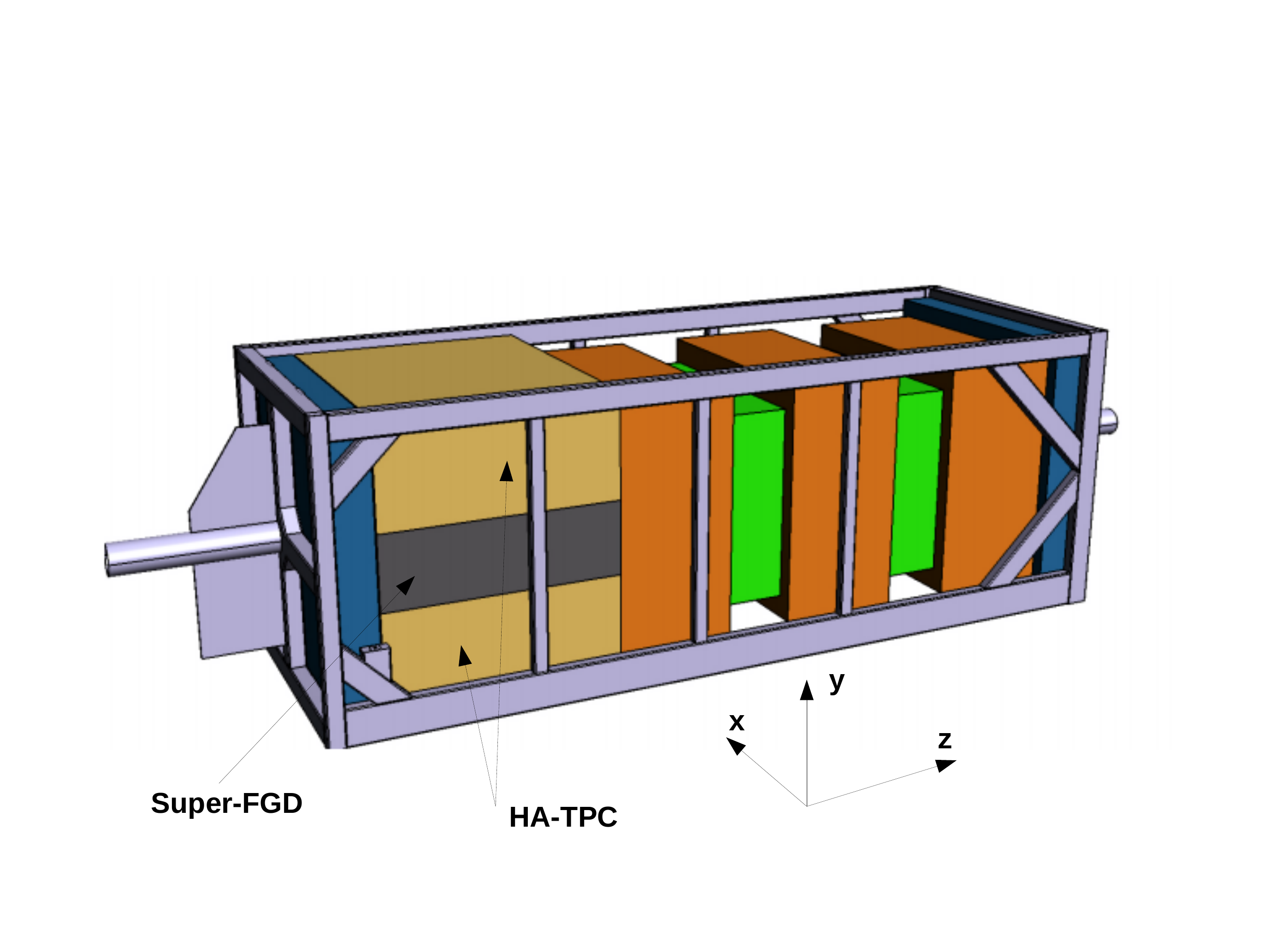}
\caption{CAD 3D Model of the ND280 upgrade detector. In the upstream part (on the left in the drawing) two High-Angle TPCs (brown) with the scintillator detector Super-FGD (gray) in the middle will be installed. In the downstream part, the tracker system composed by three TPCs (orange) and the two FGDs (green) will remain unchanged. The TOF detectors are not shown in this plot. The detector is mechanically mounted on the basket, a steel beam structure (light gray), supported at both ends. The beam is approximately parallel to the z axis, the magnetic field is parallel to the x axis.} \label{fig:nd280upgrade-intro} \end{figure}

\begin{figure} [htbp!] \centering \includegraphics[width=0.9\linewidth]{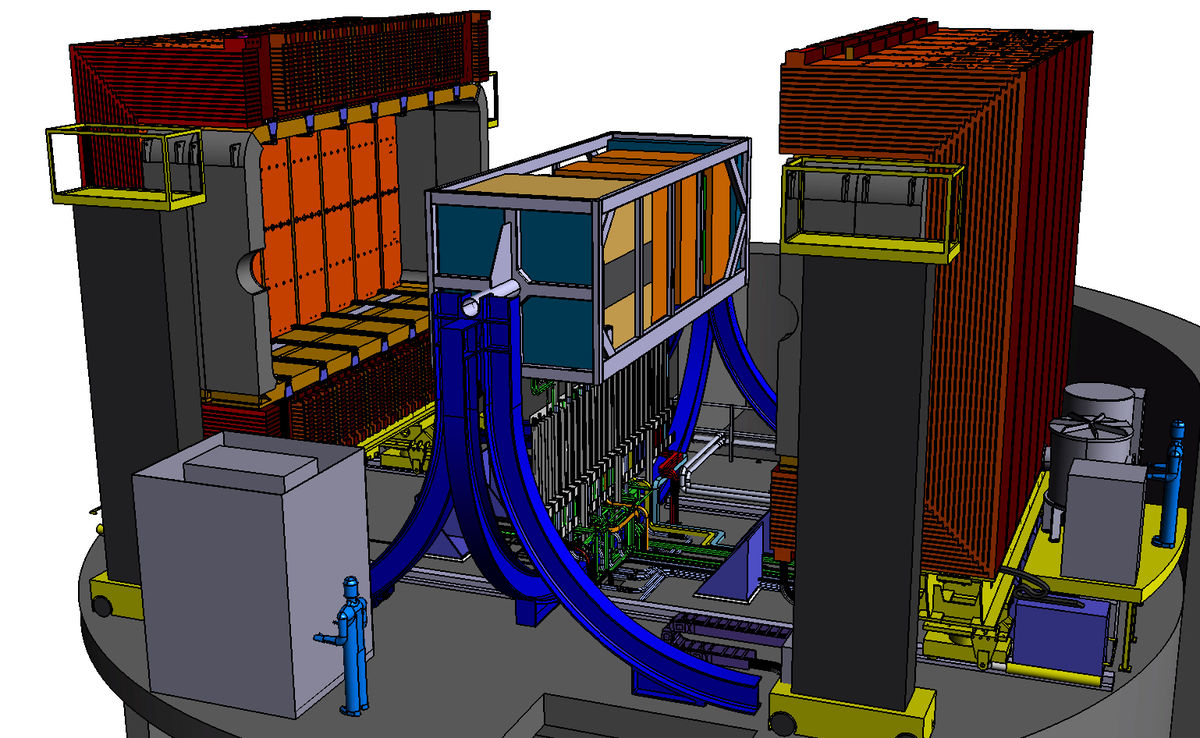} \caption{CAD 3D Model of the B1 floor of the ND280 pit. The magnet is shown in the open position with the two large magnet yokes (dark red) separated. The inner detectors are supported by the basket, a steel structure, on the basket stands (blue curved beams). } \label{fig:nd280upgrade-basket} \end{figure}

We plan to improve the performance of ND280 by adding a new highly granular, 3D scintillator detector, Super-FGD composed of small plastic scintillator cubes, read out by three WLS fibers in the three orthogonal directions. Above and below this detector are two High-Angle atmospheric pressure TPCs. These three detectors form approximately a cube with 2m-long sides (Fig.~\ref{fig:nd280upgrade-intro}). It is positioned in the upstream part of the ND280 magnet and is surrounded by six thin Time-of-Flight scintillator layers.
In the most upstream part of ND280, we will keep the P0D Upstream Calorimeter, with 4.9 radiation lengths, as a veto and to detect neutrals.
The downstream part of ND280, namely three TPCs, two scintillator detectors FGD and the full calorimeter system will remain unchanged, as well as the muon-range detector SMRD.
Figure~\ref{fig:nd280upgrade-basket} presents a general view of the B1 floor of the ND280 pit, with the magnet in the open position. 
The reference system shown in the same figure has the z axis along the neutrino beam direction (longest axis of the ND280 detector), the y axis in the vertical direction. The magnetic field is parallel to the x axis.

This configuration achieves a full polar angle acceptance for muons produced in charged-current interactions. The tracking of charged particles in the Super-FGD is also very efficient. 

\begin{figure} [htbp!] \centering \includegraphics[width=0.8\linewidth]{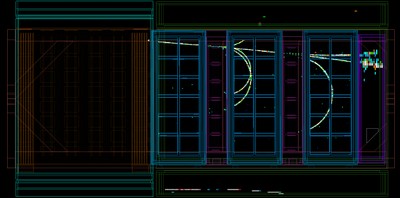}
\caption{Event display of a neutrino interaction recorded in ND280. A projection of the hits on the z-y plane transverse to the magnetic field is shown.} \label{fig:nd280eventdisplay} \end{figure}

An example of the level of information provided by the current ND280 is shown by the event display of a neutrino interaction shown in Fig.~\ref{fig:nd280eventdisplay}. 

This report is organized as follows. 
We first present in chapter~2 the design of the Super-FGD, followed by a chapter devoted to the High-Angle TPCs.
We then present the Time-of-Flight detector. In chapter 5 we present some preliminary studies on the integration of the new detectors in the existing infrastructure.
Chapter 6 present a more articulated motivation of the physics requirements and detector configuration, an update on the studies of the physics performance of the new ND280 detector suite, and new studies with transverse variables to constrain nuclear effects.

%% file: Target.tex
  \chapter{Scintillator Target Tracker (SuperFGD)}


\input{Target_intro.tex}

\input{Target_cube.tex}

\input{Target_fiber.tex}

\input{Target_MPPC.tex}

\input{Target_mechanics.tex}


\input{Target_electronics.tex}

\input{Target_DAQ.tex}

\input{Target_calibration.tex}

\input{Target_schedule.tex}

\input{Target_prototype.tex}

%% file: Target_intro.tex
\section{Conceptual Design}

In the current ND280, Fine-Grained Detectors (FGDs)~\cite{Amaudruz:2012agx} act as the active targets for neutrino interactions.
They provide measurements of charged particles generated in the neutrino interactions in combination with TPCs.
The existing FGDs consist of plastic scintillator bars aligned in either $x$ or $y$ direction perpendicular to the beam direction, which limits the acceptance to essentially the forward direction.
One of major goals of the upgrade is to improve the angular acceptance for large angle and backward-going tracks,
while keeping the basic concept of the combination of an active target and TPCs, which has been proved by ND280 to be a quite successful configuration.
Thus, a new approach is necessary for the target tracker detector.

The target detector will act as the target for the neutrino interaction 
as well as the detector to reconstruct the tracks around the interaction vertex.
It needs to have:
\begin{itemize}
\item sufficiently large mass to provide a sufficient number of neutrino interactions (comparable to the total mass of the current FGD, 2~tons),
\item acceptance for charged leptons (muons and electrons) from charged current interactions in large scattering angle, and
\item capability to reconstruct and identify short tracks of low energy hadrons around the interaction vertex.
\end{itemize}

\begin{figure}[tb]
\centering
\includegraphics[width=0.7\textwidth]{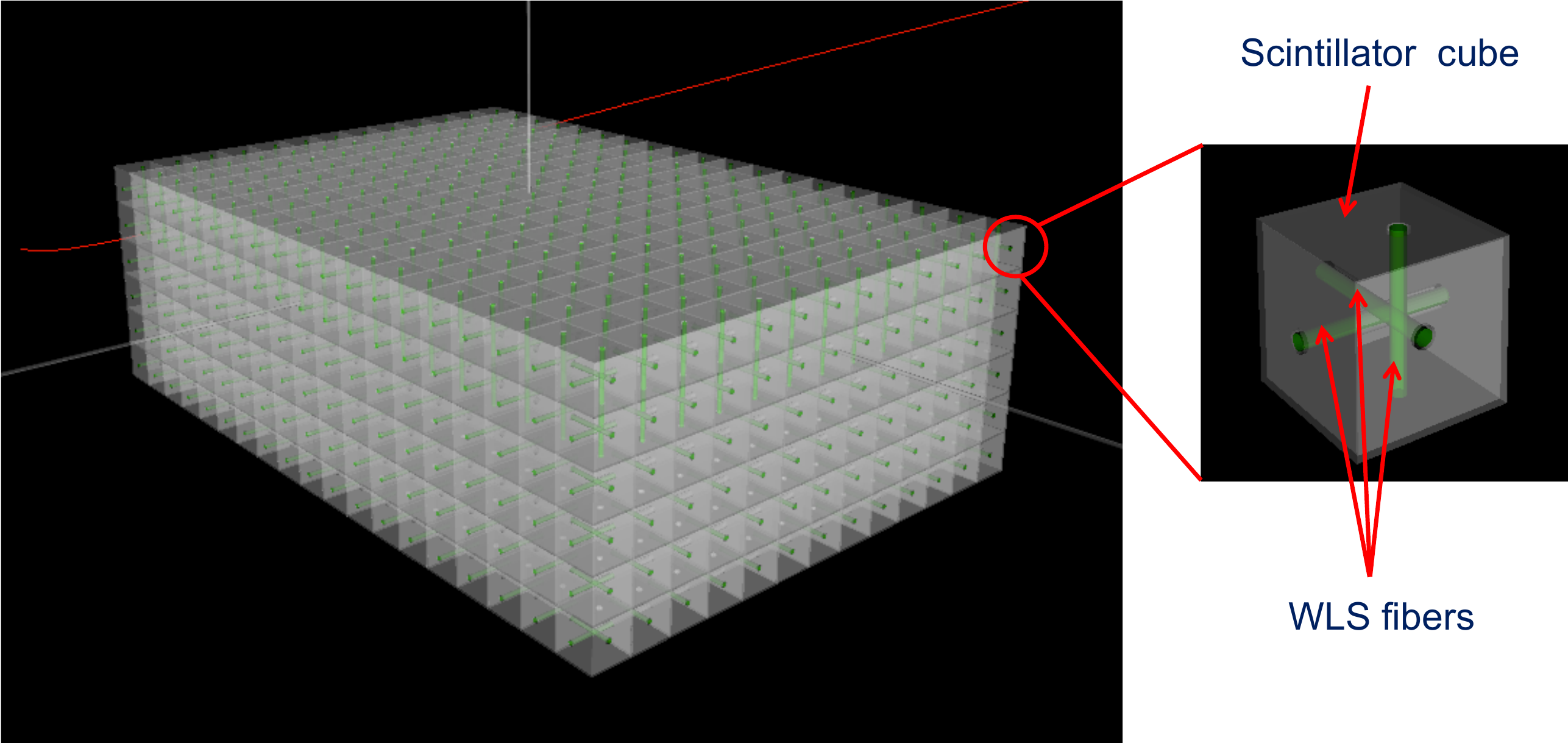}
\caption{Schematic concept of the SuperFGD structure. The size of each cube is $1\times 1\times 1$~cm$^3$.}
\label{fig:Target-SuperFGD-schematic}
\end{figure}

\begin{figure}[htb]
\centering
\includegraphics[width=0.7\textwidth]{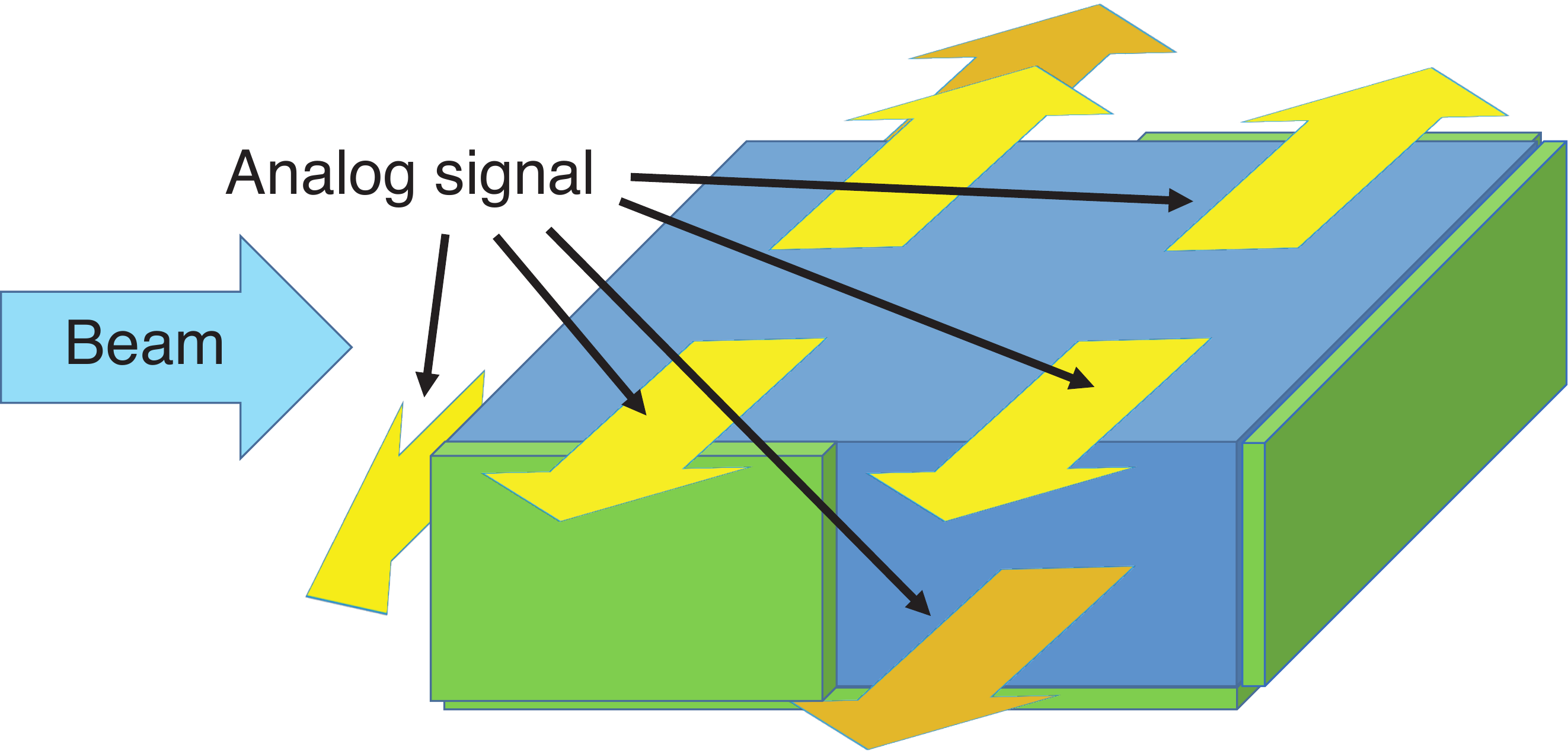}
\caption{
Schematic of the signal routing for SuperFGD.
The frontend electronics will be placed on the left and right sides of the detector.
Analog signal from the upstream and the top face will be routed to left/right.
}
\label{fig:superfgd-signal-routing}
\end{figure}

We have chosen a novel design of fine grained fully-active plastic scintillator detector, called SuperFGD,
which is a concept recently proposed by members of ND280 upgrade working group~\cite{Sgalaberna:2017khy}.
It consists of many optically independent cubes of plastic scintillator, read out along three orthogonal directions by wavelength shifting (WLS) fibers.
Figure~\ref{fig:Target-SuperFGD-schematic} shows a conceptual drawing of SuperFGD.
Each scintillator cube has three holes in $x$, $y$, and $z$ directions, where WLS fibers are inserted.
One end of each WLS fiber is instrumented with a Multi-Pixel Photon Counter (MPPC).
Because SuperFGD will provide projections of charged particle trajectories onto three planes without inactive regions, it will provide us significantly more information on the neutrino interaction compared to the existing FGDs. 

In the baseline design, the dimension of the active part of SuperFGD is $192 \times 192 \times 56$ cubes, with the size of each cube being $1\times 1\times 1$~cm$^3$.
The total numbers of cubes and readout channels will be 2,064,384 cubes and 58,368 channels, respectively.

The MPPCs will be placed on the upstream, top, left and right side of the detector.
For the readout of $y$-$z$ plane, half of MPPCs are placed on each of the left and right side in order to equalize the density of readout channels.
The analog signal from the upstream and the top face will be routed to the left or right side, where the frontend electronics will be placed, as shown in Fig.~\ref{fig:superfgd-signal-routing}.



%% file: Target_cube.tex
\section{Scintillator cubes}

\subsection{Scintillator cube production }

The scintillator cubes are produced at UNIPLAST Co. (Vladimir, Russia).
The scintillator composition is  polystyrene doped with 1.5\% of paraterphenyl (PTP) and 0.01\% of POPOP. 
After fabrication the cubes are covered by a reflecting layer by etching the scintillator surface with a chemical agent. 
The etching results in the formation of a white polystyrene micropore deposit over the scintillator. 
The thickness of the reflector layer is within 50--80~$\mu$m.
Three orthogonal through holes of 1.5~mm diameter are drilled in the cubes to accommodate WLS fibers as shown in Fig.~\ref{fig:Target-SuperFGD-schematic}. 

More than 10k cubes were produced to assemble mock-ups and prototypes. At the initial stage of R\&D the cubes were cut in size $1\times 1\times 1$~cm$^3$ out of long 1~cm thick extruded slabs.  For the real detector we plan to use  another production method of cubes, injection molding, which is  now under development.  Both methods provide the same light yield, the main differences are in the manufacturing cost of large quantities of cubes, and in the reproducibility of geometrical size.


\begin{figure}[htb]
\centering
\includegraphics[width=0.55\textwidth]{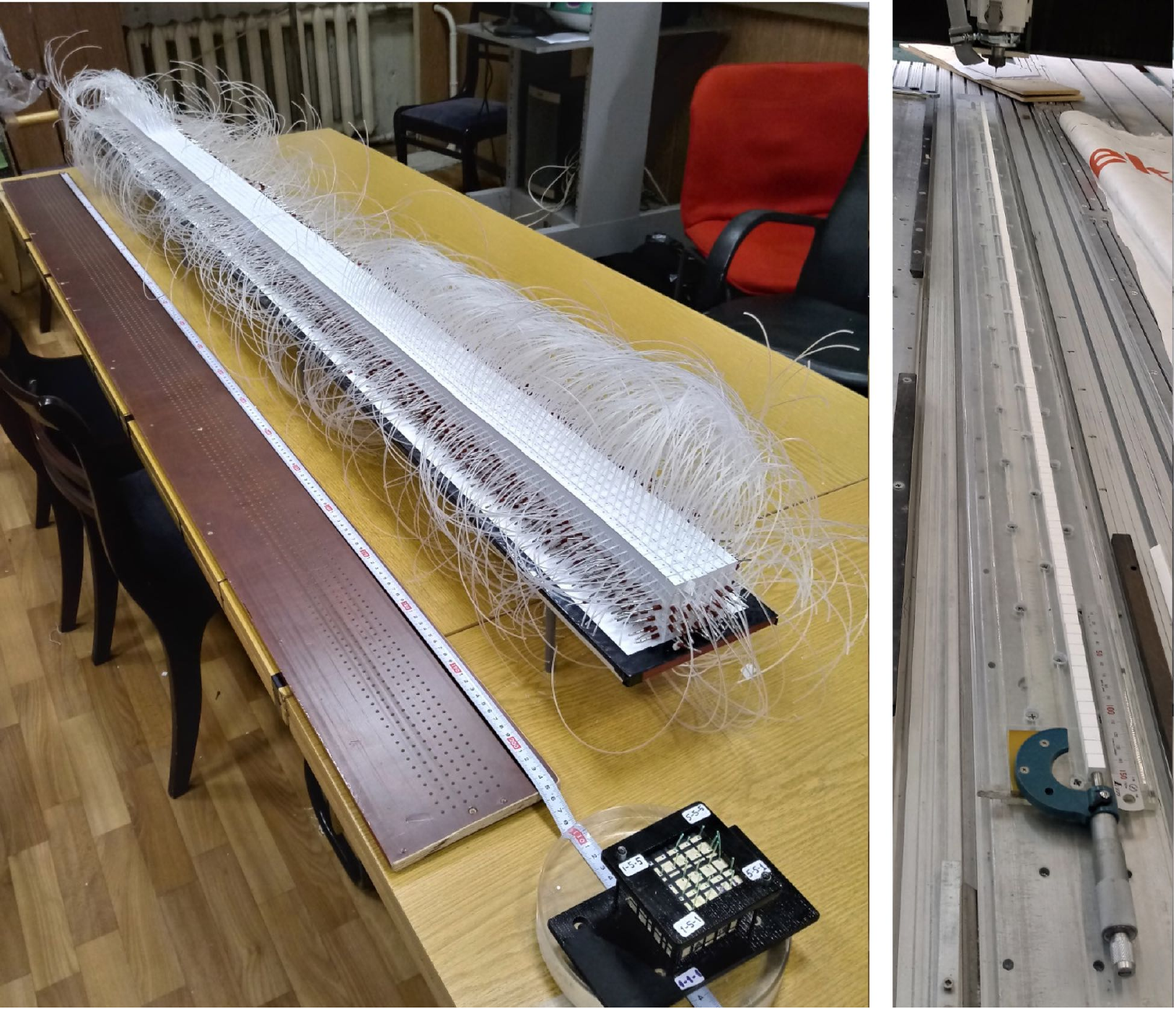}
\caption{Left: the array of $6\times 6\times 200$ cubes to check WLS fiber installation. Right: 199 cube array aligned in a single line to measure the total length variations. }
\label{fig:Target_cube_mockup} 
\end{figure}

\begin{figure}[htb]
\centering
\includegraphics[width=0.6\textwidth]{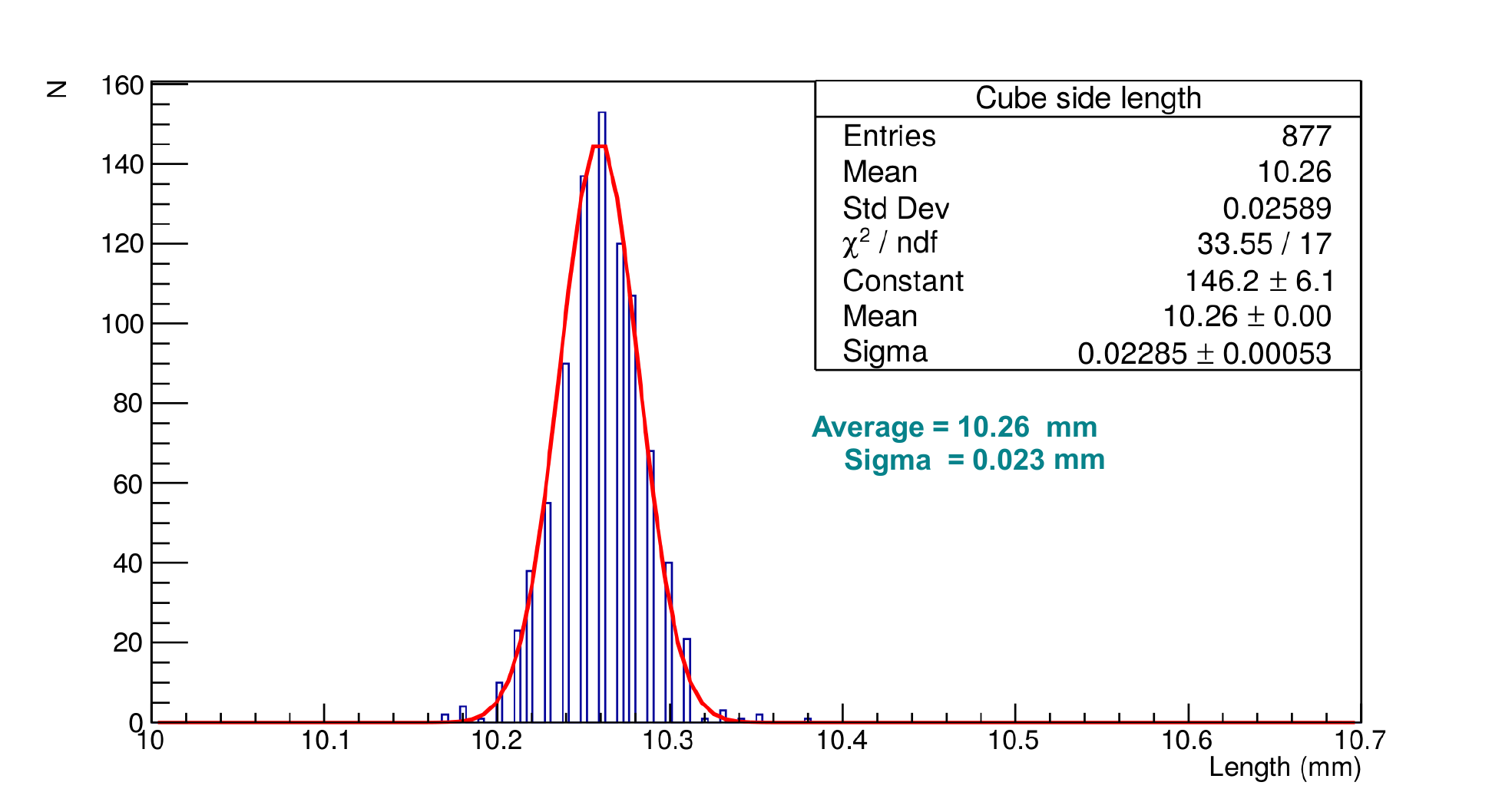}
\caption{Distribution  of the cube side length. 877 cubes were measured in random order, one side per a cube. }
\label{fig:Target_cube_cubesize} 
\end{figure}

Different size of mock-ups were assembled  to check the options for the detector construction.  Fig.~\ref{fig:Target_cube_mockup} shows two of them, each of 2~m long.  The left picture shows the array of $6\times 6\times 200$ cubes (7200 cubes). Fishing lines of 1.3~mm diameter were used to assemble the cubes into a 3D structure with a determined space position for each cube. Then the fishing lines were removed and replaced with the WLS fibers. Tests have demonstrated that 2~m long fibers can be inserted instead of the fishing lines even though this mock-up was made with the first bunch of extruded cubes of relatively variable size ($\sigma_x=100~\mu$m).

The right picture in Fig.~\ref{fig:Target_cube_mockup} shows the measurements of the length of the 199 cubes array. A groove was machined in a support base where 199 cubes were stacked as a single line in different random combinations.  The cubes were injection molded, the average width of a side was measured to be 10.26~mm with variation $\sigma_w$=23~$\mu$m, see Fig.~\ref{fig:Target_cube_cubesize}. The volume per  cube chamber in the mold  is $10.00\times 10.00\times 10.00~mm^3$. The  cube side width increases to 10.26~mm because of the diffuse reflector thickness. The total length of 199-cube line is expected to be around 199$\times$10.26 = 2041.7~mm.  The actual measured length of the array varies from 2040 to 2044~mm for 40 different sets with the average value being 2041.0~mm. 

The length  was reduced to 2038--2042~mm with the average value of 2040.0~mm under a controllable pressure limited by  ratched mechanism of a micrometer which pushes the measuring rod till it stops moving. 
The elastic diffuse reflector  works as a damper and affects the total length of the array. The total length of 199 cubes is reduced by 2~mm under weak pressure. The elasticity of the cubes can be used during SuperFGD assembly for the accurate positioning of the cubes within 2~mm over a 2~m length (0.1\%).  

\begin{figure}[htb]
\begin{centering}
\includegraphics[width=0.55\textwidth]{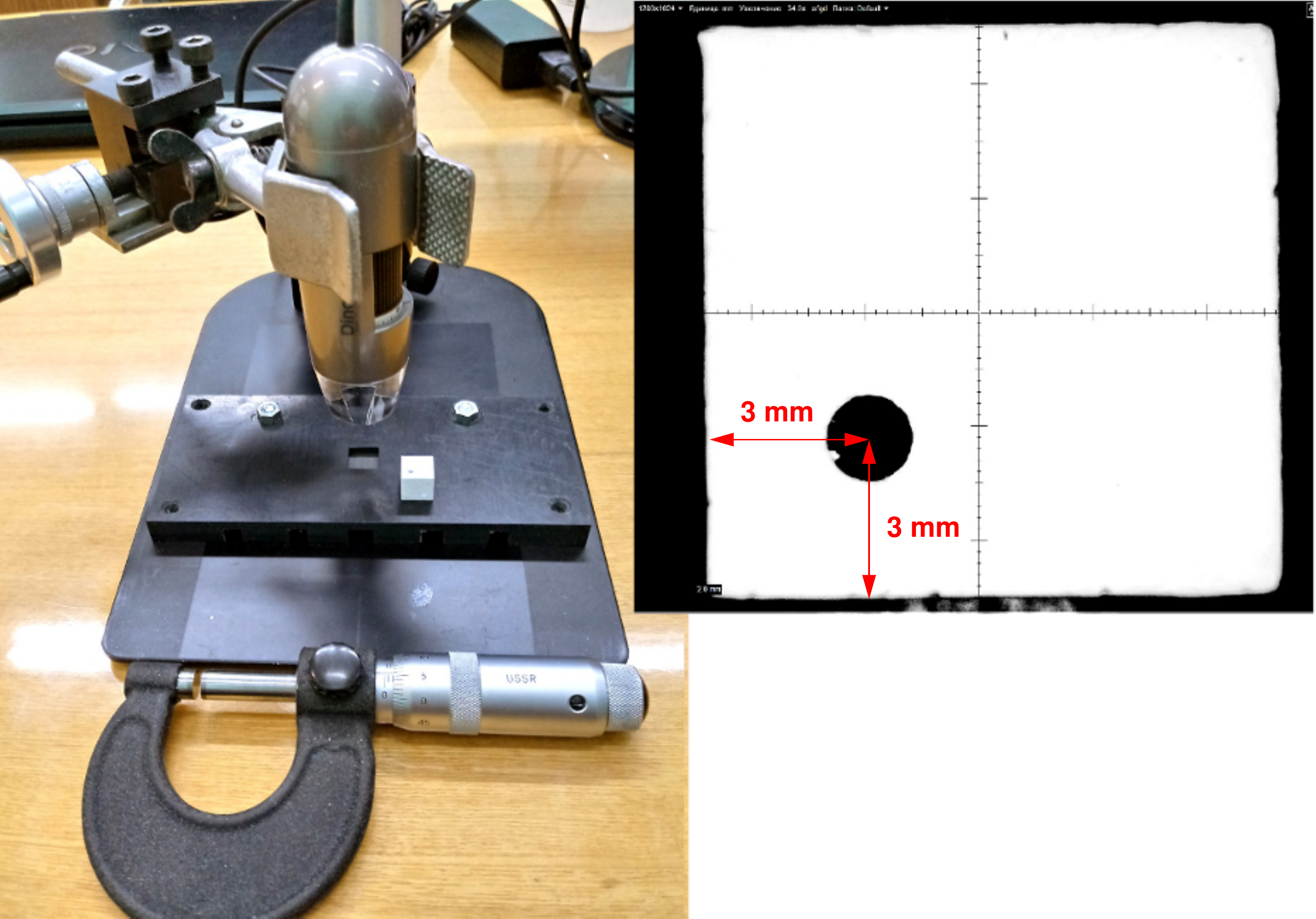}
\caption{Digital microscope setup to measure the accuracy of fiber holes drilling. The cube image shows the measured parameters (red arrows).}
\label{fig:Target_cube_microscope} 
\end{centering}
\end{figure}

A digital microscope was used to measure the position of the fiber holes relative to two cube sides.
Cubes were installed into a fixed position within a support frame, then an image was taken by the microscope.  A program  finds the hole center and calculates the distances using the image of the cube.  The microscope and sample image are shown in  Fig.~\ref{fig:Target_cube_microscope}.  The results are preliminary as the drilling technology is under development. The average distance between the hole center and the cube side was measured to be 3.11~mm that is slightly above the specified value of 3.00~mm. The variation is  $\sigma_x$=80~$\mu$m. Random deviations of the hole position less than 0.2~mm should not increase  significantly the cube position uncertainty because of the free gap between 1~mm fiber and 1.5~mm hole. The cube size stability is the key factor for the detector assembling.


The current productivity of cube injection per mold is 4 cubes each 72~s. 
A new mold form for 8 cubes is in preparation to increase the productivity, satisfying a rate of more than 4000 cubes per day.

Another challenge in cube production is the process of drilling the holes. 
We have to make 12,000 holes per day to keep up the manufacturing rate.  
We are optimizing the fabrication method to achieve a sufficient production rate while keeping the geometrical accuracy.

The preliminary schedule to produce 2 million cubes envisages the start of mass production in January, 2019. The last cubes must be delivered by January 2021.

\subsection{Scintillator cube assembly}
\label{sec:superfgd-assembly}
The main challenge in the 2 million cubes  assembling process is the variation of cube geometry that leads to the following  problems. First, if adjacent holes will be shifted more than 0.2~mm relative to each other the WLS fiber can be jammed during the insertion  into the corresponding raw or column of cubes. Another problem is the accurate positioning of WLS fibers for the correct coupling to  photosensors.  The latter issue can arise once small random fluctuations of the cube size lead to relatively large deflection of fiber position from the calculated coordinate which is fixed in the detector mechanics. A more precise cube size variation of 30~$\mu$m envisaged with the injection molding production will relax the problem, nevertheless we have to develop the technology of assembling to ensure the assembly and coupling of all the detector components: cubes, fibers, mechanics, photosensors.

\subsubsection{Fishing line method for detector assembling}
We plan to employ a ``fishing line'' method to align the cube positions into the projected geometry. The main idea is to assemble the cube arrays using a flexible plastic thread of calibrated diameter. A fishing line of 1.3~mm diameter was the natural choice for this purpose. First, the cube array is assembled on the fishing lines which form the 3D skeleton structure of specified geometry. Then the fishing lines are removed and the WLS fibers are inserted in place one by one. The fishing line diameter allows smooth insertion through the 1.5~mm cube holes while leaving some tolerance for the subsequent installation of the 1.0~mm fibers. 

\begin{figure}[htb]
\centering\includegraphics[width=0.7\textwidth]{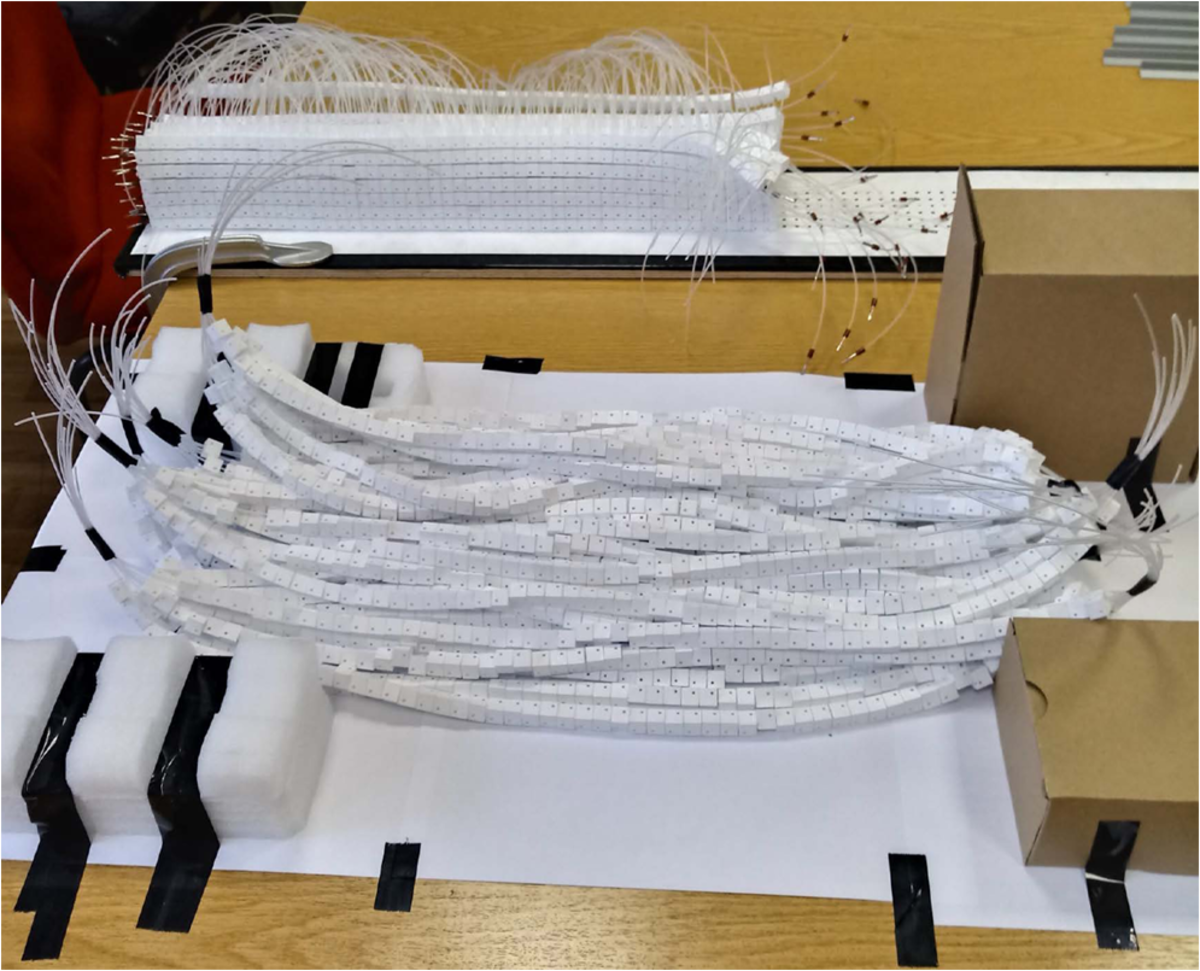}
 \caption{Linear cube arrays on the fishing lines.}
\label{fig:fishingline_chain}
\end{figure}

\begin{figure}[htb]
\centering\includegraphics[width=0.7\textwidth]{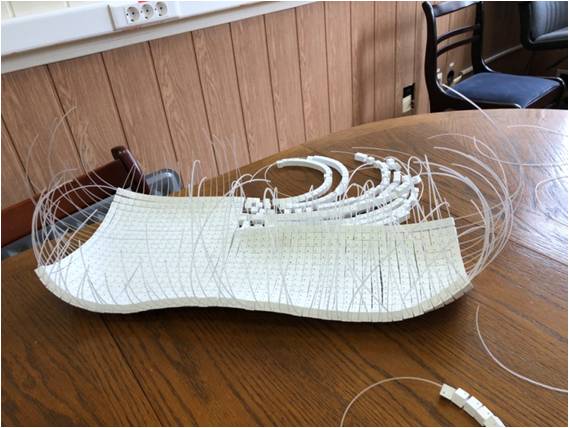}
 \caption{A plane of scintillator cubes formed with fishing lines}
\label{fig:fishingline_plane}
\end{figure}

\begin{figure}[htb]
\centering\includegraphics[width=0.4\textwidth]{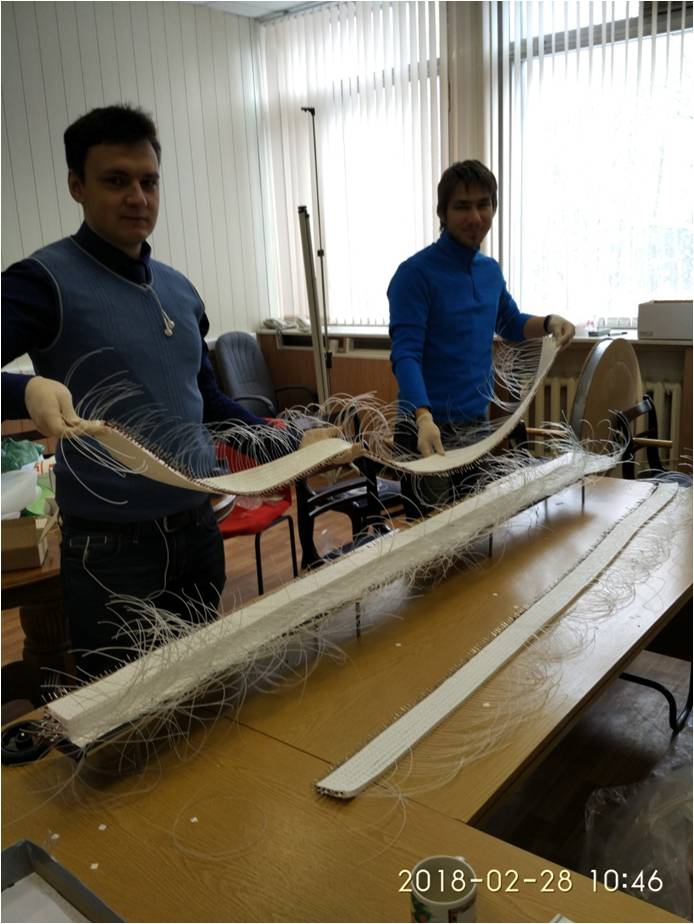}
\caption{Example of a flexible 2~m long plane with fishing lines.}
\label{fig:fishingline_long plane}
\end{figure}

\begin{figure}[htb]
\centering\includegraphics[width=0.8\textwidth]{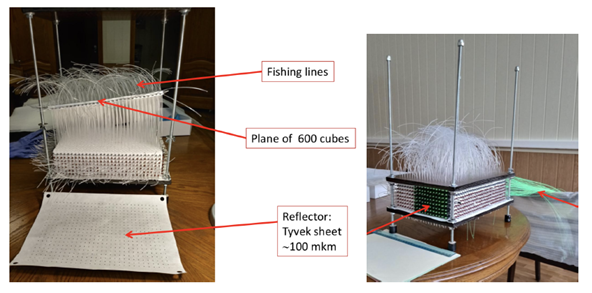}
\caption{Process of assembling of the detector prototype with vertical fishing lines (left). Some fishing lines are replaced by WLS fibers with optical connectors (right). A Tyvek reflector sheet  between planes is also shown. }
\label{fig:fishingline_vertical}
\end{figure}

Some examples of linear cube arrays with fishing lines are shown in Fig.~\ref{fig:fishingline_chain}.
A linear chain of cubes is the basic element of more complicated arrays. Then the linear chains are sewn together into 2D flat planes using also  fishing lines.  An example of a plane prepared for a detector prototype  is shown in Fig.~\ref{fig:fishingline_plane}. The most complicated stage is to merge the cube planes into a 3D body. It can be done thanks to flexibility of the cube planes. The flexibility of a 2~m long plane is demonstrated in Fig.~\ref{fig:fishingline_long plane}.
Fig.~\ref{fig:fishingline_vertical} shows the process of  merging the planes in 3D structure with  the fishing lines inserted vertically through the cubes. The right picture in Fig.~\ref{fig:fishingline_vertical} shows the detector prototype at the assembly stage when some fishing lines were replaced by WLS fibers with optical connectors.

Before the cube arrays take the final position in a mechanical box the cubes are not pressed or fixed between themselves except for the through-going fishing lines. Because of the elasticity of the whole cube array on the fishing lines we keep the possibility to adjust the positions of the cubes within a few mm at 2~m long base to build the detector into the support frame. 

We have tested in the beam the detector prototype assembled from 9200 cubes as described in Sec.~\ref{sec:superfgd-8x24x48-prototype}. Another array of 2~m length was assembled from $6\times6\times200$ cubes to check the fishing line method  and installation of WLS fibers.
No problem was found with the assembly of those prototypes.
In order to finally check the method at a larger scale, we plan a test assembly with 10--20\% of the real detector before summer 2019.

\clearpage

\subsubsection{Alternative assembly method}

\begin{figure}[htbp]
  \begin{center}
    \includegraphics*[width=0.6\textwidth]{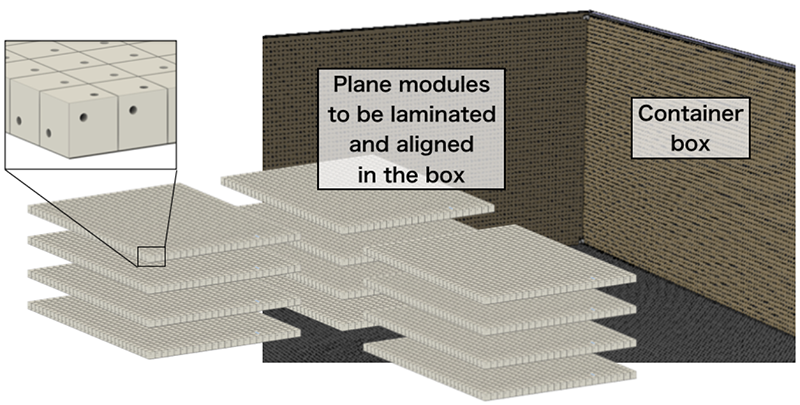}
  \end{center}
\caption{Concept of detector assembly with plane modules.}
\label{fig:Target-plane-module}
\end{figure}

We are also investigating an alternative assembly method to improve the workability and the rigidity.
The method under consideration is based on plane modules as shown in Fig.~\ref{fig:Target-plane-module}. 
The plane modules have cardboard-like structure with a cube array and are assembled to the full detector by aligning and laminating in a container box. 
Scintillator cubes are fixed on a thin sheet with controlled intervals to absorb individual variation of their size.
With such modular structure, the assembly work is divided into smaller pieces and the scalability can be assured.

In the assembly of a plane module, gluing is not preferred because of the danger of filling the holes, difficulties to control the extra material and variation of height, and the possibility of scintillator degradation.
Thus, a technique with ultrasonic welding is under development.
A white sheet of polystyrene with a few-hundred $\mu$m thickness will be welded onto cubes by an automated machine.


\begin{figure}[htbp]
 \centering
    \includegraphics[width=\textwidth]{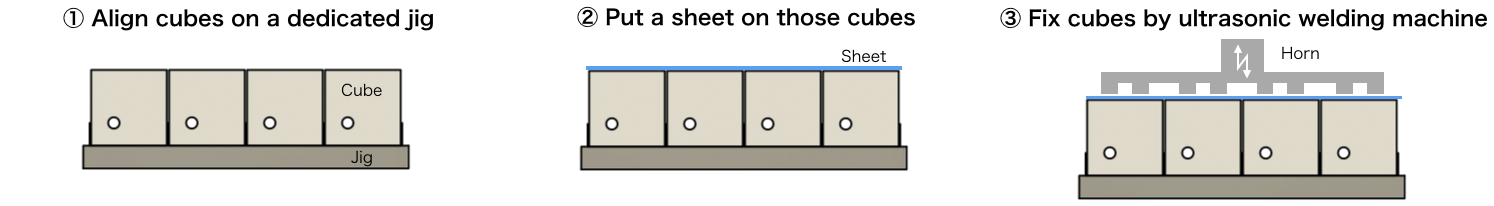}
  \caption{Procedure to make a plane module.}
  \label{fig:Target-welding-procedure}
 \end{figure} 

Figure~\ref{fig:Target-welding-procedure} illustrates the following steps to make a plane module.
\begin{enumerate}
  \item Cubes are aligned on a dedicated jig. The jig has a structure like a shallow tray with a thin grid plate to position the cubes to a predefined position with good precision. 
  \item A polystyrene sheet with holes are put on the cubes. The diameter of holes will be 2~mm to accommodate possible variations of the hole positions.
  \item The sheet is welded onto the cubes by an ultrasonic welding machine. Only a part of the contact surface of the sheet is welded, avoiding the area close to the holes and edges.
  Fast and uniform fabrication can be archived by an automated welding machine and a moving stage. A pressure monitor and a logger will be used to monitor and record the quality. 
\end{enumerate}
A unit of 24 $\times$ 32 cubes is possible with the available size of polystyrene sheet.
8 $\times$ 6 modules will make a horizontal plane of SuperFGD.
In total, 2,688 modules will be necessary to construct the full detector with 192 $\times$ 192 $\times$ 56 cubes.

\begin{figure}[htbp]
 \centering
   \includegraphics[width=0.6\textwidth]{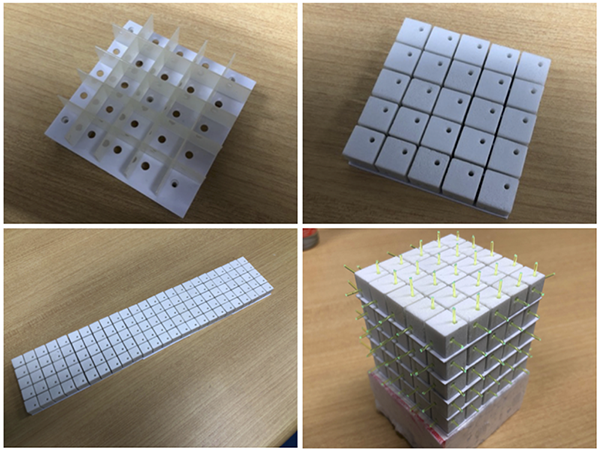}
  \caption{Demonstration of welding method using mock-up cubes, polystyrene sheets and solvent-based liquid.}
  \label{fig:Target-welding-demo}
 \end{figure} 

This concept was first demonstrated by welding mock-up cubes on a sheet using solvent-based liquid as shown in Fig.~\ref{fig:Target-welding-demo}. 
We confirmed that fibers can be easily inserted with this mock-up. 
Then, a single cube was welded on a polystyrene sheet. 
The strength of welding was confirmed to be sufficient for the handling.
The light yield was checked with cosmic rays and no significant degradation was observed.

We plan to proceed with prototyping full sheet size modules as well as development of jigs for cube alignment and welding. 
%


%% file: Target_fiber.tex
\section{Wavelength Shifting Fiber}

Wavelength shifting (WLS) fibers are commonly used to collect light from large area of scintillators. 
We use the same fiber as the current ND280, Y-11~(200) produced by KURARAY CO., LTD~\cite{Kuraray}.
The main specifications are summarized in Table~\ref{tab:Target-fiber}.
It is a multi-cladding, round shape type fiber with 1.0~mm diameter. 
Absorption spectra peaking at 430~nm is matched with the wavelength of light emitted from plastic scintillator.
The performance and quality of this fiber are very well established by many experiments.

The total length of WLS fiber will be 70~km including spares.
As an established commercial product, there is no problem foreseen for the production and quality control for this amount.
The lead time for the production of WLS fibers is estimated to be three months based on a quotation.
The procurement of WLS fibers is scheduled in 2019.

\begin{table}[ht]
  \centering
  \begin{tabular}{ll} \hline \hline
    Item & Specification \\ \hline
    Fiber type & Round shape, Multi-cladding \\
    Diameter & 1.0~mm\\
    Materials & Core: Polystyrene (PS),  \\
                     & Middle clad: Polymethylmethacrylate (PMMA), \\
                     & Outer clad: Fluorinated polymer (FP) \\
    Refractive index & Core: 1.59, Middle clad: 1.49, Outer clad: 1.42\\
    Density & Core: 1.05~g/cm$^2$, Middle clad: 1.19~g/cm$^2$, \\
    & Outer clad: 1.43~g/cm$^2$ \\
    Absorption wavelength & 430~nm (peak) \\
    Emission wavelength & 476~nm (peak) \\
    Trapping efficiency & $\sim$5.4\% \\
    Attenuation length & $>$3.5~m \\ \hline \hline
  \end{tabular}
  \caption{Main specifications of the WLS fiber, Y-11\,(200)}
  \label{tab:Target-fiber}
\end{table}

%% file: Target_MPPC.tex
\section{Multi-Pixel Photon Counter (MPPC)}
 
\begin{figure}[tbp]
\centering
\includegraphics[width=0.42\textwidth]{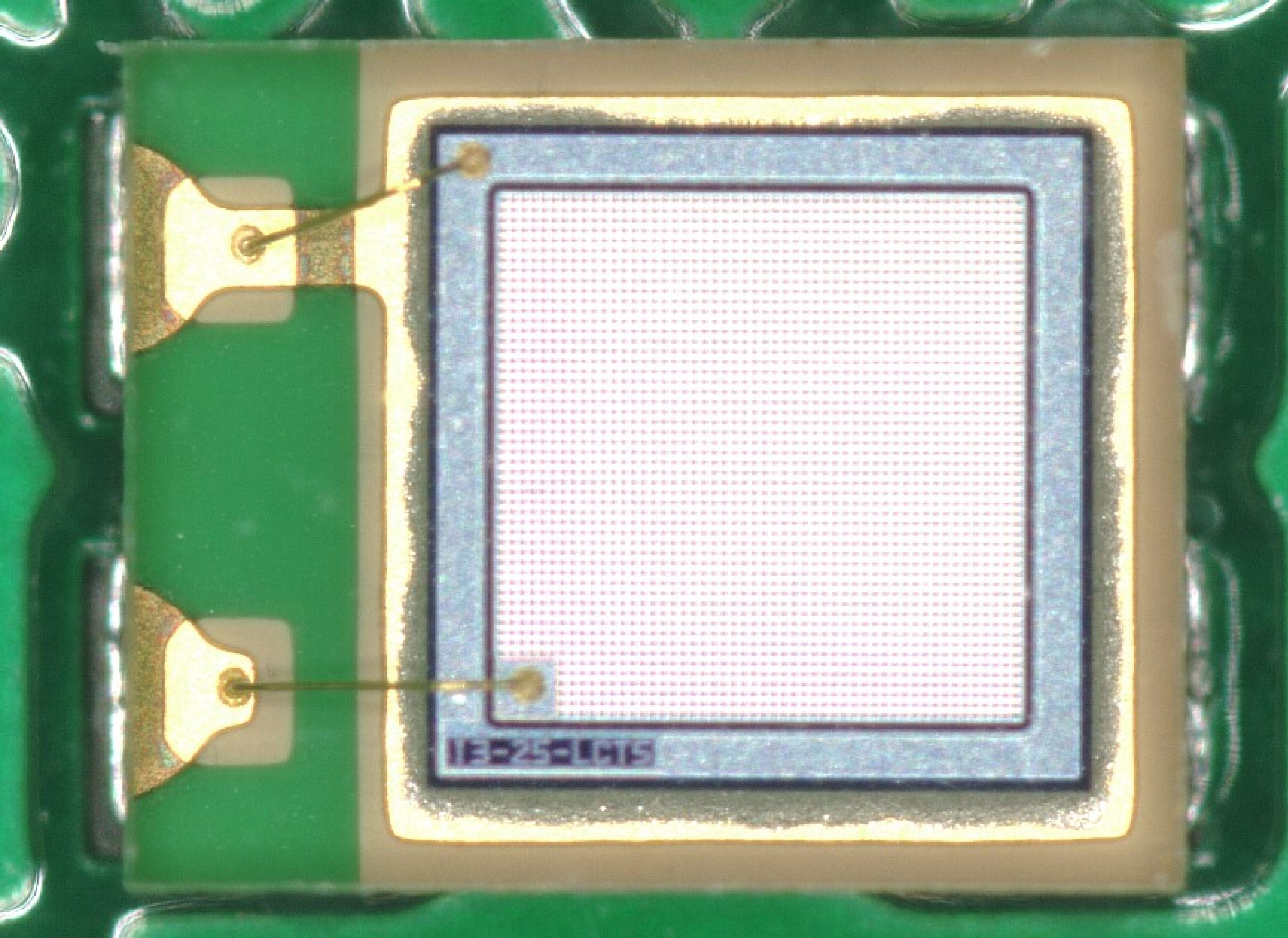}
\hspace*{0.08\textwidth}
 \includegraphics[width=0.45\textwidth]{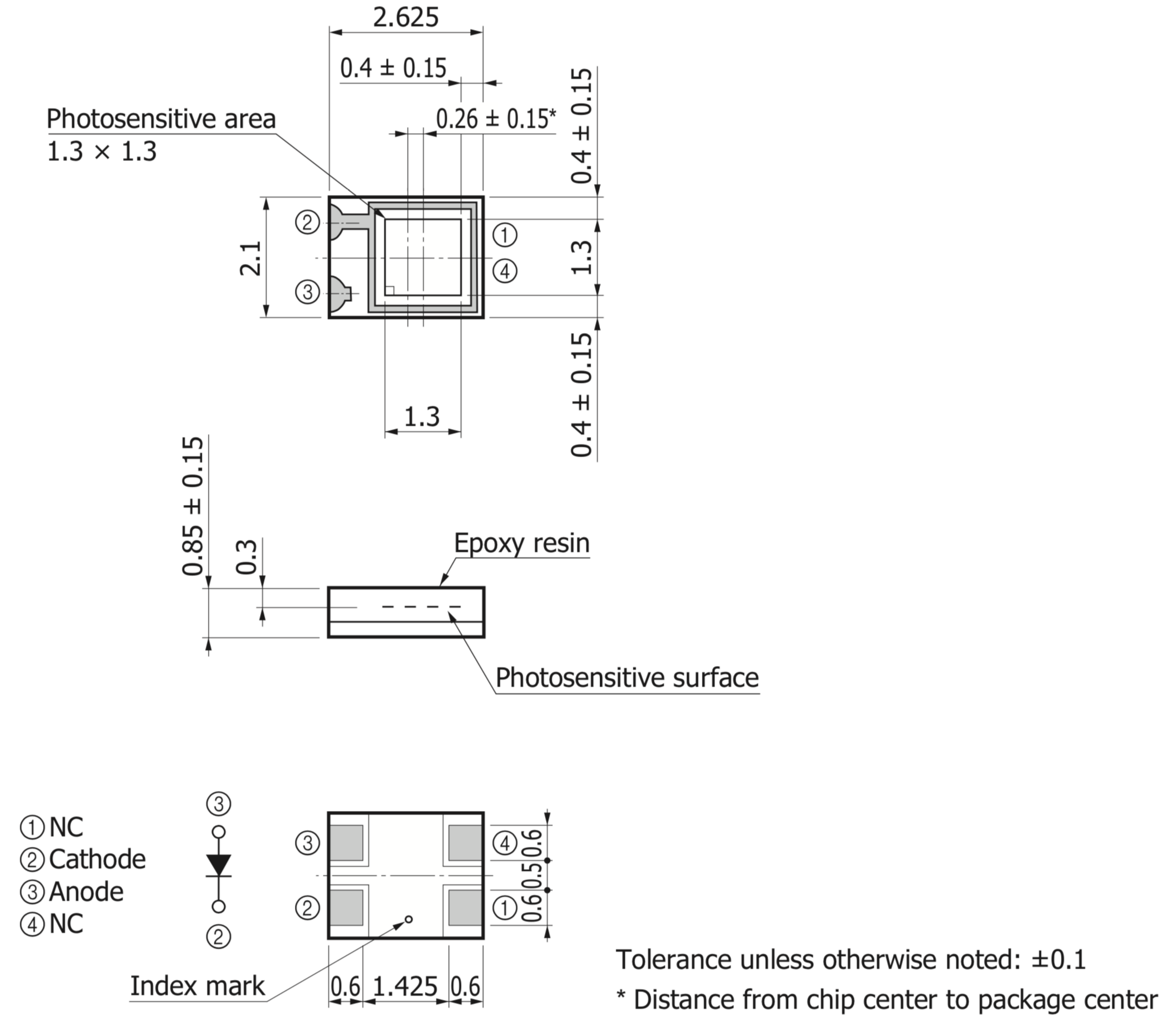}
\caption{Left: picture of the MPPC S13360-1325PE. Right: Dimensional specifications of S13360-1325PE (from MPPC catalogue of Hamamatsu Photonics).}
\label{fig:Target-mppc_picture}
\end{figure}

The photosensor is the key device to detect the scintillation light.
We adopt the Multi-Pixel Photon Counter (MPPC) produced by Hamamatsu Photonics K.K.
The MPPCs have been successfully used in all plastic scintillator detectors of the current near detectors of T2K since 2009~\cite{Abe:2011ks, Yokoyama:2010qa}.
The MPPC type chosen for SuperFGD is S13360-1325PE (Fig.~\ref{fig:Target-mppc_picture}).

The specifications of the S13360-1325PE are summarized in Table~\ref{tab:Target-mppc_spec}. 
Its sensitive area is 1.3~mm $\times$ 1.3~mm, the same as the MPPCs used for the current near detectors and designed to match the diameter of the WLS fiber.
The pixel pitch is smaller than that for the current ND280 MPPCs, S10362-13-050C (25~$\mu$m compared to 50~$\mu$m) in order to attain a larger dynamic range.
The surface mount package was chosen to minimize the space and cost.
Thanks to the development over past 10 years since the construction of original ND280, S13360-1325PE has about an order of magnitude smaller dark noise rate, cross-talk probability, and afterpulse probability compared to S10362-13-050C.
%

\begin{table}[htbp]
  \centering
  \begin{tabular}{ll} \hline \hline
    Item & Specification \\ \hline
    Effective photosensitive area & 1.3~mm x 1.3~mm \\
    Pixel pitch & 25~$\mu$m \\
    Number of pixels & 2668~pixels \\
    Fill factor & 47\% \\
    Package type & Surface mount \\
    Breakdown voltage (V$_{\rm BR}$) & 53 $\pm$ 5~V \\
    Peak sensitivity wavelength & 450~nm \\
    Photo detection efficiency &  25\% \\
    Gain &  7.0 x 10$^{5}$ \\
    Dark count & 70~kcps (typ.) \\
    Crosstalk probability & 1\% \\ \hline \hline
  \end{tabular}
  \caption{Specifications of the S13360-1325PE MPPC. The characteristics are measured at (V$_{\rm BR}$+5)~V and 25 degree C.}
  \label{tab:Target-mppc_spec}
\end{table}

The mass production plan is fixed based on the discussion with Hamamatsu, accounting for the lead time including the bidding and contract.
The first batch of 7,680 MPPCs will be delivered before March 2019.
The production of the remaining MPPCs will be completed by the end of 2019.
The MPPCs will be packed in standard reels, conforming the JEITA ET-7200 standard, to enable automatic mounting onto PCB boards.
768 MPPCs will be packed in a reel, where MPPCs with a similar operation voltage ($<\sim$0.15V) will be grouped.
The average, minimum and maximum operation voltages will be provided for each reel so that they can be arranged to minimize the operation voltage difference for a group of bias voltage supply unit.

The quality check and detailed characterization of MPPCs will be performed after they are mounted on PCB boards.
With the existing ND280 detectors, we have experience of production, test, and characterization of a large number of MPPCs~\cite{Yokoyama:2010qa, Moreau:2010zz, Vacheret:2011zza} and similar procedure is envisaged for the SuperFGD MPPCs.

%% file: Target_mechanics.tex
\section{Mechanics}
\label{sec:superfgd-mechanics-interface}



In order to maximize the acceptance of the TPCs for particles 
produced by neutrino interactions in SuperFGD, 
the dead space and material must be minimized, while keeping sufficient strength to support $\sim$2 tons of detector.

\begin{figure}[bt]
\centering
\includegraphics[width=1.0\textwidth]{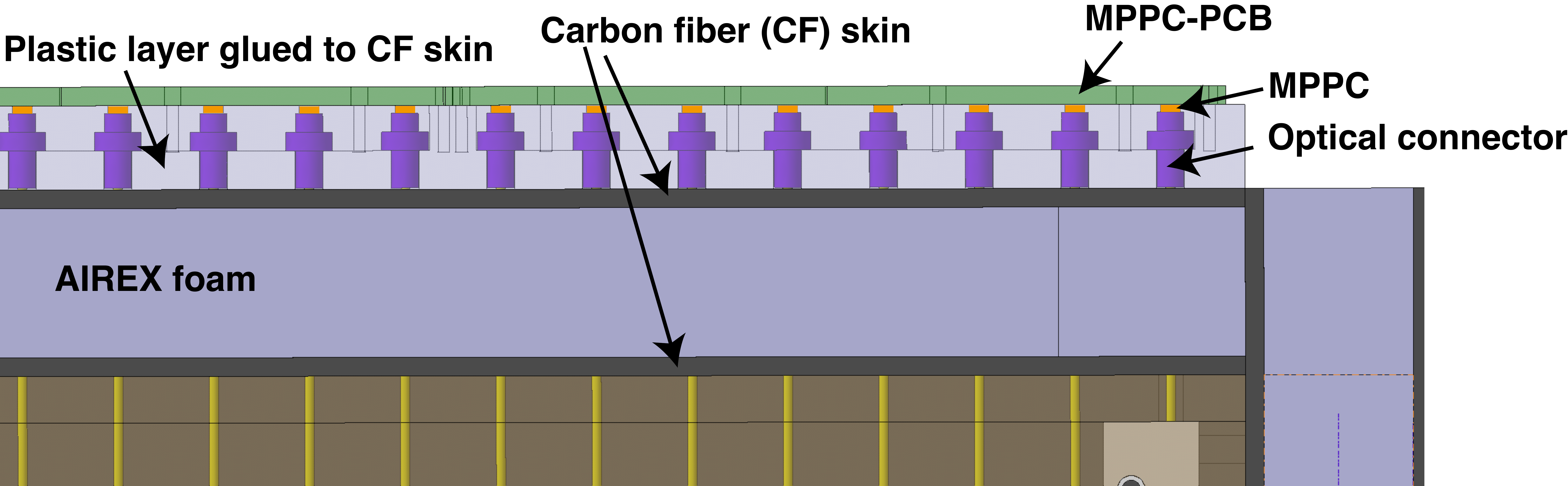}
\caption{
Cross-sectional view of the box panel and the optical interface.
The box panel is made of AIREX foam sandwiched by carbon fiber (CF) skins.
WLS fibers are brought outside the box through the holes in the panel (not shown), and glued to the optical connectors.
The optical connectors are inserted in holes of the plastic layer glued to the CF skin for the mechanical alignment to the MPPCs soldered on the MPPC-PCBs.
}
\label{fig:superfgd-box-panel-xsec}
\end{figure}
The mechanical structure for SuperFGD consists of a box that contains the scintillator cubes.
The box is made of carbon fiber (CF)-based panels with holes for WLS fibers.
The MPPCs are soldered on Printed Circuit Boards (MPPC-PCBs) that are screwed on the box. 
Four out of six panels of the mechanical box will host the optical interface, which is composed by optical connectors glued to the WLS fibers, surface-mount MPPCs soldered to MPPC-PCBs, and related mechanical structure as shown in Fig.~\ref{fig:superfgd-box-panel-xsec}.

%

\subsection{Box mechanics}
\label{sec:superfgd-box}

\begin{figure}[tb]
\centering
\includegraphics[width=0.7\textwidth]{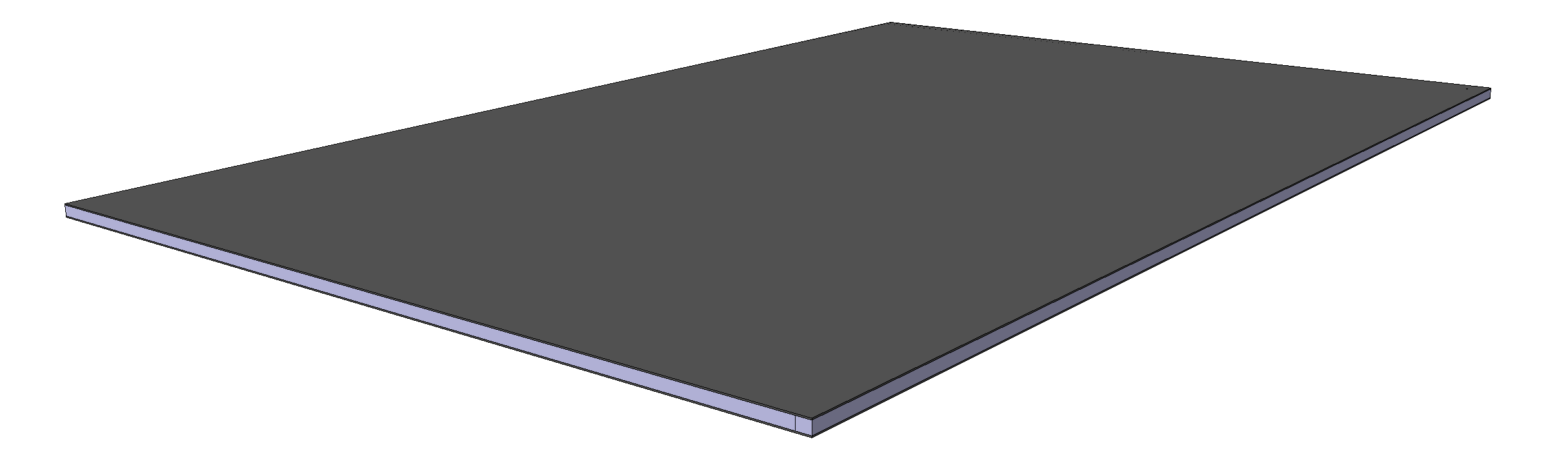} \\
\vspace*{3mm}
\centering\includegraphics[width=0.7\textwidth]{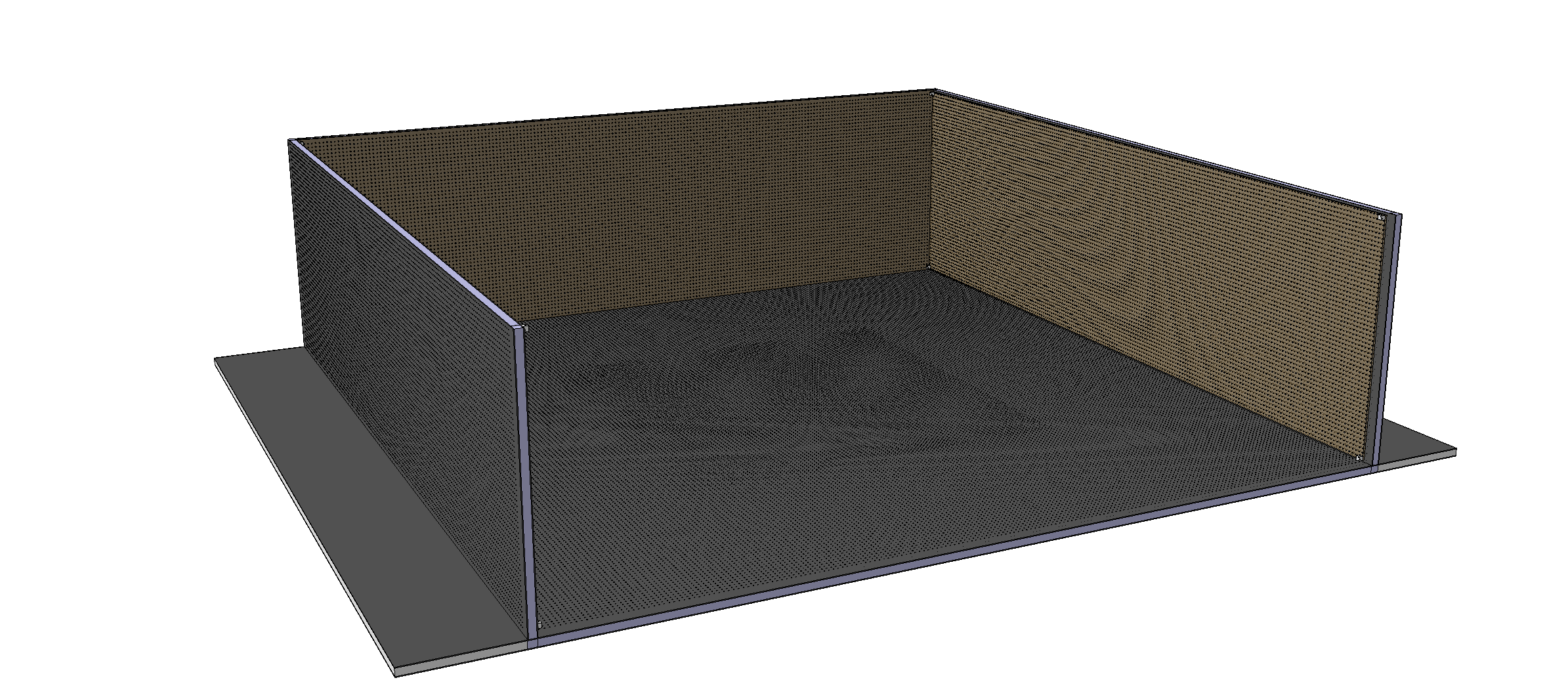}
\caption{
Top: A carbon-fiber (CF)-based panel of the box. An AIREX foam is sandwiched by CF skins.
Bottom: The SuperFGD mechanical box made of the CF-based panels. Two panels are not drawn to show the inside.
}
\label{fig:superfgd-box-sandwich}
\end{figure}


\begin{figure}[htb]
\centering\includegraphics[width=0.6\textwidth]{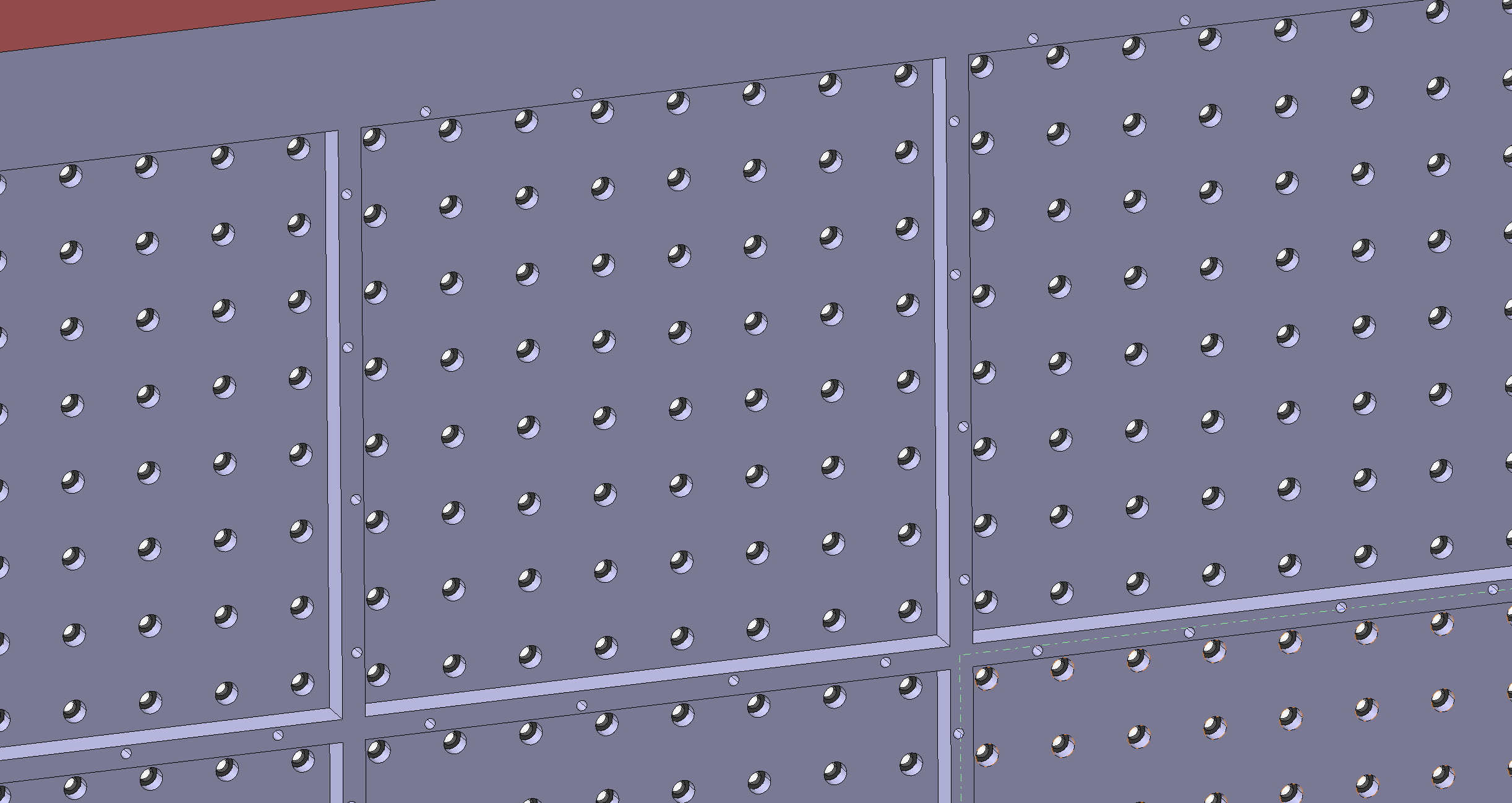}
\caption{
A zoomed view of the box surface showing the cavity of the plastic layer that host the optical interface components. 
}
\label{fig:superfgd-box-cavity}
\end{figure}

The mechanical box will be made of six carbon-fiber (CF) based panels screwed together (Fig.~\ref{fig:superfgd-box-sandwich}).
Each panel consists of a sandwich composed by a core 
of 16 mm thick AIREX spacer
and 
two 2 mm thick CF skins.
The AIREX core and the CF skins will be glued together.
In order to let the WLS fibers exit the box, the panel will have holes of 3~mm diameter spaced with a pitch of about 1 cm.  
On the external side of each CF-based panel,
an additional plastic layer with cavity structure (Fig.~\ref{fig:superfgd-box-cavity}) is glued to provide a space for the optical interface. 
The MPPC-PCBs will be screwed on this plastic layer.

AIREX is a particular type of 
low-density (about 60 $\tt{kg}/\tt{m}^2$) 
but strong (Young's modulus of 46 MPa)
foam and
 is often used for its mechanical characteristics. 
Its advantages 
is given by the material uniformity
while providing good rigidity and very low material budget.
In addition, thanks to its uniformity, it makes gluing with the CF skins less problematic against stresses. 
The candidate material of CF is Toray T300, composed by eight plies of 0.125 mm thickness with orientation of $0^{\circ}$, $+45^{\circ}$, $-45^{\circ}$ and $90^{\circ}$.
The CF sandwich structure and thickness is currently being optimized.
Preliminary studies of Finite Element Analysis (FEA) are shown in Sec.~\ref{sec:superfgd-stress-tests-fea} together with the 
stress tests performed at the CERN mechanical workshop. 



In order to assemble the CF-based box, the six panels will be screwed together. 
On the external edge of each panel, the AIREX core will be replaced by a 1~cm wide aluminum beam, glued between the CF skins, to facilitate the screwing and provide the required robustness to the box.
%
%
Between the cubes and the CF-based box, 
there will be a thin foam layer with a thickness of about 5~mm,
to compress and constrain the cubes to reduce any movement inside the box.

\begin{figure}[htb]
\centering\includegraphics[width=0.9\textwidth]{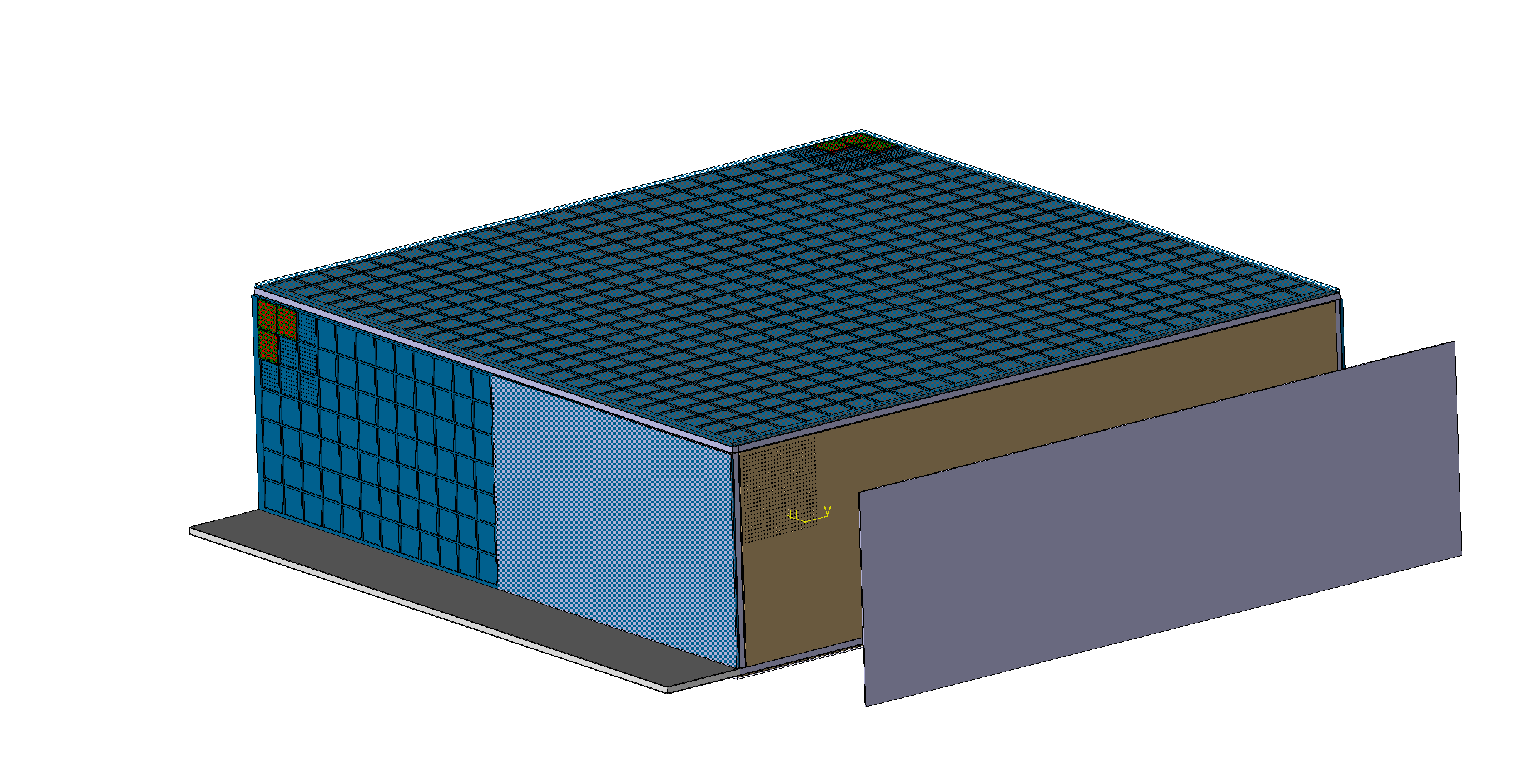}
\caption{
The SuperFGD mechanical box. 
One of side planes (the downstream plane), where MPPCs are not mounted, is shown with the cover open.
}
\label{fig:superfgd-box-fiberendview}
\end{figure}

%
%

Holes will be made also on the parts where MPPCs are not mounted, in order to take the WLS fibers outside the box and facilitate the assembly of the scintillator cubes.
The WLS fibers will be covered by a light-tight plastic cover, screwed on the plastic layer.
The space between the plastic layer and the dark cover can host the LED calibration system.
Figure~\ref{fig:superfgd-box-fiberendview} shows a model of the assembled box with a view on the face that host the not-instrumented side of the WLS fiber, with the cover open.

The left and right faces have half of their surface hosting the optical interface and the other half hosting the not-instrumented WLS fiber end, in order to balance the density of the readout electronics.

The external dimensions of the box, including the optical interface up to the MPPC-PCBs but without considering the extended bottom panel, is
$2018~\textnormal{(Width)} \times 640~\textnormal{(Height)} \times 2018~\textnormal{(Length)}~\textnormal{mm}^3$.

\subsubsection{Stress tests and Finite Element Analysis (FEA)}
\label{sec:superfgd-stress-tests-fea}

With about 60,000 holes and about two million cubes, it becomes hard to perform a full FEA simulation without  approximations in the models.
In order to validate the FEA mechanical simulations, a stress tests with a small piece of CF sandwich was performed.

\begin{figure}[htb]
\centering\includegraphics[width=0.35\textwidth]{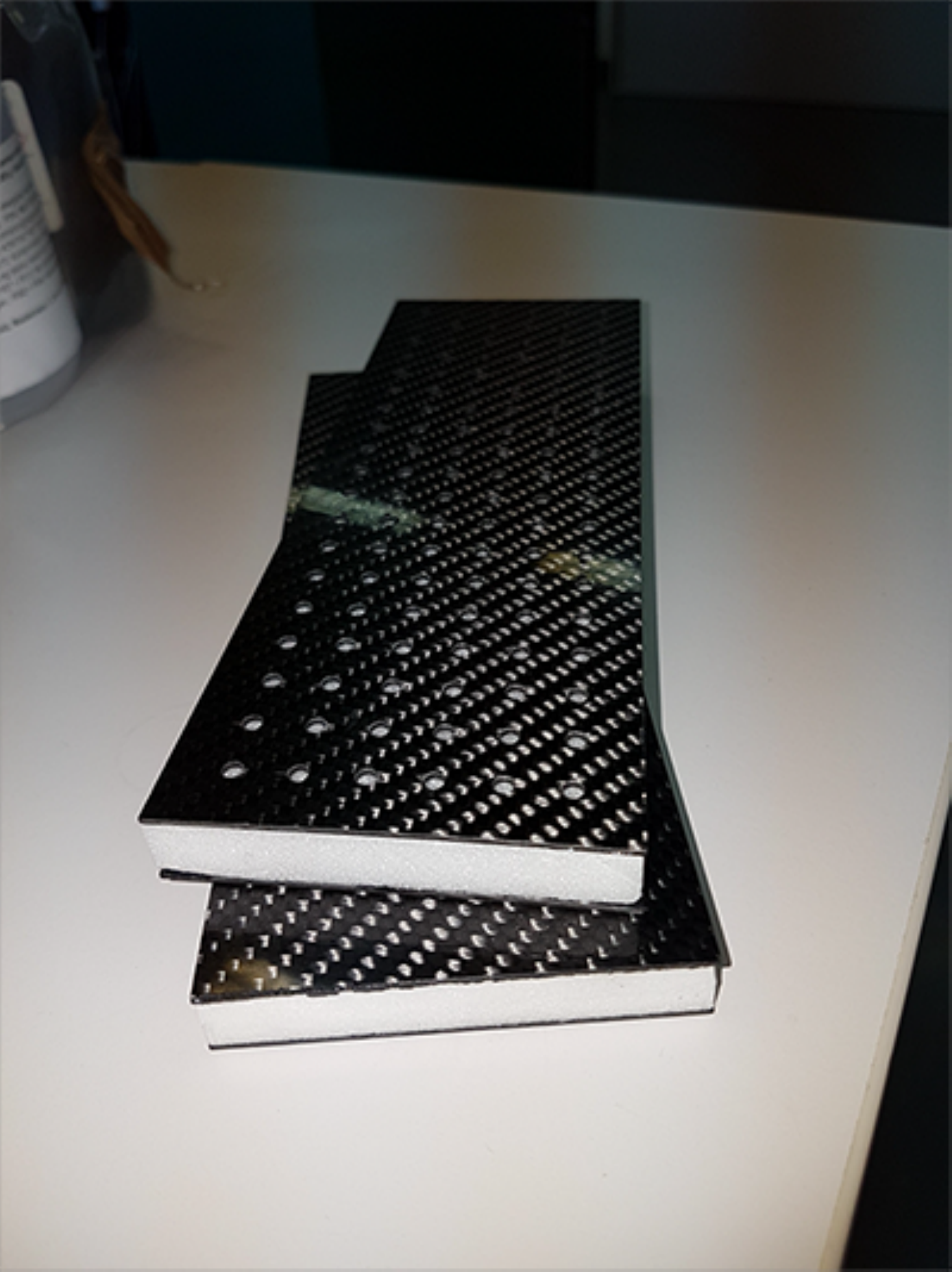}
\hspace{5mm}
\centering\includegraphics[width=0.35\textwidth]{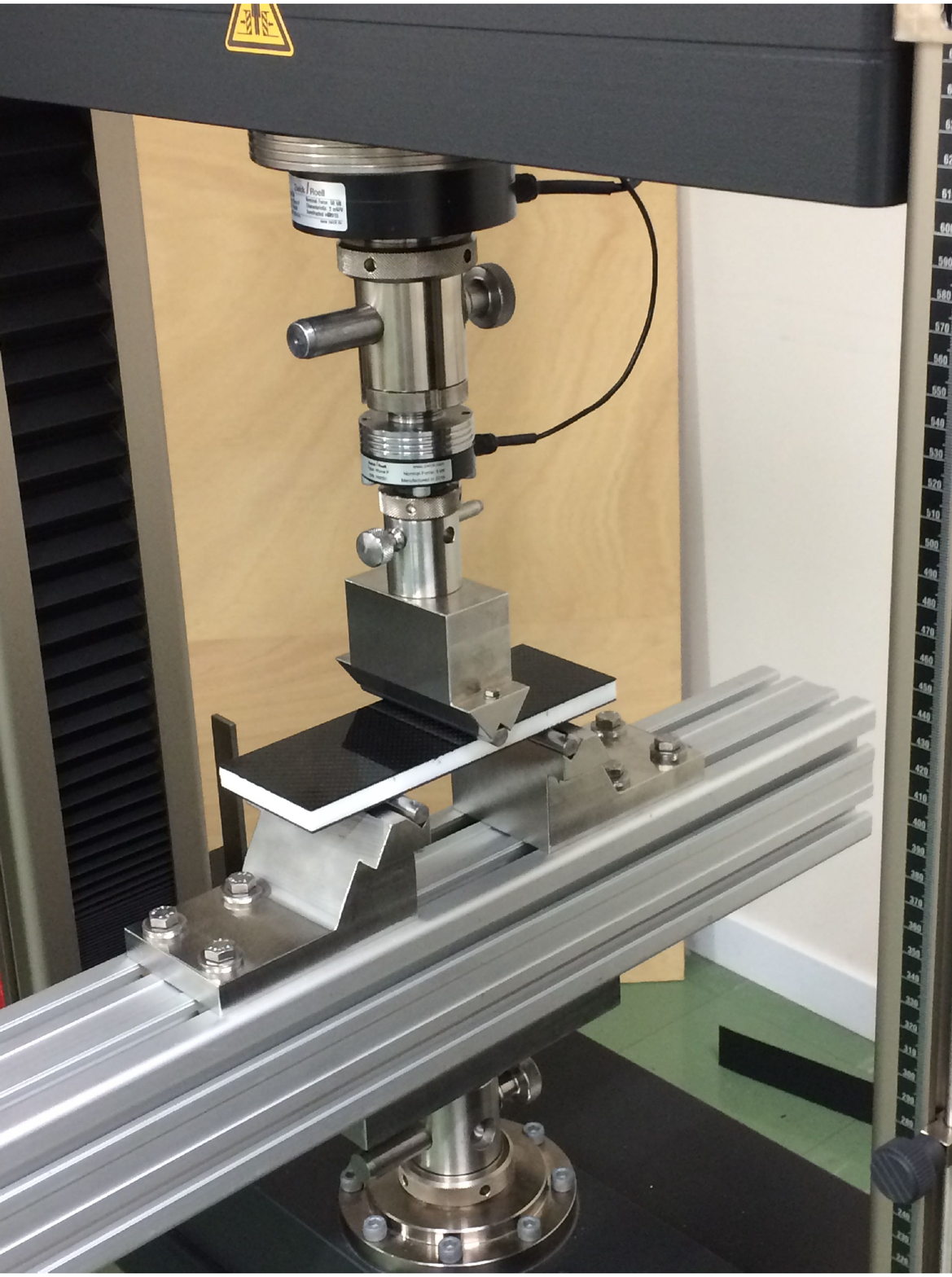}
\caption{
Left: CF sandwich samples used in the stress tests.
Right: picture of the stress tests. The span is defined by the bottom platform and a pressure is applied on the middle of the CF sandwich.
}
\label{fig:Target-stress-test}
\end{figure}

The tests were performed at the CERN mechanical workshop.
Two bars with the same sandwich structure described above, i.e. two 2~mm CF skins that sandwich a 16~mm AIREX core, were made.
In one of the bars holes with a diameter of 3~mm were drilled with a constant pitch of 1~cm.
The width of the bars was 12~cm and different spans (10 and 17~cm) and forces (300 to 800~N) were applied.
The difference in deformation between the CF sandwich with and without holes is at the level of 15-20\%. 
For example, for the case of 12~cm span and 500~N force, the maximal deformation is 0.63~mm and 0.77~mm respectively with and without holes.
The FEA simulations of the CF sandwich both with and without holes agrees with the data within 10\%.
The CF sandwich without holes never broke during the test, while
the one with holes failed when a force of 726~N was applied: the AIREX core showed a crack near the supporting platform. 
These forces, corresponding to more than 70~kg of weight, are much higher compared to the maximum static stress expected in the case of SuperFGD.
Figure~\ref{fig:Target-stress-test} shows the CF sandwich samples used in the tests and the setup.

Preliminary FEA simulations were performed with a box made by six panels with the same CF sandwich structure as described above. 
Given the difficulty to simulate about two million cubes, the studies were done by assuming 2 tons of water inside the assembled box.
The maximal deformation achieved is 2.5~cm in the middle of the bottom panel, when the a uniform load distributed over all the surface was simulated.
This simulation is considered to be conservative because water is not subject to friction as the cubes would be.
In order to obtain more reliable simulations, the plan is to perform measurements in laboratory with an assembled prototype of about 10,000 cubes. A load, uniform over the surface, will be applied on the top face of the prototype and the deformation on the bottom face will be precisely measured. The measured data will be useful to tune the FEA model and obtain results more consistent with a realistic scenario.
Based on the FEA results, the optimization of the CF-sandwich structure is ongoing in order to reduce the maximal deformation below 0.5~cm, half of the clearance required between detectors.

FEA studies were also done to evaluate the stresses on the CF sandwich. No particular issues were found.
The most problematic one is the shear stress on the AIREX core. We found it to be smaller than then failure value (0.8 MPa for AIREX), with a safety factor of four.

\subsection{Optical interface}
\label{sec:superfgd-interface}

The optical interface (see Fig.~\ref{fig:superfgd-box-panel-xsec}) is a part of the detector that brings the scintillation light outside the box, to the MPPCs.
It also serves as an interface to the frontend electronics, grouping the signal from MPPCs into a unit so that they are carried to the frontend electronics via high density cables.
Currently, two options are considered for the configuration of MPPCs on an MPPC-PCB, $8\times8$ and $8\times16$.
The drawings shown in this document are based on the $8\times8$ configuration, however the basic design of the optical interface is compatible for both options.

\begin{figure}[htb]
\centering
\includegraphics[width=0.8\textwidth]{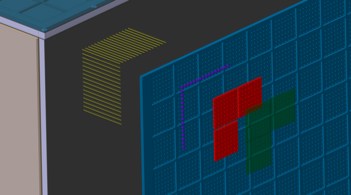}
\caption{
An expanded view of the optical interface parts.
The plastic layer (blue) is actually glued to the CF-based panel (gray).
The MPPC-PCBs (green) are screwed on the CF-based box. 
The optical connectors (purple) are inserted to 3~mm-diameter holes of plastic optical separators (red), which provide the optical separation between channels.
}
\label{fig:superfgd-interface-expandc}
\end{figure}
\begin{figure}[htb]
\centering
\includegraphics[width=0.68\textwidth]{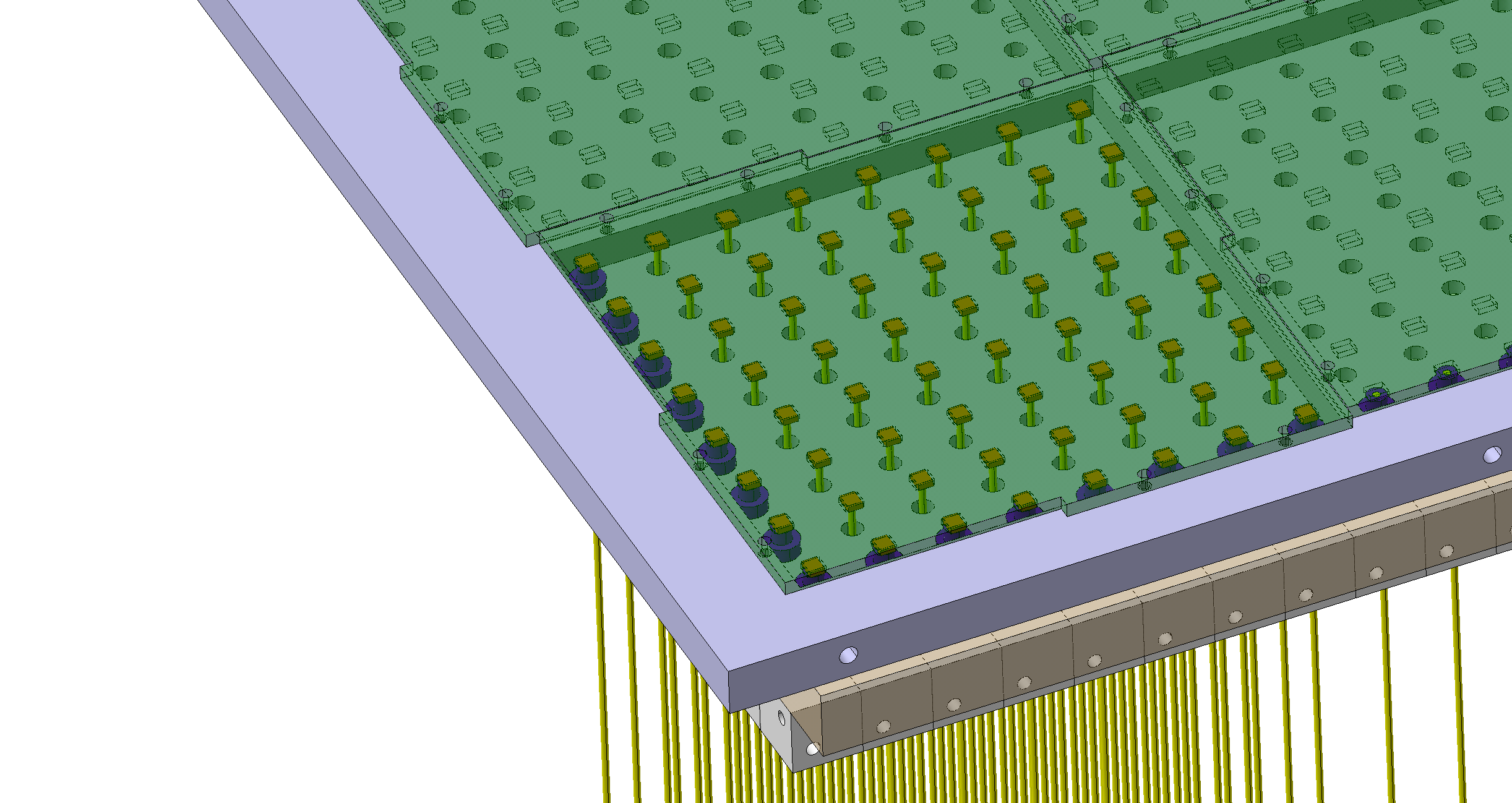}
\includegraphics[width=0.28\textwidth]{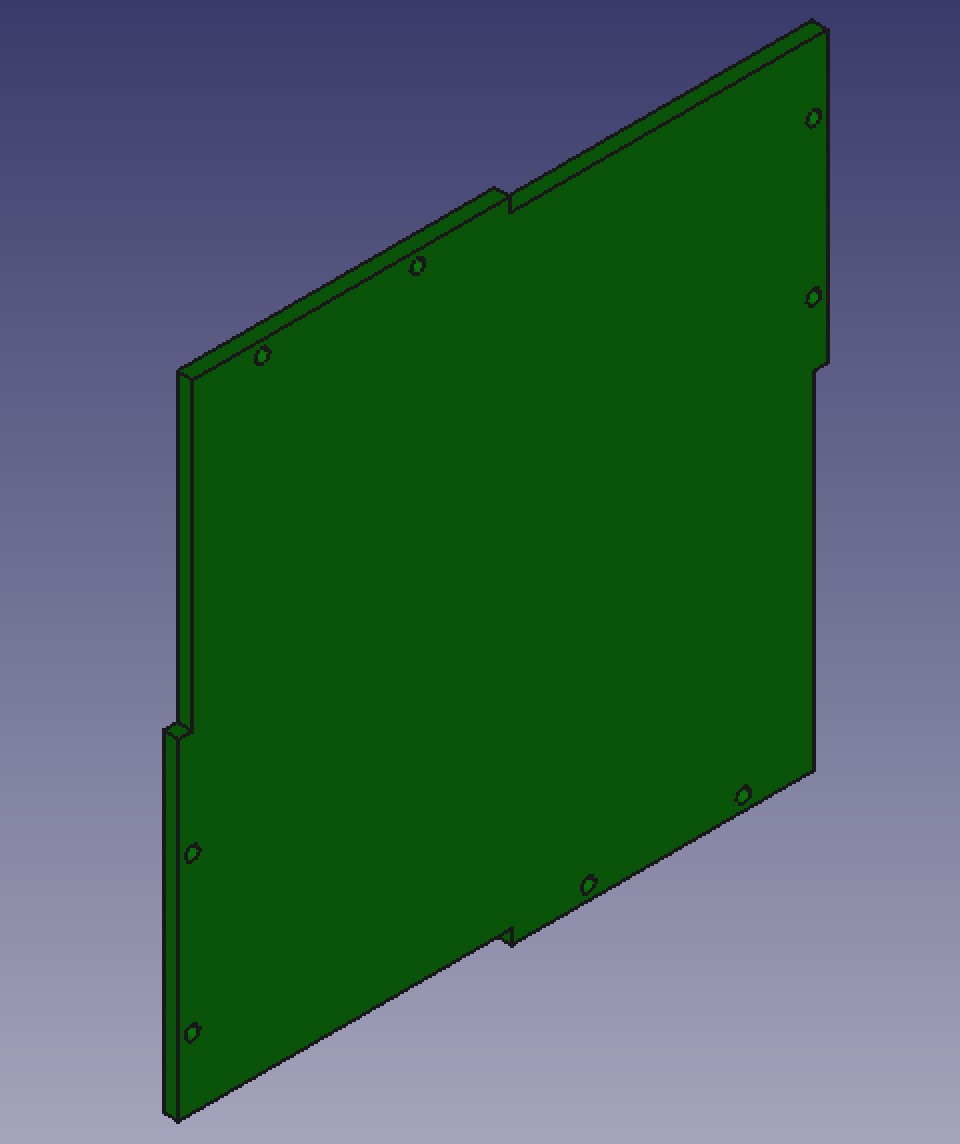}
\caption{
Top view of an instrumented panel and a MPPC-PCB.
}
\label{fig:superfgd-box-panel-pcb}
\end{figure}

Figure~\ref{fig:superfgd-interface-expandc} shows an expanded view of the optical interface parts.
The WLS fibers collect the scintillation light from the cubes
and bring it outside the box through the holes in the CF-based sandwich and the plastic layer.
The WLS fibers are glued to optical connectors (see Sec.~\ref{sec:superfgd-connectors}),
which are placed inside the plastic layer cavity and aligned with its surface, to provide a good coupling between the WLS fiber and the MPPC.
The optical connectors are inserted in the holes of the plastic layer, without touching the external part of the CF-based sandwich.
The MPPC-PCBs are screwed to the plastic layer. 
Another plastic layer is placed in the cavity to provide optical separation between channels.
The total thickness of the plastic layer is $9~\textnormal{mm}$.
A top view of an instrumented panel is shown in Fig.~\ref{fig:superfgd-box-panel-pcb}, together with the shape of a MPPC-PCB.

In order to have a good coupling between a 1~mm diameter WLS fiber and a MPPC with an active region of $1.3\times1.3~\textnormal{mm}^2$, an alignment better than $\sim$0.1~mm is required.
The positioning of the MPPC-PCB must be made with sufficient precision, since it defines the alignment between WLS fibers and MPPCs.
In order to reduce the source of misalignment, the flatness of both the MPPC-PCB and the carbon-fiber box is important.
The feasibility of the designed optical interface was confirmed with a small prototype as described in Sec.~\ref{sec:target:prototype-interface}.

The whole system must be light-tight.
A few options are being considered:
one is to use opaque soft glue or silicon
on the slots between the MPPC-PCB and the edge of the plastic layer.
If this solution does not fulfill the light-tightness requirements, the MPPC-PCBs will be surrounded by a dark cover.
The first option is preferred because of the complexity due to the large number of cables.


On the side opposite to the optical interface, fibers exit the box and contained in cavities of the plastic layer covered by a layer for light tightness. 

%

\subsubsection{Optical connectors}
\label{sec:superfgd-connectors}

\begin{figure}[htb]
\centering\includegraphics[width=0.55\textwidth]{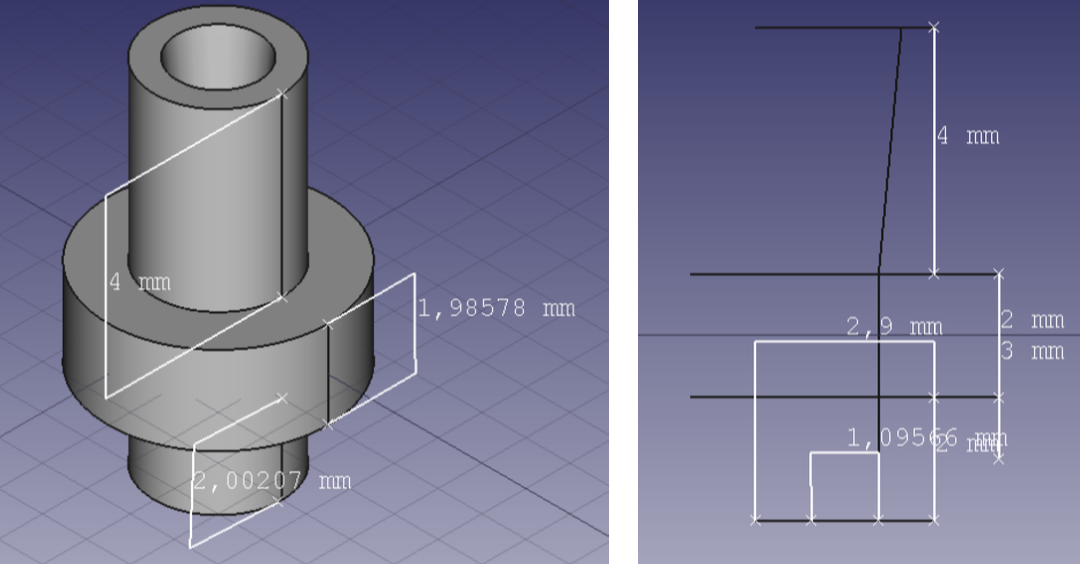}\\
\centering\includegraphics[width=0.4\textwidth]{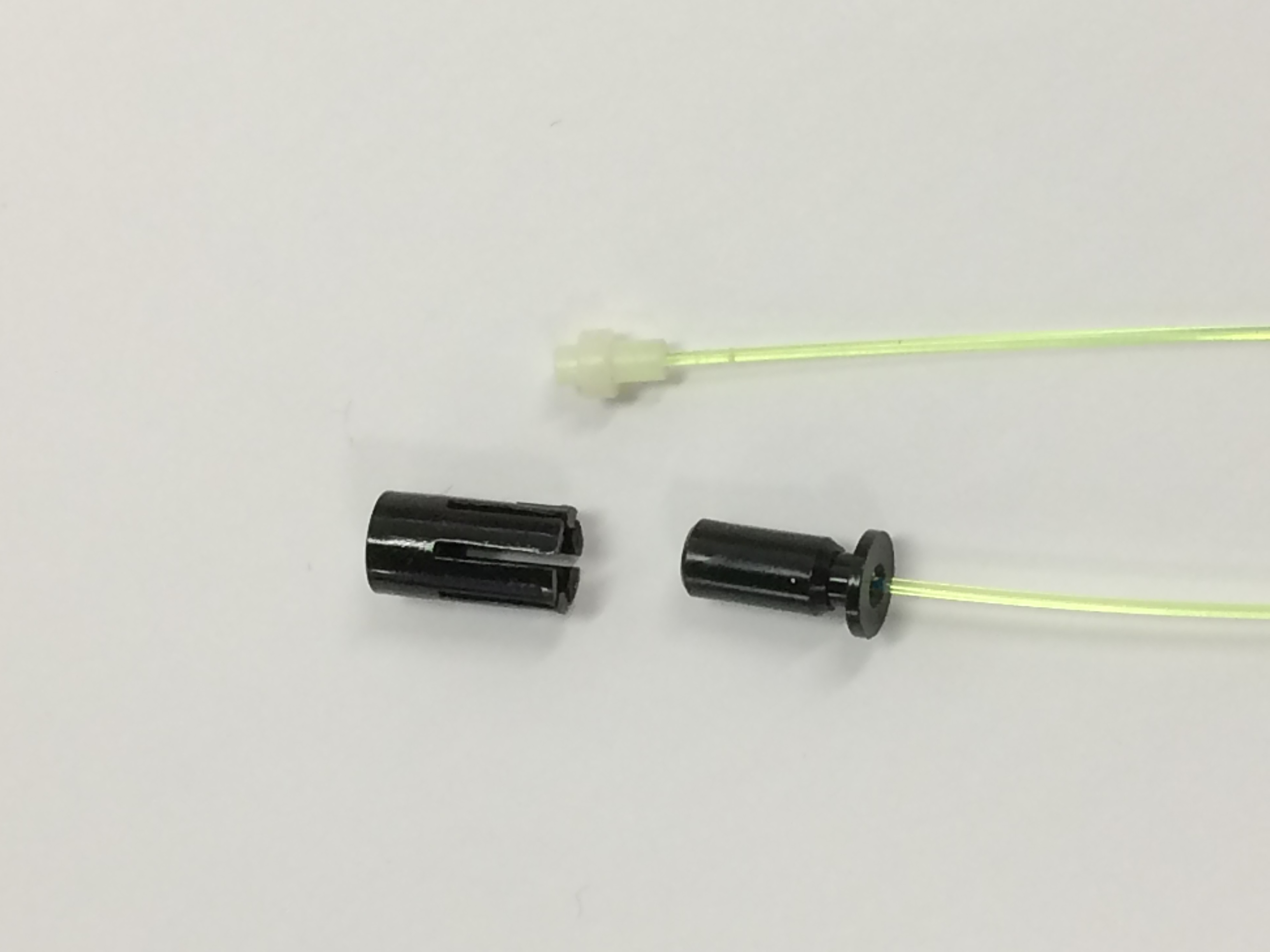}
\caption{
Top: CAD model of the optical connector with the corresponding dimensions.
Bottom: picture of 3D printed optical connector 
glued with the WLS fiber. 
The connector designed for SuperFGD (white, top) is compared to the one currently used in the ND280 detectors (black, bottom). 
}
\label{fig:superfgd-connectors}
\end{figure}

A CAD model and a picture of the optical connector, compared to the one currently used in ND280, are shown in Fig.~\ref{fig:superfgd-connectors}.
The lid of the connector has a step-like shape to facilitate the polishing
with a diamond polisher e.g. FiberFin~\cite{FiberFin}.
%

The WLS fibers are glued to the plastic connectors using the EJ-500 epoxy optical cement, also used for all the already existing ND280 detectors.
It was found that, even though the designed connectors are very small,
the contact with the WLS is mechanically strong enough against possible stresses.

%
%
%

\subsubsection{Prototype of the optical interface}
\label{sec:target:prototype-interface}

The optical interface is designed to minimize the necessary space and material budget. 
On the other hand, dimensional accuracy is required to have a good optical coupling.
We therefore made a small prototype to check the feasibility of the design. 

\begin{figure}[htb]
  \begin{center}
    \includegraphics*[width=0.8\textwidth]{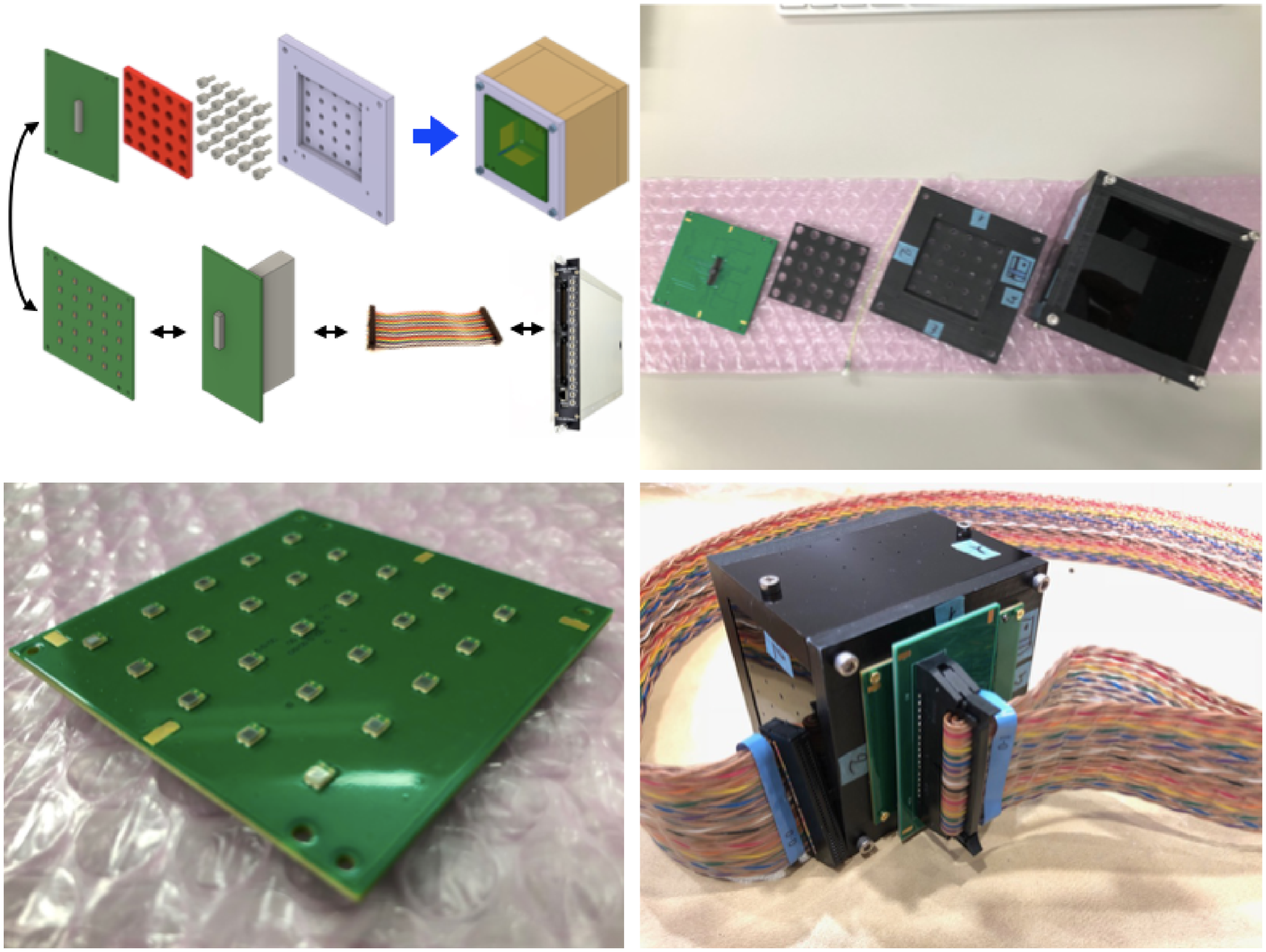}
  \end{center}
\caption{Prototype of the interface. CAD design (top left), components of the prototype (top right), MPPC-PCB prototype (bottom left) and assembled prototype are shown.}
\label{figure_interface_prototype}
\end{figure}

The first small prototype is designed to have 5 $\times$ 5 channels to demonstrate the feasibility.
Larger prototypes, with expected dimensions of 8 $\times$ 8 and 8 $\times$ 16, are foreseen as steps towards the final design.
A CAD design and pictures of those components are shown in Fig.~\ref{figure_interface_prototype}. 
Components of the prototype detector are a box surface, a plastic plate, WLS fibers, fiber connectors and a printed circuit with 25 surface-mount MPPCs (MPPC-PCB).
The box surfaces, a plastic plate and fiber connectors for this prototype were made by a 3D printer.
We plan to fabricate these elements by machining to achieve accuracy of $<$100~$\mu$m for the actual detector. 

The prototype MPPC-PCB was manufactured with a standard 4 layer PCB of 60~mm $\times$ 60~mm $\times$ 1.6~mm.
The MPPCs were mounted by KE-2060M (JUKI CORPORATION), which has $<$50~$\mu$m precision.
The soldering was done with a metal mask and a reflow oven (SOLSYS-6310IR, ANTOM CO.,LTD). 
A surface-mount connector, SAMTEC ST4-40-3.00-L-D-K-TR, is used for this prototype.



The 3D shape of MPPC-PCB prototype was measured with KEYENCE VR-3000 One-Shot 3D Measuring Macroscope, which has $<$5~$\mu$m precision. 
The positions of mechanical holes for screws and alignment pins with respect to the MPPCs were confirmed to be within $\pm$ 60~$\mu$m.
The flatness of the PCB was measured and the deformation was within 50~$\mu$m.
The MPPC alignment was evaluated by measuring the pitch between neighboring MPPCs.
With the design value of 1~cm, the mean and RMS of measured pitch was 9993.6~$\mu$m and 31.3~$\mu$m, respectively.
The height for 25 MPPCs after soldering was measured and the RMS was 7.3~$\mu$m.

%


We have tested the prototype by using the readout setup with a NIM module designed for multi-channel MPPC readout~\cite{NAKAMURA2015376}, utilizing the EASIROC ASIC chip. 
%
Using prototype scintillator cubes,
we observed about 70 photoelectrons per MIP for a channel in average. The uniformity of light yield was $<$10\% in RMS. 
Optical crosstalk was also checked by injecting LED light to a single fiber and looking at the output of neighboring channels. 
No optical crosstalk was observed up to 1 MIP level of light yield injection.

In conclusion, we validated mechanical, electrical and optical performance with the prototype and found no problem in the basic concept.


%% file: Target_electronics.tex
\section{Electronics }

Given the relatively short period of time for the development of a full electronics and DAQ chain for SuperFGD, 
we adapt systems for which some experience exists in design and operation. 
The baseline design is structured around the CITIROC (Cherenkov Imaging Telescope Integrated Read Out Chip) readout chip used by the Baby MIND collaboration in electronics deployed for the WAGASCI experiment T69~\cite{Noah:2016ikh}, with an alternative based on the SPIROC (Silicon PM Integrated Read-Out Chip) readout chip also used in the WAGASCI experiment~\cite{Chikuma:2017eed}. 


\begin{figure}[htbp]
  \centering
    \includegraphics*[width=0.45\textwidth]{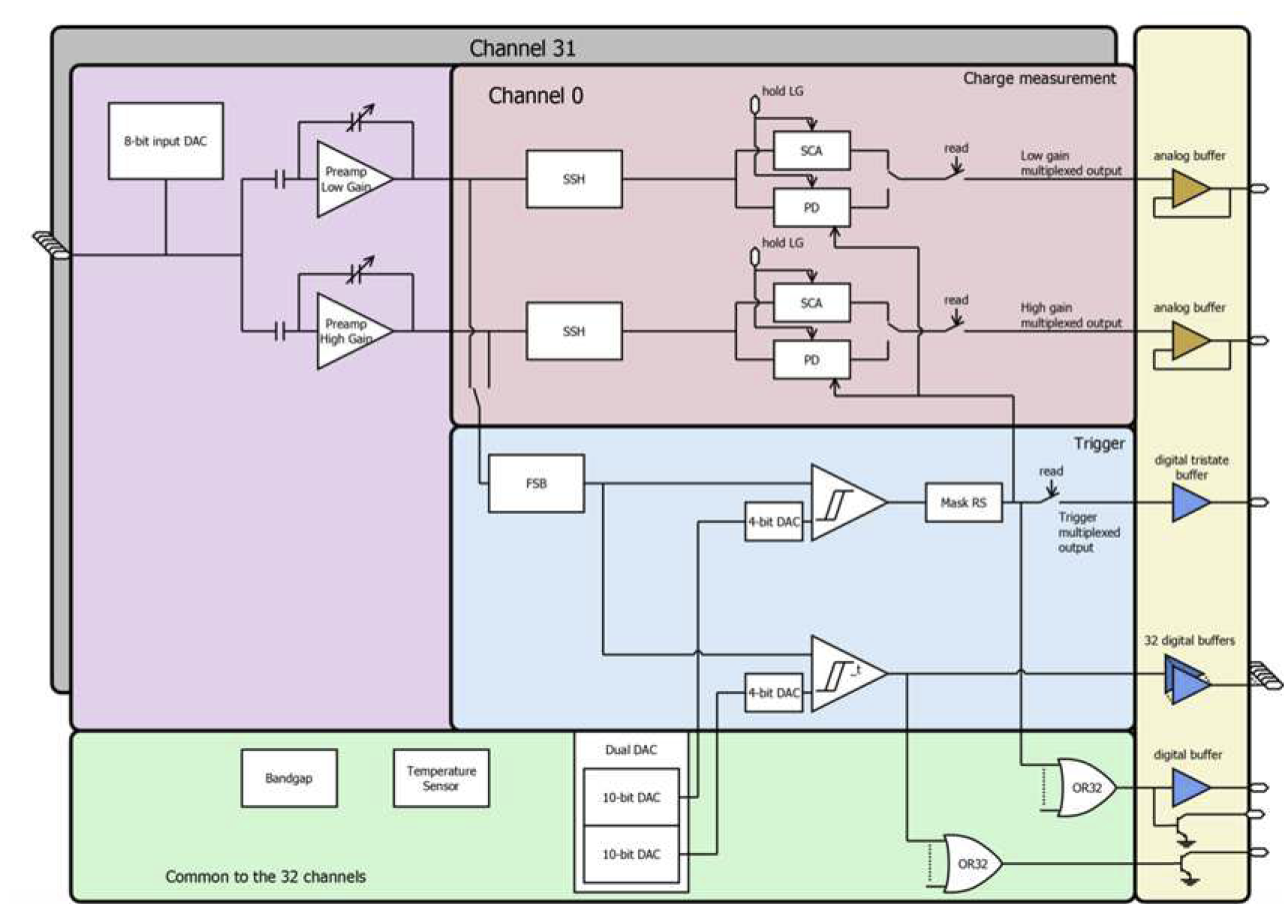}
    \hspace{2mm}
    \includegraphics*[width=0.5\textwidth]{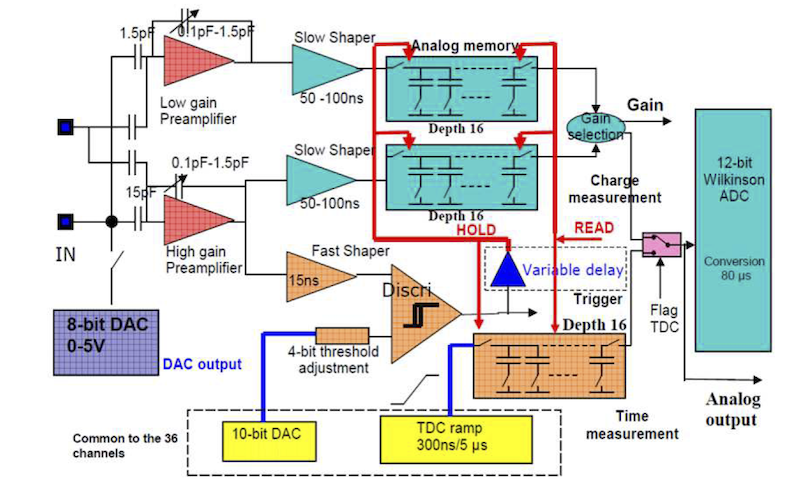}
\caption{Block diagrams of CITIROC (left) and SPIROC (right).}
\label{fig:Target-ASIC-diagrams}
\end{figure}

CITIROC and SPIROC are frontend ASICs developed by Omega laboratory at Ecole Polytechnique~\cite{Omega}.
Both are designed for the readout of large number of SiPM devices.
Figure~\ref{fig:Target-ASIC-diagrams} shows the block diagrams of CITIROC and SPIROC.
The first stage of the architecture is very similar for both chips.
For an input channel, there are two preamps with different gain (high and low gain), slow shapers for charge readout, and a fast shaper together with a discriminator for timing information.
While SPIROC has an on-chip ADC to provide digitized information of charge and timing, CITIROC uses external electronics for digitization.
The number of channels is 32 for CITIROC and 36 for SPIROC.


\subsection{Requirements and system overview}
\begin{figure}[htbp]
  \begin{center}
    \includegraphics*[width=0.95\textwidth]{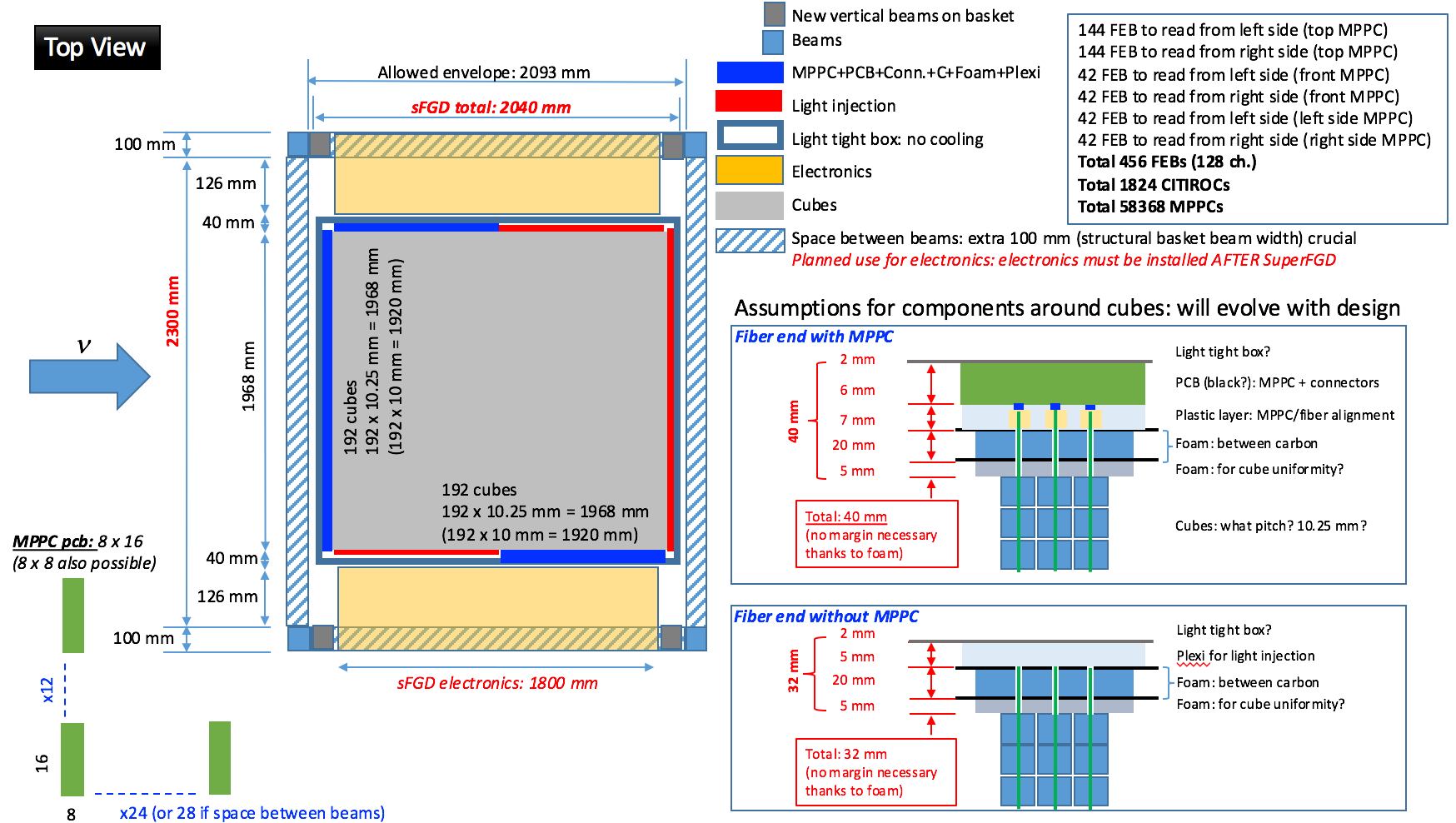}
  \end{center}
\caption{SuperFGD dimensions, indicating the location and space for the frontend electronics.}
\label{fig:Target-top-view}
\end{figure}

The requirements and constraints are summarised in Tab.~\ref{tab:elec_req}. 
One of the key points is that the electronics will have to be installed from the sides, after the SuperFGD is dropped into position from above the pit. This is in order to use the space between the structure beams of the basket, as shown in Fig.~\ref{fig:Target-top-view}.


\begin{table}[htbp]

\centering

\begin{tabular}{llllcccc}
\toprule
\textbf{Item}  	&	\textbf{Unit} & 	\textbf{Nominal}	\\

\hline
\multicolumn{3}{l}{\textit{System}}\\
\hline
Number of channels			&				[]			&	58368			 \\
Life term: experiment phase			&				[yr]			&	20		\\
Life term: test/QC phase			&				[yr]			&	0.2	\\
Expected operation fraction			&				[hr/year]			&	4000		\\
Power requirements			&				[W]			&	$<$2000	 \\
\hline
\multicolumn{3}{l}{\textit{Beam parameters}}\\
\hline
Bunches per spill			&				[]			&	8		\\
Bunch width (separation)			&				[ns]			&	80 (581)	\\
Spill duration			&				[us]			&	5			\\
Spill rate 2018 (for design)			&				[Hz]			&	0.4 (1.0)			\\
Beam power 2018 (for design)			&				[kW]			&	500 (1300)			\\
\hline
\multicolumn{3}{l}{\textit{Readout chip}}\\
\hline
Readout window beam/calibration/cosmics			&				[ms/spill]			&	0.020/100/300			\\
Deadtime	(within beam readout)		&				[$\mu$s/spill]			&	0			\\
Deadtime	(outside beam readout)		&				[ms/spill]			&	0to500			\\
Hit amplitude dynamic range		&				[pe]			&	1500			\\
Hit amplitude resolution 1 MIP (10 MIPs)	&				[pe]			&	2 (100)			\\
Hit detection threshold		&				[pe]			&	0.5			\\
Hit time resolution (1 cube)		&				[ns]			&	1			\\
Hits per channel per spill (beam window)		&				[/ch/spill]			&	0.01			  \\
Hits per channel per spill (noise, b.w.)		&				[/ch/spill]			&	1			\\
Hits per ROC per spill (b.w.)		&				[/ROC/spill]			&	50			 \\
\hline
\multicolumn{3}{l}{\textit{Material budget}}\\
\hline
FEE (if direct mount)			&				[\% x/X$_0$]			&2				 \\
MPPC PCB, cables, connectors						&	[\% x/X$_0$]			& 3		 \\
\hline
\multicolumn{3}{l}{\textit{Environmental conditions}}\\
\hline
Operating temperature (storage)			&				[C]			&	20 {(0-40)}			 \\
Operating humidity (storage)						&	[\% RH]			&	10			 \\
Magnetic field		&				[T]			&	0.2 \\
\hline	
\bottomrule
\end{tabular}

\caption{\em ND280 upgrade SuperFGD electronics requirements, input design parameters.}
\label{tab:elec_req}
\end{table}


\begin{figure}[htbp]
  \begin{center}
    \includegraphics*[width=14 cm]{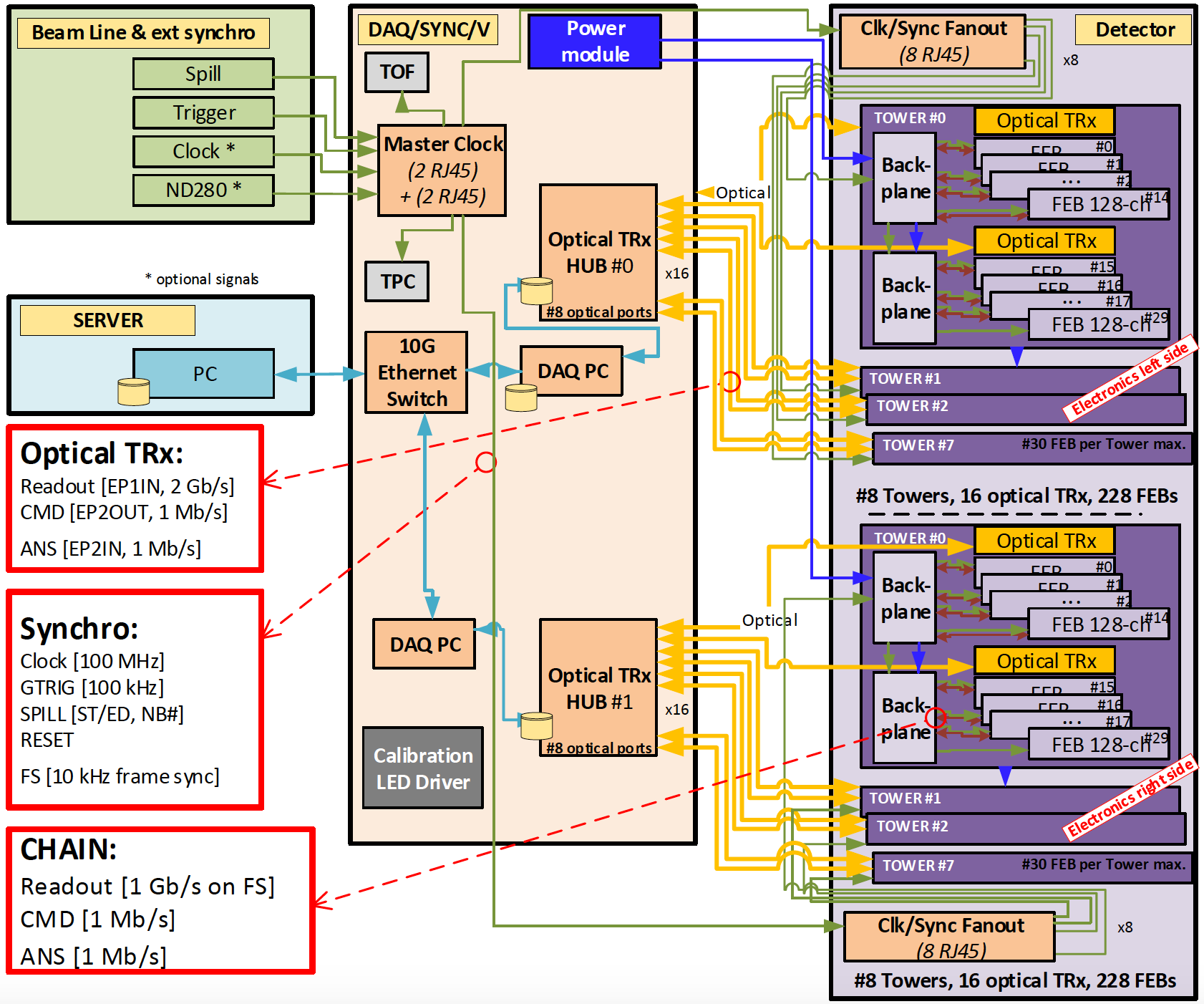}
  \end{center}
\caption{SuperFGD readout general architecture.}
\label{fig:Target-SuperFGD_readout}
\end{figure}

The general architecture of the SuperFGD readout baseline design is shown in Fig.~\ref{fig:Target-SuperFGD_readout}. 
%
%
Because of space limitations around the ND280 detector basket, it was decided to place the readout electronics on the left and right sides of the allocated SuperFGD detector volume. Accessibility is good for most elements in the electronics chain, with good prospects for maintenance, repair, exchange with spares when required. The MPPC PCBs and associated connectors will not be easily accessible. By design, these must be made reliable, so they can survive the life term of the experiment with little probability of failure. One key element in the chain will be the system of connectors and cables carrying HV to the MPPC PCB, and signals from the MPPC PCB to the Front End Boards (FEBs). These are organized in towers, 8 either side of the basket, with up to 30 FEBs per tower linked together via a Backplane.

Synchronisation with the T2K beam and other ND280 systems will be handled by a Master Clock Board (MCB)and a network of fanout boards. The MCB will most likely be located outside the ND280 detector basket, along with DAQ components such as optical transceiver hubs and PCs, power modules designed to supply the various voltages required to power components in the basket, calibration electronics.

\subsection{Detailed Description of the main components}

\begin{figure}[htbp]
  \begin{center}
    \includegraphics*[width=14 cm]{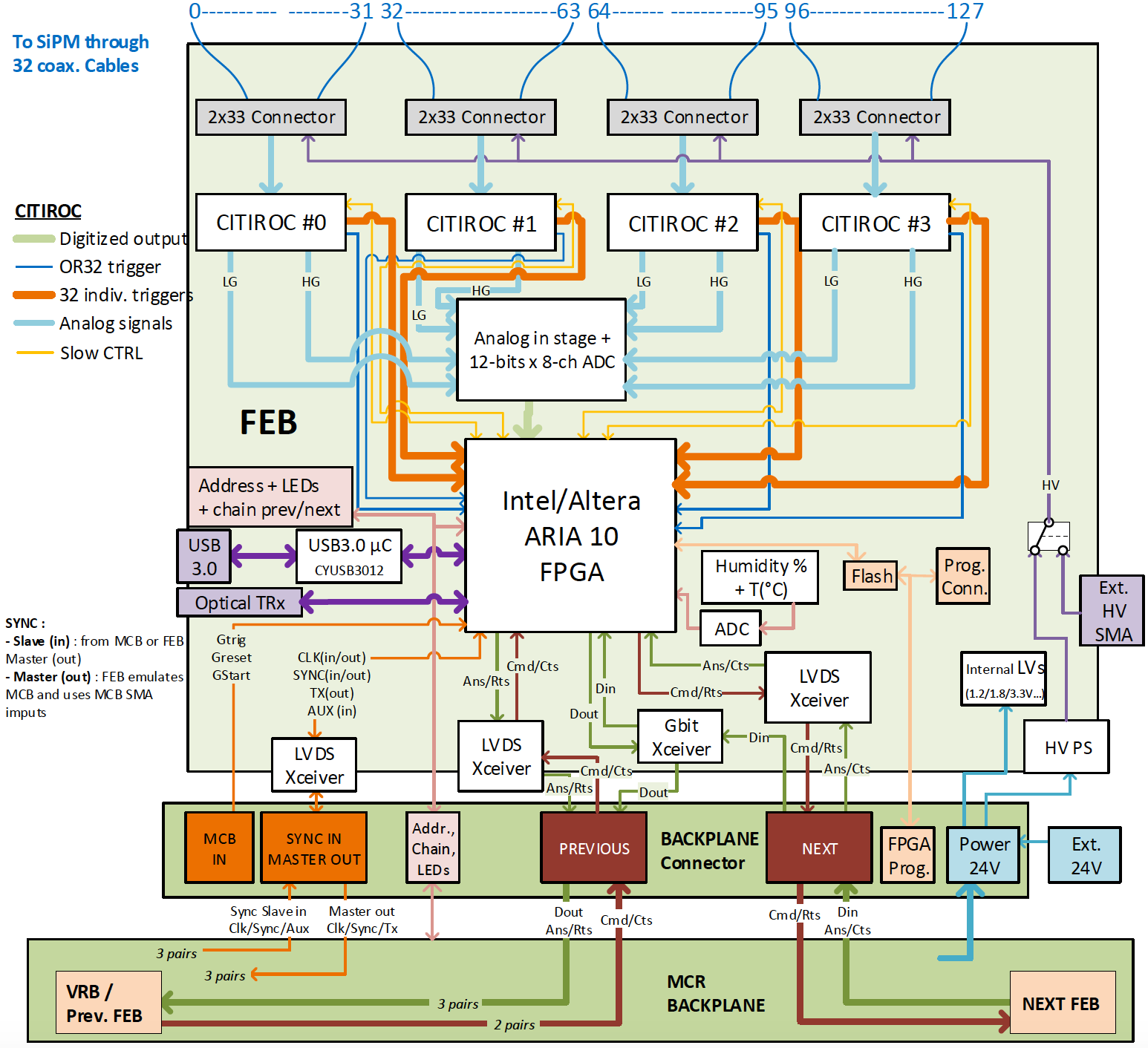}
  \end{center}
\caption{SuperFGD Front End Board sketch.}
\label{fig:Target-FEB}
\end{figure}

At the heart of the system is the electronics Front End Board (FEB). Its main features are illustrated in Fig.~\ref{fig:Target-FEB}. Daisy chaining and synchronisation functions are carried out by two ancillary boards, the Backplane and the Master Clock Board (MCB).

The FEB architecture is based on 4 CITIROC chips that can each read signals from 32 MPPCs, one FPGA Altera Aria X to control and manage the timing and data flow from the CITIROCs, one 8-channel ADC for the digitisation of the CITIROC analogue output and data transmission to a data acquisition system either via a USB3 or optical interface.

Within the CITIROC, each signal input is processed by two main adjustable signal paths: a high gain (HG) path, and a separate low gain (LG) path. Each of these two signal paths has a dedicated slow shaper, the output of which can be sampled using one of two modes: a mode with an externally applied delay, and a peak detector mode. After sampling, the analogue information from both signal paths is sent off-chip via 32-channel multiplexers, one for each path, towards an external ADC on the FEB. Timing information for each signal is provided by an independent fast shaper that can either be switched to the HG or LG path pre-amplifier output. The fast shaper is followed by a discriminator with adjustable thresholds. The $4\times32$ individual trigger outputs are sampled by the FPGA at 400~MHz, which records both rising and falling edges of these outputs and assigns time stamps. The difference in time between rising and falling edges (Time-over-Threshold) gives some measure of signal amplitude. It is used in addition to the analogue charge information and proves useful if there is more than one hit per fiber within the 9~$\mu$s deadtime due to the processing of the multiplexed charge outputs.

The internal 400~MHz clock on the FEB can be synchronised to a common 100 MHz clock. The synchronisation subsystem combines input signals from the accelerator beam line on the Master Clock Board, including a pre-beam trigger  issued 30 us before the beam, into a digital synchronisation signal (SYNC) and produces a common detector clock (CLK) which can eventually be synchronised to an external experiment clock. Both SYNC and CLK signals are distributed to the FEBs via the backplane. Tests show the FEB-to-FEB CLK (SYNC) delay difference to be 50 ps (70 ps). 
The accelerator beam spill number will also be recorded as a 16-bit signal.

\begin{figure}[htbp]
  \begin{center}
    \includegraphics*[width=14 cm]{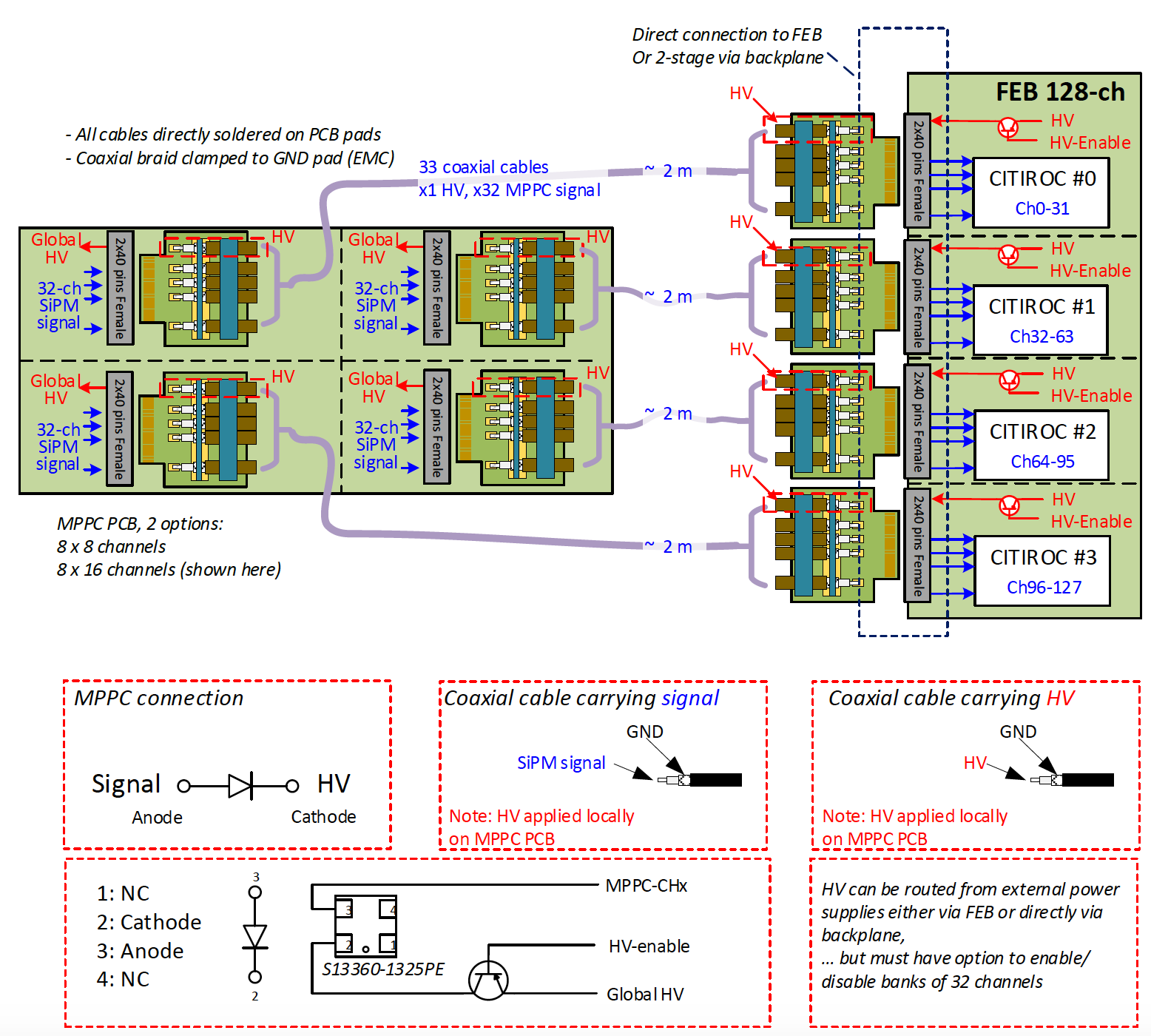}
  \end{center}
\caption{SuperFGD electronics connectivity scheme, MPPC PCB to FEB.}
\label{fig:Target-connectivity}
\end{figure}

The scheme selected for MPPC PCB connectivity is shown in Fig.~\ref{fig:Target-connectivity}. The cables are organised in bundles of 32 channels at the FEB end to match the 32 inputs of a CITIROC. Because a malfunction of a single channel could potentially affect all other channels in a bundle, one HV line is drawn separately on the 2 m extension cable bundle from the FEB. The HV is then applied to each channel at the MPPC-end of the cable bundle. The HV is then transmitted locally via the MPPC PCB to each MPPC. The 2 m extension coax cable copper braid must be connected to ground for noise immunity of the MPPC signal carried in the inner conductor wire. Having the amplifier on the FEB side and not locally close to the MPPC ensures a current-mode signal transmission from the MPPC up to the FEB through the coaxial cable, with good noise immunity. 

\begin{figure}[htbp]
  \begin{center}
    \includegraphics*[width=14 cm]{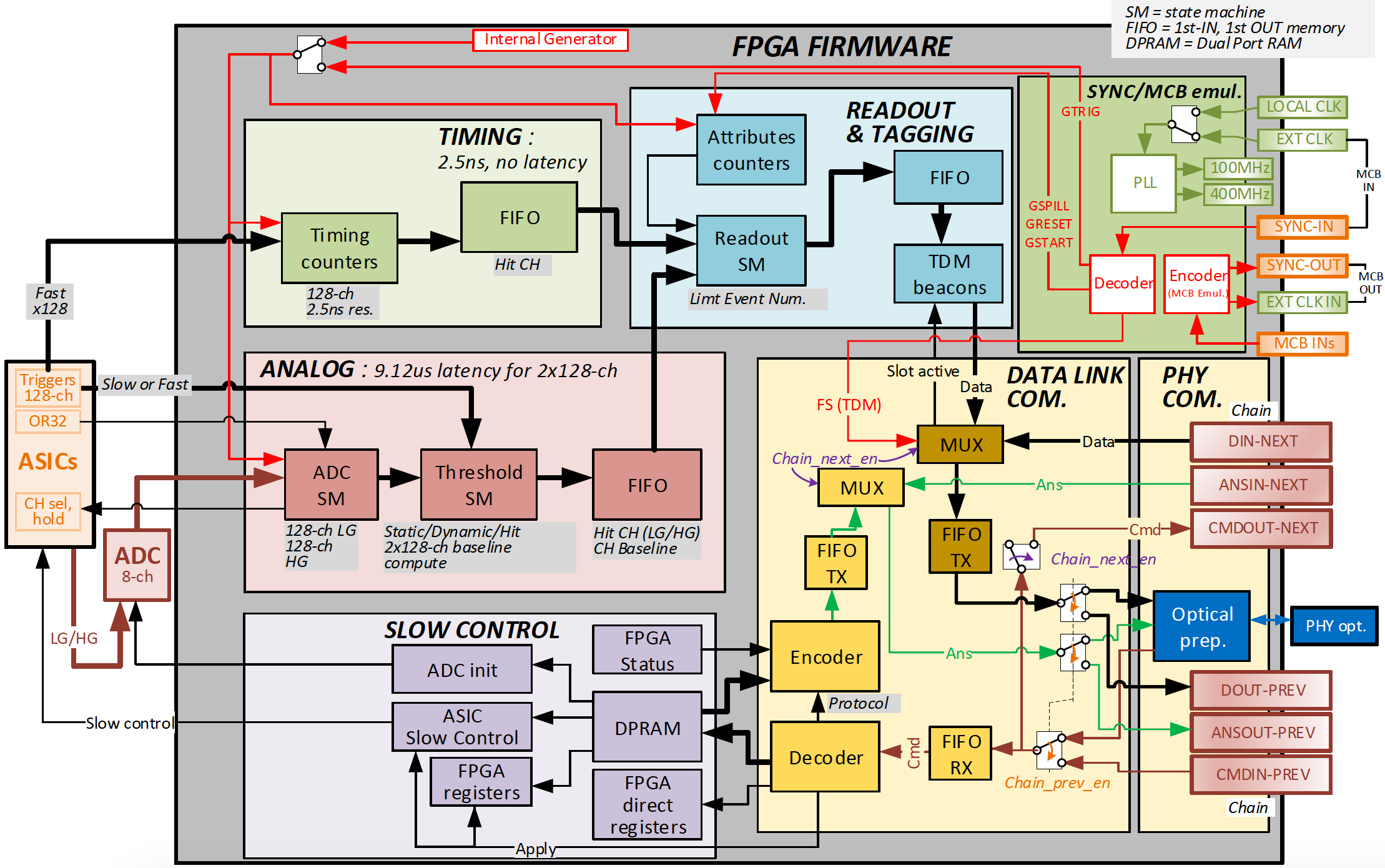}
  \end{center}
\caption{SuperFGD firmware main blocks.}
\label{fig:Target-firmware}
\end{figure}

The firmware blocks for the FPGA on FEB are shown in Fig. \ref{fig:Target-firmware}.

\subsection{Schedule}
\begin{figure}[htbp]
  \begin{center}
    \includegraphics*[width=1.0\textwidth]{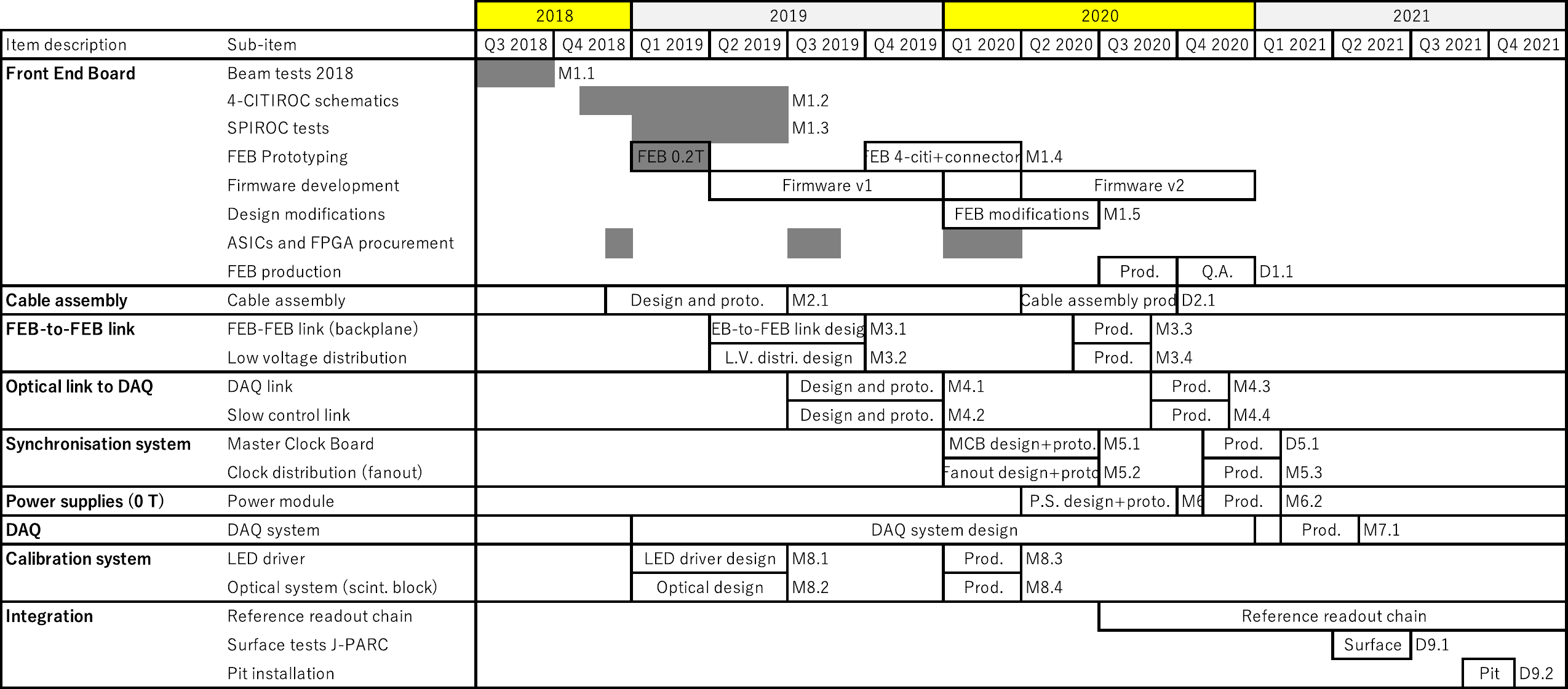}
  \end{center}
\caption{Preliminary schedule of SuperFGD electronics development.}
\label{fig:Target-elec-shcedule}
\end{figure}

Figure~\ref{fig:Target-elec-shcedule} shows a tentative schedule of the SuperFGD electronics development.
The electronics is designed based on the existing Baby-MIND architecture which has been successfully tested in CERN and also at J-PARC neutrino beamline.
The layout of FEB will be modified with much compact design to fit into a limited space available for superFGD.

Because the production phase for electronics components should not be lengthy, at most 3 months for all FEBs for example, much of the project time will be dedicated to design and prototype evaluation, with full production reserved for the end of the project, allowing for QC tests and integration. 

Production of the first prototype FEB is foreseen for Q4 2019. With feedback from tests of this first prototype, full production for all 456 FEBs and spares is expected to take place Q3 2020, with the procurement of the main components (FPGA and CITIROC ASICs) complete by start of production. Prototyping of most of all other systems will be carried out in 2019. The main exception is the MCB, whose development could be initiated in Q1 2020.

\subsection{Quality Control}
A quality control plan will be established to cover sub-systems, and integrated systems. 

Sub-systems such as the FEB, MCB, MPPC PCB, optical TRx modules, cable assemblies, calibration system, power supply modules, can be tested on dedicated test benches by the institutes responsible for their production. The QC approach is to test every component before assembly, rather than small samples per batch. This is due to the relatively long chain of components, for which troubleshooting of faults to find the malfunctioning element will be challenging once the chain is assembled. 

The definitive testing of the full integrated readout electronics will be carried out at J-PARC, on the surface.
Subsets of the full readout chain are likely to be made available to several institutes for characterisation and testing beforehand.

QC procedures will cover testing at various levels such as visual inspection, passive and active tests where necessary, as well as steps to be taken to handle components that do not pass the QC tests such as repair and re-evaluation. A database will be established to record the QC process and track changes in time.

\subsection{Alternative design based on the SPIROC readout chip} 

\begin{figure}[tb]
\begin{center}
\includegraphics[width=0.8\columnwidth]{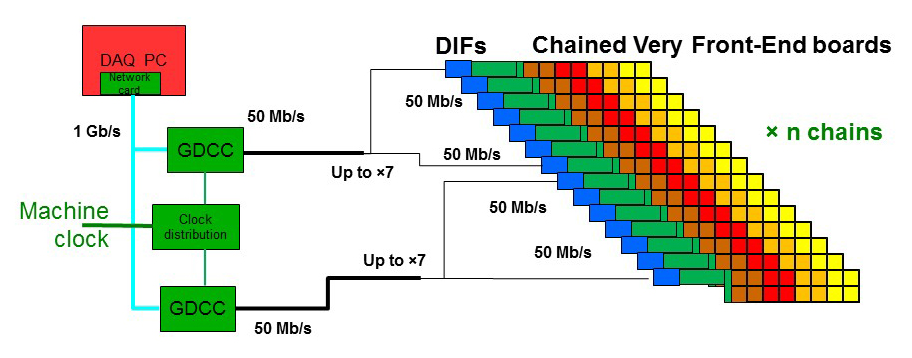}
\end{center}
\caption{Schematic view of the readout and DAQ system developed for the ultra-granular calorimetry at the future International Linear Collider (adapted from \protect\cite{TIPP-Calice}). For the proposed Super--FGD detector, the original silicon slabs will be replaced by chains of about 20 boards hosting 4 SPIROC(2E) chips, each chain corresponding to the readout of $\sim 3000$ MPPCs.}
\label{fig:DAQ-Calice}
\end{figure}

An alternative readout system is being studied in case constraints on compactness and power consumption are found to prevent from using the baseline design. 
It is based on the SPIROC(2E) chip developed by Omega, used together with a generic readout system that has been developed for the ultra-granular calorimetry part of the ILD/ILC project (CALICE)\cite{TIPP-Calice}, and which is shown in Figure~\ref{fig:DAQ-Calice}. 
This design has been used for the water modules of the WAGASCI detectors. It is scalable to a huge number of channels and comprises the very front-end electronics that can in principle be made of up to $\sim 200$ chained readout chips, each chain being read out by ``Detector InterFace (DIF)'' boards. Up to 7 DIF boards are read out by a ``Giga Data Concentrator Card (GDCC)'' board which directly communicates with the DAQ computer.

For the very front-end electronics, the 36 channel SPIROC readout chip combines high precision time measurement and dual-gain charge measurement with power pulsing capabilities, and is therefore well adapted to the T2K beam time structure (one beam spill made of 8 bunches spaced by 580~ns, with a repetition cycle of 2.5~s. In this design, the front-end boards could comprise up to 16 SPIROC chips and would be placed in towers on each side of the detector, as already described in the baseline design. Chains of $\sim 8$ of these boards will be read out by one small DIF board, also located on the side of the detector. The DIF board signals will in turn be conveyed by standard HDMI cables to the GDCC boards, located out of the magnet. This feature is important, as the GDCC is where the main FPGA chip resides, and a failing board will be easier to replace if it is not enclosed inside the magnet. Furthermore, the GDCC board has not been designed to be compatible with the 0.2~T magnetic field. Each GDCC board can control up to 7 DIFs, so 2 to 4 GDCC boards would be sufficient to read the entire detector.

The SPIROC chip stores the signal in a 16 column analog memory array, each column comprising 2$\times$36 cells for the charge (one for each gain) and 36 cells for the time measurement, also stored as a charge by means of a dedicated internal voltage ramp. A column is used every time one of the 36 channels presents a signal above a given programmable threshold during a cycle of the master clock. 
The master clock frequency will be synchronized with the beam bunch spacing of 580 ns. This way, each one of the 8 beam bunch fills at maximum one column, and several columns are left available to record signal arising in the few microseconds after the beam, allowing therefore to record tracks from the so-called ``Michel electrons'' produced by muon decays.

The signals are then digitized and sent to the DIF boards in between beam spills. The total time needed to read the data might vary from a few milliseconds to a few tenths of a seconds, depending on the detector occupancy, and therefore on the exact value we choose for the programmable chip threshold. Due to the very low interaction rate, only a few tracks are present in the whole detector during the entire beam spill, and the signal pile-up is therefore very unlikely, given the very high granularity of this detector.

The total power dissipation does not exceed 700~W with no use of the power-pulsing capability of the chips, which then allows a light cooling system. This alternative readout system is however not able to properly register two hits occurring in the same channel within the same 580 ns time window. This is the main reason why the baseline design is preferred at present. Tests of this alternative design reading the signals from the SuperFGD prototype will be performed in the beginning of year 2019. 

%% file: Target_DAQ.tex
\section{DAQ}
\begin{figure}[tbp]
  \begin{center}
    \includegraphics*[width=1.0\textwidth]{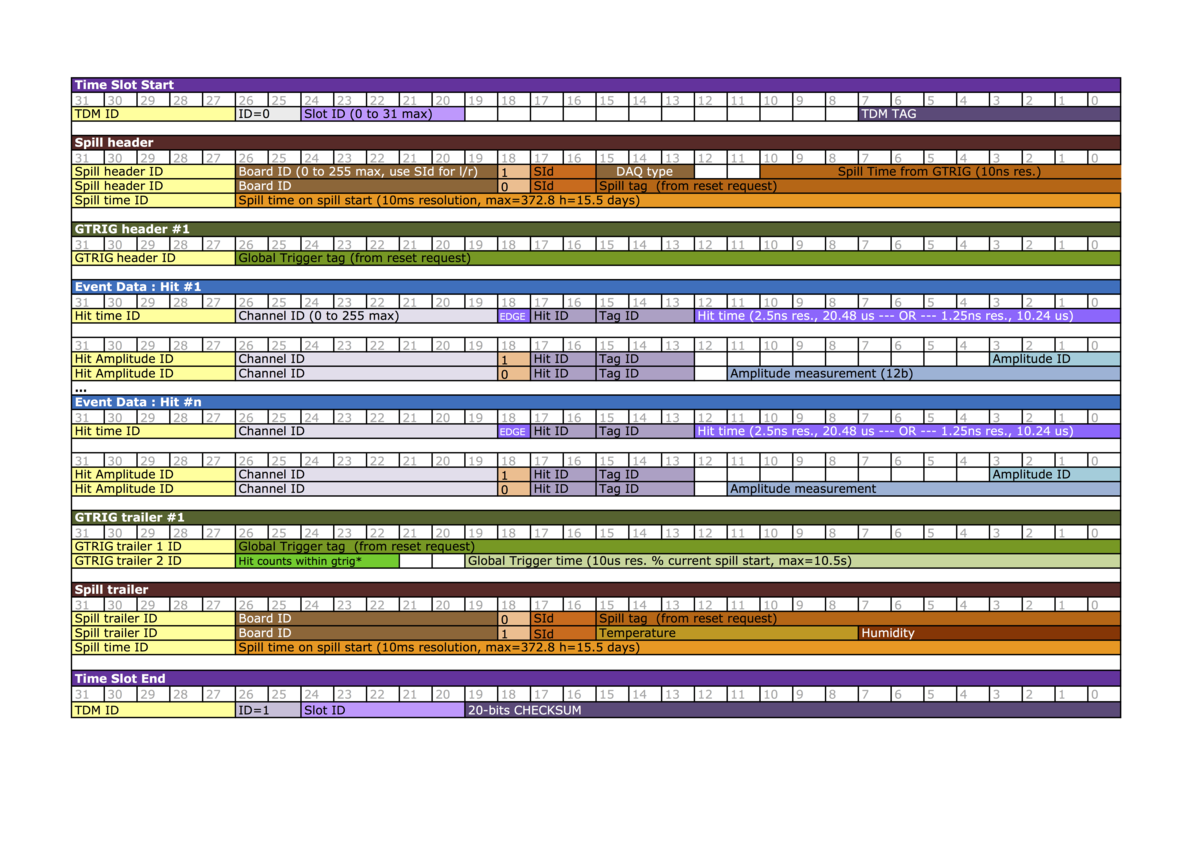}
  \end{center}
\caption{SuperFGD DAQ binary data output protocol.}
\label{fig:Target-daq}
\end{figure}

The SuperFGD DAQ binary data output protocol, adopted from what was developed for Baby-MIND, is outlined in Fig.~\ref{fig:Target-daq}. 

In order to provide realistic estimates of data rates for the SuperFGD, several measurements were carried out with electronics settings close to those that are likely to be applied at T2K. These settings were those chosen after optimization of the full readout chain during beam tests at CERN in August/September 2018 and include the high-gain and low-gain preamplifier settings, discriminator thresholds and MPPC operating voltages. 

The measurements were done on the $8\times24\times48$ - cubes SuperFGD prototype. Two types of measurements were carried out with FEBs instrumented with 96 MPPC S13360-025 each: cosmic rays at surface level with 1 FEB, then 5 FEBs, and LED measurements with an LED frequency of 15.6 kHz. The cosmics measurements at surface level are taken to be conservative in terms of data rate with respect to operation in the pit, despite the factor $\times4$ shorter WLS fiber length used (48 cm). The LED measurements are taken to be representative of measurements with a calibration system. 

Cosmic rays measurements with 1 FEB (Time Division Multiplexing (TDM) mode with a single FEB) return a raw binary file size of 0.04~kB, 0.16~kB and 1288~kB for acquisition windows of 10~$\mu$s, 100~$\mu$s and 1~s, respectively. Between 10~$\mu$s and 100~$\mu$s the rates do not scale linearly due to the large overhead of 32-bit headers and trailers for data encapsulation. The measurements with 5 FEBs in a readout chain show exactly the same data rates for averages per FEB.

LED measurements with 1 FEB (again, TDM mode with a single FEB) return a raw binary file size of 0.11~kB and 0.88~kB for acquisition windows of 10~$\mu$s and 100~$\mu$s, respectively. 

The data size estimated with various assumption on acquisition windows are summarized in Table~\ref{tab:daq_file_size}. 
 They include a further $\times2$ safety factor for beam/cosmics. This safety factor is taken to cover the larger channel count per FEB in the final configuration and a slight increase in recorded hits if thresholds are lowered. 

\begin{table}[ht]

\centering

\begin{tabular}{llllcccc}
\toprule
\textbf{Readout block}  	&	\textbf{10 $\mu$s window} & 	\textbf{100 $\mu$s window} & \textbf{1 s window}	\\

\hline
\multicolumn{4}{l}{\textit{Beam or cosmics}}\\
\hline

1 FEB			&			0.1 KB			&	0.3 KB & 2 MB 			 \\
30 FEB (1 tower)			&			3 KB			&	9 KB & 60 MB 			 \\
228 FEB (1 side)			&			22.8 KB			&	68.4 KB & 456 MB 			 \\
456 FEB (2 sides)			&			45.6 KB			&	136.8 KB & 912 MB 			 \\

\hline
\multicolumn{4}{l}{\textit{LED calibration}}\\
\hline

1 FEB			&			0.2 KB			&	2 KB & 20 MB 			 \\
30 FEB (1 tower)			&			6 KB			&	60 KB & 600 MB 			 \\
228 FEB (1 side)			&			45.6 KB			&	456 KB & 4560 MB 			 \\
456 FEB (2 sides)			&			91.2 KB			&	912 KB & 9120 MB 			 \\

\hline	
\bottomrule
\end{tabular}

\caption{\em SuperFGD DAQ binary output data sizes as a function of acquisition window duration. The acquisition window is defined for continuous self-triggering mode. The SuperFGD will likely have a total of 456 FEBs.}
\label{tab:daq_file_size}
\end{table}

As can be seen, the data rate is very dependent on the definition of the calibration sequence. A system based on a total data throughput of 10 GBit/s writing raw binary to DAQ PC disk (level 1 DAQ before compression) would be sufficient for the operation of the SuperFGD. But the exact rate to the level-1 DAQ PC must be estimated based not only on the FEB throughput capability, but the DAQ PC writing to disk rate.

As an example, a 100 $\mu$s acquisition window for the beam spill (factor $\times20$ more than the beam width), a 500 ms acquisition window for cosmics in self-triggering mode, and a 50 ms acquisition window for LED calibration would lead to data file sizes of 912 MB, and an acquisition rate for a proton beam operating at 0.8 Hz of 730 MB/s. 

In practice, there are constraints on the maximum rate at which data can be written to tape via the MIDAS back end. The average maximum rate is 40 MB/s. The current implementation of the ND280 detector uses approximately 7 MB/s. It seems reasonable to assume that 10 MB/s can be reserved for the SuperFGD. 

The cosmic and LED triggers must be designed to be compatible with the exiting ND280 trigger and DAQ sequence.
For cosmics, SuperFGD could receive triggers generated by the Cosmic Trigger Module (CTM) based on information from surrounding detectors. 
For each such cosmic trigger, the SuperFGD can send back data corresponding to a 100 $\mu$s acquisition window. For standard running, with no LED calibration, one 100 $\mu$s window for the beam spill, and $10\times100$ $\mu$s window corresponding to 10 cosmic event triggers from CTM via Master Clock Module (MCM) between spills, the data rate is 1.2 MB/s. The raw binary data from all FEBs can be written to tape via MIDAS without compression.

The LED calibration mode poses a different type of challenge. It may be more efficient to have a separate local DAQ PC to handle the much larger data rate, and produce calibration histograms that can then be pushed to tape. A single LED trigger from the MCM would be read by the local MCB, which would then issue a command to the LED driver to start emitting pulses at the required frequency. 
Since all sub-detectors must respond to the initial LED trigger, the SuperFGD could respond with an empty or dummy event. The processing of the raw binary data to calibration histograms would then proceed independently of the main ND280 DAQ, and could be pushed to MIDAS independently. 

To dimension data storage, if only acquiring beam data with 100 $\mu$s acquisition window per spill at 0.8 Hz, 10 GB/day would be written to tape. This is acceptable for a standalone system, though the overall ND280 DAQ architecture must be considered. With a quasi-continuous acquisition of cosmics data in self-triggering mode, it is a requirement to unpack the raw binary data and carry out a first level analysis to form track candidates and compress the data 
before writing to tape.

%% file: Target_calibration.tex
\section{LED light injection system}
\label{sec:superfgd-calibration}

In order to provide a good measurement of the scintillation light produced by the charged particles,
the response of the MPPCs must be well known.
\label{sec:superfgd-calibration-led}

A calibration system based on LED light injection is under consideration.
As shown in Fig.~\ref{fig:superfgd-box-panel-xsec} of Sec.~\ref{sec:superfgd-interface}, 
the WLS fibers exit the mechanical box 
through the holes and 
about 1~cm is available for the integration of the LED system. 


A very compact system, with a limited number of LEDs per channel,
has been proposed and developed by the CALICE collaboration~\cite{Kvasnicka:2012hwa}.
It consists of distributing the calibration light from the LED to the MPPC via optical fibers,
that have notches in coincidence of each MPPC (notched fiber).
When the light propagating through the fiber encounters the notch,
it is scattered perpendicularly toward the opposite side.
This concept could fulfill the requirements for SuperFGD detector,
since with only one LED more than 200 MPPCs could be calibrated. 

\begin{figure}[htb]
\centering\includegraphics[width=0.8\textwidth]{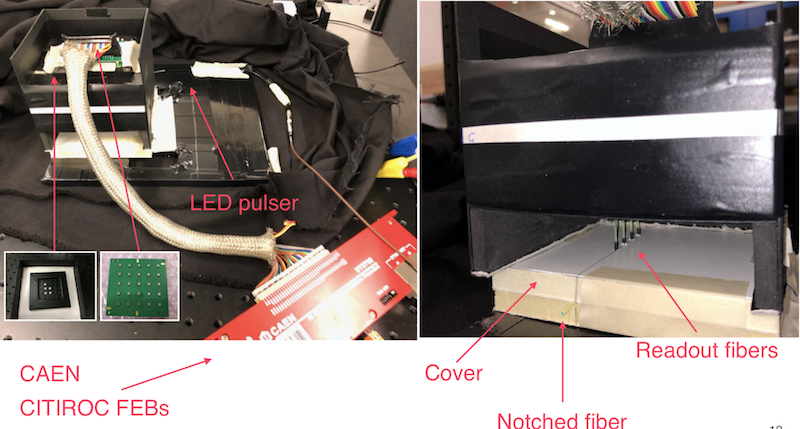}
\caption{
The prototype used for the LED calibration tests.
}
\label{fig:Target-LED-notched-fiber-prototype}
\end{figure}

\begin{figure}[htb]
\centering\includegraphics[width=0.8\textwidth]{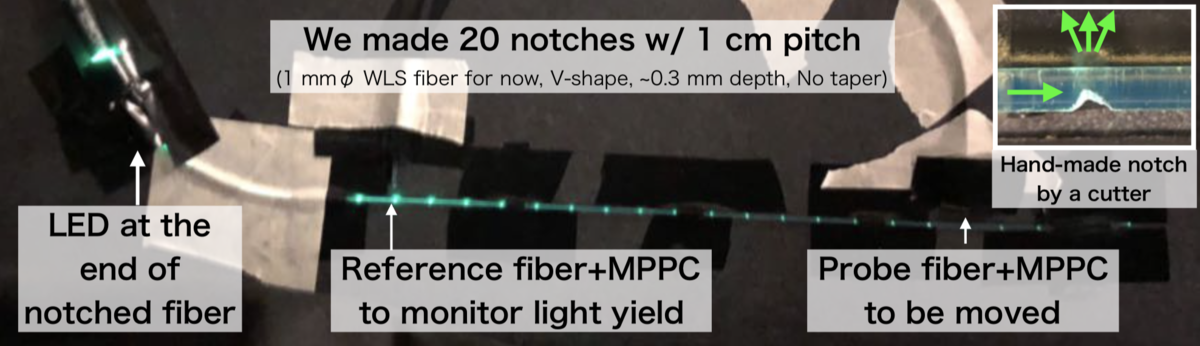}
\caption{
The notched fiber used to test the calibration system. 
The notches were hand-made with a cutter, so the light uniformity cannot be achieved.
}
\label{fig:Target-LED-notched-fiber}
\end{figure}

Some tests were performed with a small prototype box
that was instrumented with all the components of the SuperFGD optical interface, 
described in Sec.~\ref{sec:superfgd-interface}.
As shown in Fig.~\ref{fig:Target-LED-notched-fiber-prototype},
on the bottom side of the box an empty volume was used to couple the notched fiber with the WLS fibers.
The notched fiber is shown in Fig.~\ref{fig:Target-LED-notched-fiber}.
The notches were hand-made with a cutter, so it was not possible to obtain the same light yield for each MPPC,
but we can successfully observe the photoelectron peaks in the ADC distributions and proved the concept of this system.

While the concept should work, optimization of the light source and distribution system is necessary.
More studies of a LED-based calibration system are ongoing and a few other variations will be tested.

%% file: Target_schedule.tex
\section{Schedule}
\begin{figure}[htbp]
  \begin{center}
    \includegraphics*[width=1.0\textwidth]{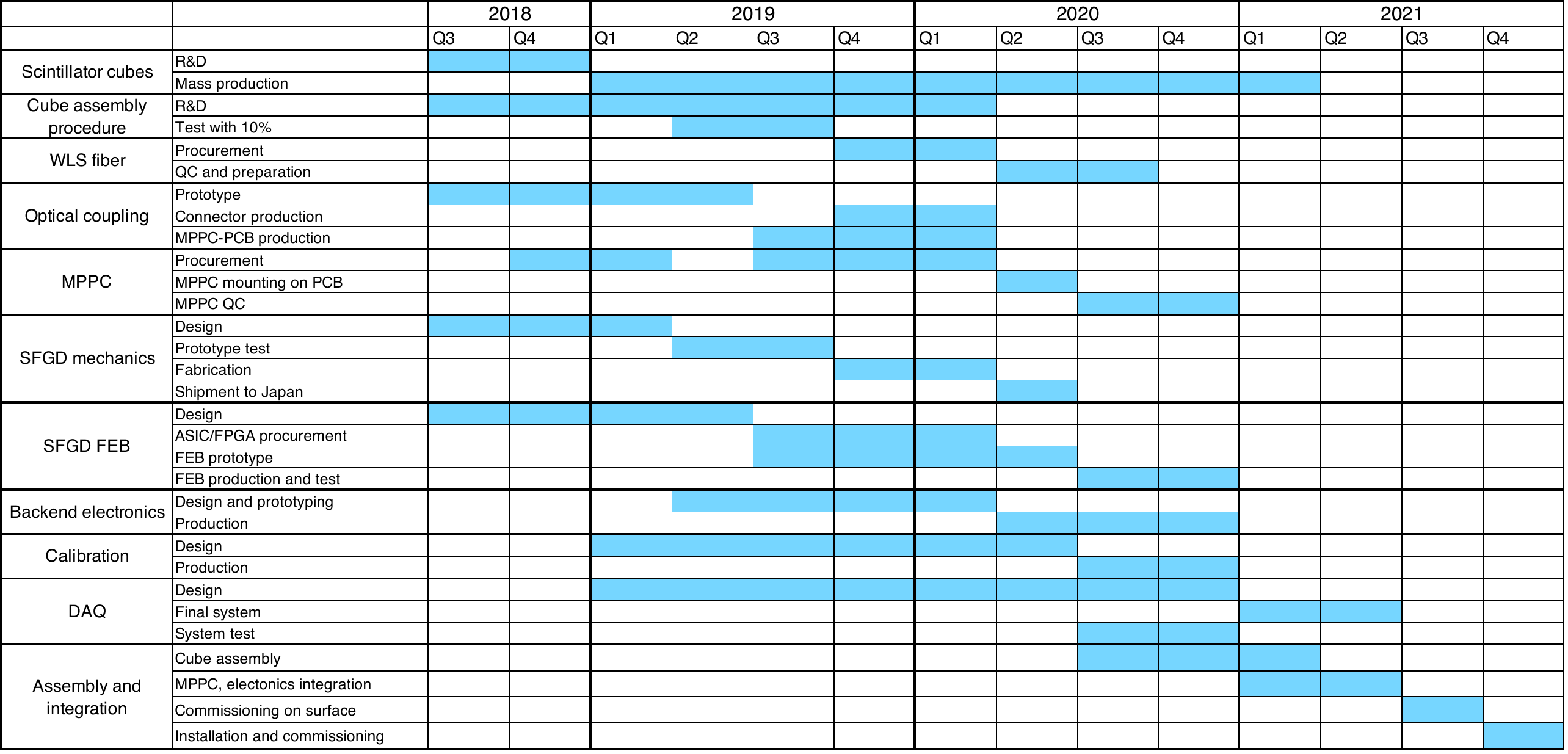}
  \end{center}
\caption{Preliminary schedule of SuperFGD construction.}
\label{fig:Target-shcedule}
\end{figure}

Figure~\ref{fig:Target-shcedule} shows the preliminary schedule of SuperFGD development and construction.

%% file: Target_prototype.tex
\section{Prototype Results }

We have constructed and tested two prototypes of SuperFGD .
The first one consisted of 125 cubes and arranged in $5\times 5\times 5$ array, which was made in the fall 2017.
The second one was made from 9216 (8$\times$24$\times$48) cubes in summer 2018.
Both were tested with test beams at CERN.

\subsection{Prototype with 125 ($5\times 5\times 5$) cubes}
\label{sec:superfgd-5x5x5-prototype}

\begin{figure}[htbp]
\centering
\includegraphics[width=0.6\textwidth]{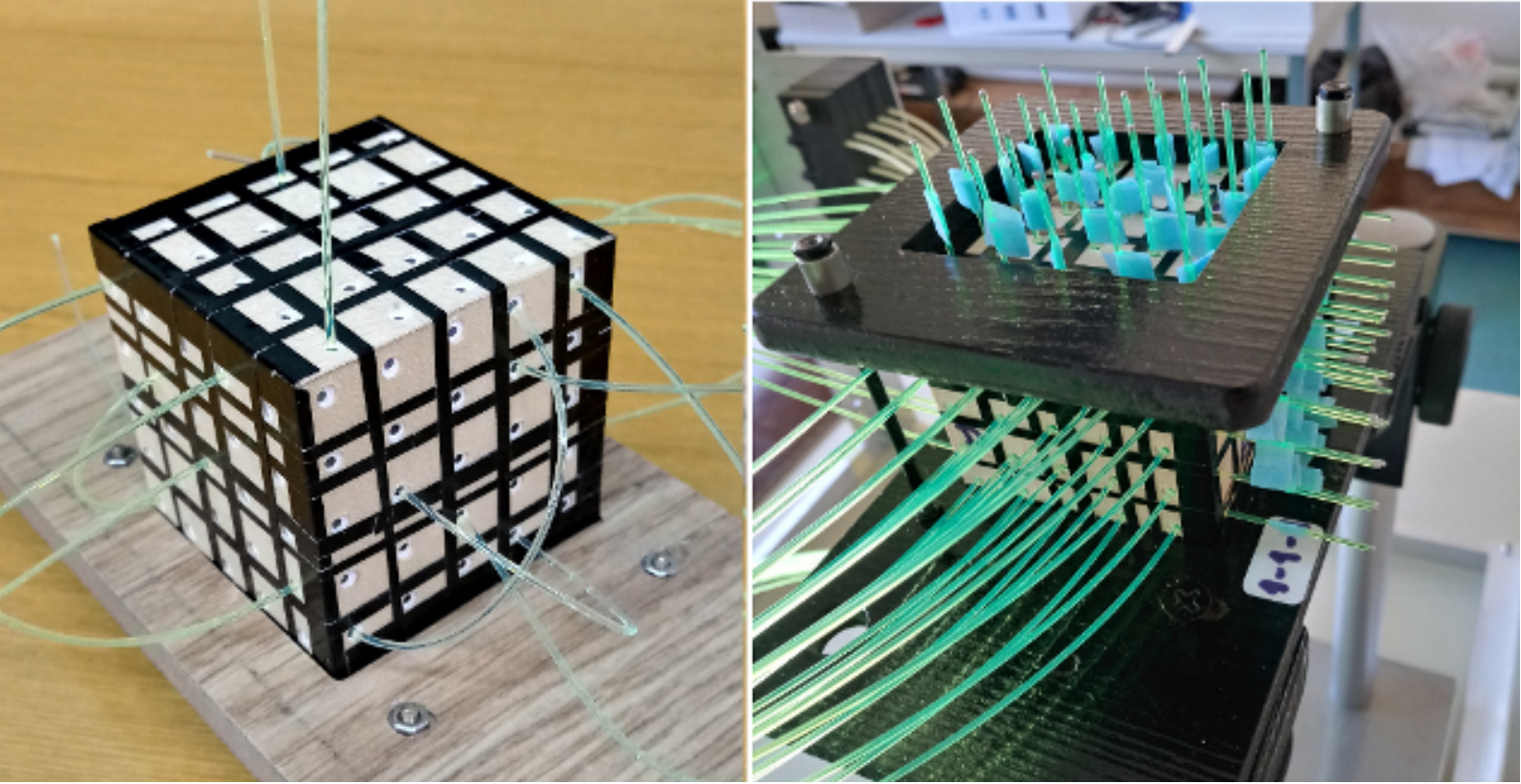}
\caption{The array of 125 cubes with 3D fiber readout. Right photo shows the array with 75 fibers mounted for the beam test. }
\label{fig:Target_cube_cube} 
\end{figure}

An array of $5\times 5\times 5$ cubes (125 cubes in total), shown in Fig.~\ref{fig:Target_cube_cube}, was assembled for a beam test with charged particles~\cite{Mineev:2018ekk} in 2017. 

75 WLS fibers were inserted through the cubes, protruding 3--4~cm out of the scintillator. The fibers are 1~mm diameter  Y11(200) Kuraray S-type of 1.3~m length. One end of the fiber is attached to a photosensor, the other end is covered by a reflective Al-based paint (Silvershine).
The photosensors in the beam test were Hamamatsu MPPCs 12571-025C with a $1\times1~mm^2$ active area and 1600 pixels. 
In order to measure the main parameters of the prototype with a high time resolution, we used custom made preamplifiers and the 16-channel CAEN digitizer DT5742 with 5~GHz sampling rate and 12-bit resolution. 

\begin{figure}[htbp]
\centering
\includegraphics[width=0.4\textwidth]{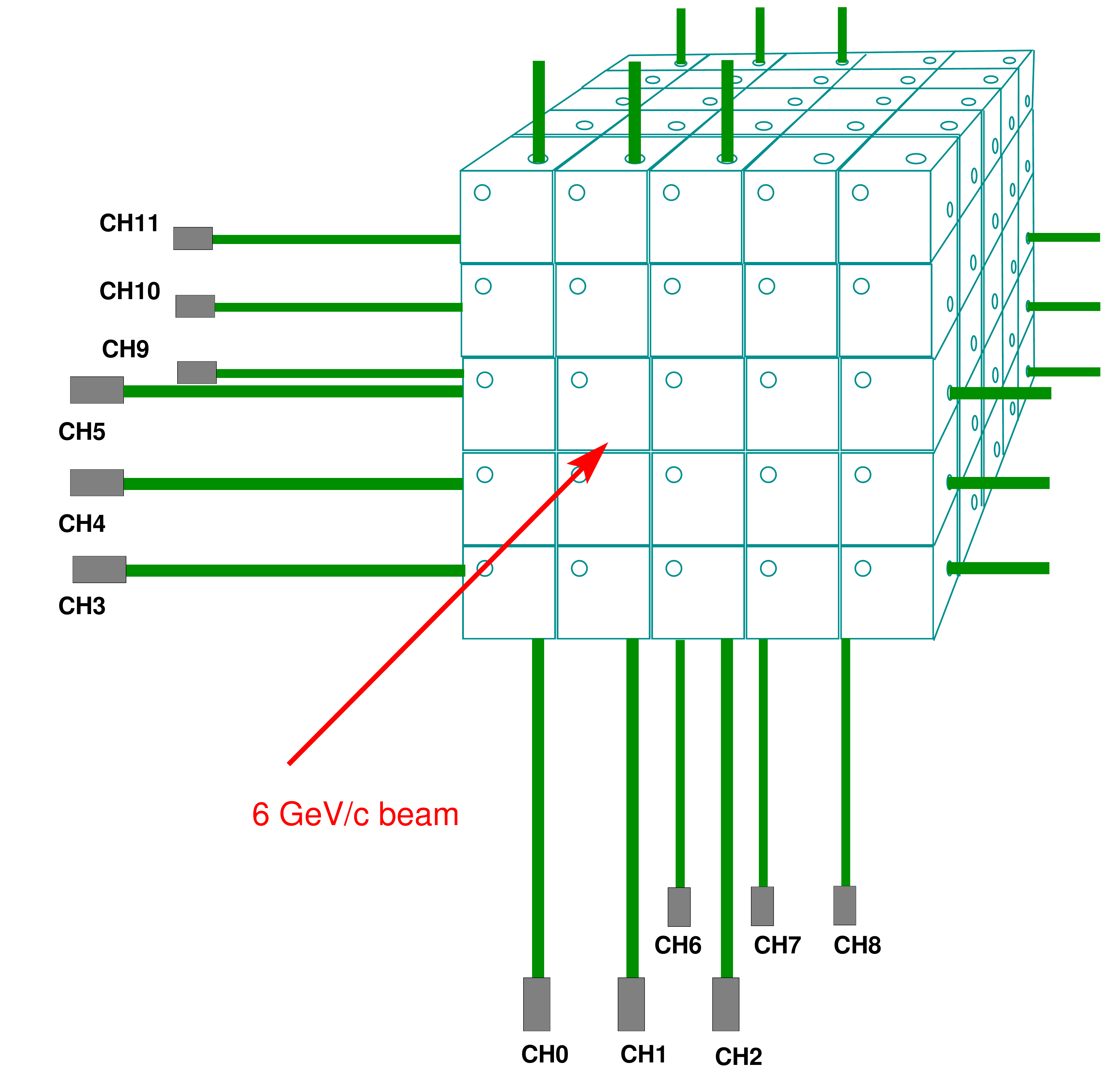}
\caption{ Readout of WLS fibers by the digitizer  and channel labeling. Inactive fibers are not shown.}
\label{fig:Target_cube_readout} 
\end{figure}

Two small  scintillator trigger counters of $3\times 3\times 10$~mm$^3$ size spaced at distance of 26~cm were installed before and after the prototype. Thus we were able to select minimum ionizing particles (MIPs) from the beam with the  position accuracy of about 3~mm. 
One of the trigger counters sent a signal to start the digitizer, signals from another one were measured by the digitizer and used in off-line analysis. Also an anti-coincidence (AC) scintillator counter with $10\times 10$~cm$^2$ area and a 9~mm aperture for the beam entrance was mounted in front of the prototype to remove accidentals. Trigger and AC counters were read out by the same MPPCs  12571-025C. All MPPCs were selected to have close values of the bias voltage, so we were able to fix it to 67.5~V recommended by Hamamatsu in the specification. 
In total, the digitizer reads out 12 WLS fibers, as shown in Fig.~\ref{fig:Target_cube_readout}, a small trigger counter and two channels from the AC counter.  The layout of readout fibers allow us to measure the parameters of 9 cubes in the front layer and 9 cubes in the back layer of the prototype. All other fibers were in place but idle for analysis.

The test beam was held at T10  area of the CERN Proto-Synchrotron (PS) in October 2018.  The line transported 6~GeV/c positive  particles  of mixed composition (mainly positrons and protons) with a momentum resolution of $\sim$0.5\%. A trigger rate of around 100~Hz has been set by closing the beam collimators in order to maximize the fraction of single-hit events.


\subsubsection{Light yield}
\begin{figure}[htbp]
\centering
\includegraphics[width=0.4\textwidth]{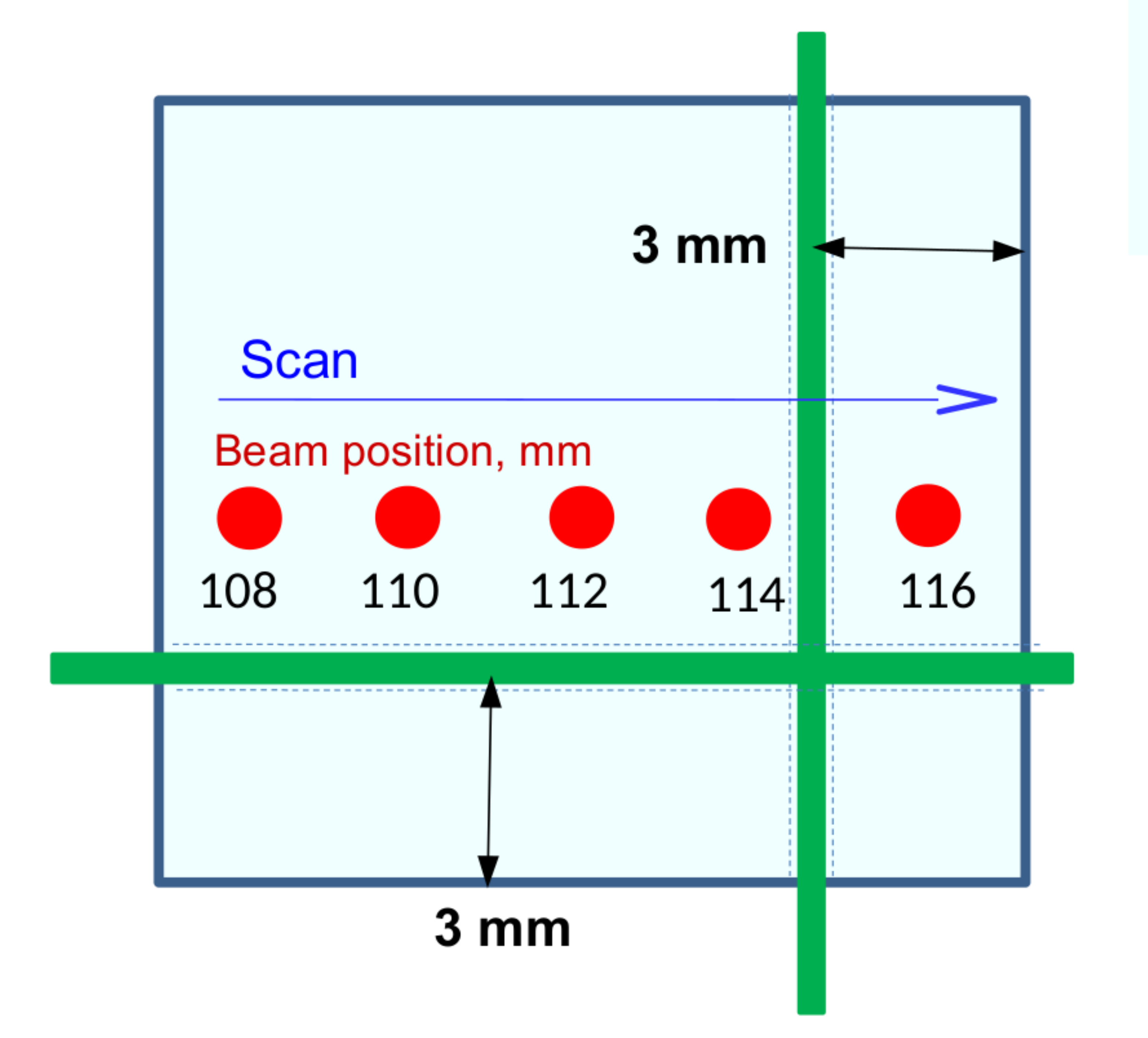}
\caption{Beam scan across a single cube. Fiber positions (in green) are shown relative to the beam hit points (in red).}
\label{fig:Target_cube_scancube} 
\end{figure}


\begin{figure}[htbp]
\centering
\includegraphics[width=0.6\textwidth]{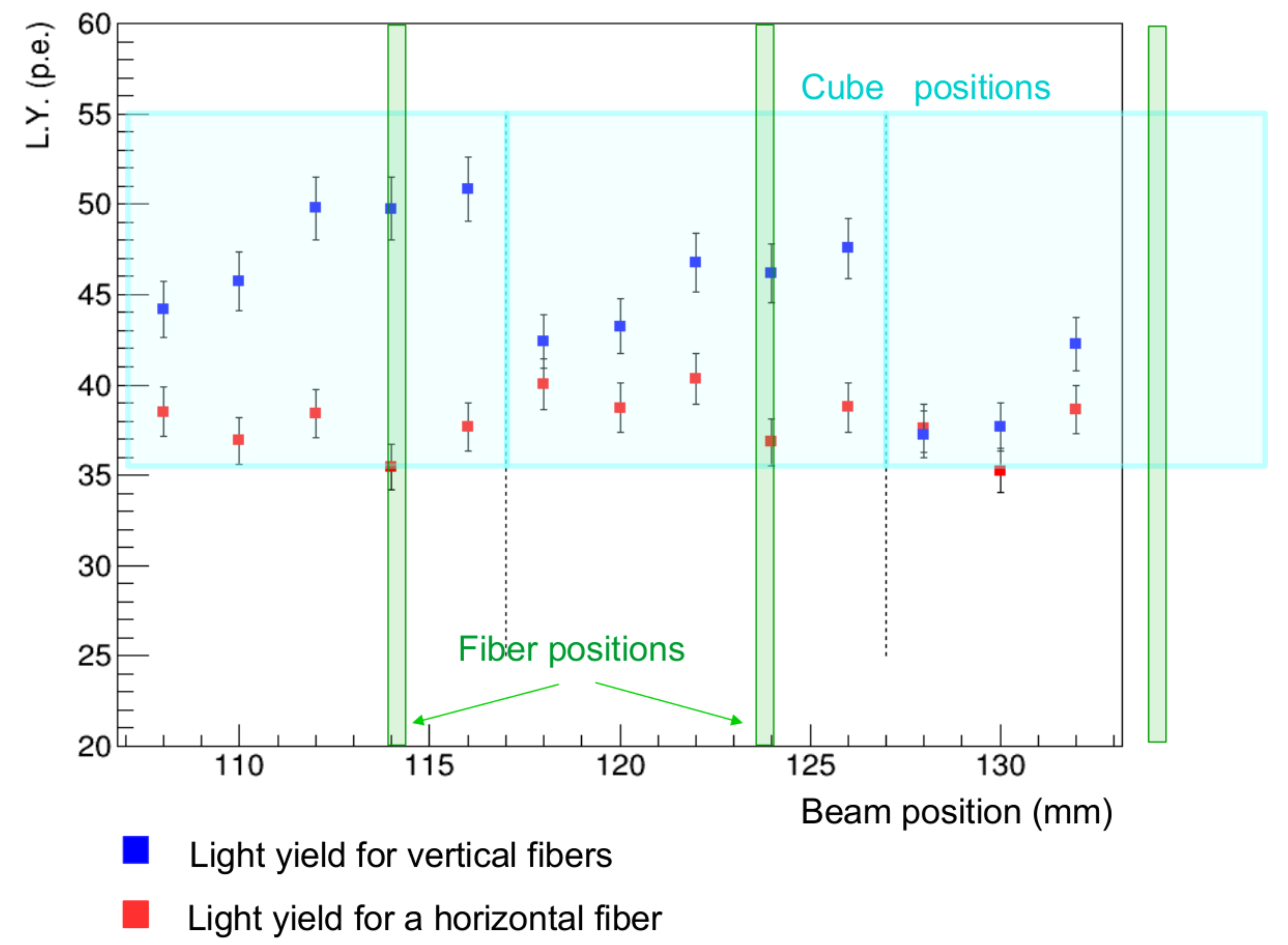}
\caption{Light yields for  horizontal and vertical fibers vs beam position. Also vertical fibers  and cube coordinates  are shown at the horizontal scale. }
\label{fig:Target_cube_scan} 
\end{figure}
A beam scan with a step of 2~mm was done across 3 cubes in the horizontal direction.  Fig.~\ref{fig:Target_cube_scancube} shows the position of the beam center for the scan points, with respect to the position of vertical and horizontal fibers in a cube. Beam particles were localized by the trigger counters within the spot of about $3\times 3$~mm$^2$. The events were selected if the light yield in a small trigger counter was larger than 50~p.e.\ and the time difference between both trigger counters did not exceed 1~ns.

We have measured 13 scan points over a span of 25~mm, 3 cubes were scanned in the front layer along with 3 cubes in the back layer.  Each cube was read out by two fibers to measure the response as a function of the beam position. The signal charge was calculated as the area of signal waveform normalized to the signal obtained for a single photoelectron (p.e.). Calibration coefficients were calculated for each run thanks to the excellent single photoelectron response of MPPCs. The result of the scan for the front layer is presented in Fig.~\ref{fig:Target_cube_scan}. 

Edge effects at the cube boundaries were minimized by selecting events with  a light output exceeding the average crosstalk in both vertical and horizontal fibers. Although the beam spot is comparable to the cube size, we can observe a systematic increase in light yield when the beam point gets closer to the vertical fiber.  The horizontal fiber demonstrates fluctuations of the light signal within measurement accuracy.

The light yield for different channels varies from 36 to 50~p.e./MIP for a single fiber.  The typical light yield was close to 40~p.e./MIP/fiber, and the total  light yield from two fibers in the same cube was measured on an event-by-event basis to be about 80~p.e., as expected.

\subsubsection{Optical crosstalk}

Since  the white chemical reflector, like any reflector of the diffuse type, does not fully isolate the scintillation light, the leakage of light from a fired cube to the neighboring ones was investigated. Crosstalk was measured on an event-to-event basis as the ratio of signals in adjacent cubes to the signal in the fired cube. The average of the distribution of these ratios was defined as the average crosstalk. 
Accidentals and induced electronic noise increase the pedestal fluctuations and create a false crosstalk. To suppress this contribution we considered the signal less than 0.5 p.e.\ as a zero value (pedestal).
The dark noise of the MPPCs generates accidental single p.e. signals.  We have measured that the dark noise adds  less than ~0.2\% to the total value of crosstalk, thanks to the low level of dark rate of MPPCs S12571-025 ($\sim$100~kHz typical value).

\begin{figure}[htbp]
\centering
\includegraphics[width=0.5\textwidth]{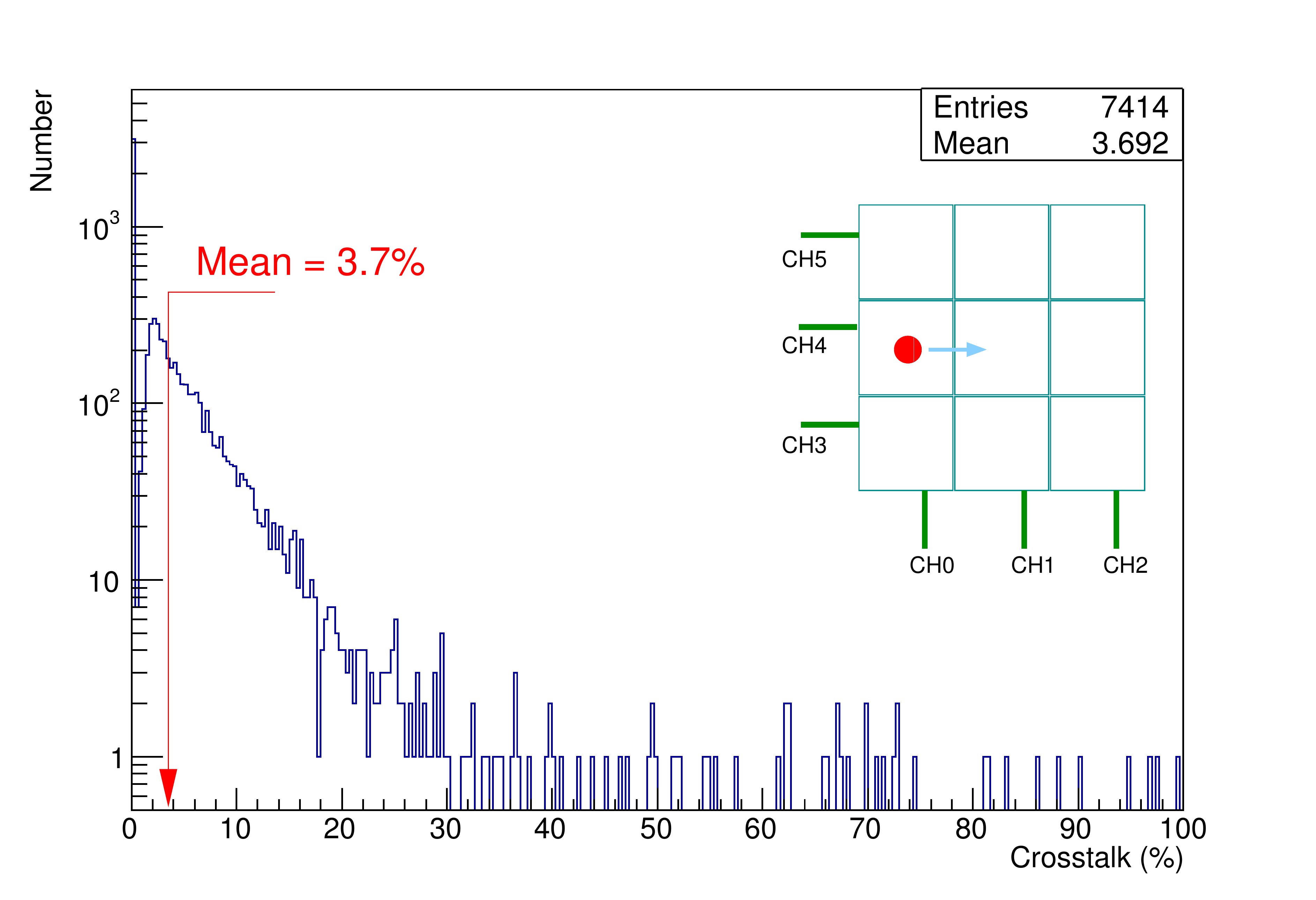}
\caption{ Crosstalk from the cube ch0/ch4  into the cube ch1/ch4. Beam hits the central part of the  ch0/ch4 cube.}
\label{fig:Target_cube_cross_spec} 
\end{figure}

\begin{figure}[htbp]
\centering
\includegraphics[width=0.35\textwidth]{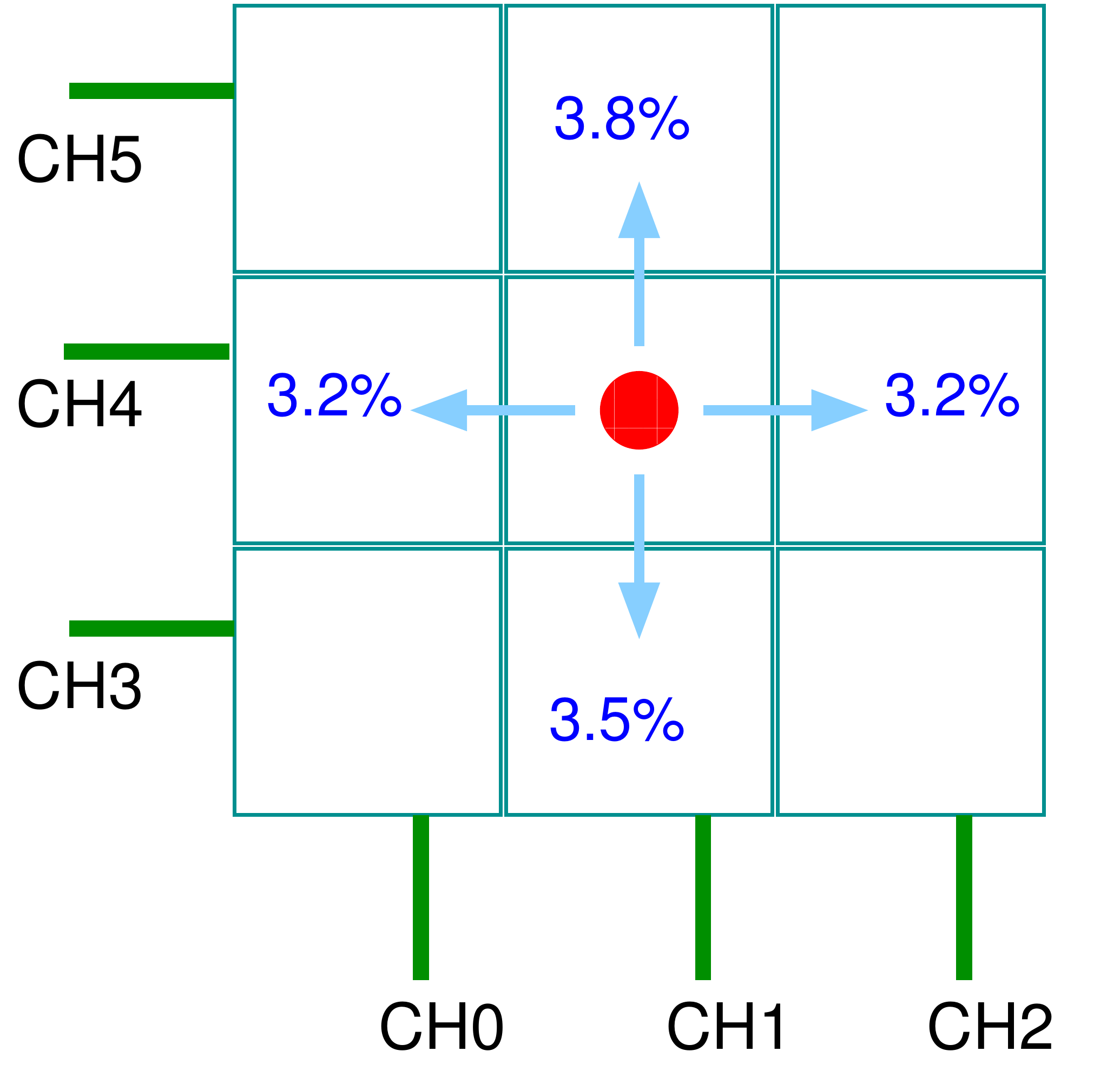}
\caption{ Crosstalk from the central cube in four directions.}
\label{fig:Target_cube_cross4} 
\end{figure}

Fig.~\ref{fig:Target_cube_cross_spec} shows the crosstalk distribution when the light from the cube CH0/CH4 leaks into the cube CH1/CH4 (see Fig.~\ref{fig:Target_cube_readout} for the channel labelling). The crosstalk was calculated as the ratio $L.Y._{CH1}/L.Y._{CH0}$. The crosstalk average value is 3.7\%, while the average of $L.Y._{CH0}$ is 41~p.e.  

The crosstalk with values higher than 30\% can be explained by shower events.
The crosstalk in four directions is shown in Fig.~\ref{fig:Target_cube_cross4}, when the beam hits the central cube CH1/CH4 in a $3\times 3$ array.  The average crosstalk is 3.4\% per side. The average of the total crosstalk into all 4 sides on an event-to-event basis was measured to be 13.7\%. From this we can conclude that ~20\% of the detected scintillation light escapes the fired cube into adjacent cubes through the cube reflective walls.

\subsubsection{Time resolution}

We have applied the constant fraction method to obtain the timing mark of the signal waveform. The preamplifiers extend the signal front to 7~ns  (measured between 0.1--0.9 fractions of the amplitude), so that we have up to 40 digitizer sample points spaced at 200~ps at the front. The following procedure was used for each waveform to obtain the timing.  First, the baseline was determined by fitting the first sample points before a signal with a horizontal line. Then a maximum amplitude of the waveform was measured.  We have found that a fraction of 10\% of the maximum amplitude at the signal front provides the best timing. 
\begin{figure}[htbp]
\begin{centering}
\includegraphics[width=0.95\textwidth]{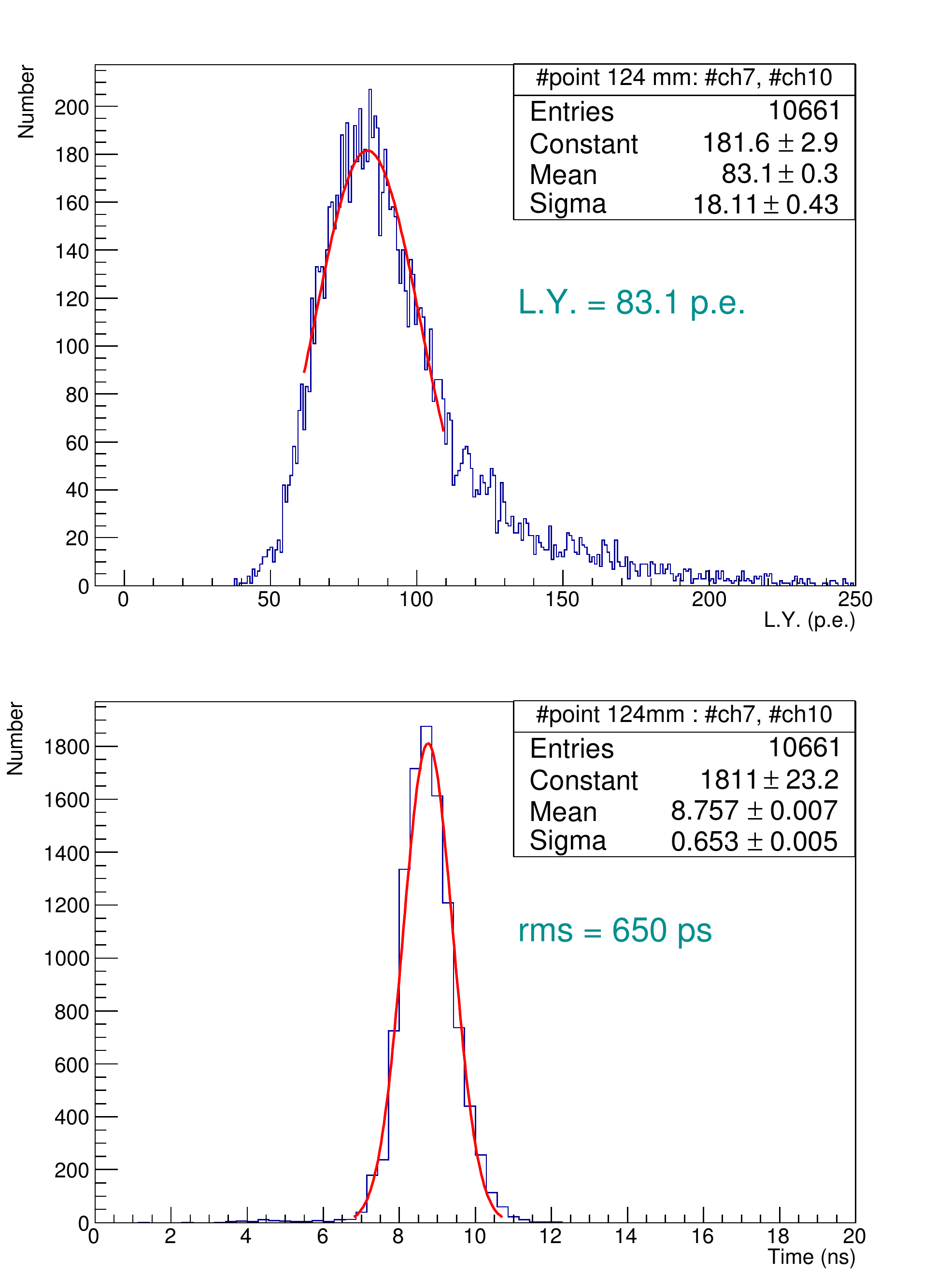}
\caption{ Charge and time spectra for a single cube. Charge signal is a sum from two fibers, the time is an average time between two fibers. }
\label{fig:Target_cube_spectra} 
\end{centering}
\end{figure}

Fig.~\ref{fig:Target_cube_spectra} shows the time and charge distributions for one of the cubes. 

Typical time resolution ($\sigma_t$) for a single fiber was around 0.95~ns.  A cube with two readout fibers gives $\sigma_t$=0.65--0.71~ns. Two cubes combined produced typical $\sigma_t$=0.52~ns for the first method of the time mark calculation, and  $\sigma_t$=0.48~ns for the second method.

\subsection{Prototype with 9216 (8$\times$24$\times$48) cubes}

\label{sec:superfgd-8x24x48-prototype}

A larger prototype, 24 cubes wide by 8 cubes high by 48 cubes in length, was produced at INR and shipped to CERN in May 2018 where it was equipped with photosensors and electronics.
The physical dimensions were imposed by the requirement to fit within the MNP17 magnet, a general purpose dipole magnet made available to  users at CERN.

For this prototype, the electronics developed for the Baby-MIND detector was used.
It is based on CITIROC frontend ASIC and is chosen for the basis of the baseline electronics design for SuperFGD.
The same type of MPPC to be used for SuperFGD, in a different packaging due to mechanical constraints but the same optical/electrical performance, was also used.
Hence, it will provide a good information to assess the performance of the final detector and feedback to the optimization of the design.

Several studies are possible with this prototype, in preparation for the full SuperFGD.
Basic properties for general detector performance optimization can be checked, such as channel uniformity, energy/timing resolution per hit, cross-talk and afterpulsing, and saturation.
Also, more information can be extracted towards physics studies, such as
hit clustering and track reconstruction, response to stopping protons, Michel electrons, and photon conversion.

Because the data analysis has just started, results shown in the following are still very preliminary.
More results are expected soon.

%

\subsubsection{Prototype assembly}
\begin{figure}[tbp]
  \begin{center}
    \includegraphics*[width=0.6\textwidth]{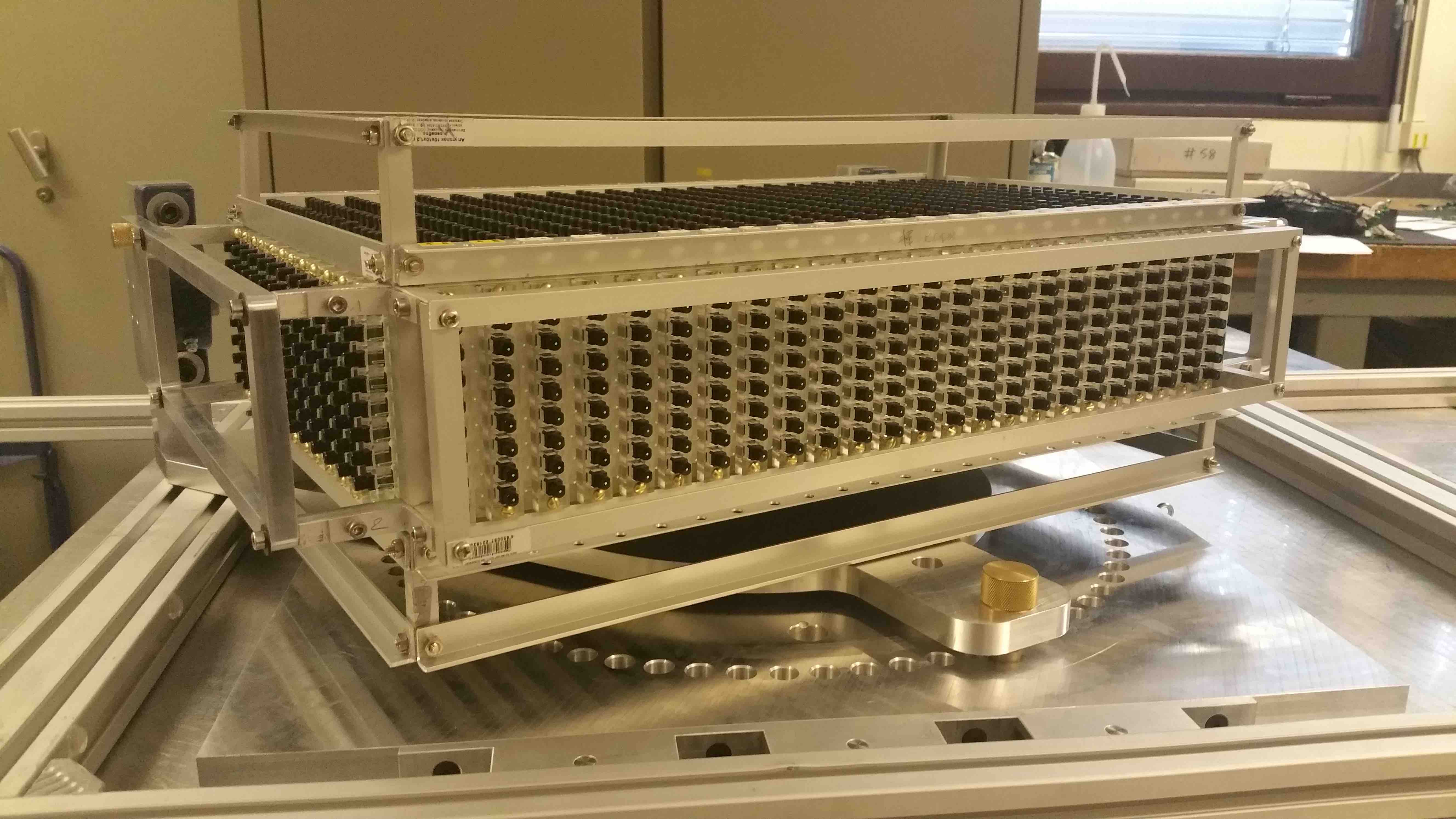}
  \end{center}
\caption{SuperFGD prototype assembled with WLS fibers and optical connectors.}
\label{fig:Target-proto_connectors}
\end{figure}

The prototype was assembled at INR 
with the fishing line method as described in section \ref{sec:superfgd-assembly}. 
When planes of $24\times48$ cubes are stacked, they are
separated by a layer of Tyvek paper reflector. 
This separation by Tyvek sheet is only for an R\&D purpose and not envisioned in the final detector.
The assembly is surrounded on all sides by plastic support plates. The fishing lines are then removed one by one and replaced by WLS fibers (Kuraray Y11) with a custom optical connector on one end of each fiber. Fig. \ref{fig:Target-proto_connectors} shows the prototype assembled with WLS fibers and connectors.

\begin{table}[htbp]

\centering

\begin{tabular}{llllcccc}
\toprule
\textbf{Hamamatsu ref.}  	&	\textbf{S13360-1325CS} & 	\textbf{S13081-050CS} & \textbf{S12571-025C}	\\
\hline
Usage			&			SuperFGD			&	WAGASCI SMRD & Baby MIND 			 \\
Prototype ref.			&			Type I			&	Type II & Type III 			 \\
Numbers in proto.			&			1152			&	384 & 192 			 \\
Package			&			Ceramic			&	Ceramic & Ceramic 			 \\
Pixel pitch	[um]		&			25			&	50 & 25			 \\
Number of pixels			&			2668			&	667 & 1600 			 \\
Active area	[mm$^2$]		&			$1.3\times1.3$			&	$1.3\times1.3$ & $1.0\times1.0$			 \\
Operating voltage [V]			&			$56\rightarrow58$			&	$53\rightarrow55$  & $67\rightarrow68$			 \\
PDE	[\%]		&			25			&	35 & 35 			 \\
Dark count rate [kHz]			&			 70			&	90 & 100 			 \\
Gain			&			$7\times10^5$			&	$1.5\times10^6$ & $5.15\times10^5$ 			 \\
Crosstalk probability [\%]			&			 1			&	1 & 10 			 \\
\hline	
\bottomrule
\end{tabular}
\caption{\em Summary of main parameters for the three types of MPPCs installed on the prototype.}
\label{tab:mppcs_on_prototype}
\end{table}

\begin{figure}[tbp]
  \begin{center}
    \includegraphics*[width=0.9\textwidth]{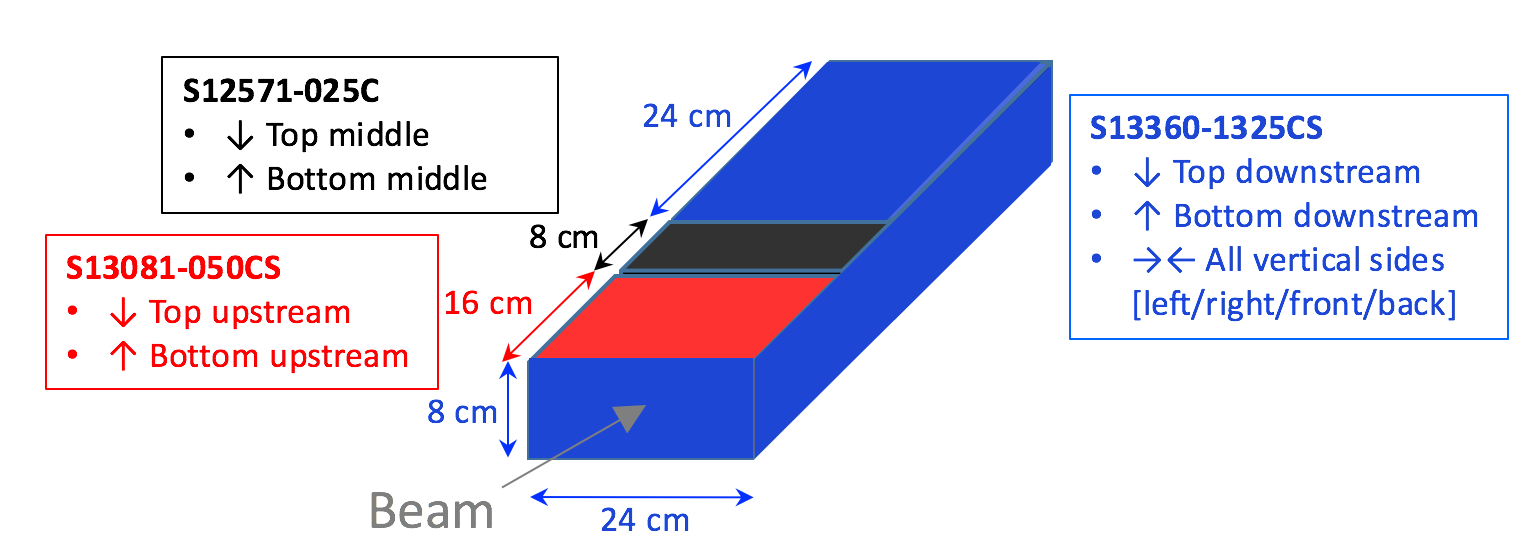}
  \end{center}
\caption{Distribution of the three types of MPPCs around the 6 faces of the SuperFGD prototype, $\times1152$ Type I (S13360-1325CS), $\times384$ Type II (S13081-050CS), $\times192$ Type III (S12571-025C).}
\label{fig:Target-mppc_distribution}
\end{figure}

\begin{figure}[htbp]
  \begin{center}
    \includegraphics*[width=0.6\textwidth]{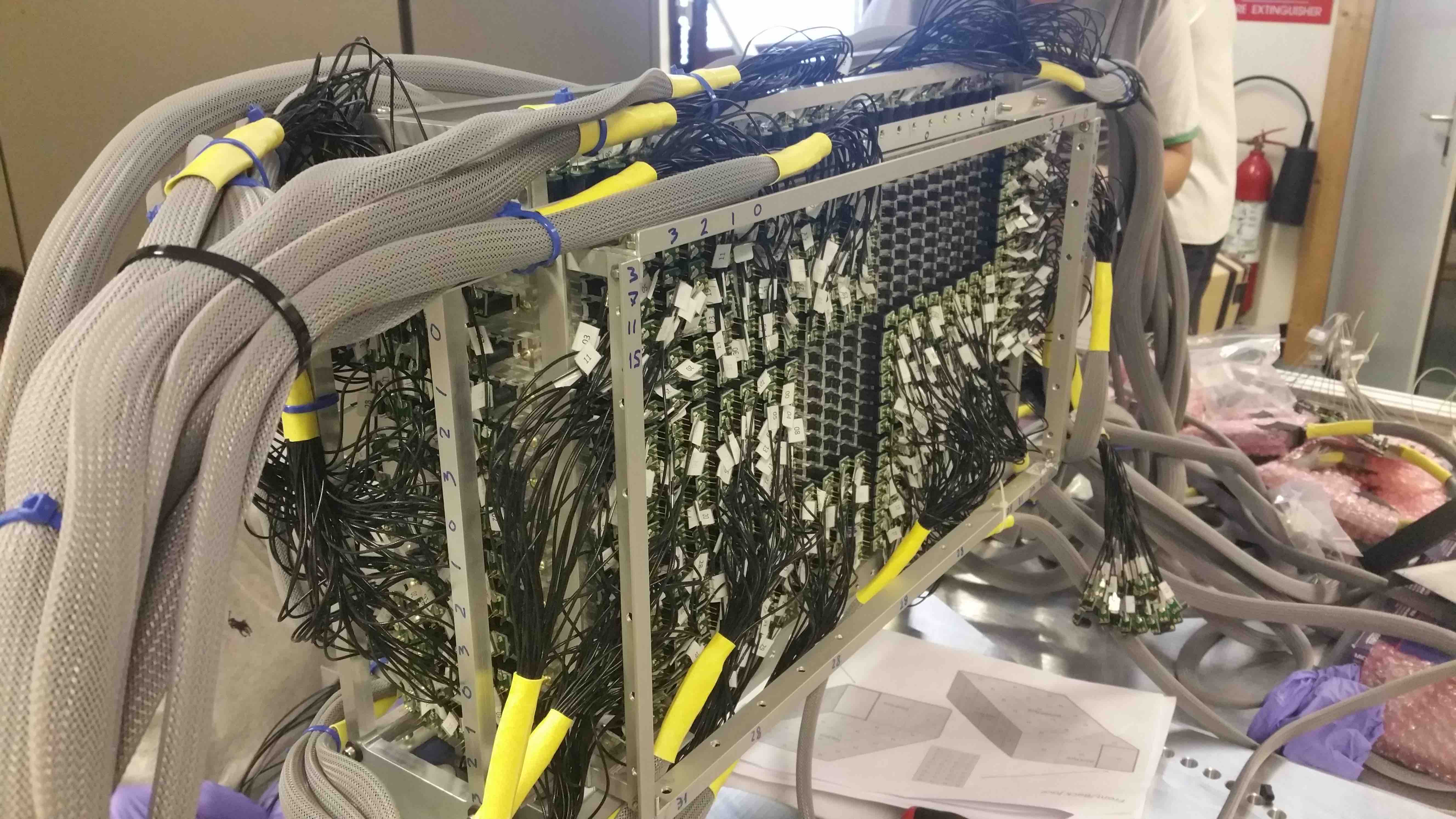}
  \end{center}
\caption{Photo of the SuperFGD prototype showing a partially instrumented bottom face, the mechanics are rotated by 90deg to enable access to the bottom face.}
\label{fig:Target-proto_photo_vert}
\end{figure}

The prototype was shipped to CERN where it was equipped with three types of photosensors, the majority of which were the type that has been chosen for the ND280 upgrade, the S13360-1325CS, though in a different package, ceramic rather than surface mounted.
Other two types of MPPC are also used for comparison.
Table~\ref{tab:mppcs_on_prototype} summarizes the specification and numbers of three types of MPPCs used for the prototype.
The distribution of MPPC types around the 6 faces of the detector is shown in Fig.~\ref{fig:Target-mppc_distribution}. 
MPPCs were pre-selected and sorted in batches of 32 to have an operating voltage spread no greater than $\pm 100$ mV per batch.
A photo taken during the assembly of photosensors on the bottom face of the detector is shown in Fig.~\ref{fig:Target-proto_photo_vert}.

\subsubsection{Layout and installation on the PS T9 beamline}
\begin{figure}[htbp]
  \begin{center}
    \includegraphics*[width=0.8\textwidth]{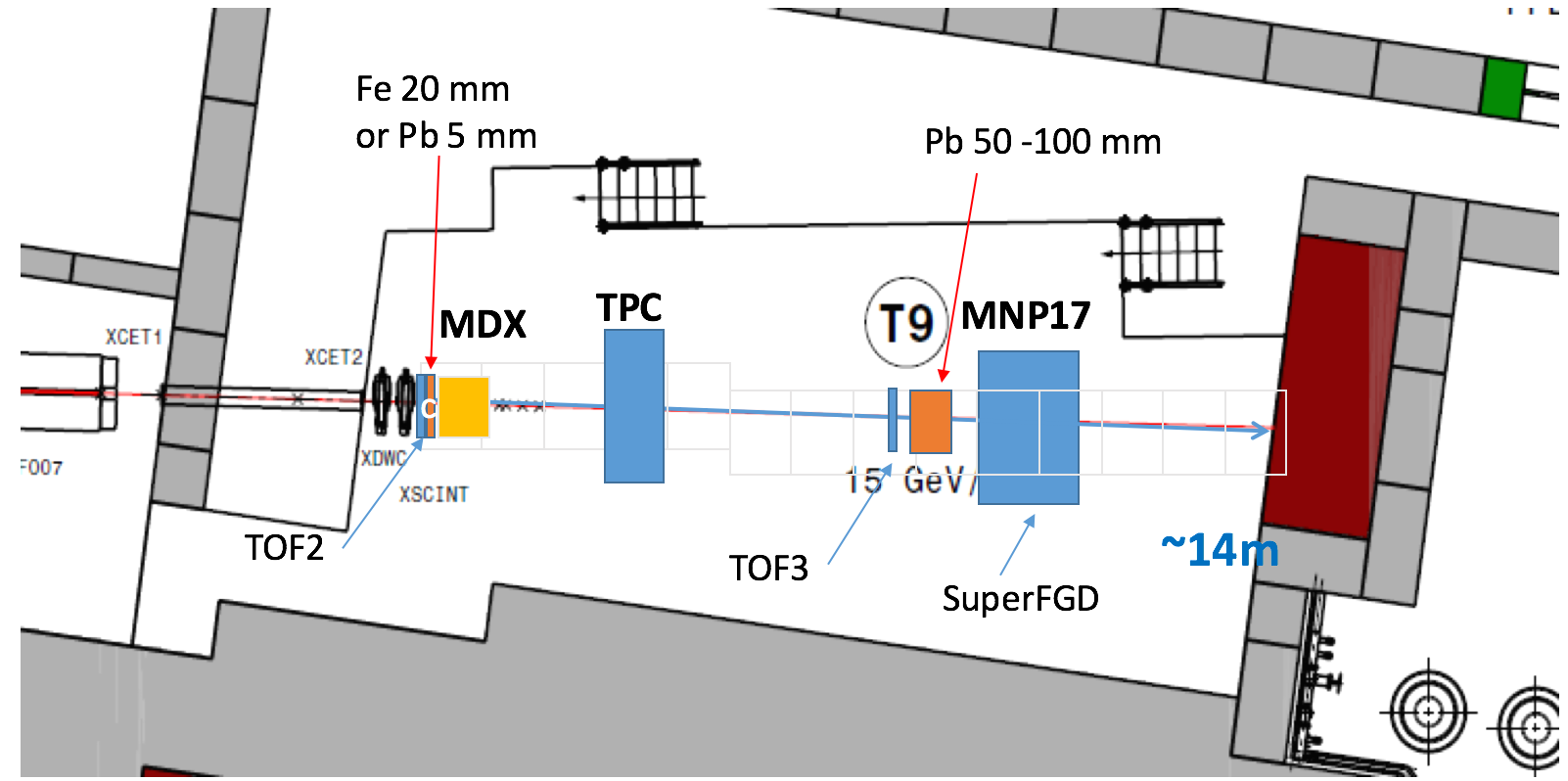}
  \end{center}
\caption{Layout showing the main components on the PS T9 beamline platform.}
\label{fig:Target-t9_layout}
\end{figure}

\begin{figure}[htbp]
  \begin{center}
    \includegraphics*[width=0.8\textwidth]{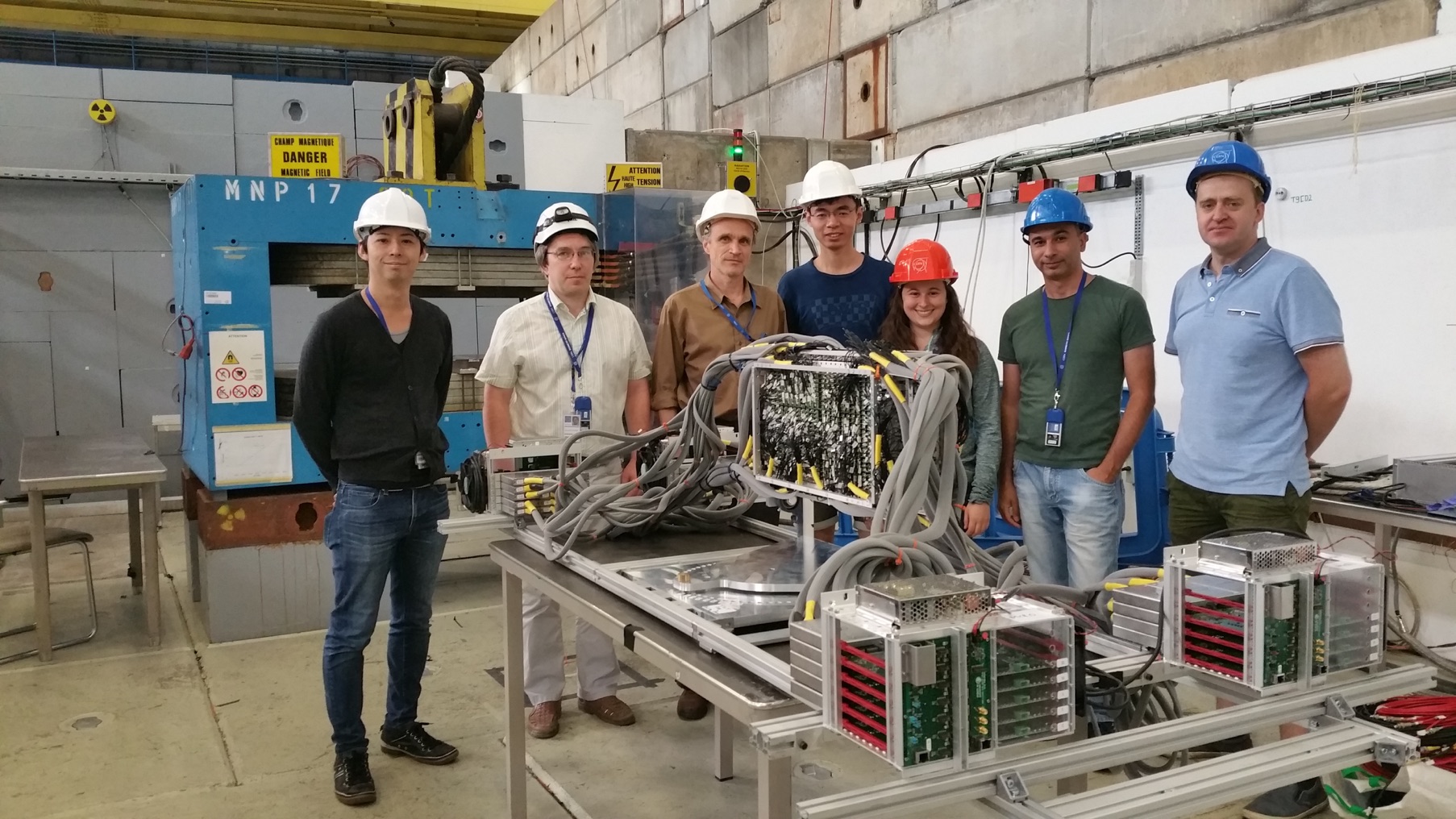}
  \end{center}
\caption{Prototype before insertion into the MNP17 magnet.}
\label{fig:Target-proto_outside_mnp17}
\end{figure}

The layout of the T9 beamline is shown in Fig.~\ref{fig:Target-t9_layout}. During the June-July 2018 test phase dedicated to the SuperFGD prototype, there was no TPC on the T9 beamline platform. Time-of-flight counters were installed along the beamline to provide particle identification. Depending on the beam required, Fe or Pb converters were inserted in the beamline upstream of the prototype. Thicker Pb degraders were also used for a few runs in an attempt to collect a sample of stopped muons in the prototype for Michel electron studies. The TPC was on the beamline during the August-September test phase.
The prototype before 
insertion into the MNP17 magnet is shown in Figs. \ref{fig:Target-proto_outside_mnp17}. 
The MNP17 magnet was operated for the vast majority of the time with a field of 0.2 T, and occasionally up to 0.7 T. The MDX magnet was operated for very short periods at 1 T, during the photon beam runs described a few sections further. 


\subsubsection{Tuning the readout and calibration}



\begin{figure}[htbp]
  \begin{center}
    \includegraphics*[width=0.8\textwidth]{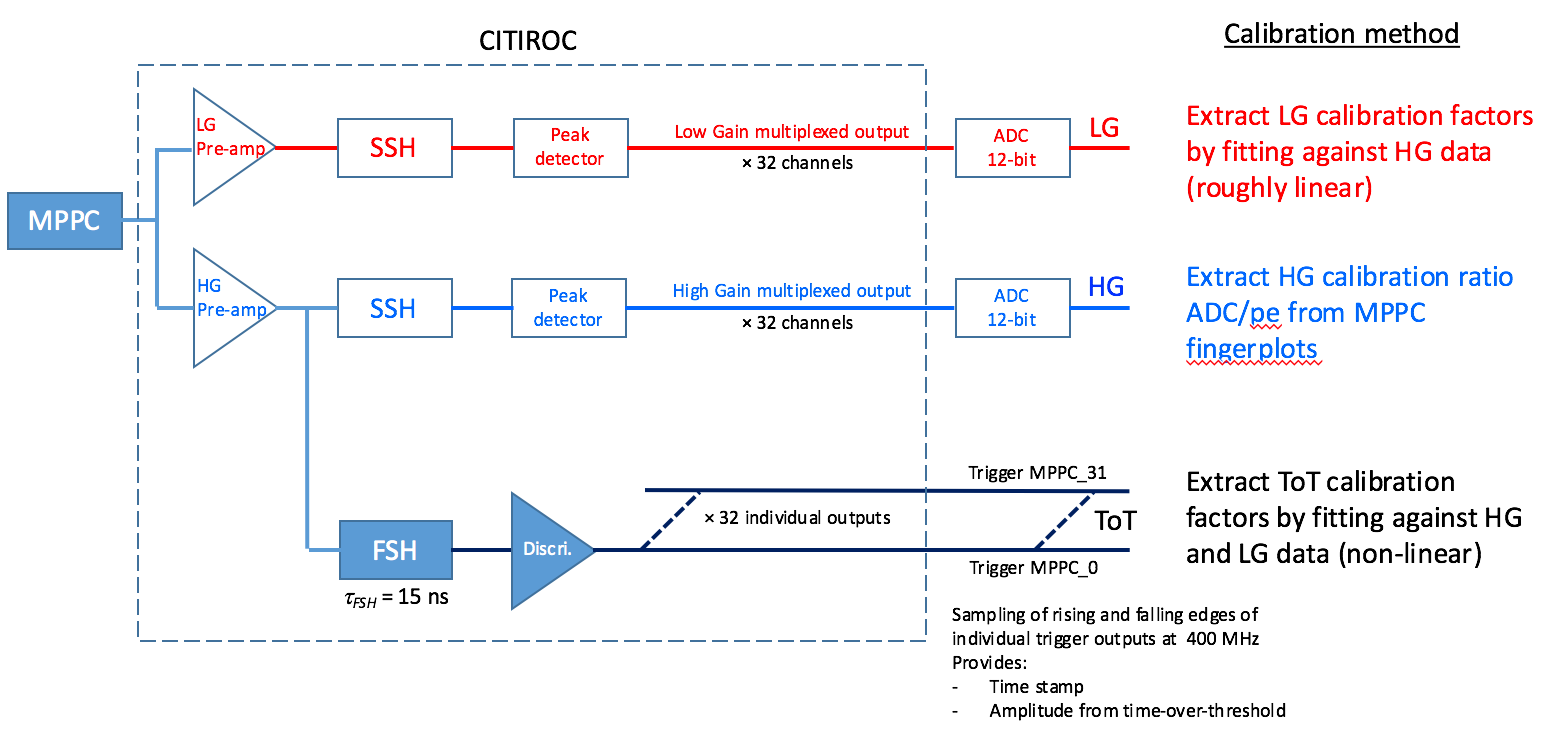}
  \end{center}
\caption{Sketch of the three signal outputs providing amplitude information for an event.}
\label{fig:Target-3signal_outputs}
\end{figure}

There are three different signal readout paths that provide a measurement of amplitude, the HG and LG signal paths, whose output from the CITIROC is digitised by a 12-bit ADC, and the Time over Threshold (ToT) obtained by sampling the rising and falling edges of the CITIROC trigger lines at 400 MHz by the FPGA on the FEB (Fig \ref{fig:Target-3signal_outputs}). All three have been calibrated:
\begin{itemize}
    \item HG calibration: done by obtaining the ADC/p.e. gain ratio from either dark counts or LED signals, for each MPPC/channel.
    \item LG calibration: done by comparing LG data against HG data for the same channel, same events.
    \item ToT calibration: done by comparing ToT data against HG data for signals up to 100 p.e. and against LG data for signals above 100 p.e.
\end{itemize}

\begin{figure}[htbp]
  \begin{center}
    \includegraphics*[width=1.0\textwidth]{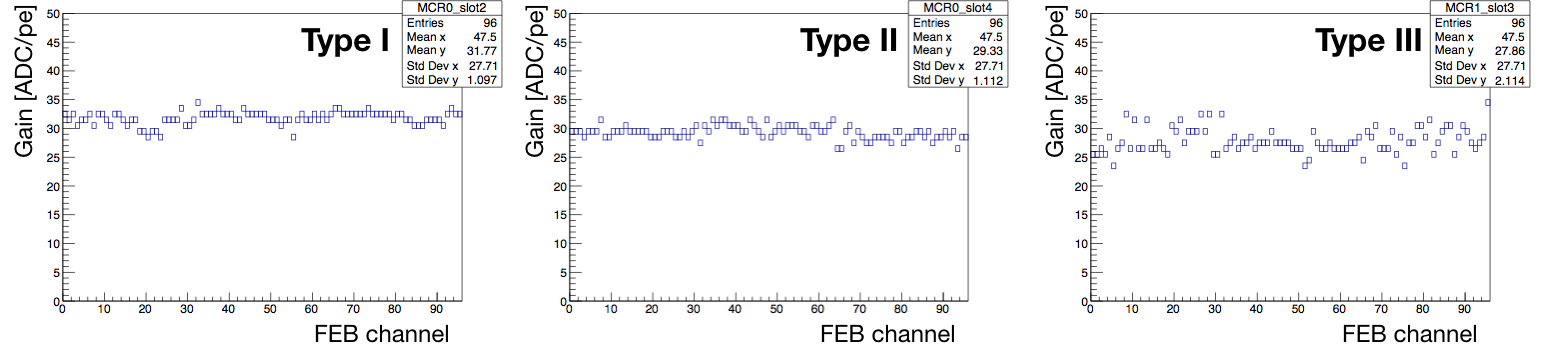}
  \end{center}
\caption{Calibration measurements: gain distributions in units of ADC/p.e. for $\times1728$ channels separated by MPPC type.}
\label{fig:Target-calib_gain_spread}
\end{figure}

\begin{figure}[htbp]
  \begin{center}
    \includegraphics*[width=1.0\textwidth]{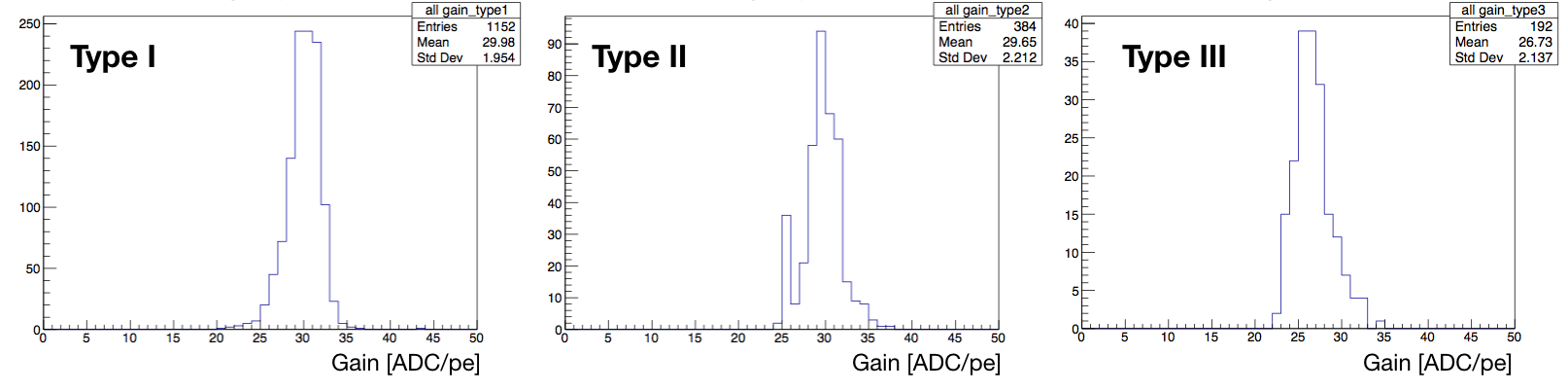}
  \end{center}
\caption{Calibration measurements: gain in units of ADC/p.e. for $\times96$ channels of one FEB each for Type I, II and III MPPCs.}
\label{fig:Target-calib_gain_distri_1728ch}
\end{figure}

The gain spreads across 96 channels for representative FEBs for each of the three types of MPPCs used in the prototype are shown in Fig. \ref{fig:Target-calib_gain_spread}. Gain distributions for all FEBs are shown in Fig. \ref{fig:Target-calib_gain_distri_1728ch}, separated by MPPC type. These distributions have relatively large spreads but can be further tuned on an individual channel basis by adjusting the CITIROC 10-bit DAC that trims the MPPC operating voltage. 

\subsubsection{Response to minimum ionising particles}

A beam of 2 GeV/c muons was used to study the response of the prototype to MIPs. There are 3 different fiber lengths, 8, 24, and 48 cm fibers. 
Only results for 
24 cm fibers are presented here. 

\begin{figure}[htbp]
  \begin{center}
    \includegraphics*[width=0.8\textwidth]{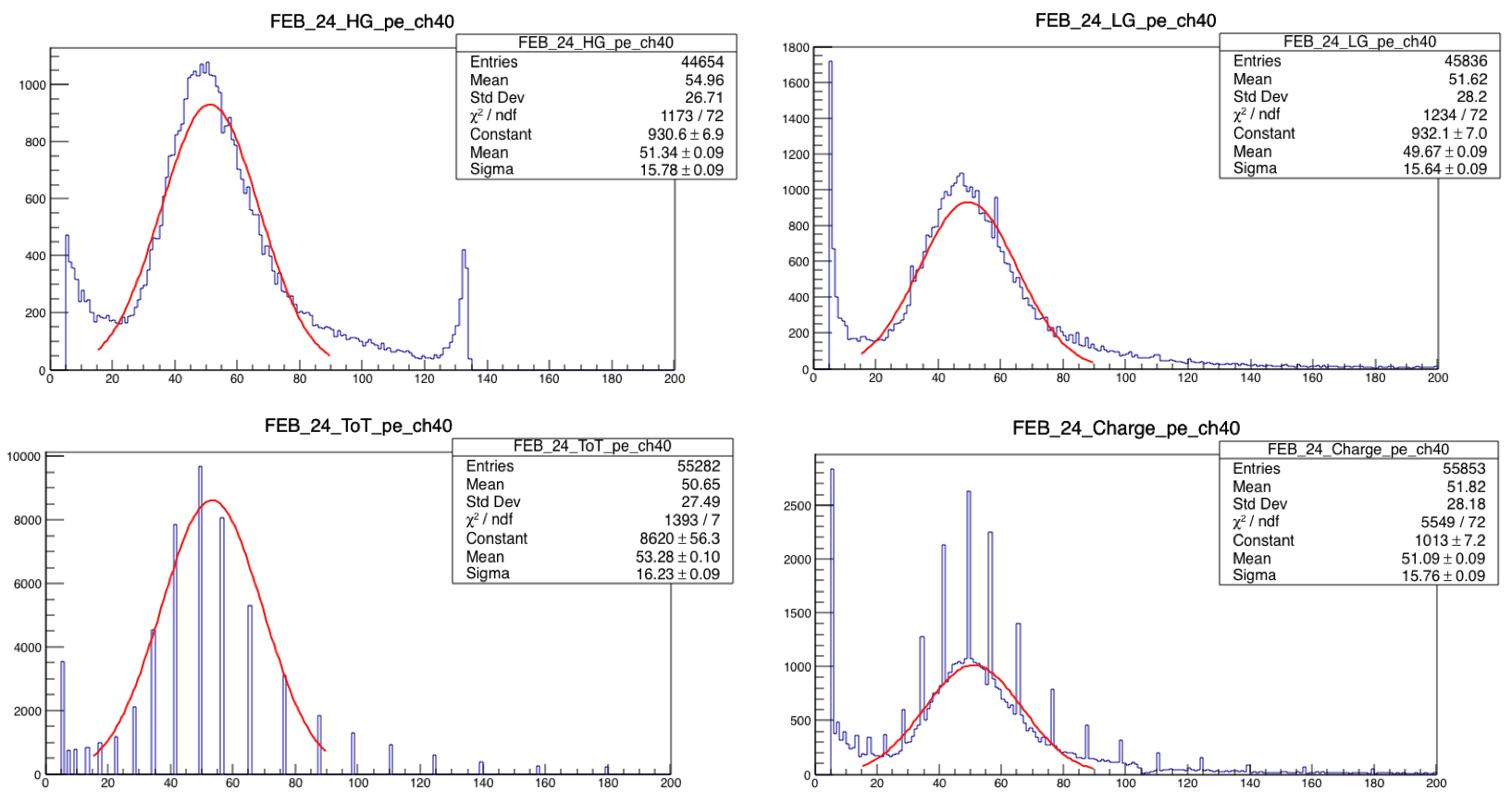}
  \end{center}
\caption{Response to 2 GeV/C muons for one MPPC reading out the light output from one 24 cm WLS fiber oriented vertically, perpendicular to the beam axis. The plots show the light yield recorded by the HG (top left), LG (top right) and ToT (bottom left) signal paths, as well as a combination of all three (bottom right). The horizontal axes are all p.e.}
\label{fig:Target-mip_24cm_1ch}
\end{figure}

The light yield for one channel is shown in Fig. \ref{fig:Target-mip_24cm_1ch}. 
The 3 different signal paths generate a value for light yield after calibration. Whenever there is a HG signal, there is a LG signal. The HG signal path records approximately 50 \% of all signals, whereas the ToT records close to 100\% of all signals. The ToT signal is discretised due to its origin as a calculated difference between rising and falling edges that are sampled at 400 MHz as can be clearly seen. The value that is retained for event displays is a combination of all three signals: whenever an event is below roughly 100 p.e. and a HG signal exist, it is retained. If a HG signal exists but it is above 100 p.e., the LG signal is used. If neither HG nor LG signals are available, then the ToT signal is used.

\begin{figure}[htbp]
  \begin{center}
    \includegraphics*[width=0.8\textwidth]{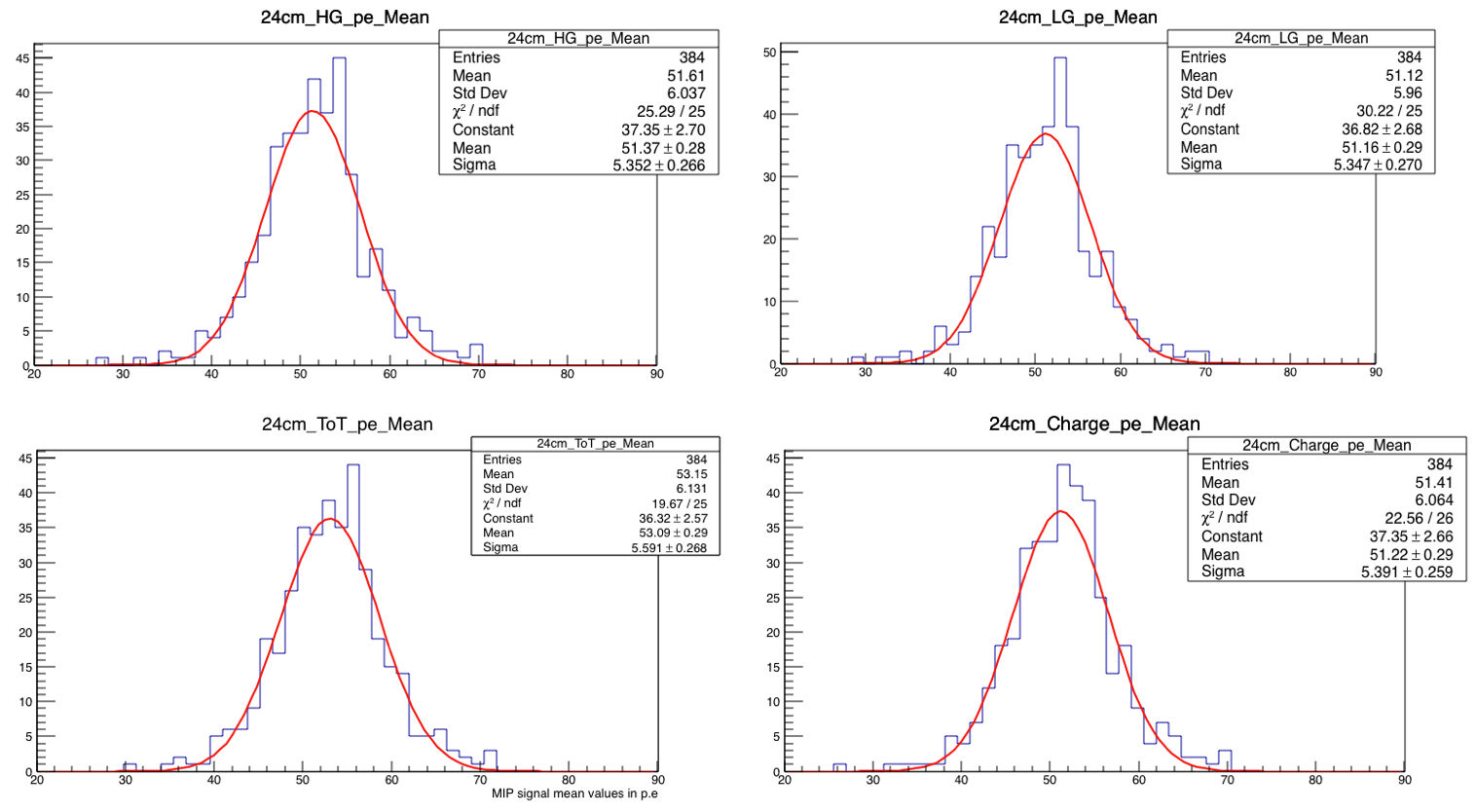}
  \end{center}
\caption{Response to 2 GeV/c muons for 384 channels, each with a 24 cm WLS fiber connected to one MPPC Type I. The horizontal axes are all p.e.}
\label{fig:Target-mip_24cm_384ch}
\end{figure}

\begin{figure}[htbp]
  \begin{center}
    \includegraphics*[width=0.8\textwidth]{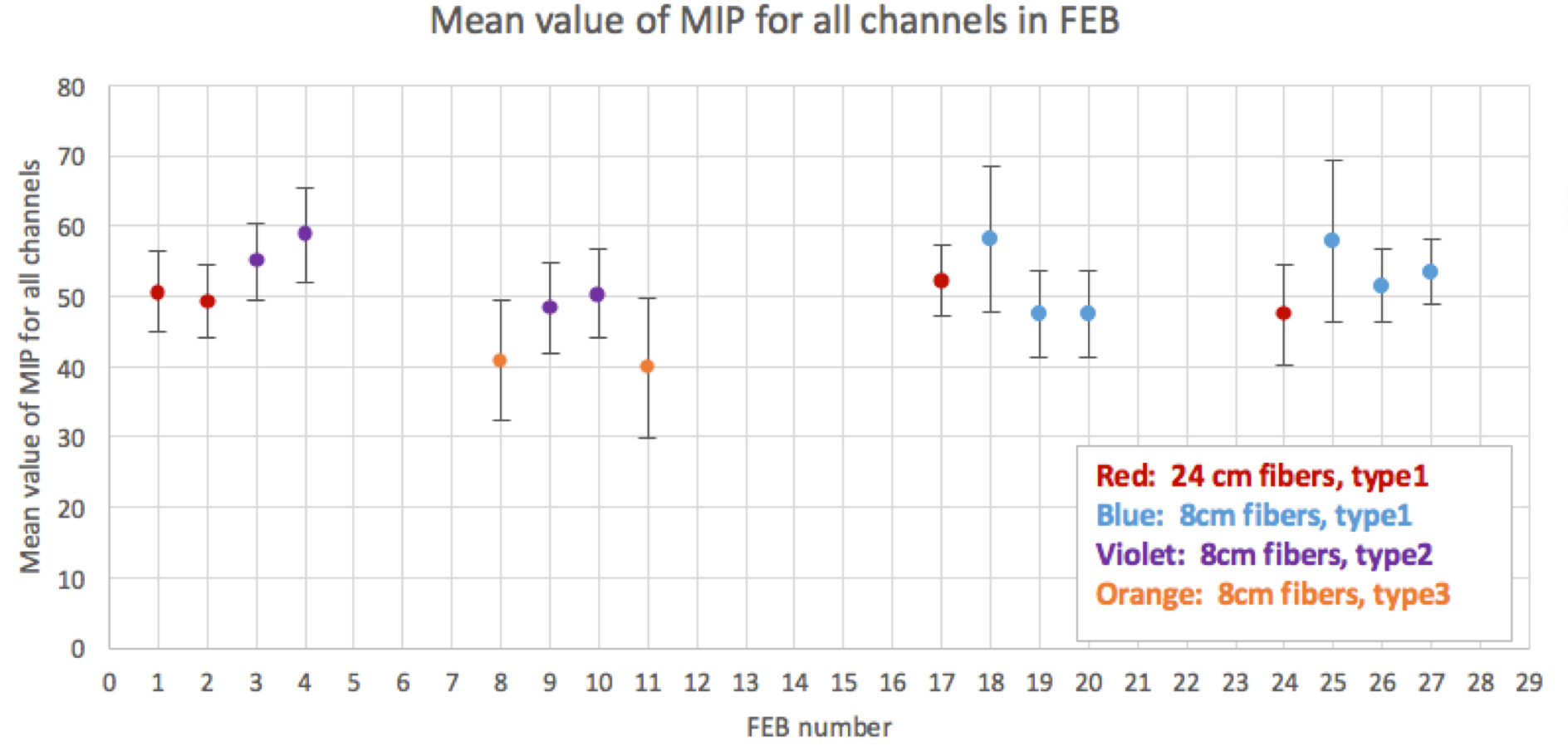}
  \end{center}
\caption{Mean value of light yield for MIPs per FEB.}
\label{fig:Target-mip_all_febs}
\end{figure}

MIP data obtained for 384 channels on 24 cm fibers with type I MPPCs is shown in Fig.~\ref{fig:Target-mip_24cm_384ch}.
Similar plots were made for other configurations of fiber length and MPPC type.
A summary of light yield per FEB is shown in Fig. \ref{fig:Target-mip_all_febs}. It should be noted that these are preliminary results, the calibration procedure in particular could be further refined, with the adoption of more precise pedestal values which may affect results by one or two p.e. 

Based on the current analysis, the following observations can be made:
\begin{itemize}
    \item Comparing 8 and 24 cm fibers: using HG data, the 8 cm fibers have a light yield of 54.73 p.e. compared to 51.37 p.e. for the 24 cm fibers.
    \item Comparing type I and II MPPCs: The Hamamatsu datasheet would suggest a much higher light yield for type II MPPCs given their higher PDE of 35\% compared with 25\% for type I MPPCs. Beam test results show very little difference: a light yield of 54.73 p.e for type I vs 54.77 p.e. for type II. This similarity in light yield between type I and II was confirmed through controlled lab tests at INR.
    \item Type III MPPCs: these show a significantly lower light yield of 43.06 p.e. compared to types I and II.
\end{itemize}
These results demonstrate the validity of the choice of Type I MPPC for SuperFGD.

\subsubsection{Stopping protons}
By selecting a beam momentum at or below 0.8 GeV/c, it was possible to obtain a small sample of protons that stopped within the SuperFGD prototype.

\begin{figure}[htbp]
  \begin{center}
    \includegraphics*[width=14 cm]{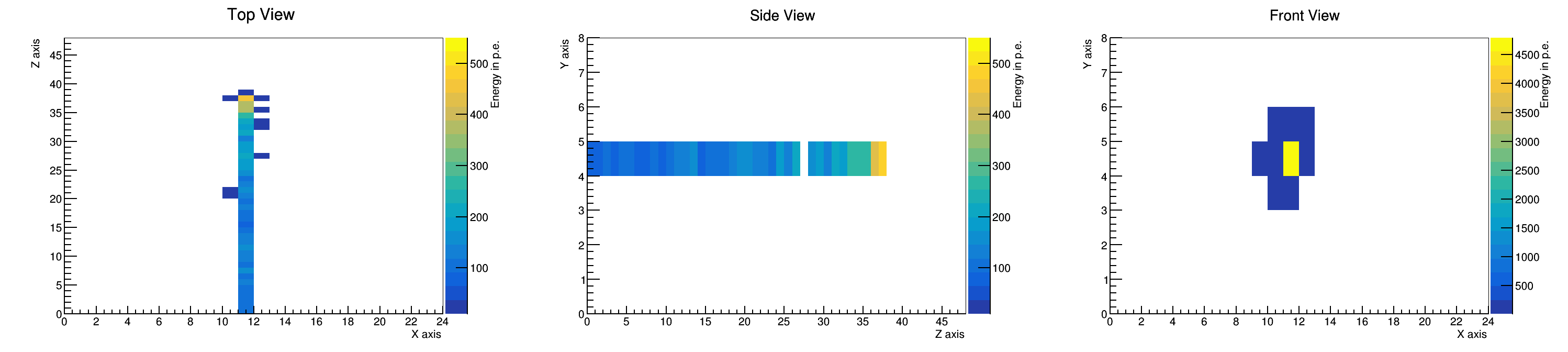}
  \end{center}
\caption{Response to a 0.8 GeV/c proton that stops in the SuperFGD prototype.}
\label{fig:Target-stopped_proton}
\end{figure}

An event display for such a stopped proton is shown in Fig. \ref{fig:Target-stopped_proton}. The top and side views show that the energy deposited in the last cube is a factor $\times10$ higher than that deposited by a minimum ionising particle. The colour scale on the front view illustrates one issue with calibration for very large energy deposition: calibration beyond 1000 p.e. is challenging due to the non-linear behaviour of the ToT in this signal region.

\subsubsection{Photon conversion}
\begin{figure}[htbp]
  \begin{center}
    \includegraphics*[width=14 cm]{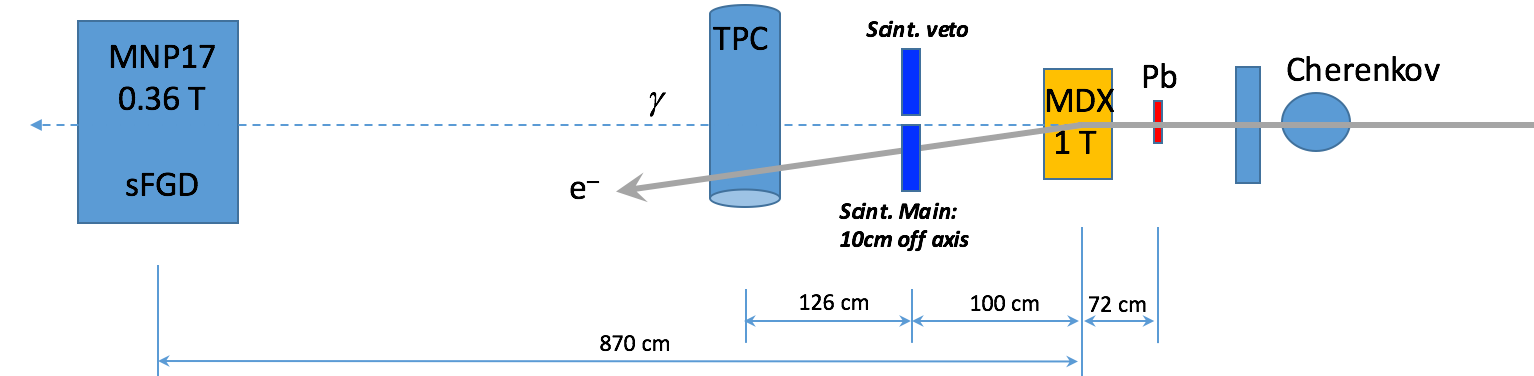}
  \end{center}
\caption{Experimental area setup for the photon conversion runs.}
\label{fig:Target-photon_setup}
\end{figure}

During the beam test period in August and September 2018 where the prototype was running in parasitic mode downstream of the TPC, a short run was dedicated to attempting to observe photon conversions in the SuperFGD. The experimental area setup is shown in Fig. \ref{fig:Target-photon_setup}. A 5 mm Pb converter was placed in front of the MDX dipole magnet operating at 1 T. On an event-by-event basis, the photon generated by interaction of the electron in the Pb target is sufficiently boosted in the forward direction to travel towards the SuperFGD prototype. 
The outgoing electron is diverted away from the initial beam axis by the MDX magnet, and recorded by an off-axis scintillator trigger.

\begin{figure}[htbp]
  \begin{center}
    \includegraphics*[width=14 cm]{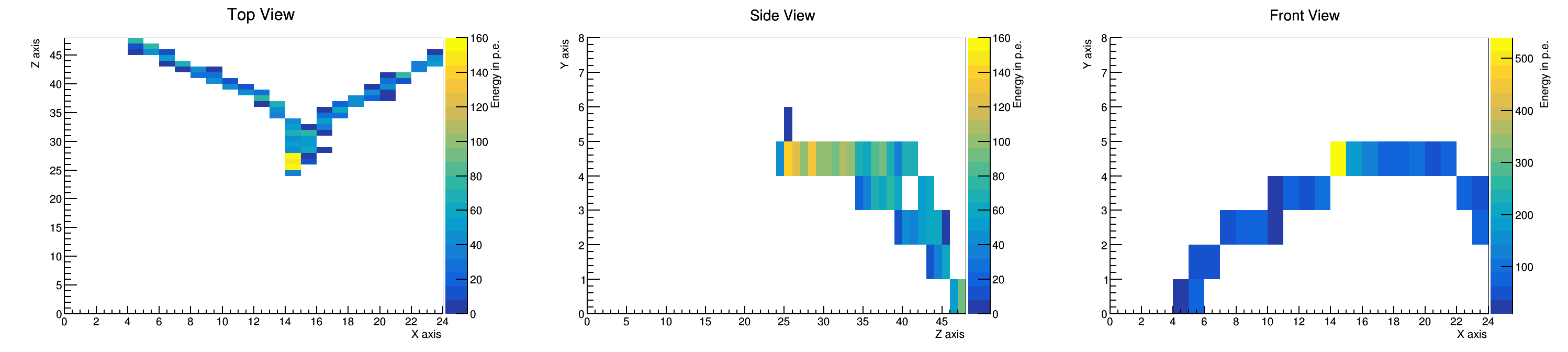}
  \end{center}
\caption{An event display showing a photon conversion in the SuperFGD prototype.}
\label{fig:Target-photon_conversion}
\end{figure}

An event display for one of the photon conversions observed is shown in Fig. \ref{fig:Target-photon_conversion}. The top view confirms that the photon, which is incident on the prototype at z=0, interacts at z=26. The photon converts to an electron-positron pair roughly in the center of the detector.

%% file: TPC.tex
\chapter{High-Angle Time Projection Chambers} 
\section{Conceptual Design}


The combination of thin active targets made of scintillators and TPCs inside the magnetized volume of the UA1 magnet is the distinctive feature of the current T2K off-axis near detector ND280. All the T2K oscillation analyses use as a constraint on the neutrino flux and cross-sections the data from these detectors. 

The ND280 TPCs~\cite{Abgrall:2010hi} have been particularly useful because they provide crucial information for the event reconstruction and the analysis: 
\begin{itemize}
 \item track reconstruction in 3D. All other detectors have coarser granularity and projected position information (mostly the in the x or y directions). Therefore TPC tracks are used as pivot in the reconstruction.
 \item charge measurement;
 \item momentum measurement;
 \item particle identification by combining dE/dx with momentum measurement.
 \end{itemize} 

We will maintain all these key features in the upgraded detectors and therefore plan to build new TPCs, called High-Angle TPCs (HA-TPC), with performances substantially similar to the performances of the existing TPCs.

Another key consideration is the fact that TPC are especially well suited to track low momentum tracks as those produced in neutrino interactions with the T2K off-axis beam: from a few GeV/c in the forward region, to a few hundred MeV/c in the high angle and backward regions.

\begin{figure} [htbp]
\begin{subfigure}[b]{.5\linewidth} 
\centering
\includegraphics[width=0.9\textwidth]{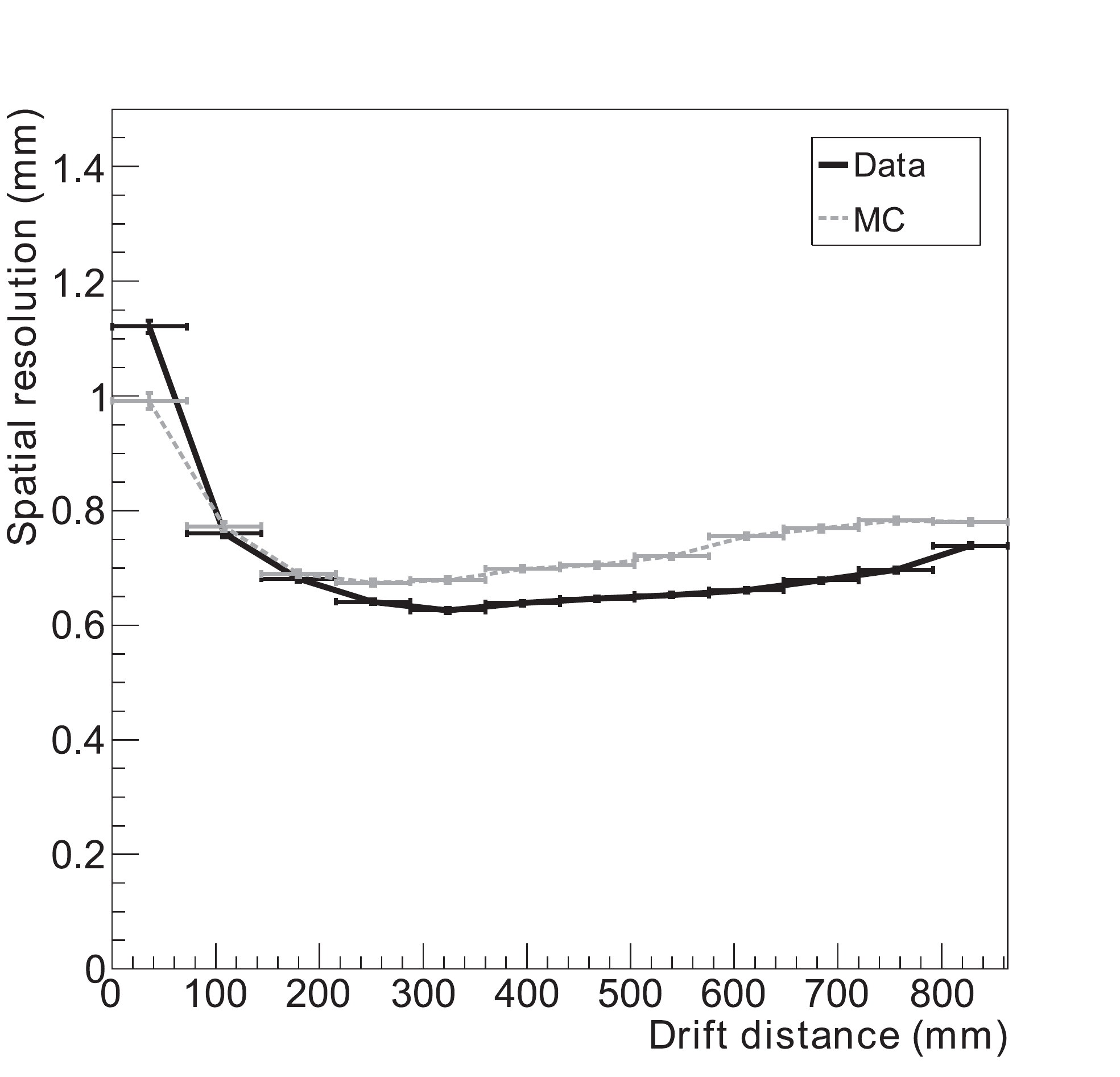} 
\caption{}
\label{fig:tpc-space:a}
\end{subfigure}%
\begin{subfigure}[b]{.5\linewidth} 
\centering
\includegraphics[width=0.9\textwidth]{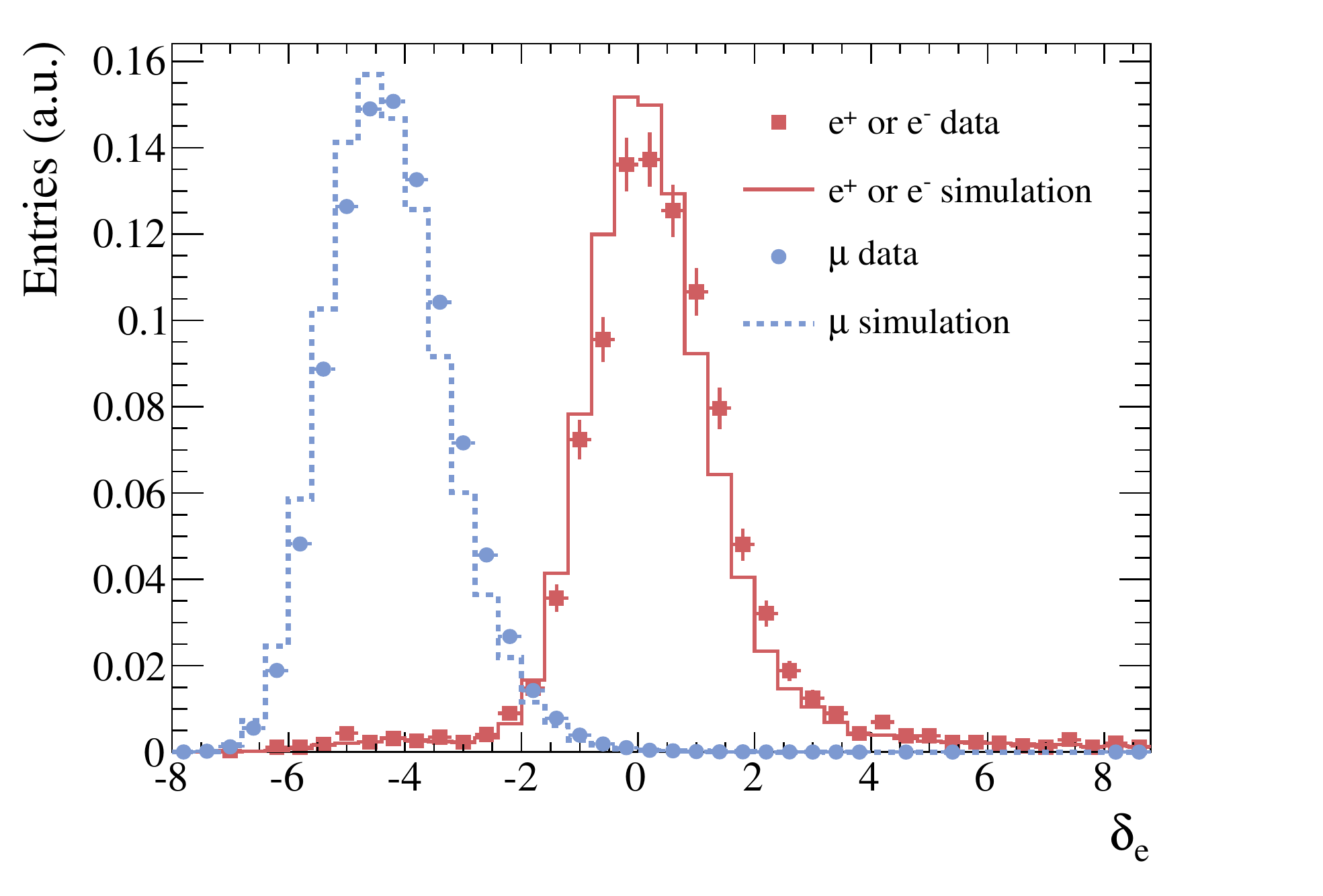} 
\caption{}
\end{subfigure} 
\caption{(a) Space point resolution for the existing ND280 TPC as function of the drift distance. 
(b) Pull in the electron hypothesis of the TPC dE/dx for a control
sample of electrons (red) and muons (blue), together with the MC
predictions 
.}
\label{fig:tpc-space}
\end{figure}

The performance obtained with the existing TPC has been completely satisfying. The requirement on the momentum resolution is 10\% at 1 GeV/c. Indeed, when reconstructing the neutrino energy, the lepton momentum is used in the Charged-Current Quasi-Elastic hypothesis: in particular the initial state nucleon is supposed free and at rest. The effect of the Fermi momentum (of the order of 200 MeV/c) introduces a smearing in the relation between the neutrino energy and the lepton momentum of the order of 10\% at 1 GeV/c. The requirement on the momentum resolution translates into a space point resolution around 800 $\mu m$ for a magnetic field of 0.2 T, 64 space points and a track length of 64 cm. Figure~\ref{fig:tpc-space:a} shows the space point resolution achieved with the existing ND280 TPC as function of the drift distance \cite{Abgrall:2010hi}.

The requirement on the momentum resolution is satisfied in the high angle and backward direction, covered by the HA-TPCs, where tracks have lower momenta, around 500 MeV/c and down to 200 MeV/c.

Another important requirement is related to the separation of electrons from muons for the  measurement of the  $\nu_e$ cross-section. Since the $\nu_e$ flux represents only approximately 1 \% of the total neutrino flux, an excellent e-$\mu$ separation is needed and the TPC particle identification is crucial to this task. We have achieved in the existing TPC a resolution of 8\% on minimum ionizing particles for the dE/dx measurement and this performance is sufficient for the $\nu_e$ studies~\cite{Abe:2014usb}, providing approximately 4 $\sigma$ separation between electrons and muons (Fig.~\ref{fig:tpc-space}). As the resolution on dE/dx is largely driven by the track length $L$ (the dependence is roughly $\sigma \propto 1/\sqrt{ L}$), we conclude that we also need a measured track length of approximately 70 cm in the vertical direction. 

The performance required for track position and angles is not critical. Indeed, what matters is a good matching 
between a track in the TPC, and either a track or hits in the Scintillator Detector, with a typical resolution at the few mm level.  

Following these considerations, the design of the new TPCs is mainly based on the design of the existing TPCs with two major changes:
\begin{itemize}
\item the Micromegas detector will be constructed with the "resistive bulk" technique, that naturally introduces a spread in the charge on the anode plane, thereby allowing in principle a lower density of readout pads. This technique allows also to eliminate the discharges (sparks) and therefore the protecting diodes on the front end cards are no longer necessary.  
\item The field cage will be realized with a layer of solid insulator laminated on a composite material. This will minimize the dead space and maximize the tracking volume. 
\end{itemize}

\begin{figure}[htbp]
	\centering
		\includegraphics[width=0.8\textwidth]{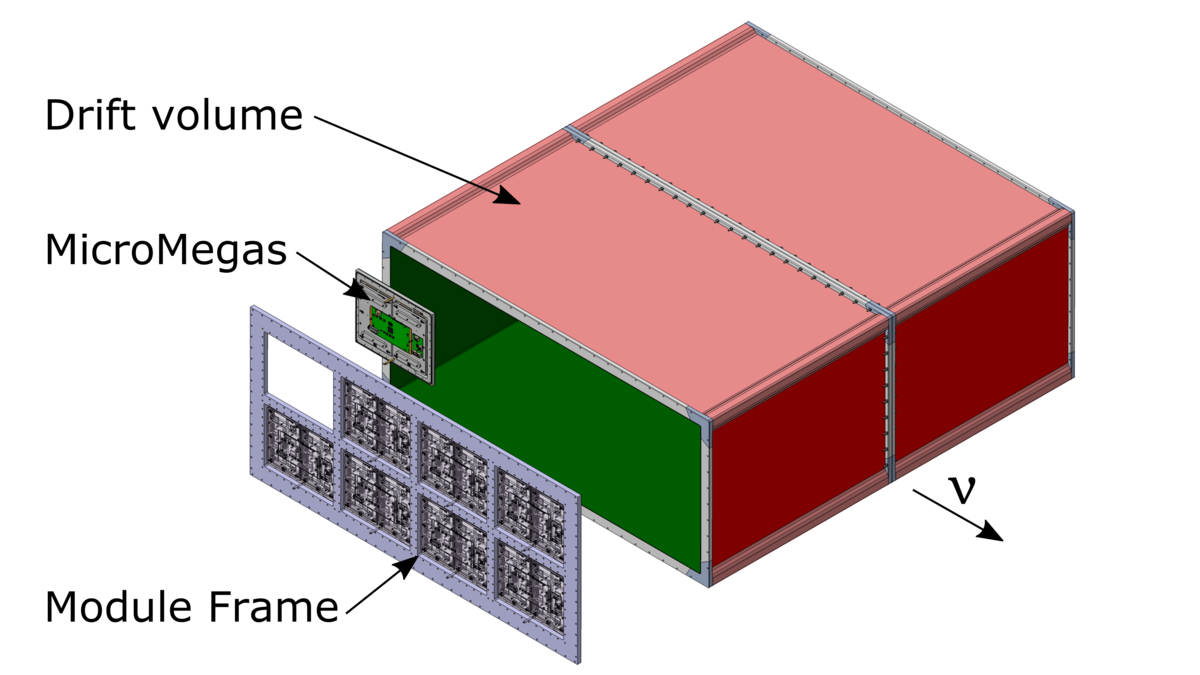}
		\caption{Schematic view of the High-Angle TPC. }
	\label{fig:hatpc}
\end{figure}

A schematic view of a HA-TPC module is presented in Fig.~\ref{fig:hatpc}.

The parameters of the High Angle TPC (HA-TPC in the following) can be found in Table~\ref{tab:vtpcpar}. 

\begin{table}[htbp]
\centering
\caption{Main parameters of the HA-TPC.}
\label{tab:vtpcpar}
\begin{tabular}{|c|c|}
  \hline
  Parameter & Value \\
  \hline
Overall  x $\times$ y $\times$ z (m) & 2.0 $\times$ 0.8 $\times$ 1.8 \\ 
Drift distance (cm) & 90 \\
Magnetic Field (T) & 0.2 \\
Electric field (V/cm) & 275 \\ 
Gas Ar-CF$_4$-iC$_4$H$_{10}$ (\%) & 95 - 3 - 2 \\
Drift Velocity $cm /\mu s$ & 7.8 \\
Transverse diffusion ($\mu m/\sqrt{cm}$) & 265 \\
Micromegas gain & 1000 \\
Micromegas dim. z$\times$y (mm) & 340 $\times$ 410 \\
Pad z $\times$ y (mm) & 10 $\times$ 11\\
N pads & 36864 \\
el. noise (ENC) & 800 \\
S/N & 100\\
Sampling frequency (MHz)& 25\\
N time samples & 511 \\  
  \hline
\end{tabular}
\end{table}


\section{TPC Structure}
\label{sec:tpc_structure}

\begin{figure}[!ht]
	\centering
		\includegraphics[width=0.65\textwidth]{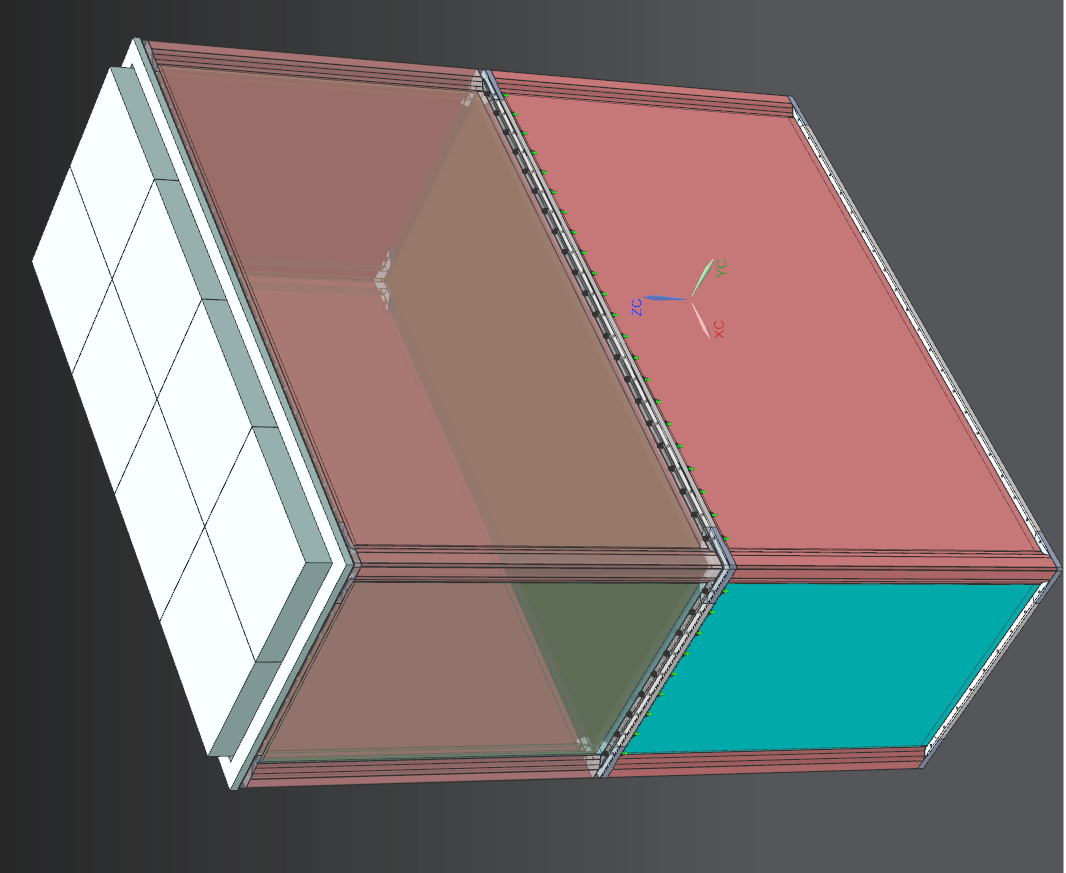}
		\caption{Model of a HA-TPC including an End-Plate. The two Field Cage boxes joined by means of their flanges in the middle of the detector are illustrated. The central cathode is also visible in transparency. The end-plates, joined at the boxes end by flanges, embed the module-frames and allow gas-tight sealing of the inner volume.}
	\label{fig:fieldcage1}
\end{figure}

The major mechanical components of the TPC are shown in Fig.~\ref{fig:fieldcage1}. The TPC consists of a gas tight rectangular prism (box) sub-divided by a common high-voltage (HV) electrode (cathode) located in its midpoint and supporting the 8 Micromegas readout modules that are located in a plane parallel to the cathode at each end of the box, where two module frame holding the Micromegas seal the TPC volume (end-plates).

The box serves as TPC Field Cage, which has to provide a highly uniform electrostatic field in the rectangular prism shaped volume containing high-purity gas to transport primary charges drifting along the magnetic field direction and over about $1\mathrm{m}$ long distance towards the readout end-plates (anode, referred to electrical ground potential). The Field Cage embeds the field shaping electrodes and provides also HV degradation towards the outside of the Cage, whose external walls are covered by a copper shielding layer, connected to electrical potential ground reference. 

\begin{figure}[!ht]
	\centering
		\includegraphics[width=0.65\textwidth]{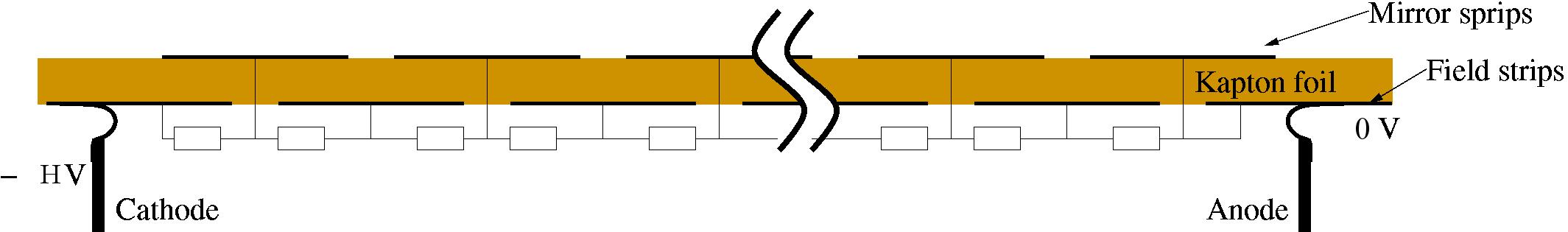}
		\caption{Field shaping electrodes located onto the inner Field Cage surface, with the related electric circuit.}
	\label{fig:strip_configuration}
\end{figure}

The field shaping electrodes consist of a series of copper strips which cover the walls joining the cathode and the two anodes on the opposite cage sides.
The normal direction of the cathode plane defines the direction of the drift field, which lays parallel to the box walls, defining the field gradient.  
The strips are joined by precision resistors forming a voltage divider (see Figure~\ref{fig:strip_configuration}). The central HV electrode and the two opposite potential degraders provide uniform drift fields of about 275 V/cm. The drift field is chosen in line with the intrinsic properties of the drift gas affecting the drift velocity and the diffusion of primary ionization electrons in that gas. Thus, given the maximum drift path of 1m, the HV at the central electrode will be as large as 27kV.

From a mechanical point of view the Field Cage is composed of two flanged boxes 
about $1\mathrm{m}$ along the drift direction 
and $1.8\times0.8\mathrm{m}^2$ in the plane transverse to the drift. 
The boxes are joined in the middle of the detector (at the cathode location) and 
are closed at the opposite ends by the end-plates including the module frames. 
The three junctions are sealed by means of O-rings in order to ensure gas tightness. The two internal volumes will be communicating via open gaps at the cathode edges in order the gas to flow trough one single volume. The maximum over-pressure allowed for the field cage will be 4 mbar.

The overall mechanical and electrical structure requirements follow:
\begin{itemize}
\item low-density and low-Z material to reduce multiple scattering and conversion processes: 
the overall Field Cage wall thickness must not exceed $4\%$ of radiation length;
\item high structural integrity against over-pressure, gravitational and thermal loads;
\item high degree of electric field uniformity: distortions resulting from imperfections in the construction of the Field Cage structure should produce distortions in the reconstructed positions of primary electrons of approximately $0.2$mm or less,
small compared to the nominal space point resolution of approximately $0.4$mm, 
and small enough to not affect the momentum scale by more than $2\%$;
\item adequate inner surface smoothness to protect against HV discharges;
\item adequate solid Field Cage walls structure such that the maximum electric field in within walls should not exceed 
the nominal breakdown by more than $30\%$;
\item very low permeability to atmospheric gas components having a negative impact on the drift of electrons (O$_2$, N$_2$ and H$_2$O);
in particular the structure must be sufficiently gas tight to keep the Oxygen level in the drift volume below about 10ppm;
\item negligible vapour pressure of contaminants emanating from material exposed to the drift volume;
\end{itemize}

\subsection{The TPC Field Cage}
\label{sec:tpc_fc}

\begin{figure}[!ht]
	\centering
		\includegraphics[width=0.45\textwidth]{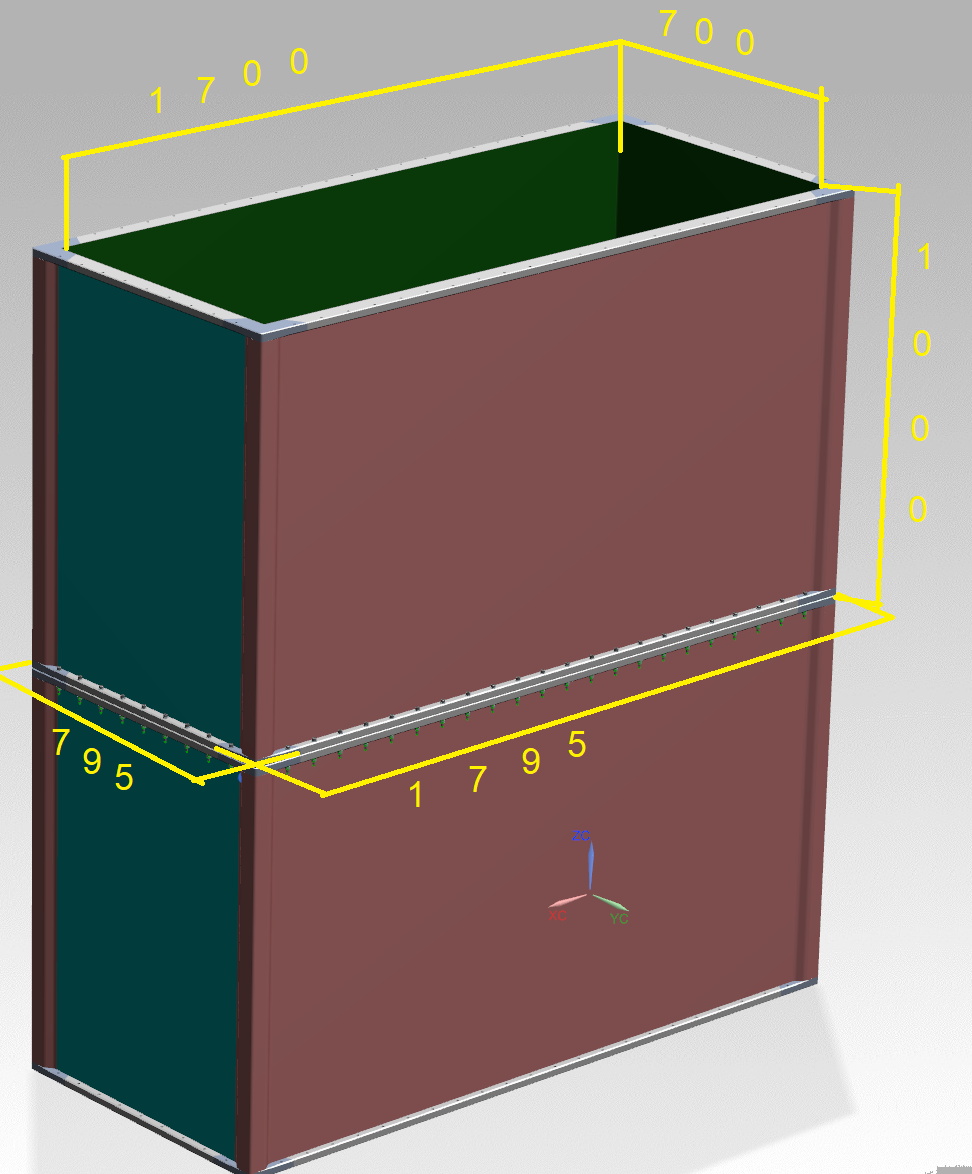}
		\includegraphics[width=0.33\textwidth]{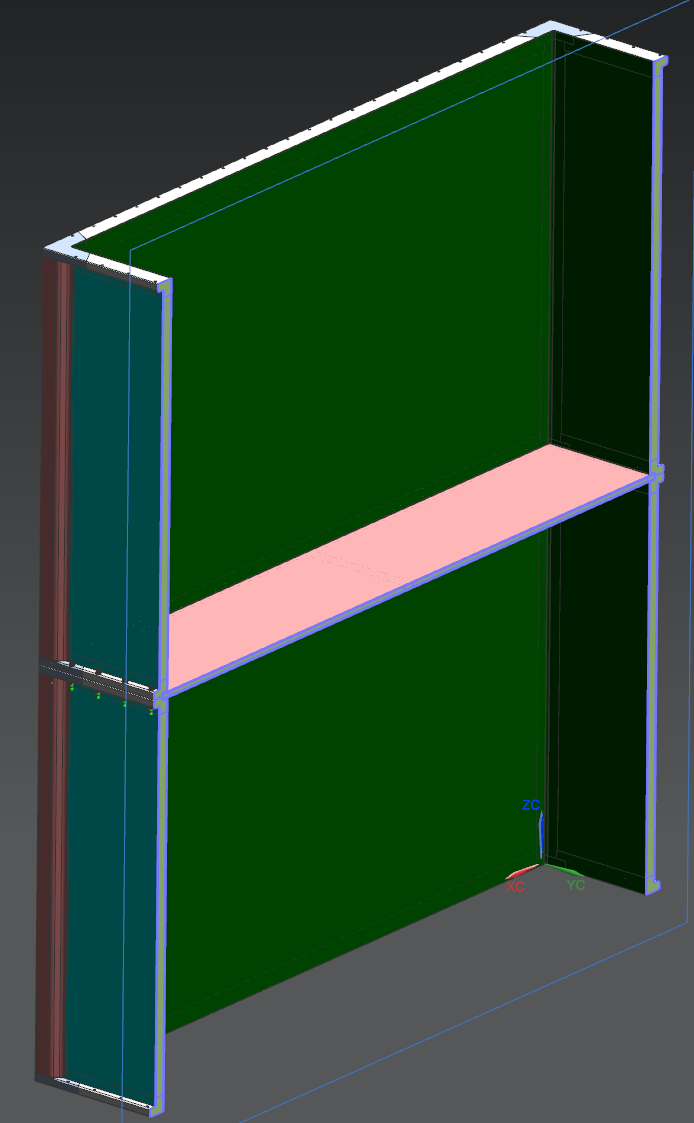}
		\caption{Left: model of a HA-TPC Field Cage made of two boxes joined in the middle of the detector, where the cathode is located. Right: Field Cage vertical cross-section. The view is such that the drift direction is vertical.}
	\label{fig:FC_1}
\end{figure}

These requirements led to the choice of composite materials for the Field Cage. 
Composite sandwich structures provide the highest stability/mass ratio and are commonly used in the industry, allowing for reduced production costs. Thus two flanged boxes composing the Field Cage consist of low-mass mechanical structures, 
having an overall hollow box shell shape (see Figure~\ref{fig:FC_1}). The Cage boxes consist of a polyimide/aramid fiber fabric sandwich with innermost and outermost surfaces embedding thin Copper superficial electrodes. 
The boxes extend $1.000$m along the drift direction (vertical direction in Figure~\ref{fig:FC_1}) and have an internal cross section of $1.700\times0.700\mathrm{m^2}$ and an external cross section of $1.795\times0.795\mathrm{m^2}$ . 

In order to optimize the Field Cage design, studies were performed with finite elements simulation software both for structural analysis and for electric field analysis. The simulation studies, which are respectively summarized in Sections~\ref{sec:tpc_sim_mech} and~\ref{sec:tpc_sim_ele}, were validated via test on small scale samples of the Field Cage structure. The distortions in tracking, resulting from combined mechanical and electric field deformations, were obtained from drifting charges from various locations. The key tolerances arising from these studies are the following:
\begin{itemize}
    \item the resistor pairs that form the voltage divider between the central cathode and the Micromegas must be matched within an rms of $0.1\%$;
    \item the central cathode should be flat to within 0.1mm;
    \item the Micromegas plane should be flat to within 0.2mm;
    \item the central cathode and Micromegas planes should be parallel to within 0.2mm;
\end{itemize}
Additional arising tolerances specific of the Field Cage are listed in the following. 

The resulting optimized design of the Field Cage structure is described as follows.

Each Cage box consists of a single piece of solid composite material. The Cage box walls (four lateral sides) consist of a single sandwich structure including a core made of Aramid honeycomb (25mm thick) and of two laminate skins (approx. 2mm thick) on opposite sides of the core. The skins incorporate, as innermost layer, a Kapton foil with Copper coated strips on both sides (Double layer strip foil) and an outermost layer consisting of an uniform Copper foil on the external face. The core of the skins consist of Aramid fiber fabric layers. The detailed stack sequence of the layers is shown in Table\ref{Table_wall}.

\begin{table}[!ht]\small
\centering
\begin{tabular}{lllll} 
\hline
Layer of the wall   & Material  & thickness d  & average      & d/X$_0$\\
                    &           & d (mm)       &  X$_0$ (mm)  & $(\%)$\\
\hline
1 (inner layer      & Double layer strip foil        & $\sim$0.05    &   143        & 0.08 \\ 
2                   & Polymide film (Kapton)         &  0.01         &   285        & $<$0.01 \\ 
3                   & Aramid Fiber Fabric (Twaron)   &  2.0          &   $\sim$240  & 0.70 \\ 
4                   & Aramid honeycomb panel (Nomex) &  25           &   14300      & 0.17 \\ 
5                   & Aramid Fiber Fabric (Twaron)   &  2.0          &   $\sim$240  & 0.07 \\ 
6 (outer layer)     & Copper foil                    &  0.01         &   143        & 0.07 \\
\hline
Total               &                               &   $\sim$30     &              & 1.7 \\ 
\end{tabular}
\caption{Composition and radiation lengths of the materials in the field cage wall}
\label{Table_wall}
\end{table}

\begin{figure}[!ht]
	\begin{center}
		\includegraphics[width=0.35\textwidth]{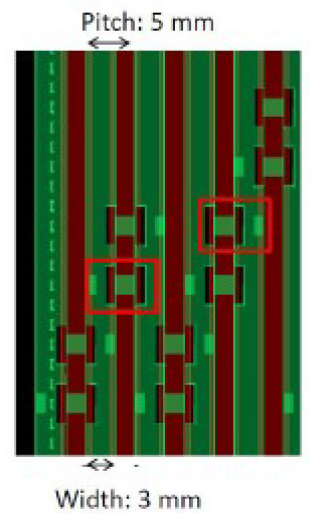}
	\end{center}
	\caption{Closeup of the Field Cage strips. Copper strips on the inner side ('field strips') are depicted in green colour, as the square $2\times2$mm pads which serve as connection between the 'mirror strips' (not visible) and the field strips. Kapton not covered by Copper is depicted in brown colour.}
	\label{Figure:strips}
\end{figure}
The main field forming element is a Kapton foil ($40\mu$m thick) covered by Copper strips ($5\mu$m thick) on both sides. The foil is embedded in the Cage boxes as the innermost layer. A close-up of the strip foil is illustrated in Figure~\ref{Figure:strips}. The pitch of the strips is 5mm. The gap between two adjacent strips is $2$mm. Strips on the opposite faces of the foils are staggered so that they overlap by $0.5$mm. One of the foil sides includes 110 strips ('field strips') on the inner side of the foil. The opposite side includes 109 strips, which will be referred to as the 'mirror strips'. Pads are present along the strips, where resistors will be soldered in order to connect electrically the strips to form the voltage divider. Pads include some vias allowing the connection of the mirror strips with the inner strips, in order to connect mirror and field strips along a common voltage divider. For redundancy reasons two voltage dividers will be used, connected in parallel.

In order to provide stiffness and strength to the Cage box, the composite sandwich incorporates 4 bars (also referred to as ``angular bars'') made of thermoplastic (POM-C type) that are embedded into the 4 vertical edges of the Cage, as an internal part of the sandwich structure. 
In order (1) to protect the edges of the honeycomb panels against transverse strain, (2) to compensate the irregular shape of honeycomb panels at their edges and (3) to provide a smooth and precision machinable top and bottom edge surfaces to the Cage, additional 8 bars and 8 corner parts (referred to as the Cage box "flanges" parts) made of POM-C are embedded into the 8 horizontal edges (and corners) of the Cage box. These surfaces constitute the top and bottom flanges and need post-production precision machining and bore drilling. As previously discussed the flanges are needed for joining the two boxes together and for sealing the Field Cage with the module-frames.

The Cage boxes composite sandwich structure must be manufactured in progressive steps by hand lay-up onto a mold the layers described in Table~\ref{Table_wall} and including the structural thermoplastic parts above mentioned. In order to obtain uniform and homogeneous quality of the detector, the curing of resins used for laying-up process require room temperature pressure based techniques with the use of vacuum bag and high pressure in autoclave. 

The mold mainly consists of 4 precision machined cast Aluminum plates (called "ALCOA" plates) $20$mm thick and with dimensions $730\times1000\mathrm{mm}^2$ (2 plates) and $1700\times1000\mathrm{mm}^2$ held by L shaped angular profiles in order to define the Cage box inner surface. Eight additional plates ("offset" plates) extend the ALCOA plates top and bottom edges for a few cm. The offset plates are removed after the composite production process for allowing the post-processing precision machining of the Cage flanges (and bore drilling). After the machining is completed the mold is dismounted in order to allow surfaces finishing and inspections.

The manufacturing steps are summarized as follows:
\begin{itemize}
    \item Application, positioning and precision alignment of the strip foils on the mold;
    \item Lay-up the polyimide film (layer $\#$2, see Table~\ref{Table_wall}) and the Aramid Fiber Fabric (Twaron) inner layer (layer $\#$3), namely the sandwich inner skin;
    \item Placement of the core of the sandwich structure (layer $\#$4), namely the honeycomb panels, angular bars and flange parts;
    \item Lay-up of the sandwich external skin (layer $\#$5, Twaron) and of the externally copper layer;
\end{itemize}

A final post-production phase is needed for cutting and treating the skins material in excess, for machining and polishing the flange surfaces, and for drilling the flange bores. 
The following tolerances are expected for Cage geometry after post-process machining: 
\begin{itemize}
\item Parallelism between the outer surfaces of the top and bottom flanges better than 0.1mm over 1m;
\item Planarity of the inner box Cage surfaces better than 0.3mm over their extension;
\item Flange surfaces roughness below 2 $\mu$m;
\item Inner surface waviness better or equal than ISO 1302 N9 grade;
\item Orthogonality between adjacent faces better than 0.25mm over 1m;
\item Orthogonality between inner surfaces and flange plane better than 0.25mm over 1m;
\item Overall thickness of the composite sandwich: -0,+5mm with respect to the nominal values;
\end{itemize}

These tolerances are compliant with the above mentioned studies concerning the effects on tracking, resulting from combined mechanical and electric field deformations. In particular
in terms of electric field component transverse to the drift direction $E_{\perp}$, these tolerances ensure that a relative variation below $\Delta E_{\perp}/E_{\parallel} < 10^{-4}$.

\subsection{Cathode}

The central cathode is a Copper-clad G10/rohacell panel constructed
from 1mm copper-clad G10 laminated onto both surfaces of
10mm (nominal) thick rohacell, giving a total nominal thickness of 12mm.
To make the panel a frame is constructed from glued and screwed G10 bars, which are machined to match the
measured rohacell thickness. The frame is then loaded with rohacell and laminated to both copper-clad G10 
sheets at the same time. The lamination is done on a granite flat table with a vacuum bag. 
The frame are milled all along the sides (except for a few cm about the Corners) 
in order to mill a wedge cross section ($45^o$~degrees)

\begin{figure}[!ht]
	\centering
		\includegraphics[width=0.45\textwidth]{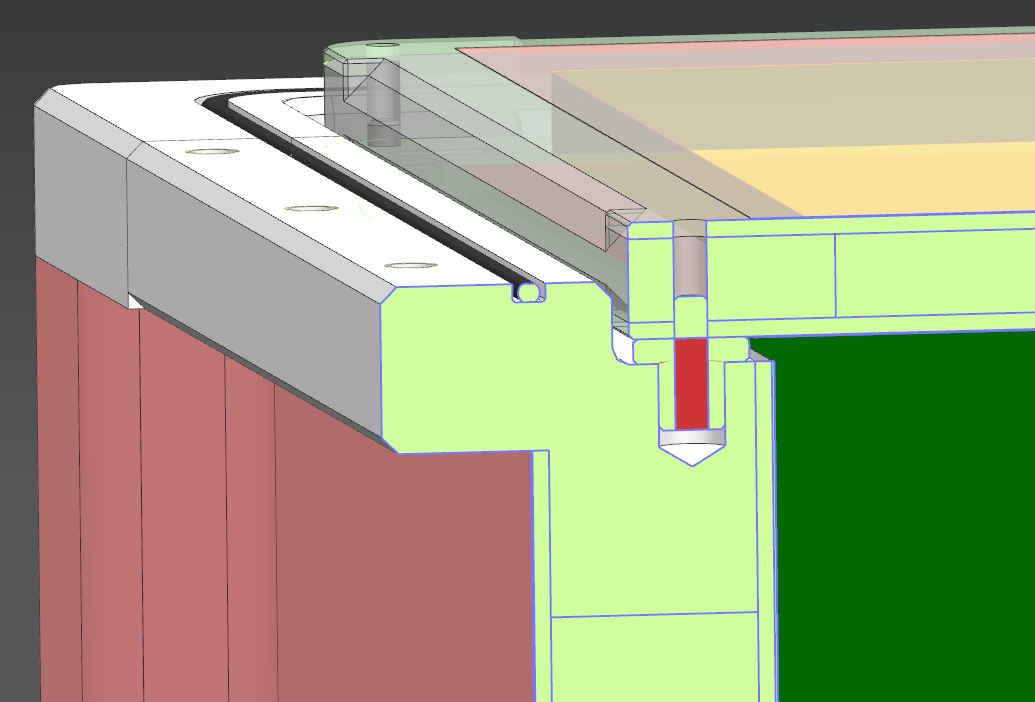}
		\includegraphics[width=0.45\textwidth]{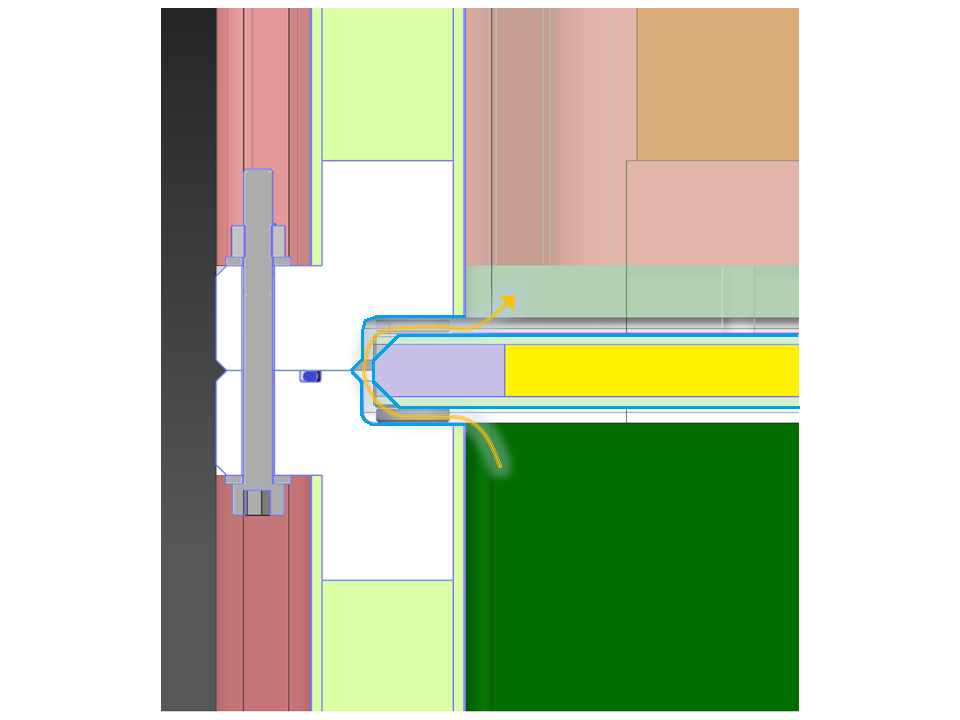}
		\caption{
		Left: closeup of the cathode supported through a spacer on the grooved flange of a Cage box.  
		Right: cathode infixed between the flanges of the two Cage boxes. 
		Gas flow between the two Cage boxes volumes is allowed by the 
		distance of the cathode edges from the inner surface of the flange and 
		by the wedged cathode edges.}
	\label{fig:TPC_CC_1}
\end{figure}

Left figure~\ref{fig:TPC_CC_1} shows a closeup of the cathode 
supported on the internal side of a Cage box flange by a spacer.
The internal side of the flanges are properly grooved ($8\times18$mm) 
in order to host the edges of the cathode, which are protruding 16mm into the groove.
Right figure~\ref{fig:TPC_CC_1} shows the cathode infixed between the flanges of the two Cage boxes. 
Precision machined spacers allow for alignment of the Cathode and the End-Plate planes at a level better than 0.1mm.

Low impedance gas flow between the two Cage boxes volumes is ensured 
by the the 2mm distance of the cathode edges from the (grooved) inner surface of the flange and 
by the wedge shape of the cathode edges.

The high-voltage connection to the central cathode is made by properly feeding the High Voltage cable through the flange in a position near to a corner on a short side of the flange. Electrical connection of the Cathode to the proper Field Cage strip is provided by soldered fine wires.  

\subsection{Module frame}

\begin{figure}[ht]
	\begin{center}
		\includegraphics[width=0.8\textwidth]{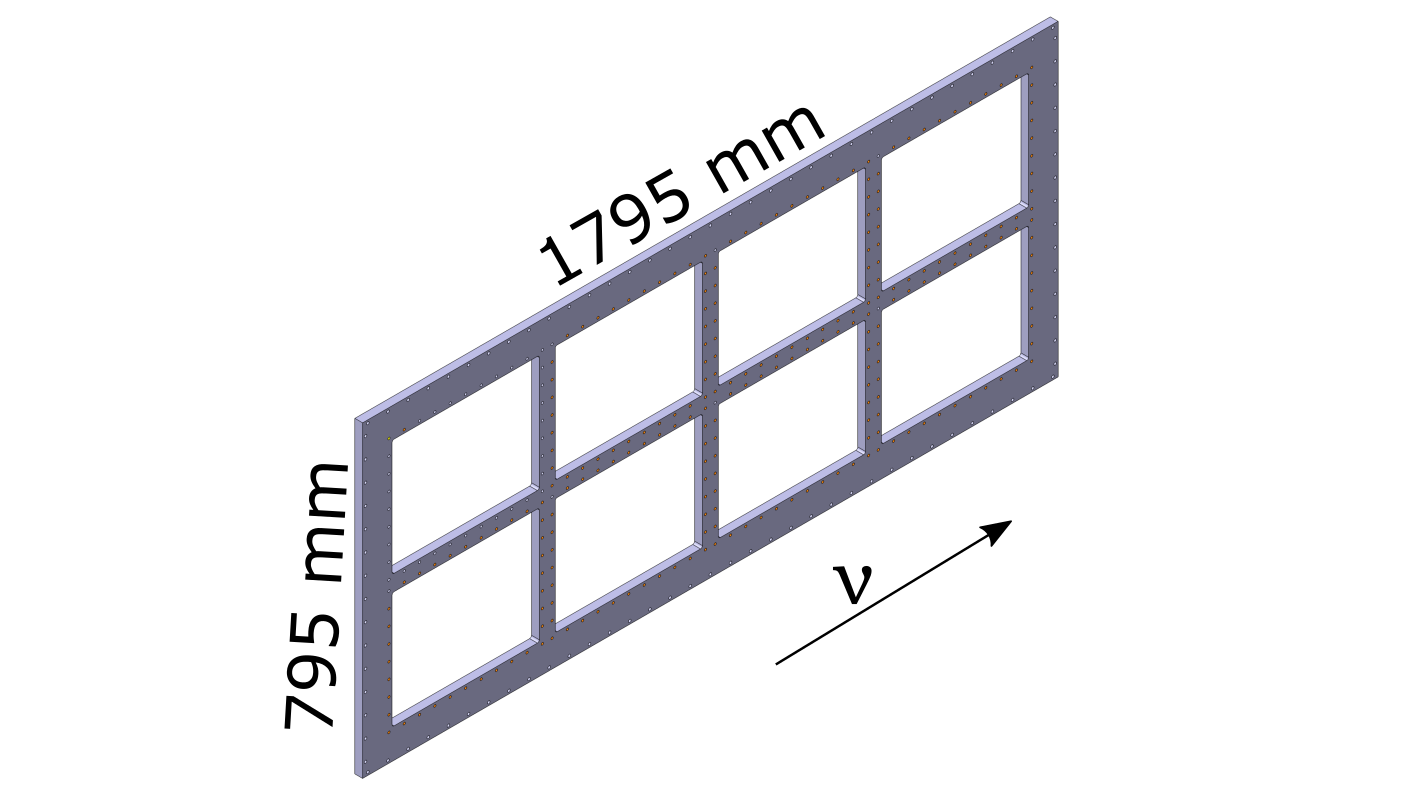}
		\caption{The TPC \MF}
		\label{fig:TPC_MF_cad}
	\end{center}
\end{figure}

The \MF~ is presented in Figures~\ref{fig:hatpc} and \ref{fig:TPC_MF_cad}.
Four such frames, together with the total number of 32 \MM~modules mounted on them, are needed to complete the two HA-TPCs.

For each \MF~the \MM~are organised in two rows of four modules and are mounted on its inner side, which allows to decrease spaces between \MM~to 1~mm and to provide a better coverage of the TPC area. For the \MM~dimensions specified in Table~\ref{tab:vtpcpar} the minimal dimensions of \MF~are \SI{795}{\milli\metre}~$\times$ \SI{1795}{\milli\metre} $\times$ \SI{20}{\milli\metre}.


The requirements for the \MF~are as follows:
\begin{enumerate}
	\item gas tightness and guarantee of a high purity of the gas mixture, which implies the use of O-ring seals and non out-gassing materials,
	\item precise positioning of the eight Micromegas~detection surface 
on a vertical plane parallel to the cathode plane with a 	precision of 100 $\mu$m,
	 which means that the global flatness must be within $\pm$~\SI{0.1}{\milli\metre} under \SI{4}{\milli\bar} overpressure -- the nominal working overpressure (the operating mode),
	\item no plastic deformation under \SI{10}{\milli\bar} overpressure -- the overpressure during gas tightness tests (the test mode).
\end{enumerate}

The new Micromegas are grounded electrically, which allows the use of an aluminium \MF~and Micromegas~stiffener. Aluminium offers good machinability and structural performance.

The mechanical simulations, proving that the aluminium \MF~meets requirements 2 and 3, have been performed using the Ansys software~\cite{ANSYS_ewbsite} and their results are presented in Section~\ref{sec:tpc_sim_mech_mf}.
In general the simulations show that the minimal configuration can sustain the over-pressure and the load from the \MM. 

The next step is to include the features that are tested on the prototype such as the high voltage connector and the gas manifold. Finally, some other elements for handling and fixation in the ND280 basket will be studied.

\subsection{Mechanical Calculations and Simulations}
\label{sec:tpc_sim_mech}

In order to optimize the Field Cage, the Cathode and the Module Frame mechanical design, finite elements (FEM) simulations were carried on for the structural analysis, which are described in the following Sections.


\subsubsection{ Field Cage Mechanical simulations} 
\label{sec:tpc_sim_mech_fc}

The Field Cage mechanics design was studied and optimized with MSC FEM simulation software ~\cite{MSCwebsite}. The simulation was validated via test on small scale samples of the Field Cage structure.

The goal of these calculations was to find a structure that is as thin as possible but nevertheless stable against gravity and over-pressure stresses. For these calculations is has been assumed that the two boxes are joined by the central flanges and that the two end-flanges are at fixed positions.

\begin{figure}[!ht]
	\centering
		\includegraphics[width=0.90\textwidth]{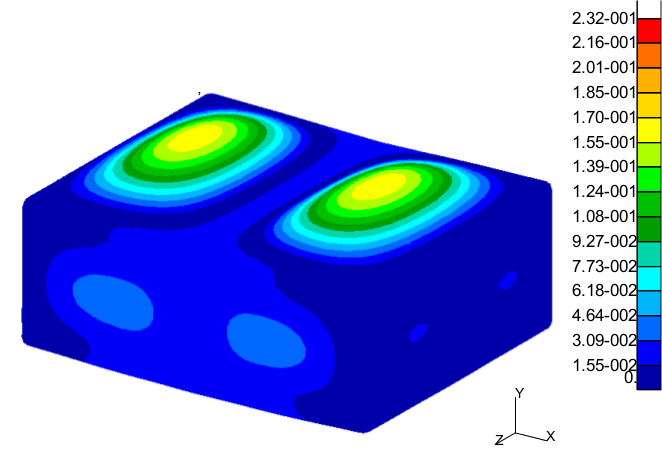}
		\caption{Result of the FEM simulation of the Field Cage with an over-pressure of 10mbar. The maximum deformation is estimated to be 0.23mm.}
	\label{fig:FC_sim_mech}
\end{figure}

Maximum deformation was estimated for various values of the over-pressure parameter $\Delta P$. For instance, in the case of $\Delta P=10$mbar (a factor of 2.5 in excess with respect to the working conditions) we find a maximum deformation of 0.23mm. This result is illustrated in Figure~\ref{fig:FC_sim_mech}.

Assuming an over-pressure of 4mbar simulations results show that the maximum deviation from planarity is well below 0.15mm. In terms of electric field component transverse to the drift direction $E_{\perp}$, this translates into a relative variation below $\Delta E_{\perp}/E_{\parallel} < 10^{-4}$. 

\subsubsection{\MF~Mechanical simulations} 
\label{sec:tpc_sim_mech_mf}

The simulations related to the studies and the optimization of the~\MF~mechanics are performed using the Ansys software~\cite{ANSYS_ewbsite} and their results are presented here. 

\begin{figure}[htbp!]
	\begin{subfigure}[b]{.5\linewidth}
		\includegraphics[width=0.9\textwidth]{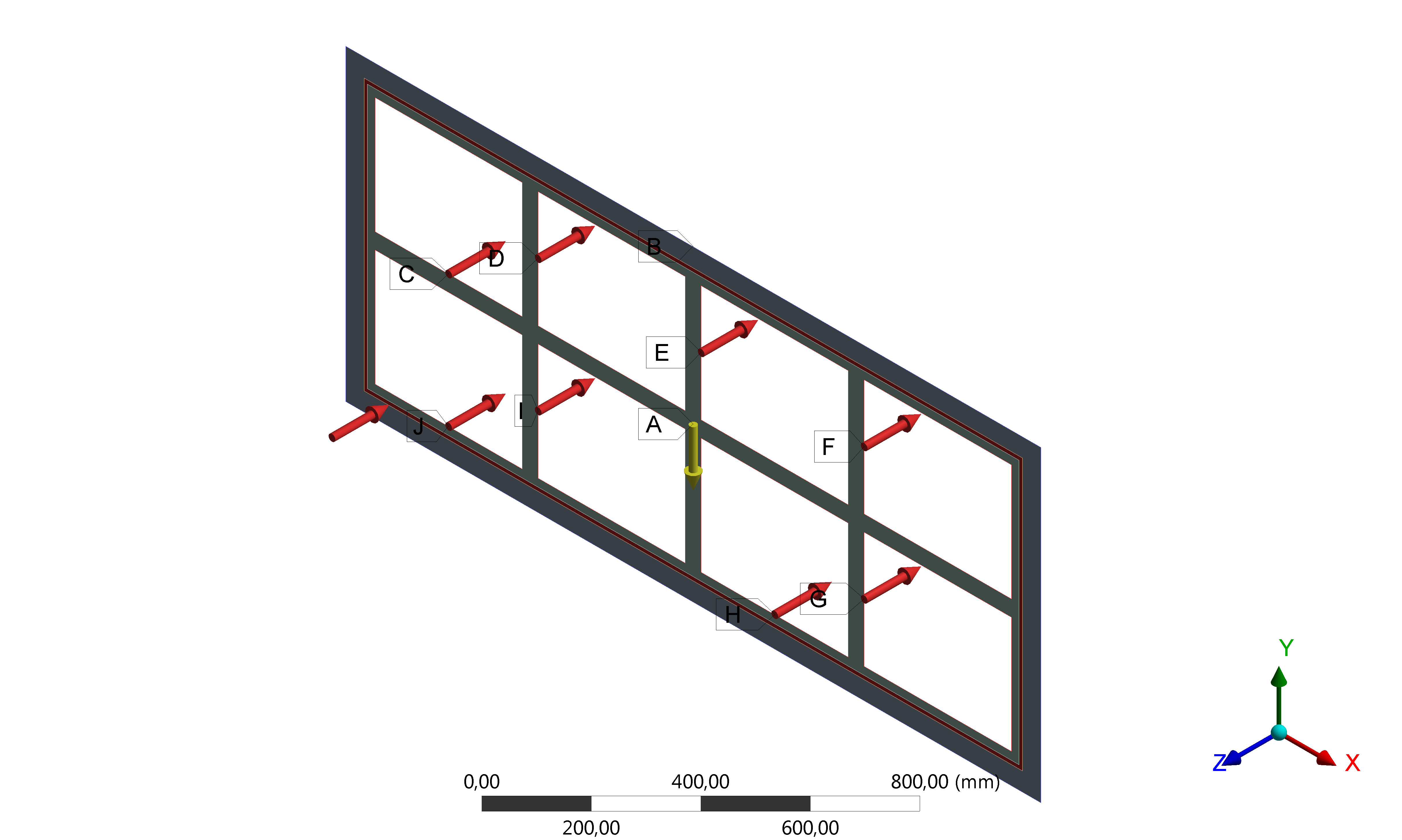}
		\centering
		\caption{Simplified model (without \MM)}\label{fig:TPC_MF_surf}
	\end{subfigure}%
	\begin{subfigure}[b]{.5\linewidth}
		\centering
		\includegraphics[width=0.9\textwidth]{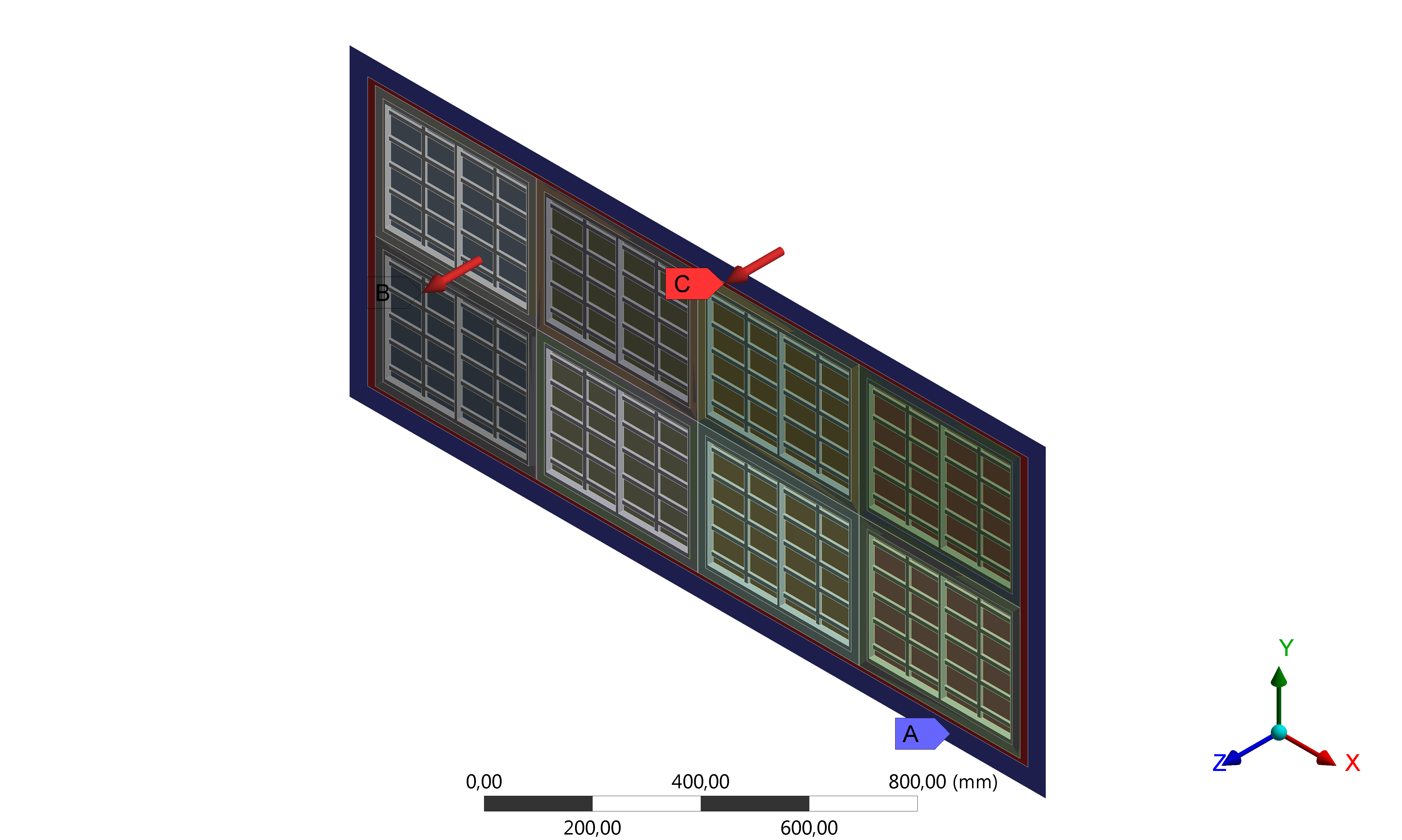}
		\caption{Realistic model (with Micromegas)}\label{fig:TPC_MF_as}
	\end{subfigure}
	\caption{Two models of the \MF~used in mechanical study.}\label{fig:TPC_MF}
\end{figure}

Two different models were applied in the simulations. At the early stage of the design, the simplified model demonstrated the feasibility of using a \MF~in Aluminum, while the realistic one rendered the results of the advanced design.
In the simplified study (Fig.~\ref{fig:TPC_MF_surf}), the pressure engendered by the Micromegas~was taken into account, but not the stiffness given by the Micromegas~body. This assumption gives deformation and stress values which are higher than ones present in reality.
For the realistic mechanical study (Fig.~\ref{fig:TPC_MF_as}), the Micromegas~design was advanced enough to take it into account.

The 7075-T6 aluminium alloy chosen for the simulations is a good structural alloy and is widely used in physics applications. For both models, two case studies were considered \begin{enumerate*}[label=(\roman*)] \item the operating mode at \SI{4}{\milli\bar} and \item the test mode at \SI{10}{\milli\bar} overpressure.\end{enumerate*} For the operating mode, the main focus is at the deformation over stress results in order to verify that the displacements stay within the $\pm$\SI{0.1}{\milli\metre} requirement. For the test mode, the stresses become more important in order to exclude any irreversible damage to the component.

\textbf{The simplified mechanical study}

The Micromegas~surface is 
\[\SI{340}{\milli\metre}\times\SI{420}{\milli\metre} = \SI{1.428e5}{\square\milli\metre}
\]

\noindent multiplied by \SI{4}{\milli\bar} ($\equiv$ \SI{4e-5}{\newton\per\square\milli\metre}), we obtain a resultant force per \MM~equals to 
\[
\SI{1.428e5}{\square\milli\metre} \times \SI{4e-5}{\newton\per\square\milli\metre} = \SI{57}{\newton}.
\]

\noindent Aluminium EN-AW 7075 T6:

\begin{center}
	\begin{itemize*}
	\item $\rho$ = \SI{2810}{\kilogram\per\cubic\metre} 
	\item E = \SI{71.9}{\giga\pascal}
	\item $\nu$ = 0.33
	\item $\sigma_{u}$ = \SI{572}{\mega\pascal}
	\item $\sigma_{y}$ = \SI{503}{\mega\pascal}	
\end{itemize*}
\end{center}

\begin{figure}[htbp!]
	\begin{subfigure}[b]{.5\linewidth}
		\includegraphics[width=0.9\textwidth]{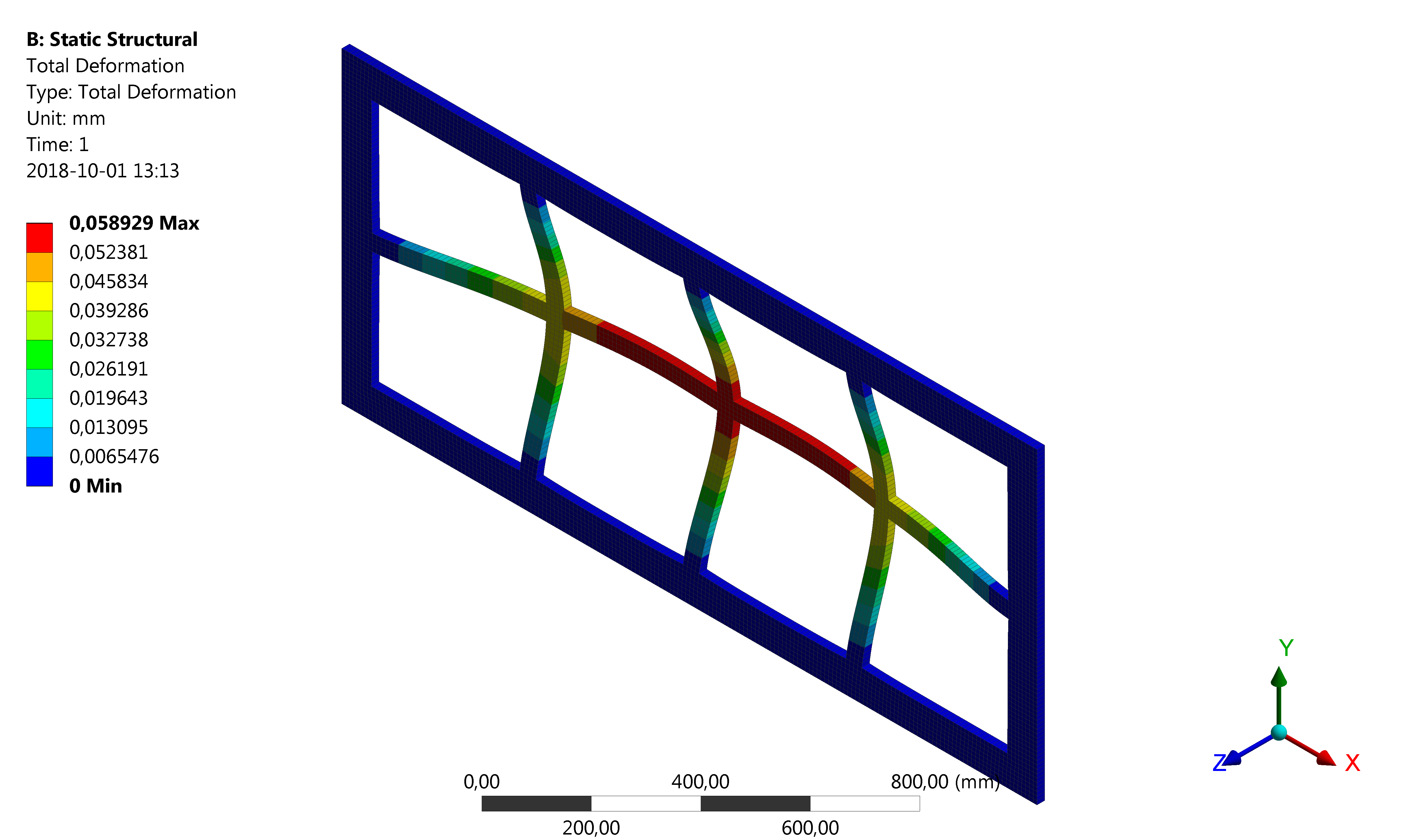}
		\centering
		\caption{Deformation [\si{\milli\metre}]}\label{fig:TPC_MF_surf_def_4mbar}
	\end{subfigure}%
	\begin{subfigure}[b]{.5\linewidth}
		\centering
		\includegraphics[width=0.9\textwidth]{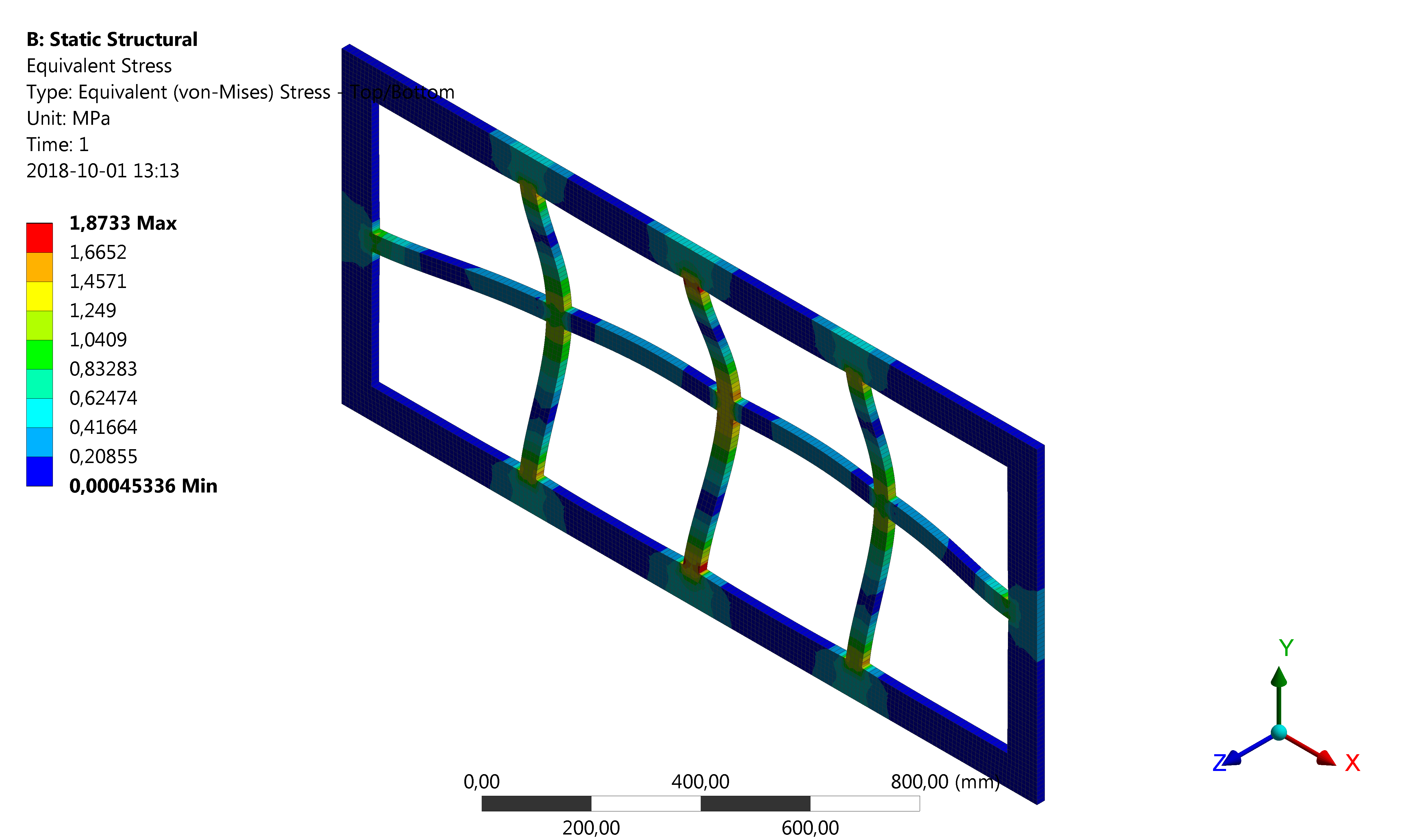}
		\caption{Von-Mises stress [\si{\mega\pascal}]}\label{fig:TPC_MF_surf_stress_4mbar}
	\end{subfigure}
	\caption{\MF~deformation and von-Mises stress at \SI{4}{\milli\bar} in simplified mechanical study.}\label{fig:TPC_MF_surf_4mbar}
\end{figure}

One can see that the maximum deformation (Fig.~\ref{fig:TPC_MF_surf_def_4mbar}) is \SI{0.06}{\milli\metre} which is below the limit of \SI{0.1}{\milli\metre}. However, the mechanical uncertainties mainly due to the machining and assembly are not taken into account. The von-Mises stresses (Fig.~\ref{fig:TPC_MF_surf_stress_4mbar}) are considered negligible.


Such a test at \SI{10}{\milli\bar} overpressure is performed to check the gas tightness of the \TPC. The von-Mises maximum stress ($\approx$\SI{5}{\mega\pascal}) is still far from the limit $\sigma_{y}$ = \SI{503}{\mega\pascal}.

\textbf{The realistic mechanical study}

\begin{figure}[htbp!]
	\begin{subfigure}[b]{.5\linewidth}
		\includegraphics[width=0.9\textwidth]{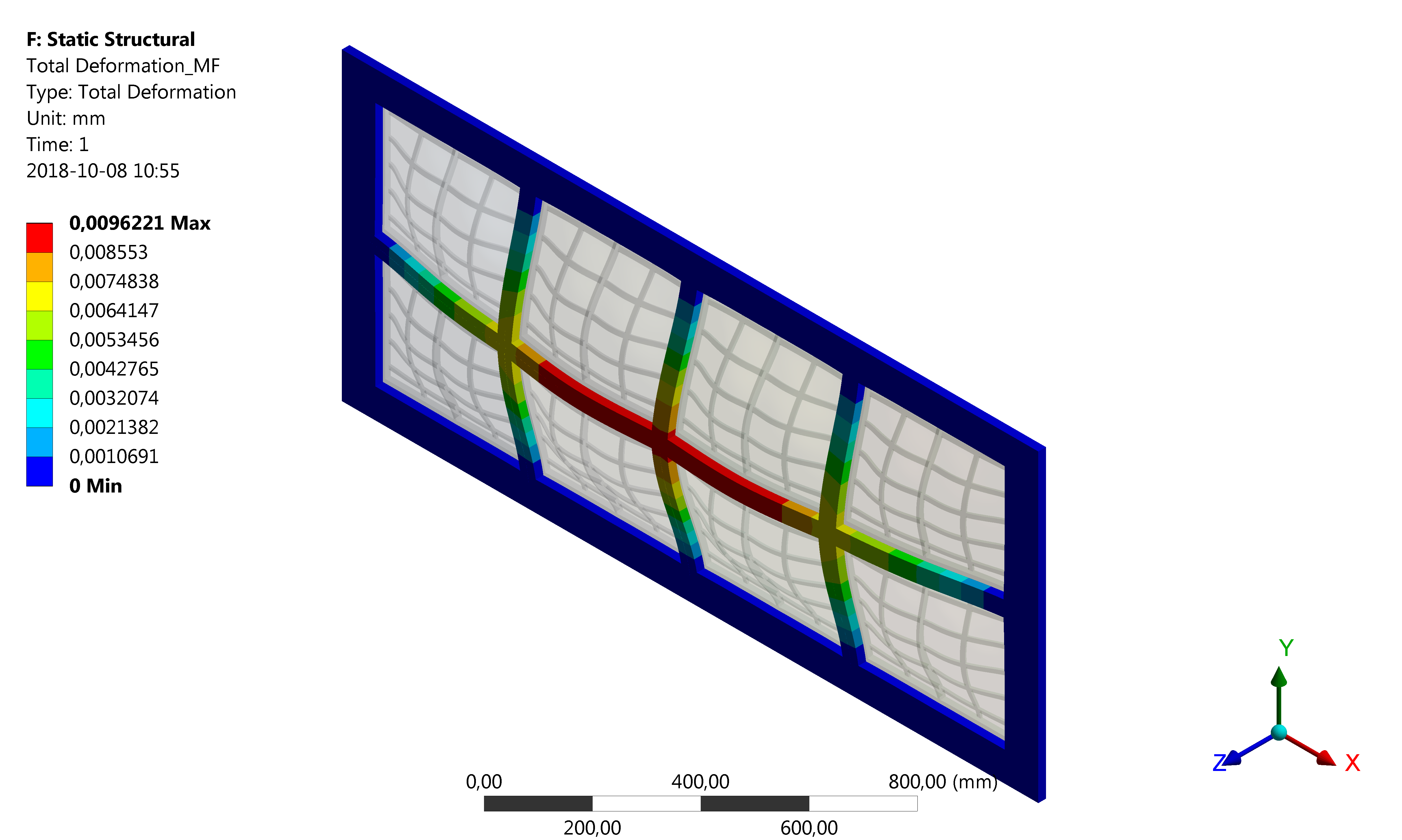}
		\centering
		\caption{Deformation [\si{\milli\metre}]}\label{fig:TPC_MF_as_def_4mbar}
	\end{subfigure}%
	\begin{subfigure}[b]{.5\linewidth}
		\centering
		\includegraphics[width=0.9\textwidth]{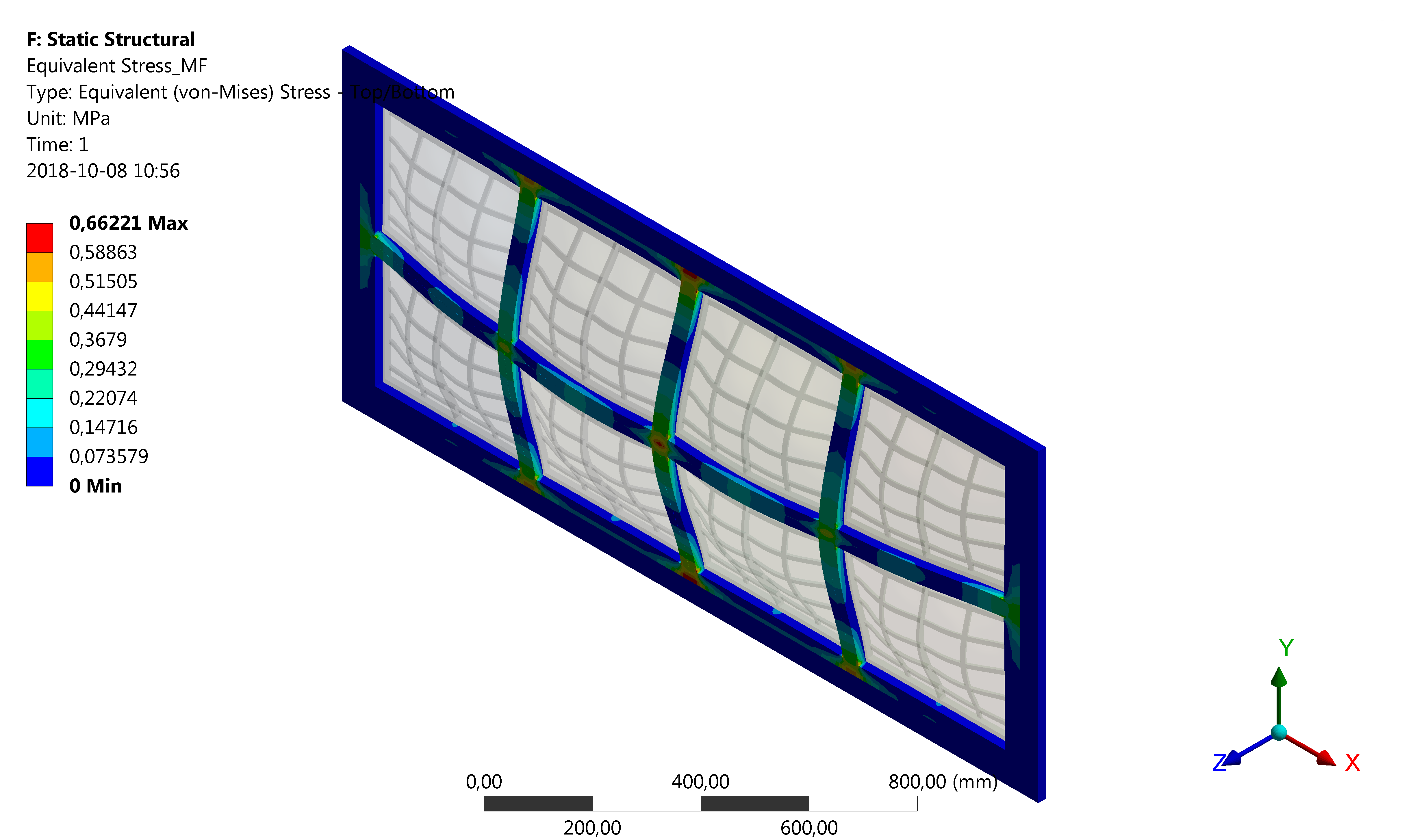}
		\caption{Von-Mises stress [\si{\mega\pascal}]}\label{fig:TPC_MF_as_stress_4mbar}
	\end{subfigure}
	\caption{\MF~deformation and von-Mises stress at \SI{4}{\milli\bar} in realistic mechanical study.}\label{fig:TPC_MF_as_4mbar}
\end{figure}

As one could expect, the displacements and the von-Mises stresses (Fig.~\ref{fig:TPC_MF_as_4mbar}) are smaller than in the previous case (Fig.~\ref{fig:TPC_MF_surf_4mbar}). The maximum deformation (Fig.~\ref{fig:TPC_MF_as_def_4mbar}) is \SI{0.01}{\milli\metre} which is well within the requirement and gives some margins for mechanical uncertainties due to the machining and assembly.


\begin{table}[htbp]
\centering
\begin{tabular}{|l|l|c|c|}
\hline
\multicolumn{2}{|l|}{Configuration}                         & \SI{4}{\milli\bar} & \SI{10}{\milli\bar} \\ \hline
\multirow{2}{*}{Simplified} & Disp. [\si{\milli\metre}]     & 0.06               & 0.15                \\ \cline{2-4}
                            & VM stress [\si{\mega\pascal}] & 1.87               & 4.67                \\ \hline
\multirow{2}{*}{Realistic}  & Disp. [\si{\milli\metre}]       & 0.01               & 0.025               \\ \cline{2-4}
                            & VM stress [\si{\mega\pascal}] & 0.66               & 1.66                \\ \hline
\end{tabular}
\caption{Summary table for the simplified and realistic mechanical studies of the module frame.}
\label{table:mf}
\end{table}

Once again, one can see that there is no problem to test the \TPC~at \SI{10}{\milli\bar}. 

The results of these simulations are summarized in Table~\ref{table:mf}.

\subsection{Electrical field calculations and simulations}
\label{sec:tpc_sim_ele}


The electrical simulations reported in this subsection have been performed with COMSOL~\cite{COMSOLwebsite} and CST~\cite{CSTwebsite} software. The goal of these calculations was to find an electrode configuration assuring an electric field homogeneity better than 10$^{-4}$ at a distance larger than 15 mm from the inner field cage walls. 
In terms of electric field component transverse to the drift direction $E_{\perp}$, this requirement translates into a relative variation below $\Delta E_{\perp}/E_{\parallel} < 10^{-4}$.

\begin{figure}[htbp!]
    \centering
    \includegraphics[width=7.8cm,height=5cm]{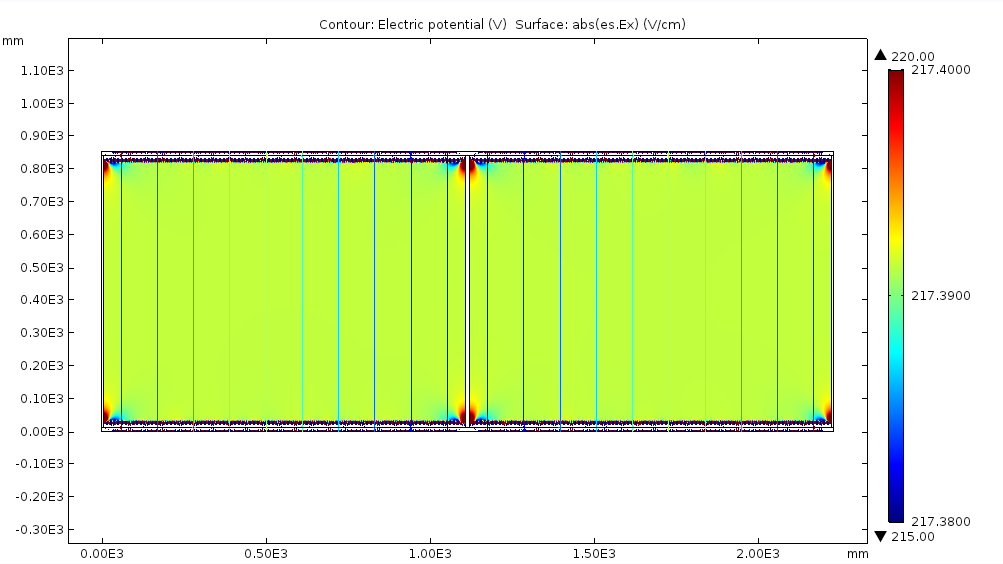}
    \includegraphics[width=7.8cm,height=5cm]{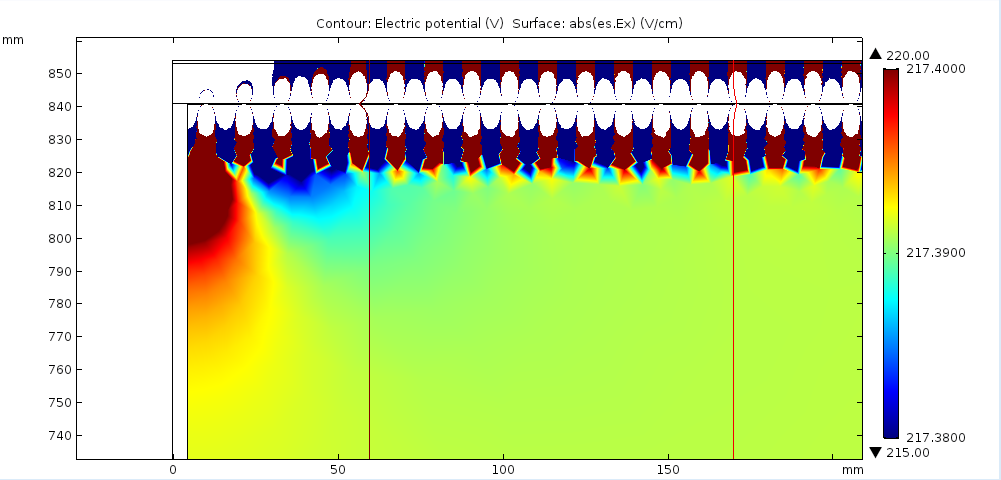}
    \includegraphics[width=7.8cm,height=5cm]{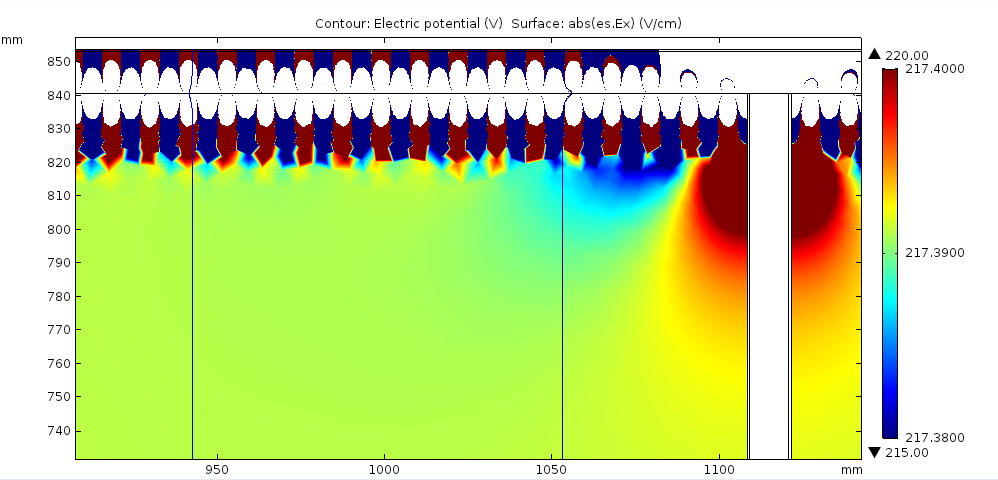}
    \includegraphics[width=7.8cm,height=5cm]{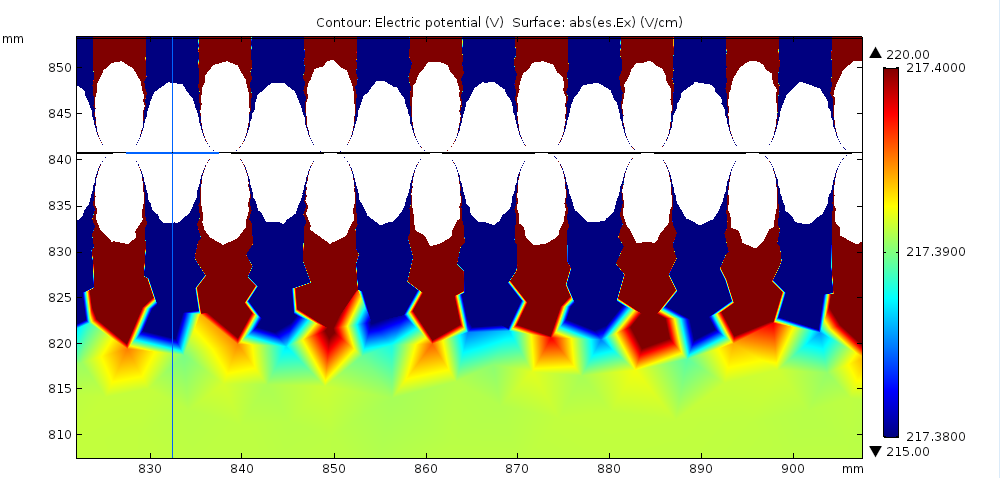}    
    \caption{Electric field simulation within the field cage equipped with a cathode foil of 13.2 mm. The strips layout is the same used for the current ND280 TPCs (11.5 mm pitch). As can be seen from the colored scale on the right of each plot, intense red and blue colors means a deviation of 0.01 V/cm respectively above and below the Electric field desired of 217.39 V/cm.
    Top: Electric field in the whole field cage (left) and close the anode (right).
    Bottom: Electric field close the cathode (left) and close the strips (right).
    The anode plane is set to 0 V.}
 \label{fig:efield_sim_old}
\end{figure}

\begin{figure}[h!]
    \centering
    \includegraphics[width=7.8cm,height=5cm]{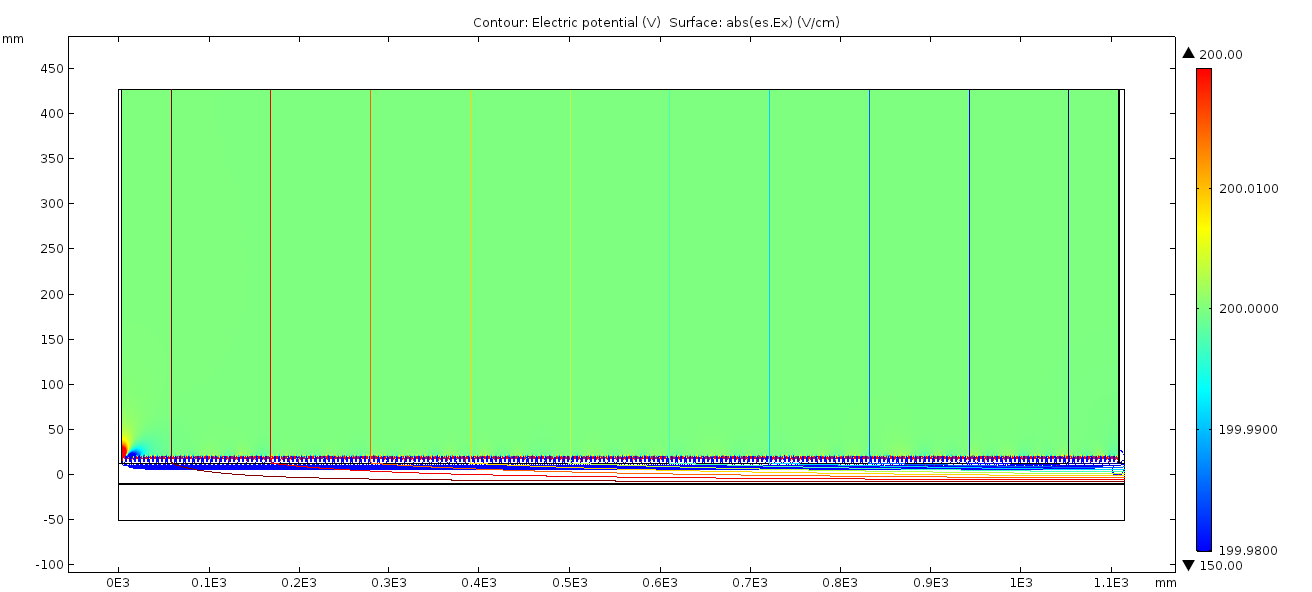}
    \includegraphics[width=7.8cm,height=5cm]{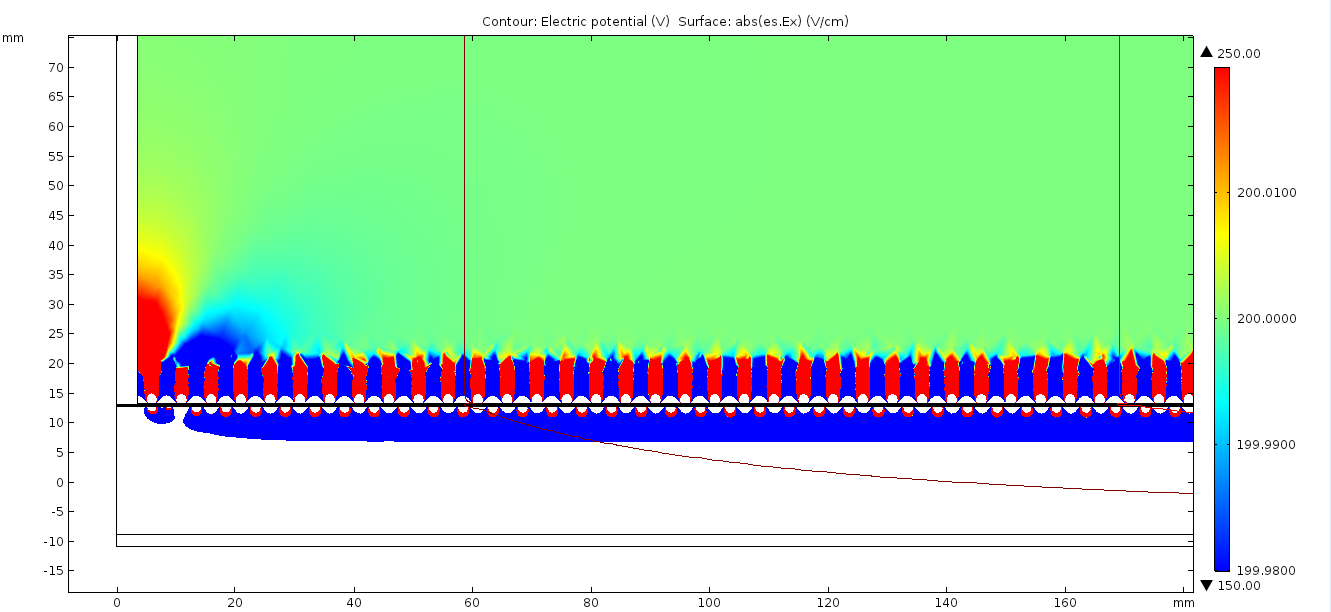}
    \includegraphics[width=7.8cm,height=5cm]{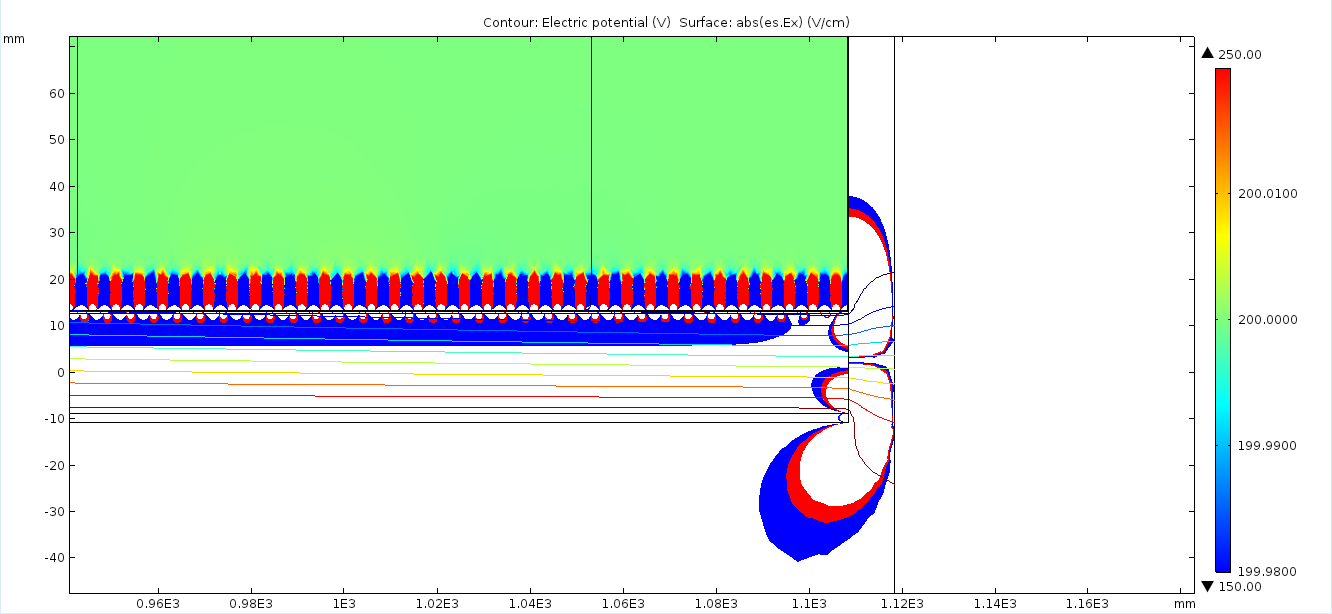}
    \includegraphics[width=7.8cm,height=5cm]{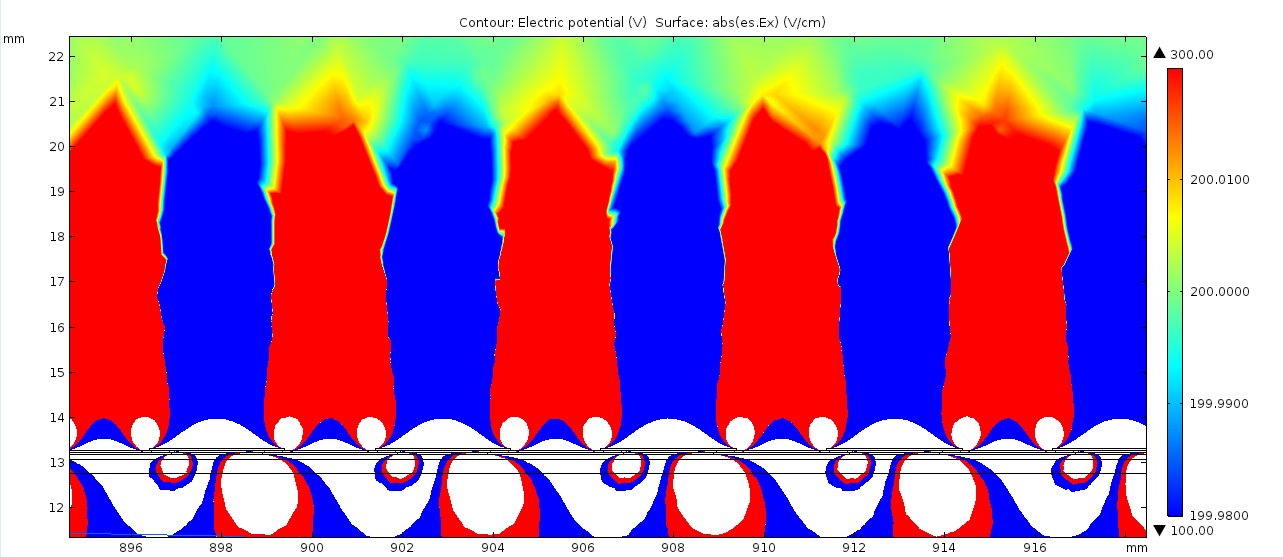}    
    \caption{Electric field simulation within a portion of the field cage equipped with a cathode of 13.2 mm. Mirror strips (3 mm) and field strips (3 mm) with a pitch of 5 mm are used. As can be seen from the colored scale on the right of each plot, intense red and blue colors means a deviation of 0.02 V/cm respectively above and below the Electric field desired of 200 V/cm.
    Top: Electric field in the whole field cage (left) and close the anode (right).
    Bottom: Electric field close the cathode (left) and close the strips (right).
    The anode plane is set to 0 V.}
    \label{fig:efield_sim_cathod_t2k}
\end{figure}

In order to achieve an electric field homogeneity better than 10$^{-4}$, the field cage (whose design is described in Section\ref{sec:tpc_fc}) is equipped with a layer of ``field strips'' and a second layer of ``mirror strips'' installed directly under the field strips (see Figure~\ref{Figure:strips}). Each mirror strip covers the gap between two field strips in front. 
Together, the two layers provide a shielding against external electrical influences on the internal field. 
With the help of finite-element field calculations several strip arrangements were investigated.
The first strips layout tested was the same of the current ND280 TPCs~\cite{Abgrall:2010hi}: 
a strip width of 10 mm with a 1.5 mm gap in between the strips, giving an 11.5 mm pitch. 
With this configuration, 95 field strip (plus two additional half field strips attached to the cathode and anode) and 96 mirror strips are used on each side of the field cage. 
This means that the field cage is equipped with 190 field strips plus 4 half field strips and 192 mirror strips in total.
The field shaping strips, together with mirror strips,
lie on stepwise decreasing potentials from the anode to the cathode and define the boundary condition for the electric field along the inside of the TPC barrel.
The cathode potential is set to -24 kV while the anode is grounded. 
Two adjacent strips have a voltage drop of 250 V, while the voltage drop between a field strip and a mirror strip is 125 V.
The simulation results are shown in Fig. \ref{fig:efield_sim_old}.
In the drift volume, the electric field obtained is $\sim$217.39 V/cm and it is  uniform to better than 10$^{-4}$ for distances larger than $\sim$25 mm from the inner side wall.
In order to increase the electric field uniformity region down to $\sim$10 mm from the inner side wall,
the layout chosen foresees a strip width of 3 mm with a 2 mm gap in between the strips, giving an 5 mm pitch for both field and mirror strips.
In this case, 220 field strip (plus two additional half field strips attached to the cathode and anode) and 221 mirror strips are used on each side of the field cage (440 field strips + 4 half field strips and 442 mirror strips in total). 
Due to the high number of strips, for this simulation, the COMSOL feature ``zero charge plane'' is used. 
This feature allows to simulate only a portion of the field cage by taking advance of symmetries of the field cage geometry.
In this case, the cathode potential is set to -22.1 kV while the anode is grounded. The high voltage has changed with respect to the previous simulations for convenience, 
in order to have an round value voltage drop between strips (of course what is relevant in these studies is the amount of transverse electric field component relative to the longitudinal component).
Two adjacent strips have a voltage drop of 100 V, while the voltage drop between a field strip and a mirror strip is 50 V.
Simulation results are shown in Fig. \ref{fig:efield_sim_cathod_t2k}.
In the drift volume, the electric field obtained is $\sim$200 V/cm, while the field is uniform to better than $10^{-4}$ for distances larger than $\sim$10 mm from the inner side of the field cage wall, more then two times better than the results obtained for the current ND280 TPCs shown previously.

Several simulations have been performed by checking different cathode thicknesses and it turns out that the electric field homogeneity is almost not influenced by the cathode width. In particular, concerning corner and edge regions next to Cathode and Anode structures, we estimated that field uniformity is degraded by larger extents with respect to other regions close to Field Cage walls (e.g. in the central part). 

\begin{figure}[h!]
    \centering
    \includegraphics[width=7.8cm,height=5cm]{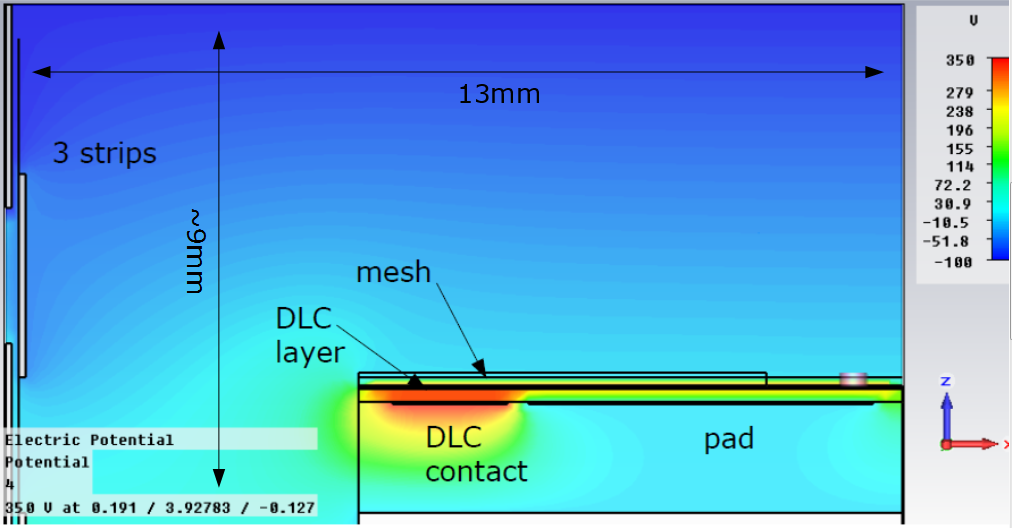}
    \includegraphics[width=7.8cm,height=5cm]{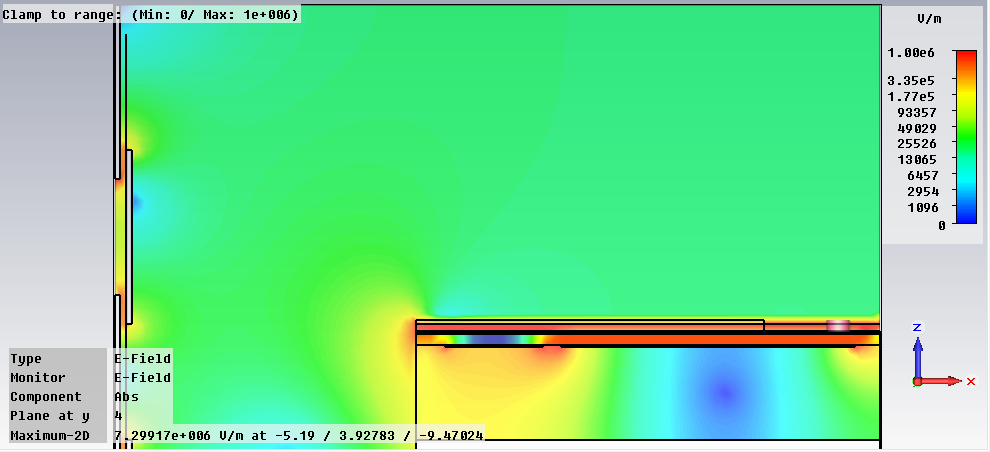}
    \caption{The Electric Potential and Field calculated in a $\sim 1\mathrm{cm}^3$ cubic region at the Anode, close to a~\MM~edge. The fields are shown on a cut-plane parallel to the drift direction and orthogonal to the Field Cage wall. Cross-sections of three strips and of the edge of a \MM are shown. The~\MM~distance from the Field Cage wall is assumed $5$mm. The transverse electrical field component is found to be limited to less than $10^{-4}$ relative to the longitudinal component at distances larger than $\sim$15mm from the Field Cage walls.}
    \label{fig:efield_sim_edge}
\end{figure}

Detailed studies were carried out to evaluate the amount of transverse electric field close to the field cage edges and corners. For instance Figure~\ref{fig:efield_sim_edge} represents the potential and the electric fields in a $\sim 1\mathrm{cm}^3$ cubic region at the Anode, close to a~\MM~edge. Two mirror strips and a field strip are represented (cross-section, in the plots' view) together with the detailed structure of a~\MM~edge (see Section\ref{sec:MM} for detailed description). Various potentials configurations were applied to the strips and the~\MM~internal anode electrode, being the~\MM~mesh set to ground potential. As a result we found optimized configurations for the field potential which allow longitudinal electric field uniformity to a level better than $10^{-4}$ for distances larger than $\sim$15mm from the Field Cage walls.

In summary we can state that the tracking quality is not affected at distances larger than 15 mm from the Field Cage walls all along the drift volume, including regions close to Cathode and Anode structures.



\clearpage

\section{Gas system}

The new TPC detector system for the T2K-ND280 upgrade proposal is composed by 5 modules. Each module has a maximum volume of about 3.68 m$^3$ (i.e. x2.3 m x 0.8 m x 2.0 m). The total detector volume is 18.4 m$^3$. The TPC detector is operated with a non-flammable three components gas mixture made of Ar-CF$_4$-iC$_4$H$_{10}$ (95\% - 3 \% - 2\%). Considering the large detector volume, the presence of CF$_4$ and the past experience, the gas system will continue to be based on gas recirculation. In order to limit at reasonable value O$_2$, CO$_2$ and H$_2$O contamination, the baseline circulation gas flow is set to 1 volume exchange every 6 hours. The injection of fresh mixture represents 10 \% of the circulation flow. These numbers are based on the experience accumulated during the operation of the existing 3 TPC modules. Therefore, during normal run conditions the total circulation flow will be 3.1 m$^3$/h (i.e. about 615 nl/h per TPC module) while the fresh injection is about 310 nl/h. 
The basic function of the gas system is to mix the three components in the appropriate proportions and to distribute the gas mixture to the 5 TPC units (that is the 3 existing TPC and the 2 HA-TPC).
The system proposed consists of several modules which profit of design experience and standards adopted for the gas systems of the LHC experiments at CERN. The gas system will be running on a Programmable Logic Controller and it will be controlled/monitored using the standard WinCC-OA SCADA interface : it is a supervisory control and data acquisition (SCADA) and human-machine interface (HMI) system. For all values (pressures, flows, mixing ratios) warnings, alarms and interlocks can be configured according to the specific detector needs. 
In the following some relevant module will be briefly introduced.

\subsection{Primary supply}

The gas system will be connected to the existing primary supply network for Ar, CO$_2$, CF$_4$, iC$_4$H$_{10}$ and N$_2$ (the latter it is mainly used to control pneumatic valves). Connection to primary supply as well as the possibility to monitor the gas levels will be investigated.   

\subsection{Mixer}

\begin{figure}[ht]
\begin{center}
\includegraphics[width=0.8\textwidth]{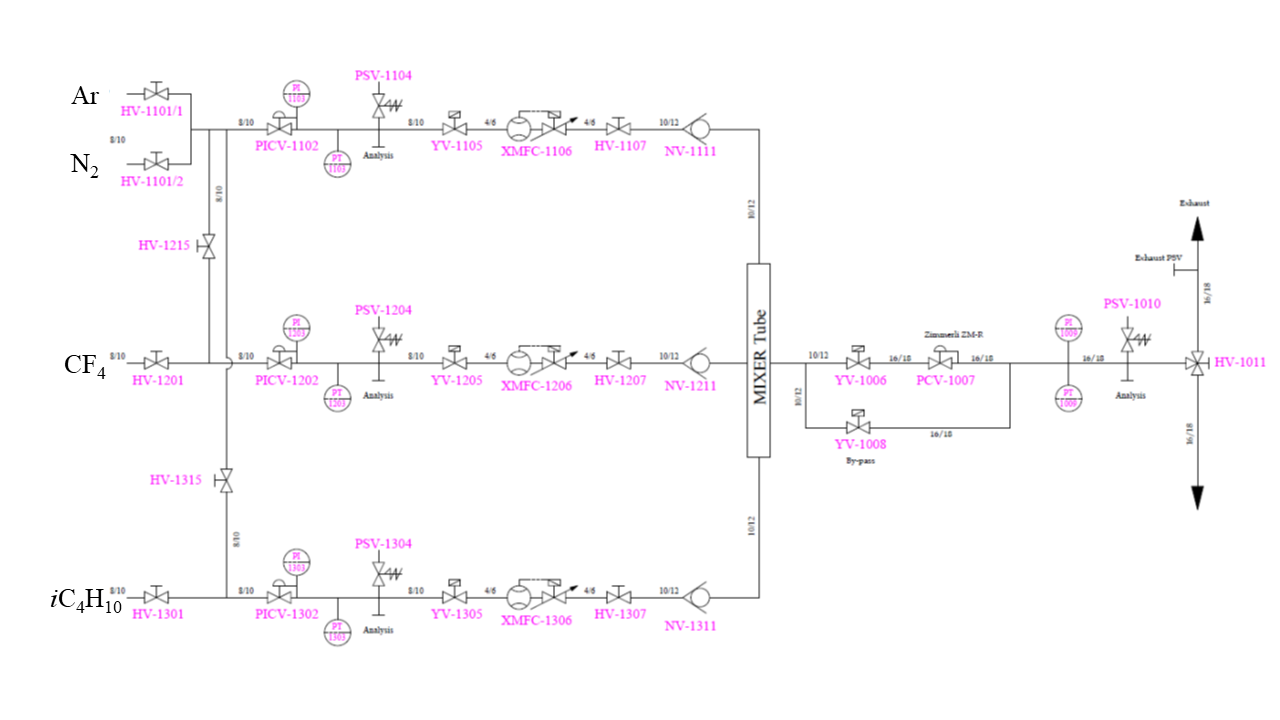}
\caption{Layout of the gas mixer unit.}
\label{fig:gasmixer}
\end{center}
\end{figure}

The flows of each gas component will be controlled by mass flow controllers (MFCs). The mixing ratio will be adjusted and monitored by the software control. The mixer unit will contain two sets of MFCs: the first called run is used for normal operation, while a second set with higher flow capacity will be employed for a fast filling of the detector with the run mixture (fill). 
In order to be operated in the optimal range for stability and reproducibility, the run set will have a total maximum flow capacity of about 1 m$^3$/h while the fill set will have a maximum capacity of about 7 m$^3$/h. Therefore, they will be normally operated at about 60-70\% of their maximum capacity. 
The possibility of a detector purge with a neutral gas (i.e. N2) is foreseen in the design of the mixer module.

\subsection{Closed-loop circulation modules}

\begin{figure}[htbp]
\begin{center}
\includegraphics[width=0.8\textwidth]{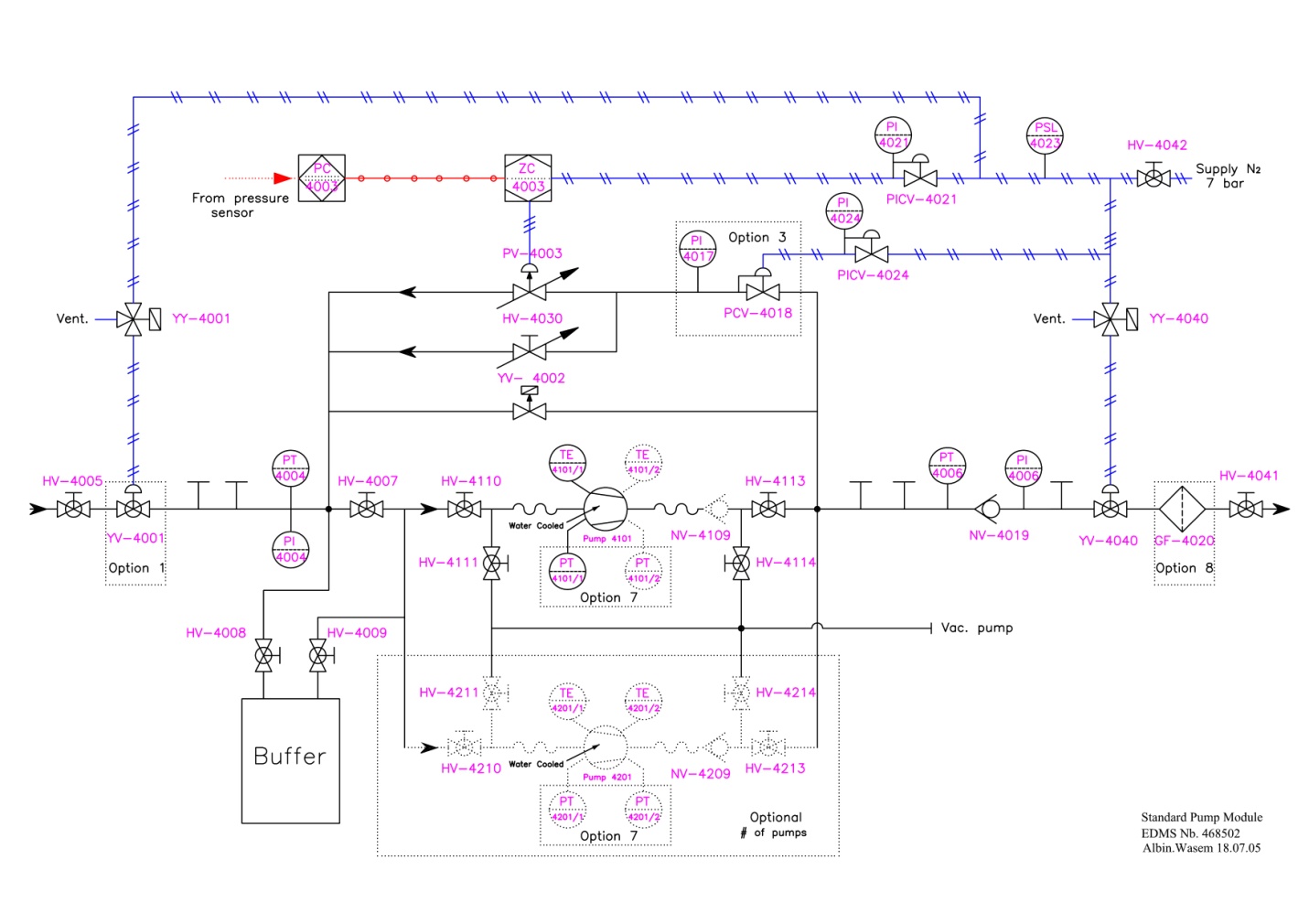}
\caption{Layout of the gas compressor unit.}
\label{fig:gasmixer-pump}
\end{center}
\end{figure}

In order to reduce the operational cost, the gas is circulated in a closed loop circuit. The circulation loop is distributed over three different areas:
Fresh mixture supply, mixture purifier and exhausted can be located in service area also far from the detector;
Circulation pump should be located close to the experimental;
Pressure controllers and final distribution module (i.e. between the 5 TPC modules) should be located very close to the TPC detector to minimize the pipe work. 
The mixture circulation in the main loop is ensured by the pump module. Figure~\ref{fig:gasmixer-pump} shows a typical design of a pump module. The flow capacity and therefore the input pressure are tuned by means of an automated regulation valve in a by-pass loop around the pump. A manual by-pass loop is also present. The pump input pressure is normally used as a setpoint for the operation of the module.
The final distribution module will contain 5 individual supply and return channels (one per each TPC module). Supply and return flows will be read by means of mass flow meters (MFMs). Manual needle valves will allow to adjust the flow in the individual channels. On the return lines, 5 automated regulation valves, will control the pressure in each TPC module. The final layout of the distribution module and the type of devices selected will have to consider the specific environmental conditions (i.e. for example the presence of magnetic field). 
Several impurities can be accumulated during operation in gas recirculation. A standard purifier module can be equipped with material able to remove O$_2$ and H$_2$O from the gas mixture. The removal capacity of other specific impurities (if any) will have to be tested and measured. The module contains two 24 liters cartridges filled with the suitable cleaning agent (normally molecular sieves and metallic catalysts are used for water and oxygen removal respectively). The purifier cycle is completely automated. During standard operation one cartridge is used for mixture cleaning, while the other is regenerated or ready and waiting for operation. Regeneration of metallic catalysts require the use of H$_2$ or mixture containing at least 3 to 5\% of H$_2$.

\subsection{Gas analysis}

A gas analysis module is used to continuously monitor with an automated cycles O$_2$ and H$_2$O concentration in the gas mixture. The module is completely automated: it can be programmed to sample all gas streams (for example supply and returns of individual TPC modules) including references or calibration gases. Expert operators can trigger remotely the analysis of specific lines at any moment. The analysis time of each stream can be tuned to compensate for the different pipe length between the analysis rack and the sampling point.
The presence of an infrared analyzer (IR) might be needed to ensure that the fresh gas mixture injected is always below the flammability level.


\section{TPC Gas Monitoring}

The gas inside the TPCs is subject to constantly varying ambient conditions that have an impact on the gas density.
To ensure consistent measurements across long periods of data-taking, those density effects have to be calibrated out.
For ND280, the temperature and pressure depended values are corrected to a standard temperature $T_0$ and pressure $p_0$.
At certain points in the gas flow, a small amount of 6 l/h gas are vented through the monitoring chambers.
This keeps the impact of the monitoring chambers on the main gas flow at a minimum and still assures a reliable measurement.

\subsection{Existing Gas Monitoring chambers}

For monitoring the supply and return gas of the existing TPCs, two
independent mini-TPCs were constructed with a design similar to
the large TPCs. The smaller Micromegas modules used in the
chambers where produced in the same way as the full size
Each of these two chambers measures both the drift velocity
and the gas amplification. To measure the drift velocity there are
two $^{90}$Sr sources above each chamber. They produce two lines of
tracks with a well-defined separation distance perpendicular to
the drift field. By measuring the time difference between the drift
times of two lines, the drift velocity can be calculated. Each drift
time measurement is triggered by signals from scintillating fibers
located directly below each chamber. For the gain measurement
there is one $^{55}$Fe source for each chamber. More details on these chambers are reported in Ref.\cite{Abgrall:2010hi}.

\subsection{Gain stability}
\begin{figure}[hbtp]
  \centering
  \includegraphics[width=.7\linewidth]{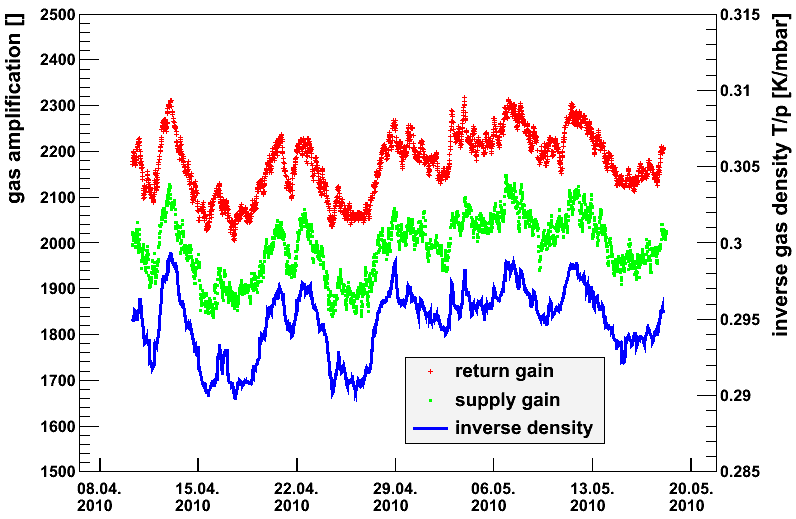}
  \caption[Gain stability over 6 weeks follows inverse gas density]{
    Gain measured by the monitor chambers over a period of 6 weeks is shown by the upper two sets of points, for the return and supply gas to the TPCs.
    The two monitor chambers have not been cross calibrated, resulting in a constant offset between the two measurements. 
    The lower curve shows the variation in the inverse gas density over the same period (using the scale on the right).
    The variation in gas gain is primarily due to atmospheric pressure changes.
  }
  \label{fig:gain_Tp_stbility}
\end{figure}

Fig.~\ref{fig:gain_Tp_stbility} shows a six-week history for gain as measured by the monitoring chambers for the gas supplied to the TPCs and returned from the TPCs.
As an overlay, the inverse gas density $T/p$ is plotted.
It can be seen that the gain variation over this period is less than $\pm10$\%, and is mostly due to gas density variations, primarily caused by atmospheric pressure changes.
This inverse gas density is used to correct the measured gain value of the gas monitor chambers and the TPCs.
A correction for density changes is given by
 $$ g_{\text{corr}}=\frac{ g_{\text{meas}} }{1 + \left(\frac{T/p}{T_0/p_0}-1\right)\cdot s} $$
with $T_0 = 298.15$ K and $p_0 = 1013$ mbar.

The relative change of the gain per relative change of $T/p$ is described by the slope $s$. It is extracted from the monitor chamber data of a given data taking period by correlating the measured gain with the gas density calculated from ambient pressure and temperature.
This correction is necessary because the temperature and pressure of the TPCs and the monitor chambers differ from $T_0$ and $p_0$ due to different barometric altitudes and climate conditions.
After applying the correction to this data, the remaining variation, due to other factors such as gas composition, is below 1\%.

\subsection{Gas Mixture Monitoring}
\begin{figure}[!ht]
  \centering
  \includegraphics[width=.85\linewidth]{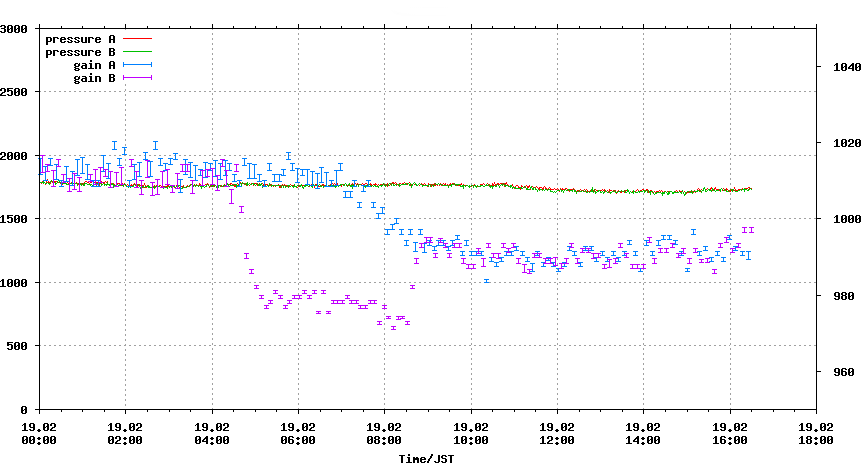}
  \caption[Change in mixture impacts gain]{
    Supply (B) and return (A) gas gain measured over a 16 h period.
    The pressure inside the gas volumes was constant.
    A drop in gas gain can be seen on the supply line gas that propagates to the return gas line after some hours.
    In this case, the cause could be traced back to a fault in the gas mixing.
  }
  \label{fig:gasmix_stability}
\end{figure}

In the current setup at ND280, both supply and return line of the TPC gas are monitored in dedicated monitoring chambers.
From the return gas flow, 90 \% are purified and recirculated with 10 \% replaced by new gas.
Since the return line is in equilibrium with the gas in the TPCs, measurements provided by the monitoring chamber on the return line are used for TPC calibration.
A monitoring of the supply line in addition provides a quick check of the stability of the supplied gas.
Fig.~\ref{fig:gasmix_stability} shows a drop in gain that is not visible on the return line at first.
After a recirculation time, the drop in gain is also visible on the return line.

The connection of the chambers to supply and return can be interchanged with a valve for added failure safety and systematics estimation.

\subsection{Integration in ND280}
The new HA-TPCs will be added into the existing gas flow to and from the current TPCs in parallel. 
Since their construction is not identical, both will have a dedicated connection to new monitoring chambers.
The proposed extension of the current gas monitoring system is to double the number of monitoring chambers.
With a total of four chambers, two will be used to monitor supply and return gas lines with the additional two chambers connected to the new HA-TPCs.

Due to the low flow requirement through the monitoring chambers, the feed lines have to be as short as possible.
Placing them above ground in the gas hut, where the current monitoring chambers are installed, would place them about 50 m of piping away from their gas source.
A suitable location would be on the service level in the ND280 pit.
About 10 U in a 19" rack would be required.
Additional exhaust lines will have to be installed for the outflowing gas, which can be made of low purity materials that are easy to install, i.e. plastic tubing.


\section{Resistive Bulk Micromegas Modules}
\label{sec:MM}

Micro-Pattern-Gaseous Detectors (MPGD) have been successfully used 
in a variety of particle physics experiments in the last two decades. They offer distinctive advantages in TPC with respect to wire chambers: while providing good gas amplification they do not suffer from the degradation of the resolution due to the E $\times$ B effect, substantially reduce the ion back-flow in the drift volume, and are free from the long-term ageing and mechanical constraints affecting wire chambers. They are therefore suited to paving large surfaces with minimal dead regions. 

The performance of the ND280 TPCs, the first large TPC built with MPGD, has in this respect been excellent. Since their installation in 2009, the 72 Micromegas are performing according to the specifications, without degradation of their response and without failures. 
In the following we will first describe the key feature of bulk Micromegas, then the resistive Micromegas and the resistive foils. Then we will describe the two detectors developed for the HA-TPC, called MM0 and MM1.

\subsection{Bulk Micromegas}

The detector modules of the TPC will be built using the bulk
Micromegas technology invented in 2004 by a CERN-Saclay
collaboration \cite{ref:bulk}. This technique provides an excellent solution
to minimize the unavoidable dead areas on the edges of a module
and allows large detection areas with excellent gas gain uniformity to be built. Moreover, such detectors can be manufactured in
a single process, reducing the production time and cost.
The bulk Micromegas technique consists in laminating a
woven mesh on a Printed Circuit Board (PCB) covered by a
photoimageable film. At the end of the process, the micromesh
is sandwiched between two layers of the same insulating material. The detector then undergoes UV exposure with an appropriate mask, followed by chemical development. A thin, few
millimeter wide border at the edge can thus be obtained avoiding
the need of an external additional frame to support the stretched
micromesh.

\subsection{Resistive Micromegas}

\begin{figure}[!htb]
\centering\includegraphics[width=0.8\textwidth]{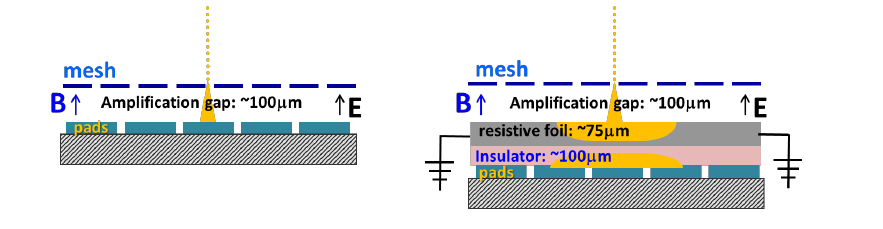}
\caption{Schematic cross-section of a normal bulk Micromegas (left) and a resistive Micromegas (right). The pads are covered by a layer of insulating material and a layer of resistive material.}
\label{fig:resistive}
\end{figure}

The ILC TPC has succesfully tested a new kind of detector, the resistive Micromegas~\cite{dixit, Bellerive}. A schematic cross-section view of this device is shown on Fig.~\ref{fig:resistive}. 

The pads are covered by a layer of insulating material and then by a layer of resistive material. The avalanche is then naturally quenched because the potential difference locally drops in presence of a high charge density. The resistive layer acts like a 2-D RC network and the charge deposited by the avalanche spreads naturally with time with a Gaussian behaviour. For a point charge deposited at $r=0$ and $t=0$, the charge density as a function of radius $r$ and time $t$ reads
\begin{equation}
\rho(r,t) = \frac {RC} {2 t} e^{- r^2 RC / (4 t)}
\end{equation}
where R is the resistivity per unit area and C the capacitance per unit area.
For our purpose, for an electronic shaping time of the order of 100 ns, the optimal resistivity is 0.4 MOhm/square, the optimal glue thickness 75 $\mu$m (controlling the capacitance C) : this will give a spread of 2.6 mm.

\begin{figure}[!htbp]
\centering\includegraphics[width=0.8\textwidth]{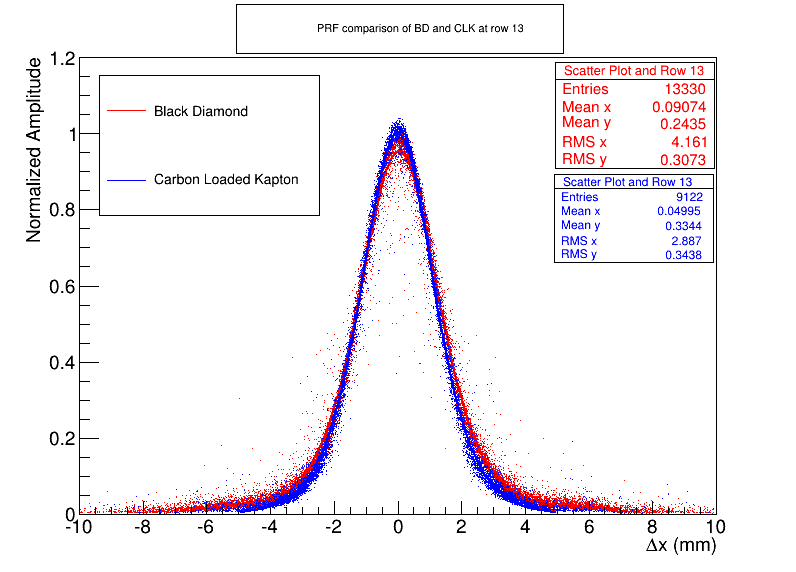}
\caption{Pad Response Function (PRF) for two types of resistive Micromegas tested by the ILC TPC collaboration  with a resistivity of 5 M$\Omega$/square.}
\label{fig:prf}
\end{figure}

In this way, even for small drifts, when the electron cloud width is small, the resistive layer will enable the charge to be detected over several pads. In the ILC TPC this configuration allowed to reach excellent spatial resolution of 70 $\mu$m even for small drifts. Examples of the Pad Response Functions (PRF) measured in ILC-TPC prototypes are shown in Fig.~\ref{fig:prf}. 

In our case, this device allows a readout structure with large pads, without compromising the space point resolution. In addition, the natural quenching properties naturally suppress Micromegas discharges (so-called sparks) and therefore no protection diodes are required for the front-end electronics.

We have developed the Micromegas to be used in the ND280 Upgrade TPC in two steps. In the first step, named MM0, we have used the same PCB as the one used for the Micromegas of the present TPCs, but covering the pads with a resistive foil. This Micromegas has been tested in the Summer 2018 test beam.

In the second step, called MM1, the PCB has been designed for the new TPCs and the pad number and size correspond to the final dimensions.

\subsection{Resistive foils}

Two techniques have been used for the production of resistive foils to be used in MPGD devices: carbon sputtering and screen printing with resistive ink.

The sputtering technique has proved to have some advantages like uniform resistivity and strong attachment to the substrate. It is the technique chosen for the Micromegas used in the New Small Wheel of the ATLAS experiment. 
Therefore, sputtering is our baseline option, however we will also investigate the performance of screen printing on a few prototypes.

Since several years some of us, involved in the RD51 collaboration at CERN, have been in contact with the Be-Sputter Company in Japan. They have a large chamber where a large foil (up to 1 $\times$ 4.5 m$^2$ foil) can be sputtered. The structure of the sputtered carbon is amorphous Diamond-Like-Carbon (DLC).

Special tests have addressed the mechanical robustness and found that the carbon surface does not suffer develop peeling and can be bent without changing the resistivity.
Chemical robustness to the chemical agents (acid and alkali) used in the PCB processing was also verified.

For low resistivity of 400 kOhm/square, pure carbon sputtering would require a thick layer and therefore a long processing time. It was found that a few percent of Nitrogen introduced in the sputtering chamber could act like a doping agent for the carbon. 
In this way the processing time could be reduced by an order of magnitude. For instance, for 3.2 \% nitrogen content, a layer of 700 Angstrom could be deposited in 42 minutes to reach 700 kOhm/square.

We will use a kapton (Apical polyimide) foil as substrate.

\subsection{Development of MM0}

The Micromegas detector MM0 was developed on the basis of the already existent Micromegas PCB used in the present TPC. It has a sensitive area of 36 x 34 cm$^2$, covered by 36x48 pads with 0.98x 0.70 cm$^2$. 
The thickness of the PCB is 2.2 mm and
comprises three layers of FR4 with blind vias in the inner layer.
This solution avoids the gas-tightness problems arising from the
conventional two-layer structure with vias sealed with epoxide
resins. The top conductive layer forming the anode pad plane is
made of 25 $\mu$m thick copper deposited on FR4. The other three
conductive layers are used for the routing network, grounding
and pad-readout connectors.

The pad surface was covered by a 200 $\mu$m insulating layer acting as the capacitance, and then a 50 $\mu$m kapton (Apical) with a  thin Diamond-Like-Carbon layer (DLC). In the first two detectors the resistivity was 2.5 MOhm/square.

On top of this surface, a bulk Micromegas was built, with a 128 $\mu$m amplification gap. The mesh is a 400 LPI stainless steel woven mesh, with 19$\mu$m wire diameter.

One feature of this design is that we had to adapt the PCB to bring the electrical contact to the DLC and that we needed to provide a sufficient resistance to ground even in the pads on the perimeter of the structure. To realize this, the peripheral pads are partly covered by the photoimageable Pyralux and only a part of their surface is available for gas amplification.

The test of MM0 was first done in Saclay on the test bench shown in Fig.\ref{fig:testbench}. Then it was mounted on the ex-HARP TPC field cage at CERN and further tested there both with cosmic rays and on the PS T9 testbeam as described later.

\subsection{Development of MM1}

\begin{figure}[htbp]
\begin{center}
\includegraphics[width=0.8\textwidth]{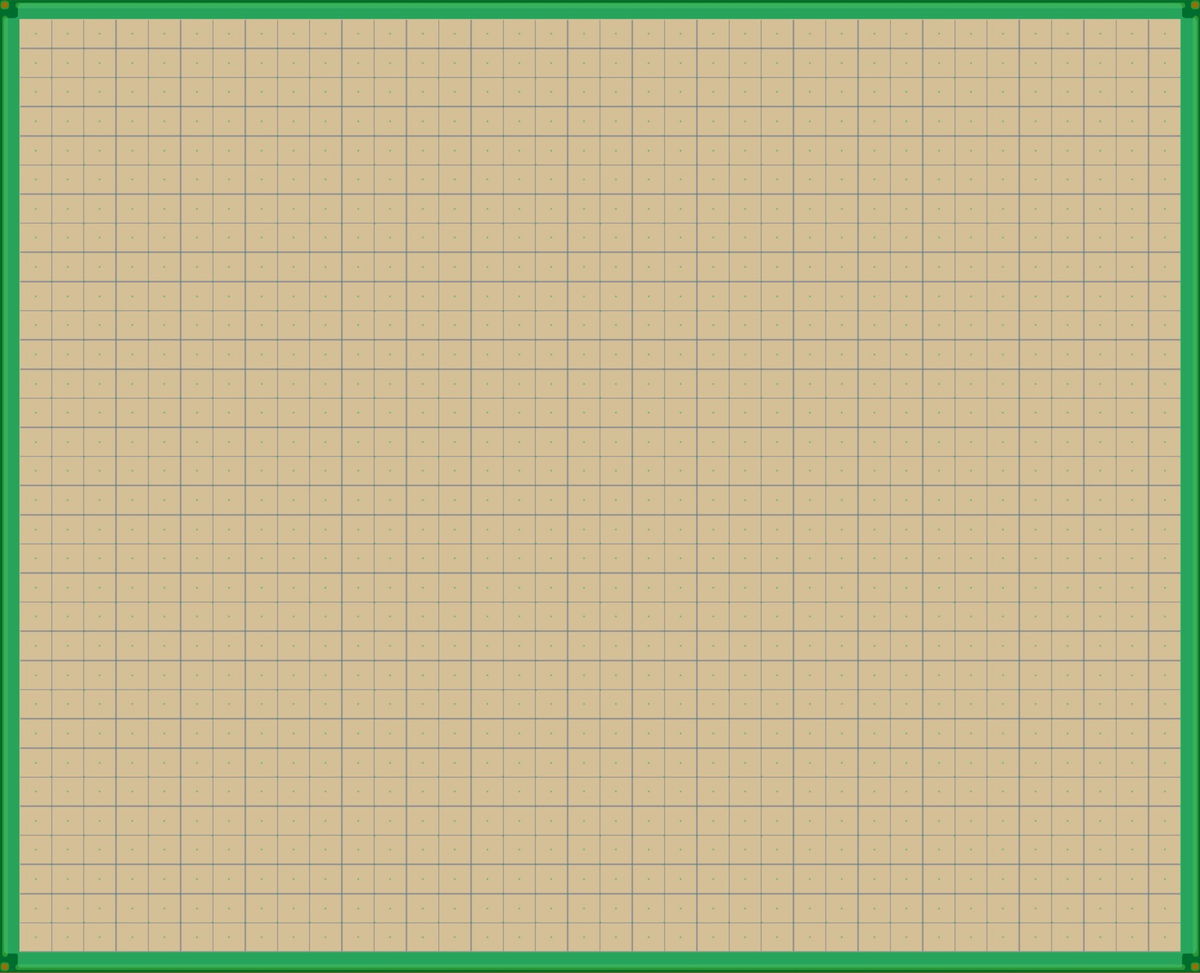}
\caption{Pad side of the MM1 Printed Circuit Board. The micromesh is connected to ground through silver pasted connections in the 4 corners of the PCB.}
\label{fig:mm1-padside}
\end{center}
\end{figure}

\begin{figure}[htbp]
\begin{center}
\includegraphics[width=0.8\textwidth]{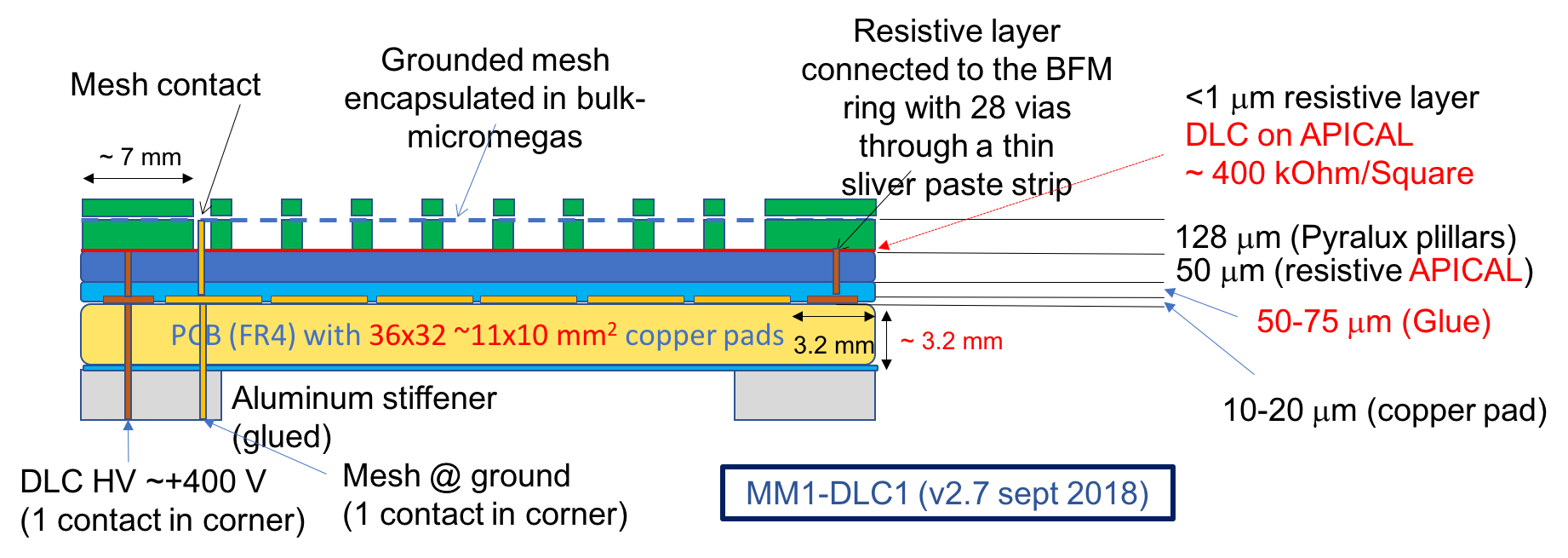}
\caption{Transverse cut of the MM1 module.}
\label{fig:mm1-transversecut}
\end{center}
\end{figure}

The Micromegas detector MM1 is being developed for the ND280 Upgrade TPCs. It has dimensions of 34 x 42 cm$^2$, covered by 32x36 pads with 1.1x 1.0 cm$^2$ (Fig. \ref{fig:mm1-padside}).
The pad surface is covered by a 75 $\mu$m insulating layer acting as the capacitance, and then a 50 $\mu$m kapton (Apical) with a thin Diamond-Like-Carbon layer. The resistivity corresponds to the design value of  0.4 MOhm/square. A transverse cut of MM1 is shown in Fig.\ref{fig:mm1-transversecut}. A 7 mm frame will be necessary to keep the mesh in position, to provide a sufficient resistance to ground for the DLC layer, and to guarantee a safe HV insulation between the grounded mesh and the few hundreds volts polarizing the DLC (see Fig.\ref{fig:mm1-zoomcorner}).

On top of this surface, a bulk Micromegas is built, with a 128 $\mu$m amplification gap. The first MM1 are under construction at the EP-DT-EF workshop at CERN and we expect the first detectors to be available at the beginning of 2019.

\subsection{Production of the Micromegas modules}

\begin{figure}[htbp]
\begin{center}
\includegraphics[width=0.6\textwidth]{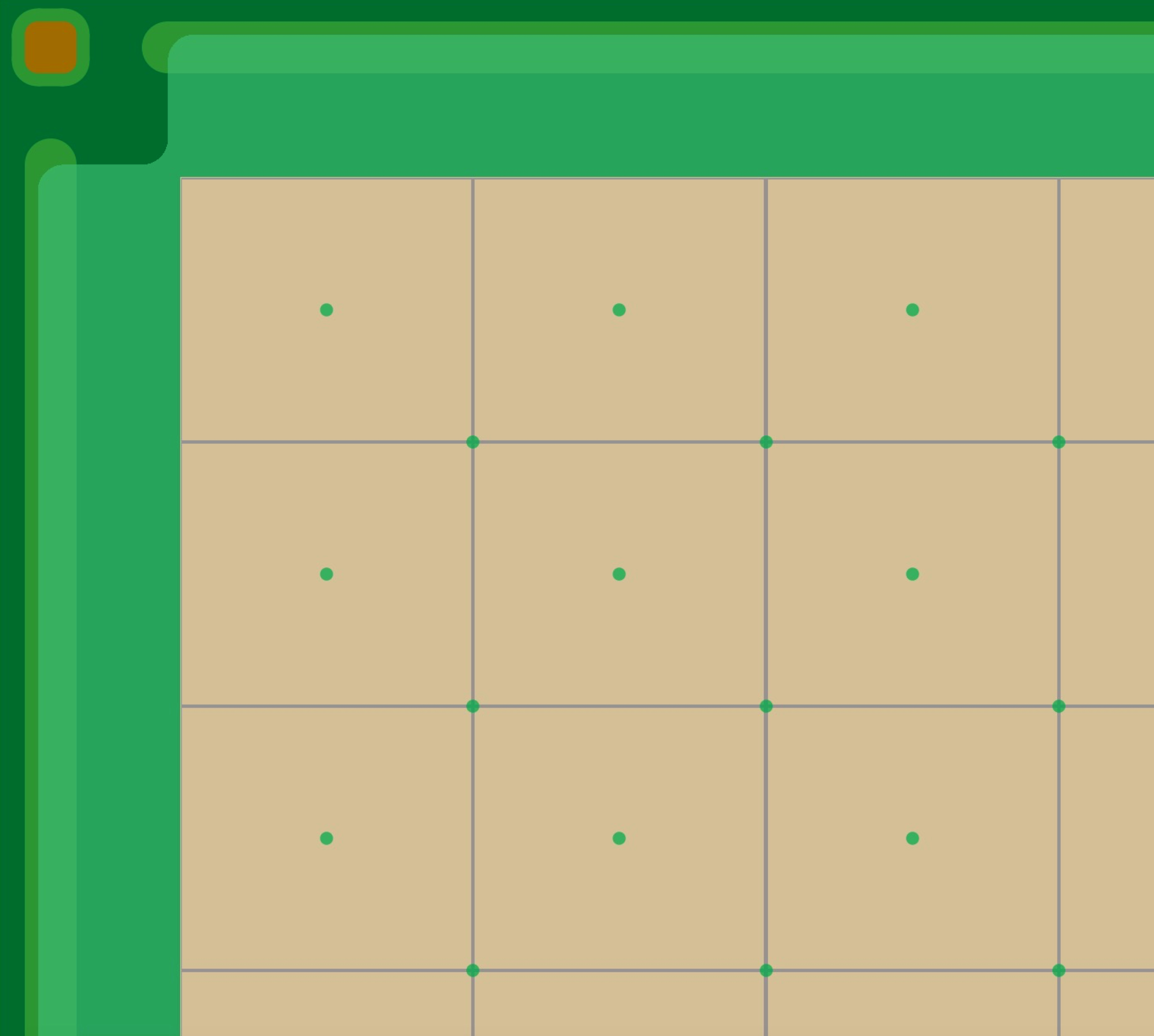}
\caption{Zoom on a corner of the MM1 Printed Circuit Board. The micromesh is grounded through a silver paste connection in the corner (brown orange pad). The Pyralux layer covers the 7 mm large border of the PCB and makes the spacing pillars in the active area (brown green). The DLC-clad Kapton foil is in light green.}
\label{fig:mm1-zoomcorner}
\end{center}
\end{figure}

The 32 bulk Micromegas modules instrumenting the two
HA-TPCs will be produced over a period of approximately 12 months 
starting in Fall 2019 by
CERN/EP-DT-EF. First a layer of insulating material and the DLC foil will be laminated onto the PCB. Then a sandwich of two layers of 64 m m Pyralux
PC1025 photoimageable polyimide by DuPont, 14 a woven micro-
mesh and finally a layer of Pyralux were laminated on the PCB.
The micromesh  manufactured by the BOPP company (Switzerland)
is made of 18 $\mu$m
diameter 304 L stainless steel wires. After weaving, its thickness
is reduced by 20-30\% by lamination. The wires are spaced with
a pitch of 63 $\mu$m (400 LPI). During the manufacturing process, the
micromesh is held on an external frame with a tension of about
12 N. This procedure guarantees sufficient flatness of the micro-
mesh during lamination and thereby a uniform amplification gap
over the entire sensitive area of the detector module. At the end of
the photoimaging process, the micromesh is held in place by a
7 mm Pyralux border and by  2237 regularly distributed pillars,
maintaining the amplification gap of 128 $\mu$m. The pillars 
are cylindrical with a diameter of about 0.5 mm. 
They are placed in the center and at each corner of the pads (see Fig.\ref{fig:mm1-zoomcorner}). The active area represents about 95\% of the module surface.
After development, the bulk Micromegas detector undergo cleaning and
baking processes to achieve complete polymerization of the
Pyralux material.

After the lamination, the Hirose connectors (16 per module) will be soldered on the detector in an oven reaching $220^\circ$ C.

\subsection{The Micromegas and associated electronics and mechanical structure}

\begin{figure}[ht]
\begin{center}
\includegraphics[width=0.8\textwidth]{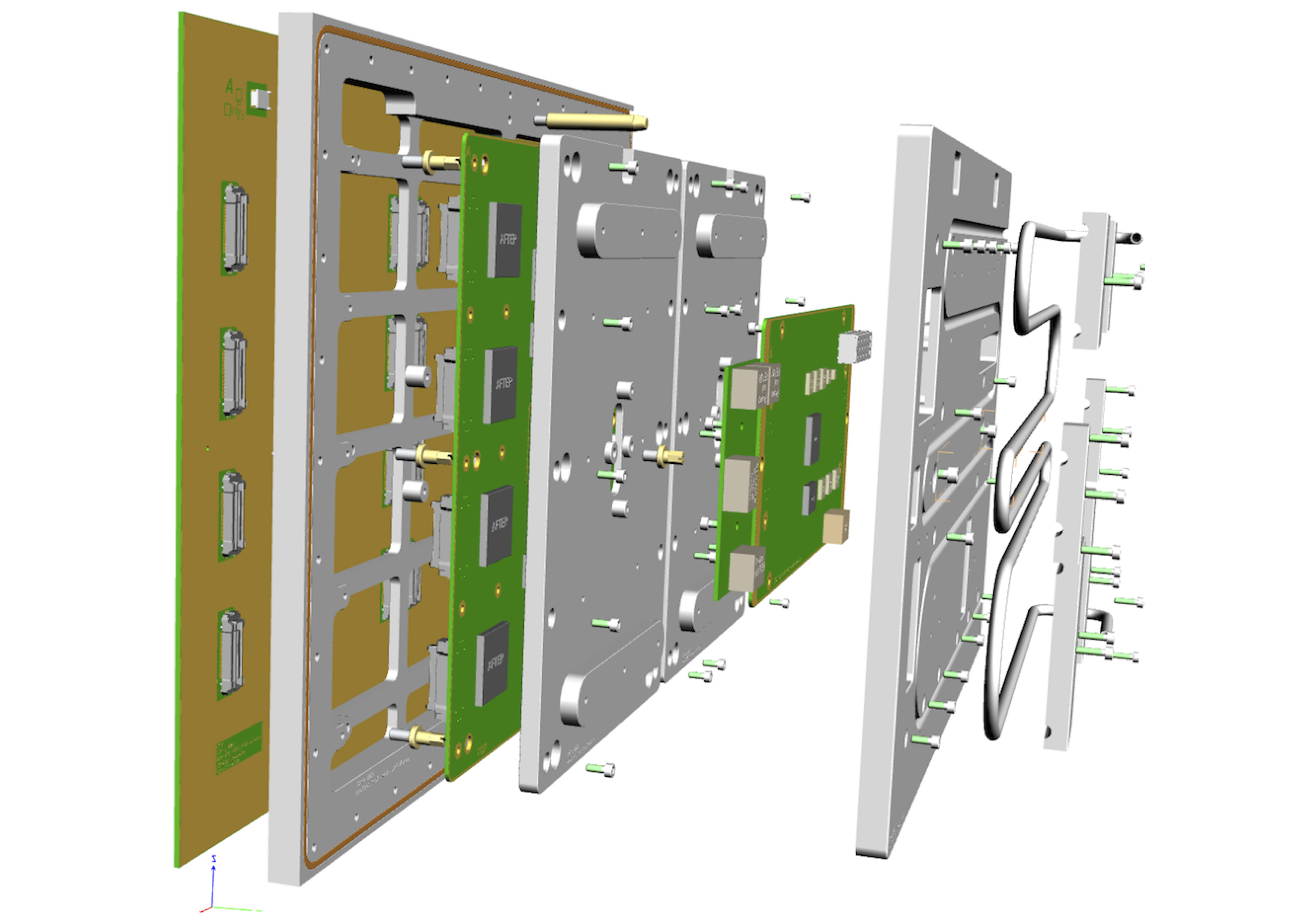}
\caption{The structure of a Micromegas module with its associated electronics and mechanical structure.}
\label{fig:mm1-exploded}
\end{center}
\end{figure}

Each Micromegas will be first glued to a support structure hosting the O-ring, called stiffener and machined out of an 20 mm thickness aluminum plate.
The thickness of the glue layer will be adapted in order to 
reach the required tolerance of 100 $\mu$m on the position of the Micromegas mesh with respect to the drift field, in the direction of the electric field.
 Then it will be mounted on the module frame and then be connected to the various electronics module and their cooling plates.
An exploded view of the full structure is shown in Fig.\ref{fig:mm1-exploded}.

It comprises:\begin{itemize}
\item the Micromegas module and its stiffener;
\item two Front-End cards (FEC); 
\item two FEC cooling plates; 
\item the FEM and PDC cards; 
\item the FEM cooling plate, connected to the water cooling piping. 
\end{itemize} 

\begin{figure}[htb]
\centering\includegraphics[width=0.8\textwidth]{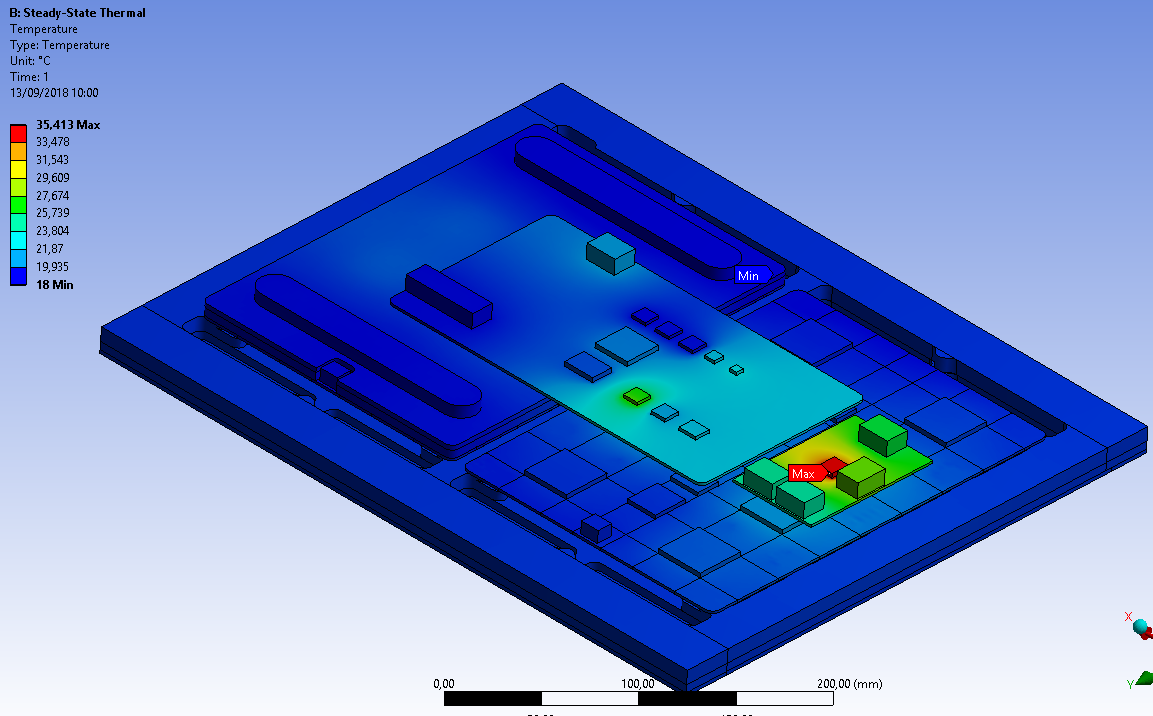}
\caption{Simulation of the cooling of the front-end card.}
\label{fig:cooling1}
\end{figure}

The thermal behaviour of the module has been investigated by developing a finite element model which considers the heat generated by the front-end electronics and dissipated into the cooling system. The power dissipation of each FEC board is about 10.5 W,  8.5 W for the FEM, whereas it is 7 W for the PDC power supply board. Each board is mounted on a cooling plate which conducts the heat away and transfers it by conduction to the cooling channel; the system is cooled down by water at room temperature and the flow rate is about 0.7 l/min. The resulting temperature field is shown in Fig.~\ref{fig:cooling1} ; it is worth noting that the maximum temperature is occurring on the DC/DC chip of the power supply board and it is $35^\circ$C. Further finite element analysis will be carried out in the next future to optimize the current mechanical design and improve the overall thermal performance of the system.

\subsection{Quality control}

Each Micromegas will be validated with a three steps procedure.
First, after the fabrication it will be tested in air at a voltage of 900 V. The requirement is that the current should be less than 10 nA. This assures that the module has no major short-circuit or other defects.

Then each Micromegas will be mounted on a dedicated test bench, described in the following section, where its amplification properties will be tested.

Finally, each TPC will be tested at CERN with cosmic rays before shipping to Japan. This will allow to be mounted and then tested it in its experimental environment.

\subsection{ Test bench for the production of the Micromegas detectors}

The test bench for the HA-TPC Micromegas detector is mainly used to 
qualify the gain and performances and to provide an absolute calibration
for the signals detected by the TPC. The architecture of the test bench will be similar to the one used in the T2K experiment in 2008~\cite{1742-6596-65-1-012019}. 
It consists of a plastic gas chamber of dimensions 50$\times$50 $\times$ 15 cm$^3$. The cathode side is closed by a thin mylar foil. Copper strips glued inside and connected through a resistor chain will provide the drift field over a distance of 15 cm.
A robotic arm equipped with a $^{55}$Fe source will be used to provide a narrow, collimated beam of X rays generating an input signal for each pad of the detector. 

\begin{figure}[htbp]
\begin{subfigure}[b]{.5\linewidth} 
\centering
\includegraphics[width=0.9\textwidth]{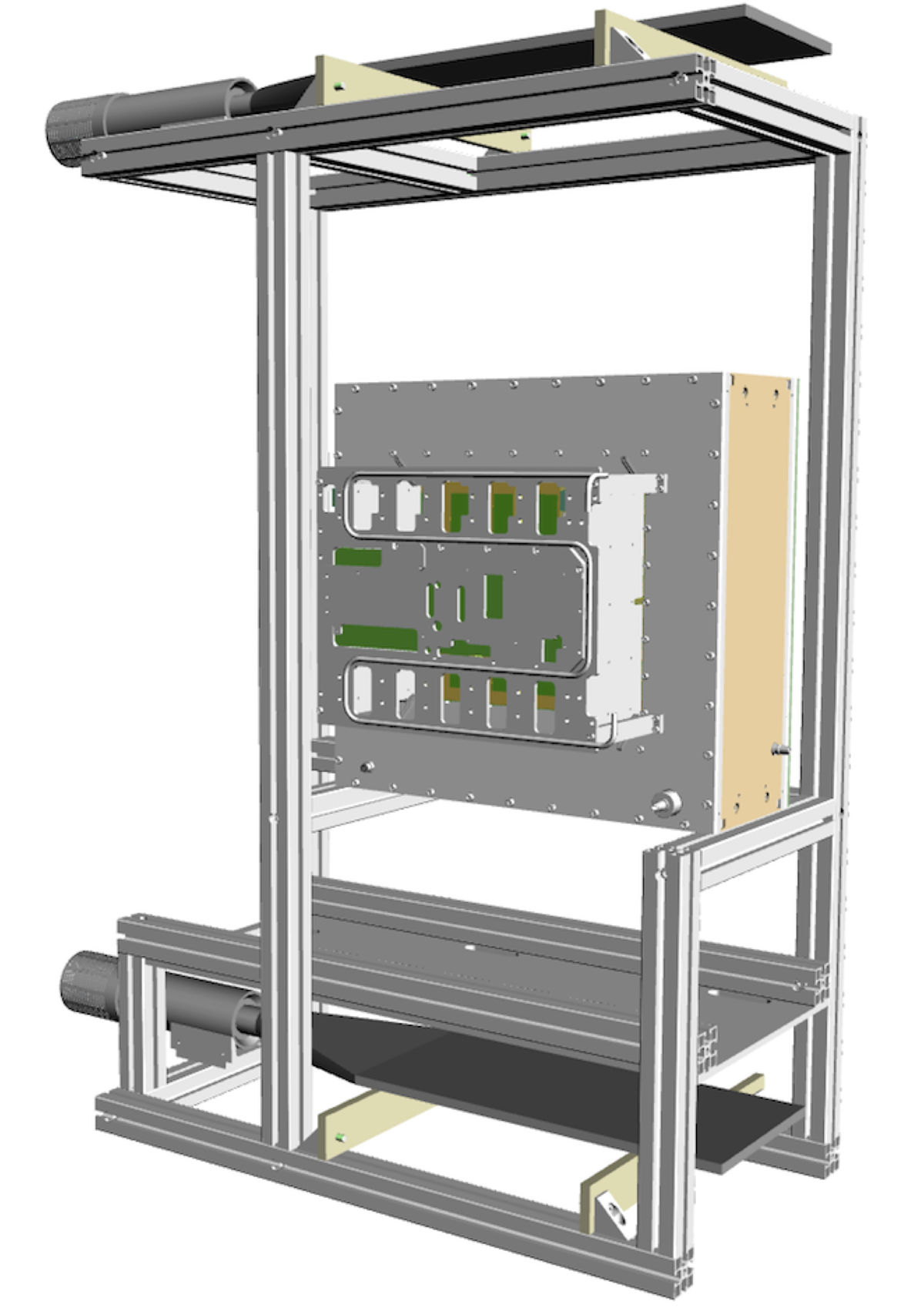} 
\caption{}
\end{subfigure}%
\begin{subfigure}[b]{.5\linewidth} 
\centering
\includegraphics[width=0.9\textwidth]{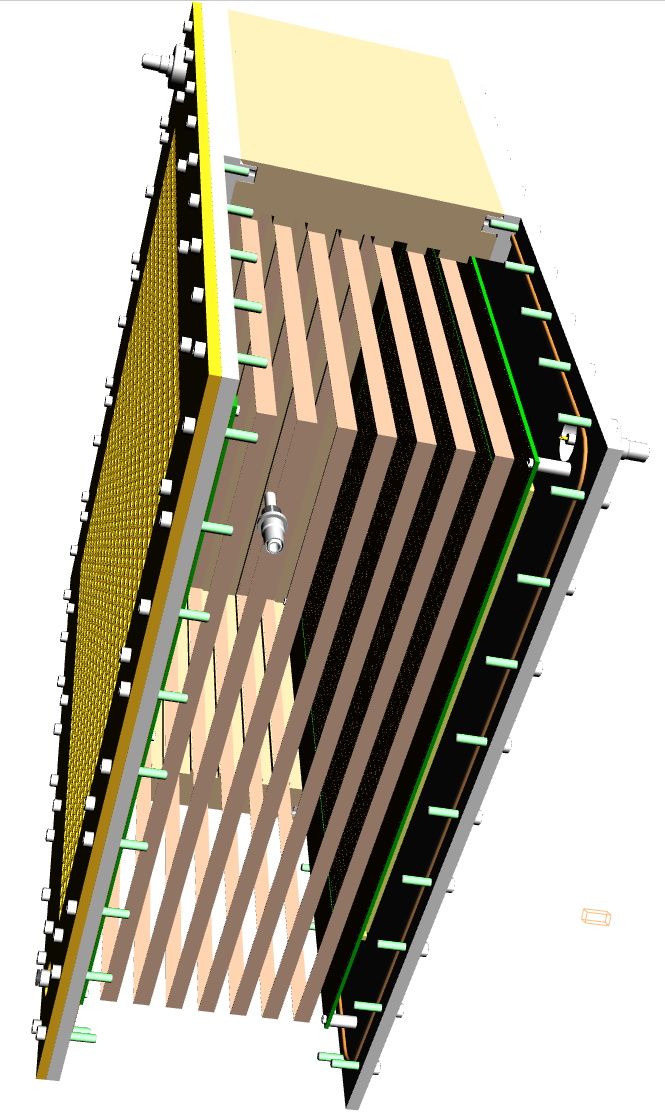} 
\caption{}
\end{subfigure} 
\caption{
Test bench used in the validation of MM0 prior to the test beam and which will be used for the tests of some of the MM1 modules.
On the right hand side a view of the inner structure of the test-bench gas chamber. The copper rings defining the drift electric field are visible.
}
\label{fig:testbench}
\end{figure}

Each produced Micromegas will be mounted on this setup and thoroughly tested.
The scan of the whole detector in the XY directions will provide the following information: uniformity, dead pads, gain map and energy resolution. As the X-ray conversion region will be narrow it will be possible to measure the spread of the signal and verify the spatial resolution, which is an important factor, especially with resistive Micromegas detectors.
A view of a similar system used in tests of MM0 and MM1 is shown in Fig.~\ref{fig:testbench}.

\section{Electronics}
\subsection{Outline of the architecture}
\begin{figure}[ht]
\begin{center}
\includegraphics[width=0.8\textwidth]{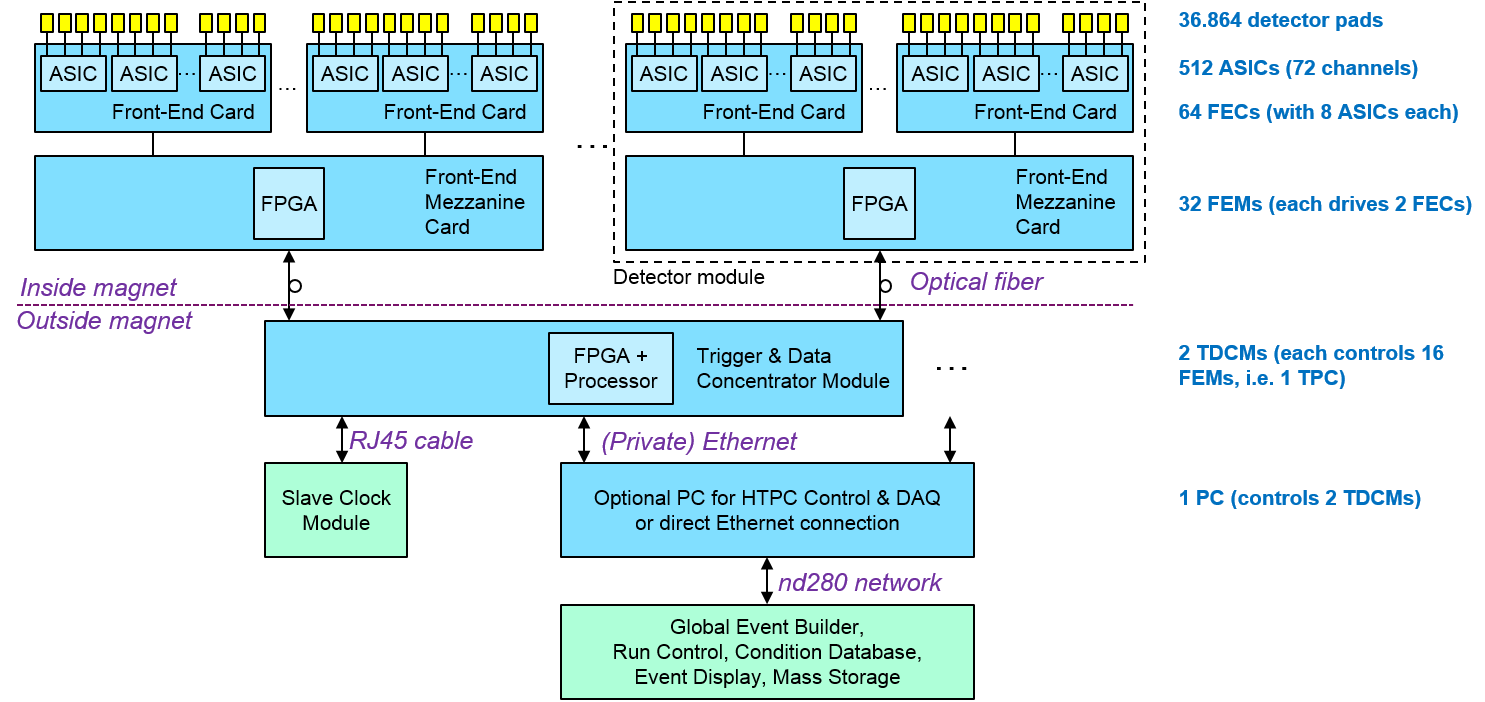}
\caption{HA-TPC readout architecture.}
\label{fig:htpc_readout_architecture}
\end{center}
\end{figure}

\begin{figure}[ht]
\begin{center}
\includegraphics[width=0.8\textwidth]{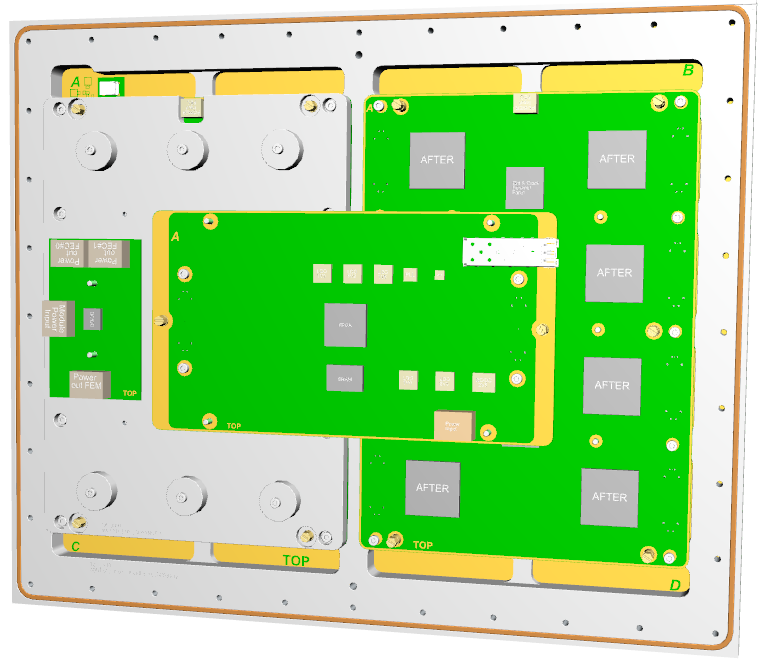}
\caption{Preliminary drawing of the detector module with its front-end electronics.}
\label{fig:htpc_front_end_electronics}
\end{center}
\end{figure}

The architecture of the complete readout system of the HA-TPCs is schematically shown on Fig. ~\ref{fig:htpc_readout_architecture}. It is based on the replication of the modular structure used to read out each of the 32 Micromegas detector modules that comprises the system. Each detector module holds three types of electronic boards:
\begin{itemize}
\item Two Front-End Cards (FECs), with 576-channel each. These capture the analog signals of the 1152 pads of the detector module and convert the acquired samples in digital format using an octal-channel analog to digital converter (ADC).
\item A Front-End Mezzanine card (FEM). This controls the two FECs and performs some elementary data processing such as baseline offset correction, zero-suppression and temporary data storage. 
\item A Power Distribution Card (PDC). This performs the local conversion of the externally higher supplied voltage (e.g. 24 V) to 4-5V used by the FECs and the FEM. Instead of being a separate dedicated board, the PDC could also be integrated on the FEM.
\end{itemize}
In order to minimize the degradation of the highly sensitive detector analog signals and avoid the high cost of cables, the FECs and the FEM are directly mounted at the back of detector modules. This is comparable to what was done on the existing TPCs, except that the reduction of channel count and the suppression of the anti-spark protection circuits which are no longer needed with resistive Micromegas detectors allow mounting the FECs parallel to the detector plane, rather than perpendicular. This leads a significantly more compact layout. A preliminary drawing of the detector module with its front-end electronics installed is shown on Fig. ~\ref{fig:htpc_front_end_electronics}.
The two FECs are side-to-side at the back of the Micromegas detector. The FEM plugs on-top of the FECs. The PDB is the small card close to the FEM. Each card has an aluminum carapace for shielding, mechanical protection, and for conducting the dissipated heat to a water pipe serpentine (not shown). The digitized and pre-processed data of each detector module is transported outside of the detector magnet via an optical fiber to a back-end unit called the \guillemotleft Trigger and Data Concentrator Module\guillemotright,  TDCM. Each TDCM aggregates the data of the 16 modules of one HA-TPC and distributes the global clock and common trigger signal to the FEMs using the return path of the corresponding optical link. Each TDCM runs locally a MIDAS front-end program and interfaces to the global run control and DAQ of the nd280 detectors via a standard Gigabit Ethernet link. Alternatively, an intermediate PC that bridges the command interpreter program running on the TDCMs to the nd280m network and the MIDAS software environment could be used.
\subsection{Detailed Description of the main components}
\subsubsection{Readout ASIC}
Several options have been considered for the readout ASIC of the HA-TPC: the AFTER chip ~\cite{ref:after}, designed for T2K and used in the current TPCs and FGDs, its successor, the AGET chip ~\cite{ref:aget}, and some other chips. However, the improvements and additional features of the newer devices would not bring any real benefit compared to the original AFTER chip given the requirements of T2K. Therefore, the AFTER chip, which is a proven solution, is retained for the readout of the HA-TPCs.
The AFTER chip is a 72-channel device that includes preamplifiers and shapers with programmable gain and shaping time (4 values of gain from 120 fC to 600 fC and 16 values of peaking time from 100 ns to 2 \(\mu\)s) coupled to a 511-time bucket switched capacitor array (SCA). The AFTER chip requires an external trigger signal. An external ADC is required to digitize the samples captured in the SCA. The maximum write speed of the SCA is 100 MHz and the maximum readout speed is 25 MHz.
\subsubsection{Front-End Cards}
Each FEC holds 8 AFTER chips leading 576 readout channels. The AFTER chips amplify and shape detector pad signals, sample them at 10-100 MHz in a 511-bucket SCA which is digitized upon trigger by a commercial octal-channel 12-bit ADC (e.g. Analog Devices AD9637 or equivalent) which is clocked at 12.5 MHz. All AFTER chips are digitized in parallel and the minimum incompressible dead-time for acquiring the 72 \(\times\) 511 time buckets of the complete SCA is approximately 3.3 ms. The inputs of the AFTER chips are connected via small surface mount capacitors to connectors that mate to their counterpart on the detector side. Eight surface mount 80-pins connectors are used to connect to the detector. To reduce the insertion and extraction forces and ensure good contacts, floating type connectors are chosen. Two candidate products have been considered: Iriso 9827 series and Hirose FX23L/FX23 series. Decision was made to retain Hirose connectors which have a higher floating range and are easier to procure than Iriso connectors. The other elements of the FEC are a 14/16-bit DAC for injecting pulses for test and individual channel calibration, small logic, voltage regulators and passive components. Similarly to the FECs of the current TPCs, the FECs for the HA-TPC do not have any programmable component and neither require firmware nor embedded software.
The FEC is expected to draw 2.5 A from a 4.2 V power supply leading to about 10.5 W of dissipation. The FEC is normally powered by the FEM, through the 4 dedicated contacts of the FEM to FEC interface connector. Alternatively, the FEC can also be powered via a cable from the PDC, or an external source (for laboratory tests for example).
\subsubsection{Front-End Mezzanine Card}
Each FEM performs the control, synchronization and data aggregation of the two FECs of a detector module. The FEM is connected to each of its two FECs via an 80-pin Hirose FX23L/FX23 floating type connector. A mid-range FPGA (Xilinx Artix 7), coupled to a memory buffer (Cypress 1 M \(\times\) 36 bit SRAM with NoBL architecture) and ancillary logic, implements all the required functions in the FPGA fabric. The FEM does not incorporate any soft or hard core processor and consequently no software runs locally. The memory buffer has sufficient capacity and write access speed to store up to three complete events interspaced with the digitization time of the SCA of the AFTER chips (i.e. 3.3 ms). Retrieving events from the buffer memory is done at lower speed (50 MSps), and processing a complete event in the FEM takes approximately 12 ms. The data that remains after zero-suppression, or complete raw data during pedestal runs, are transferred to the back-end data concentrator over a medium speed optical serial link (200 Mbps bandwidth for the transport of event data). A standard small form factor pluggable transceiver (SFP) will be used, either a bi-directional single fiber model, or a classical dual-fiber model. Current, voltage and temperature monitoring of the two FECs and the FEM are supervised by the local FPGA of the FEM and the corresponding data are time multiplexed over the optical link along with detector data. A fraction of 100 Mbps over the total 400 Mbps bandwidth of the FEM to TDCM optical link is reserved to that end. The remaining 100 Mbps of link bandwidth available from the FEM to the TDCM is reserved for fast traffic related to the trigger: acknowledge, set busy flag and release busy flag.
The FEM is expected to draw 2 A from a 4.2 V power supply, leading to about 8.5 W of dissipation. The total current supplied at the input of the FEM is expected to be 7 A because it includes the supply current of the two FECs.
\subsubsection{Power Distribution Card}
The role of the Power Distribution Card is to convert the externally supplied input voltage, selected from 12 V to 24 V and possibly more, to the 4-5 V supply voltage needed for the FEM and the two FECs of a detector module. The PDC mostly include a DC/DC converter and the I/O connectors for power supply cables. The DC/DC converter must be compatible with the 0.2 T magnetic field of T2K magnet. Derating the current of a standard DC/DC converter, using air-core inductors, or placing a small magnetic shield around a common ferrite inductor are the options being considered. Some of the components under investigation include Texas Instruments LMZ13610, \guillemotleft10A Simple Switcher Power Module with 36V Maximum Input Voltage\guillemotright and Linear Technology LTM4641, \guillemotleft38V-10A DC/DC micro-module regulator with advanced input and load protection\guillemotright. Assuming 80\% conversion yield, the power dissipated on the PDC is expected to reach around 7 W when supplying the required 7 A \(\times\) 4.2 V to the front-end electronics. The line current at the input of the PDC is expected to be 1.5 A at 24 V. This leads to a substantial reduction of the cross section of cable required to bring power inside the magnet from the external low voltage power supply unit. If it is found advantageous, the PDC may be integrated to the FEM.
\subsubsection{Back-end Electronics}
The TDCM is a generic clock and trigger distributor and data aggregator module designed for several projects, including the upgrade of T2K. A complete description of the TDCM is given in~\cite{ref:backend}.
\begin{figure}[ht]
\begin{center}
\includegraphics[width=0.8\textwidth]{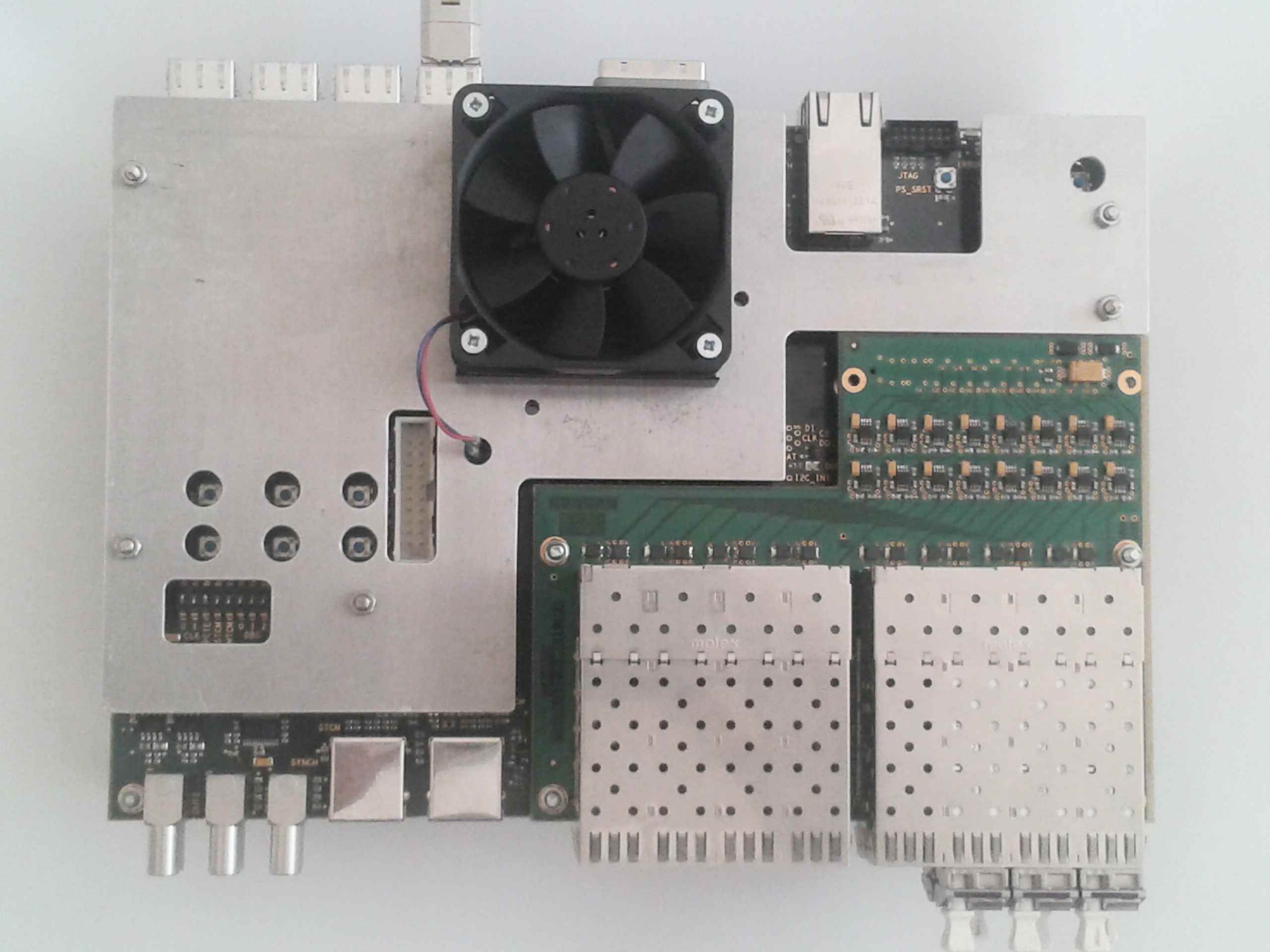}
\caption{Prototype of the TDCM.}
\label{fig:tdcm_1mezz}
\end{center}
\end{figure}
This module is composed of a commercial System-On-Module (SoM), Enclustra Mercury ZX1 series~\cite{ref:enclustra} plugged on a custom made 6U form factor carrier board. The SoM is built around a Xilinx ZYNQ 7030-7045 FPGA that contains ample programmable logic resources and integrates a multi-core 800 MHz ARM processor. The module incorporates up to 1 GB of DDR3 memory, 512 MB of NAND flash, offers more than 150 user I/O pins and 4-8 multi-gigabit per second capable transceivers. The TDCM supports up to 32 serial links to connect to the FEMs. One fully equipped TDCM would be sufficient to read out the two HA-TPCs, but to keep each HA-TPC independent, one TDCM per HA-TPC will be used. A picture of a prototype of the TDCM with a 16-port optical link mezzanine card installed is shown on Fig.~\ref{fig:tdcm_1mezz}.
The TDCM receives the primary 100 MHz clock and the common trigger signal of the T2K near detector from a Slave Clock Module (SCM) via a RJ45 copper cable and makes a fanout of these signals, along with some other information, via the set of optical links connected to the FEMs at a nominal rate of 100 Mbps (200 Mbaud after encoding). Alternatively, a single higher power optical transceiver coupled to a 1:16 passive optical splitter may be used to implement the fanout. In the upstream direction, each FEM is linked to the TDCM via a point-to-point fiber link. All the functions required to control and read out the front-end electronics of one HA-TPC are implemented in the fabric of the FPGA of the SoM of the TDCM and dedicated software running in the embedded dual-core ARM processor. This is described in more detailed in the DAQ section.
\subsection{Production and Test}
\subsubsection{Front-end ASICs}
The remaining stock of encapsulated and tested AFTER chips is $\sim$700 units (i.e.$\sim$50,000 channels) which is expected to be sufficient to build prototype FECs, produce the 64 FECs required for the two HA-TPC (512 AFTER chips are required) and still provide a sufficient number of spares. If required, more AFTER chips could be produced, but extra time and resources would be needed in that case, and the obsolescence of the plastic package of the AFTER chip would probably require to use a ceramic package which is more expensive and delicate.
\subsubsection{Test bench for the production of the Front-end Cards}
The role of this test stand is the quick validation of every FEC at the end of the assembly line: verification of all input channels, assessment of the noise level and measurement of the crosstalk level between neighboring channels. Calibration pulses will be injected with the built-in pulser of the FEC. A custom PCB will make a capacitive load representative of a Micromegas detector. The test bench will also comprise one FEM and a portable computer used for control and DAQ. A user-friendly interface will allow a non-expert technician to run a pass or fail test at the production factory. Detailed tests and the analysis of eventual defaults will be performed by the designers of the FEC in a laboratory environment.
\subsubsection{Test bench for the production of the Front-end Mezzanine Cards}
This test stand is required for the validation of every FEM at the production site. All the analog and digital functions and interfaces of this card have to be tested. The hardware part of this test stand comprises a motherboard that mimics the function of two FECs, and one TDCM, or the equivalent, to connect to the optical port of the FEM. Using the appropriate dedicated software on a laptop computer, a technician at the board factory will run a pass or fail test for every FEM. Deeper analysis on potential defects will be conducted by the designers of the FEM if that is needed.

\begin{figure}[ht]
\begin{center}
\includegraphics[width=0.8\textwidth]{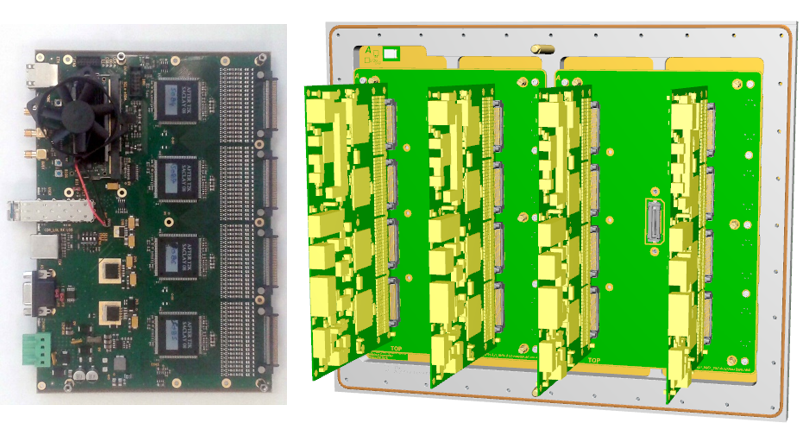}
\caption{The ARC and the front-end electronics of the detector test stand.}
\label{fig:htpc_detector_test_bench_elec}
\end{center}
\end{figure}


\subsection{DAQ and slow control}
Two options are being considered for the control and data acquisition software. In the first scheme, which is unchanged compared to the existing TPCs, the embedded processor of TDCMs execute a simple \guillemotleft bare-metal\guillemotright command interpreter program. An intermediate PC running the MIDAS framework performs the translation of the instructions emanating from the global data acquisition system to the required series of commands interpretable by the TDCMs. In the second scheme, the two CPU cores of the TDCM will be used: one CPU core runs the Linux operating system and executes the MIDAS processes locally. The second CPU core runs the command interpreter of the TDCM. An inter-process communication mechanism ensures the correct exchange of control messages and event data between the two CPUs. The advantage of the first scheme is that it decouples the development of the DAQ software from the environment of the TDCM (Xilinx development tools for ARM processor) at the expense of an additional PC and network hop. The second scheme is more integrated, but requires additional engineering to run heterogeneous software on the multi-core processor embedded in the TDCM.

\section{Detector construction and assembly}

The TPC construction will take place in several phases.

\begin{itemize}
\item Design and prototyping. At the moment (November 2018), the design of the final system is quite advanced but still not fully finalized. In this phase, extending roughly to the first half of 2019, we will produced and test the first full HA-TPC prototype, called Prototype-1. We will mount on it MM1 and test it first at CERN with cosmics, then at DESY in June 2019 with an electron beam. This phase will validate the design of these two crucial sub-components of the TPC.
\item Production of the field cages, module frames, Micromegas, electronics and associated mechanics. The field cages will be produced in Europe, with most of the components machined in INFN mechanical workshops in Italy. The Micromegas will be produced by CERN EP-DT-EF and then tested as described above on a test bench at CERN. 

\item Assembly at CERN. The two HA-TPC will be first assembled at CERN in a clean room. After integration of the module frame and the Micromegas, it will be possible to do first tests of electrical continuity and gas tightness. After mounting the from-end electronics we will also do a first system test of each of these devices, with a radioactive source and cosmic rays.

\item Integration in J-PARC. After shipment to Tokai, the TPC will undergo a short test on surface in the Neutrino Monitor Building to verify that nothing has been damaged during the shipment. Then they will be lowered in the ND280 basket and connected to the gas system, the cooling system, the high and low voltage etc.

\end{itemize}

\begin{figure}[ht]
\begin{center}
\includegraphics[width=0.98\textwidth]{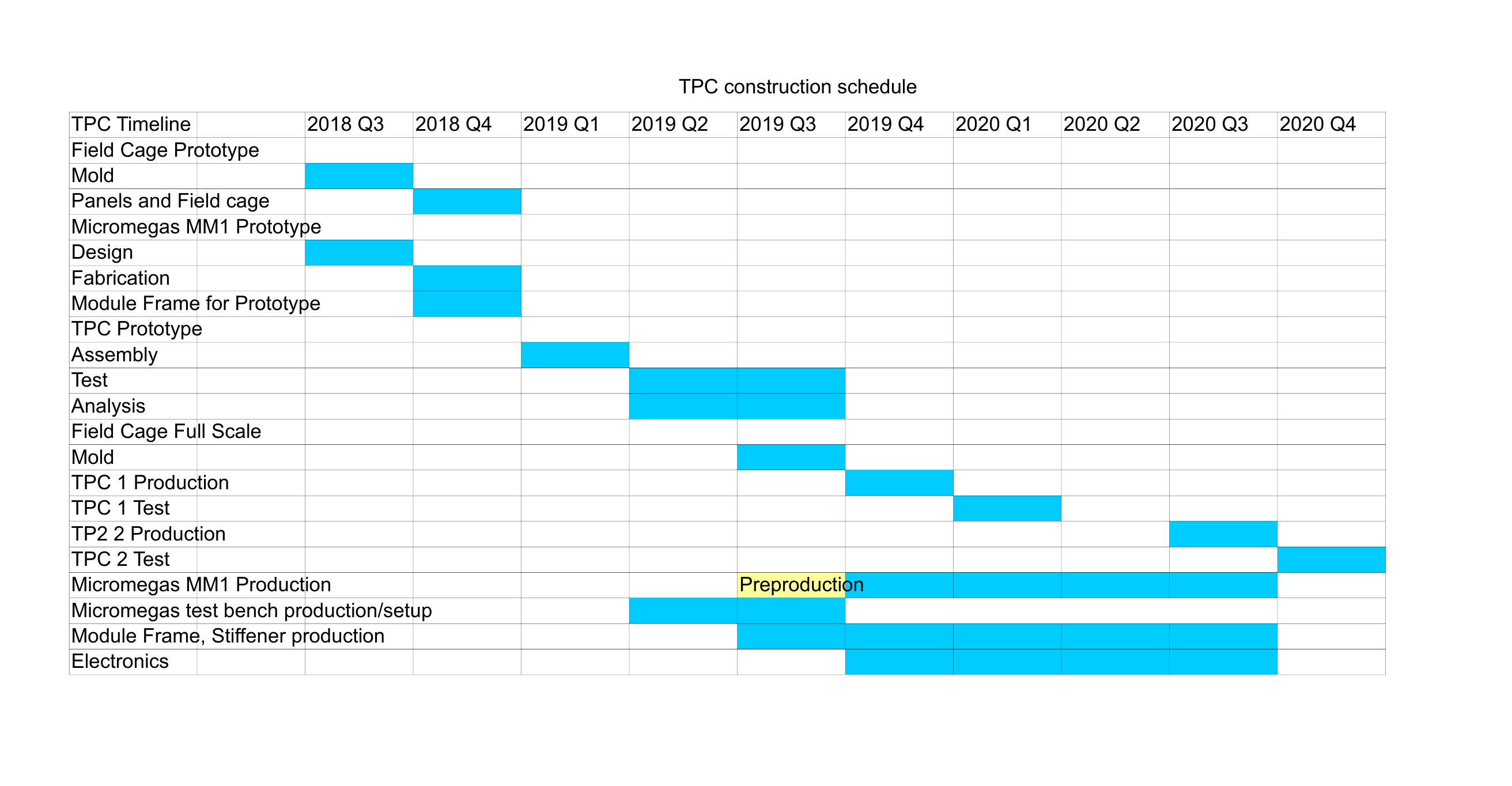}
\caption{Schedule of the construction of the HA-TPC.}
\label{fig:tpctimeline}
\end{center}
\end{figure}

These phases are summarized in Figure~\ref{fig:tpctimeline} and the most important milestones are shown in Table~\ref{tab:tpcmilestone}.

\begin{table}[htbp]
\centering
\caption{Milestones of the HA-TPC.}
\label{tab:tpcmilestone}
\begin{tabular}{|c|c|}
  \hline
  Date & Milestone \\
  \hline
March 2019 & First full TPC prototype ready \\
June 2019 & Testbeam at DESY \\
October 2019 & Start Production (Field cage, Micromegas)\\
June 2020 & HA-TPC 1 ready \\
January 2021  & HA-TPC 2 ready \\
  \hline
\end{tabular}
\end{table}

\clearpage

%% file: TestBeam.tex
\section{Test Beam}
\label{sec:testbeam}

During Summer 2018 a TPC has been assembled and operated at CERN using tagged beams from the CERN T9 beam line.
For this tests the HARP TPC field cage has been used with mounted on it the first version of resistive Micromegas (MM0). The goal for these tests being the characterization of the response of the MM0. 

A number of samples of data have been taken under different conditions of the beam (particle types and momentum) and also of the detector working conditions. The response to muons, electrons, pions and protons at different momentum but also entering in the TPC active volume at several drift distances has been studied. Both high voltage bias and sampling time of the Micromegas has been varied have been varied during the data taking to study the response of the detector at different working points.  Cosmic ray events  have been also recorded to study the Micromegas response with respect to different track orientations. Finally a radioactive source of $^{55}$Fe have been positioned at the cathode to have a reference point for  energy calibration. 

In this section we present a  description of the experimental setup (see Section~\ref{sec:testbeam_setup}) used for the test beam as well as preliminary results obtained from a first analysis of the collected data (see Section~\ref{sec:testbeam_res}).

\subsection{The setup}
\label{sec:testbeam_setup}

A cylindrical volume of 2~m long and $\sim$~0.8~m diameter host the drift volume. The field cage is made by Stesalit  with a double interleaved strip pattern to avoid electric field
inhomogeneities and high field gradients.
A foil of individual aluminized Mylar strips has been glued inside
the cylinder, and an aluminium foil has been glued onto the
outside surface.
A detailed description of the HARP field cage can be found in~\cite{HARP_fieldcage}.
The cathode is at one extremity of the field cage and at about 50~cm from the edge of the external cylinder. 
On its rear holes it hosts calibration sources. During the TPC operation a voltage of 25 kVolts have been applied to the cathode generating an electric field in the drift volume of 167~V/cm.  
On the extremity opposite to the cathode a circular flange close the cylinder where the Micromegas MM0 is installed. The description of the MM0 and of the readout electronics has been detailed in Section~\ref{sec:MM}.  

The TPC has been operated using a premixed gas with 95\% Ar , 3\% CF4 and 2\% Isobutane, which is the same mixing used by T2K for the existing ND280 TPCs. 
A simple gas system with only one line has been set up to operate the detector. Before starting the data taking the TPC has been extensively fluxed with Nitrogen gas first. Later it has been flushed with the gas mixture for several hours (3 or 4 times the volume of the TPC) to remove impurities. During normal operations the gas flux was kept of about 25 litres/hours.    Temperature measurement on the exhaust line were taken to monitor environmental condition which might induce a difference in the electron drift velocity and thus on the performances of the detector.  

To trigger on cosmic rays going through the TPC two scintillator plastic panels have been positioned on the top and the bottom of the TPC. Panels are made of 3 200(L)x20(W)x2(H) cm$^3$ bars of Polystyrene doped with 2\% PTP and 0.05\% POPOP. The surface is coated with reflective paint and two grooves are made to host fibres. The readout is done at one side with Hamamatsu MPPC while the other end-side of the fibres is mirroed. Plastic bars are installed inside aluminum boxes to get good light tightness conditions\footnote{The cosmic ray system has been kindly provided by the Neutrino Platform and the design comes from R\&D for the ICARUS cosmic ray tagger.}.

\begin{figure}[htbp]
\centering
\includegraphics[width=.7\linewidth]{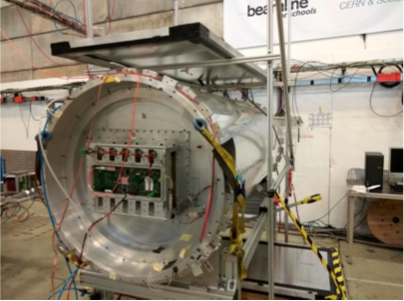}
\caption{\label{fig:setup_tpc} The experimental setup during the TPC test beam in T9 at CERN. The Micromegas MM0 is mounted on the HARP field cage. Two alluminium boxes containing plastic scintillator bars read out by SiPM are positioned on the top and the bottom of the field cage in order to select through going cosmic rays. }
\end{figure}

Figure~\ref{fig:setup_tpc} shows the experimental setup during the data taking in T9. The beam is entering from the left of the picture. As previously mentioned, data were taken with the beam entering at a number of distances from the anode to test different drift distances (e.g. 10, 30 and 80 cm ). To facilitate the displacement, the setup has been positioned on a table equipped with wheels.

\subsubsection{The trigger}

\begin{figure}
\centering
\begin{subfigure}{.40\textwidth}
  \centering
  \includegraphics[width=.9\linewidth]{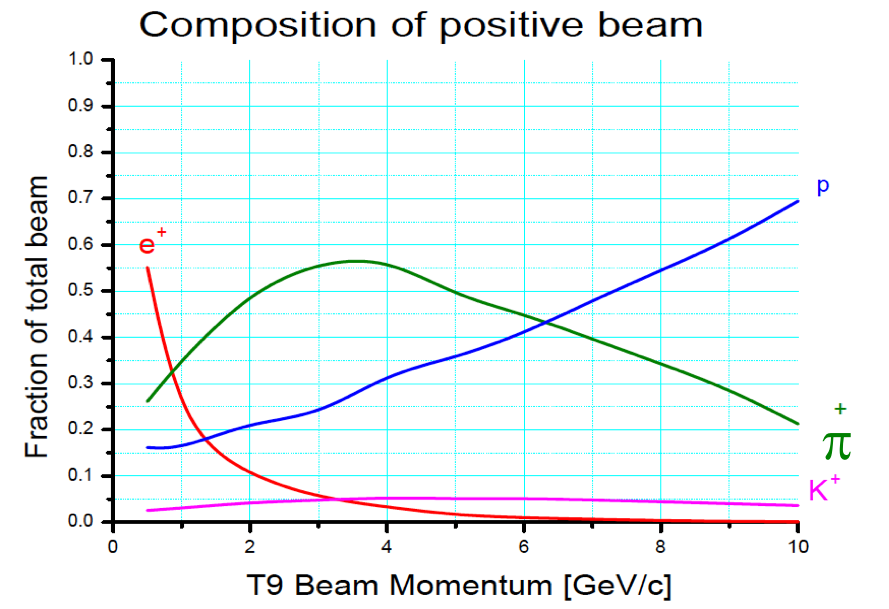}
  \caption{ \label{fig:BeamPos} }
\end{subfigure}%
\hspace{.2cm}
\begin{subfigure}{.40\textwidth}
  \centering
  \includegraphics[width=.9\linewidth]{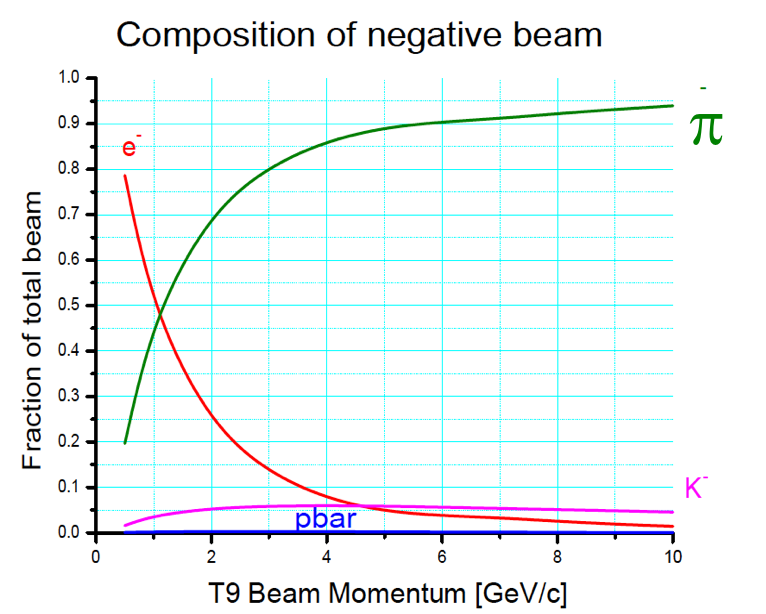}
  \caption{\label{fig:BeamNeg} }
  \end{subfigure}
\caption{\label{fig:T9beam} Breakdown of the beam composition as a function of the energy for T9 for electron enriched beams. The electron composition during hadron enriched  beam configuration is strongly reduced (by about a factor 8).   }
\end{figure}

Test beam data were taken using the T9 beamline with copper target to have an \" hadron enriched  beam configuration\". The breakdown of the beam composition as a function of the energy is shown in Figure~\ref{fig:T9beam}. The beam composition at low energies is largely dominated by electrons. The use of copper target allow to reduce the electron contribution of about a factor 8. 

\begin{figure}
\centering
\includegraphics[width=.7\linewidth]{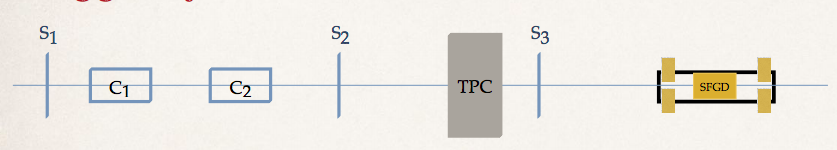}
\caption{\label{fig:trigger} Sketch of the detectors used to tag the particles from the beam and set up trigger selections.  }
\end{figure}

Particles coming from the beam have been tagged using three plastic scintillator detectors coupled with PMTs called respectively S1, S2, S3\footnote{The numbering of the detectors always start from the most upstream.} and two Cherenkov detectors called C1 and C2. 
Figure~\ref{fig:trigger} shows a cartoon of the locations for those detectors along the beamline.   
The selection for the different particle types have been done by the combination of the NIM signals coming from those detectors. A summary of the different options is presented in Table~\ref{tab:triggermenu}.

\begin{table}[htbp]
\centering
\begin{tabular}{l  l }
\hline
\hline

Particle  & Selection \\
\hline
Electrons & Scintillators + Cherenkov    \\
Protons (+Kaons) & S1(delayed) * S2 (delay ~ proton TOF between S1 and S2)   \\
Pions (+ muons) & Scintillators * $\overline{\mathrm{protons}}$ * $\overline{\mathrm{electrons}}$ \\ 
Cosmic ray &  from the scintillators panels (only out of spill) \\ 
\hline
\hline
\end{tabular}
\caption{\label{tab:triggermenu} Summary of the different signal combinations to tag the different particle types coming from the beam}
\end{table}


\subsection{Test beam data analysis}
The main goal of this section is to present some results obtained with the data collected during TPC test beam to discuss the performance of the new resistive Micromegas and evaluate the potential improvement of their usage in the HA-TPCs in contrast of the currently used Micromegas installed in the forward TPCs of the ND280 detector.
\label{sec:testbeam_res}
\subsubsection{Tracks selection}
\begin{figure}[hbtp]
    \begin{center}
    \begin{minipage}{0.49\linewidth}
        \centering{\includegraphics[width=0.9\linewidth]{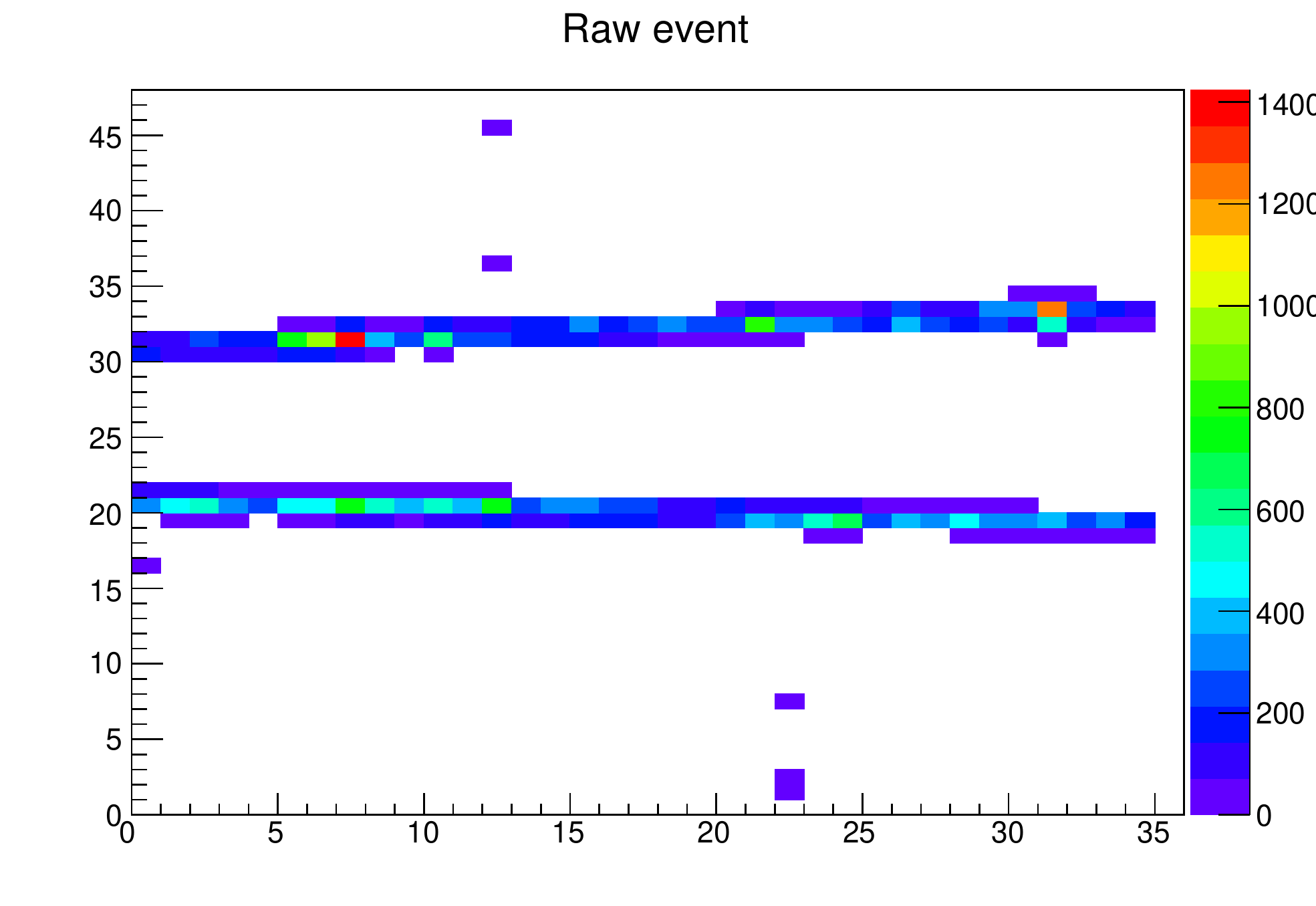} \\ a) }
    \end{minipage}
    \begin{minipage}{0.49\linewidth}
        \centering{\includegraphics[width=0.9\linewidth]{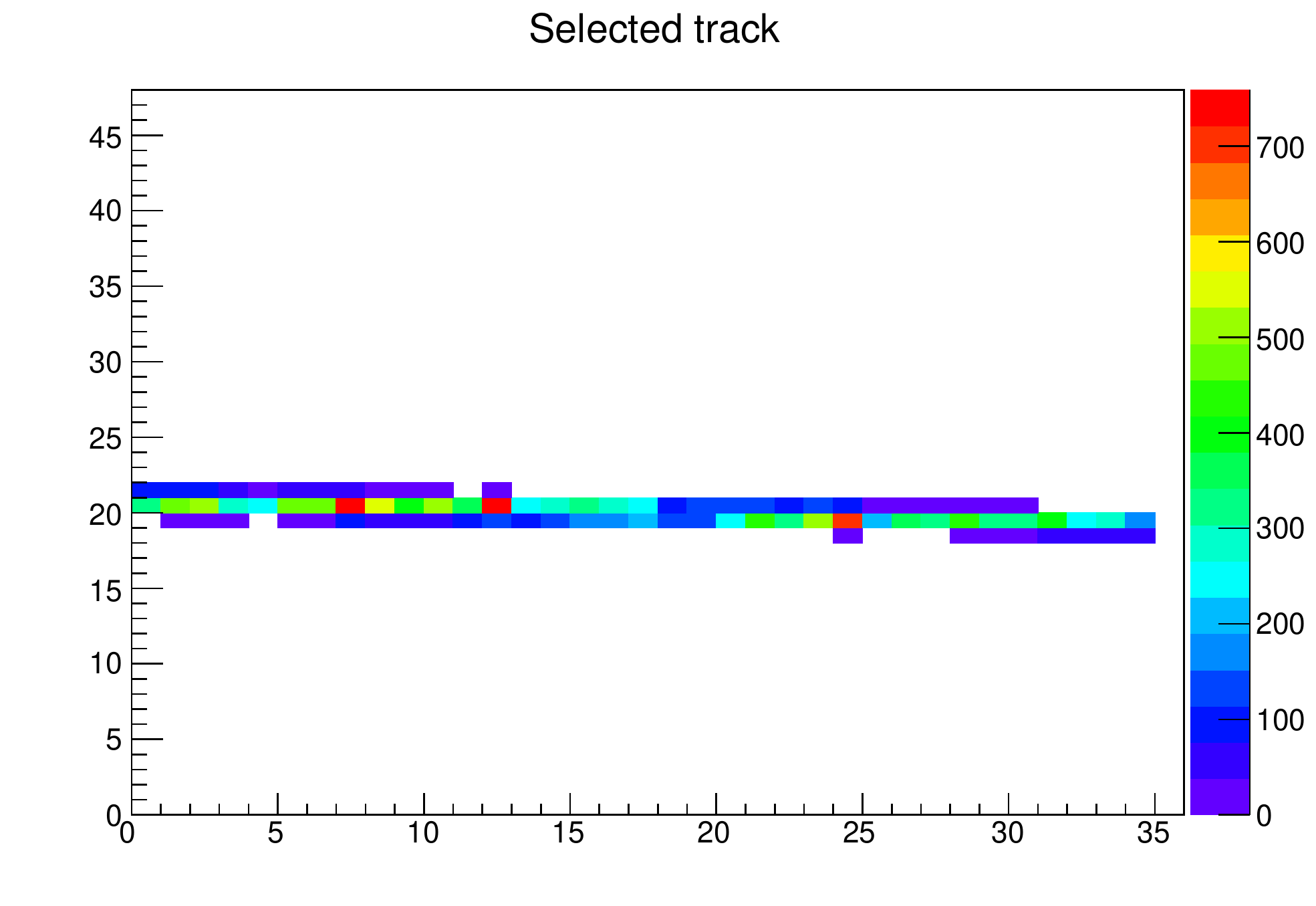} \\ b) }
    \end{minipage}
    \caption{Views of the projection of the raw event a) and one of the selected tracks b) on the read-out plane of the MM. The colormap shows the charge deposited in each pad in arbitrary units.}
    \label{fig:sel}
    \end{center}
\end{figure}

\begin{figure}[hbtp]
    \begin{center}
    \begin{minipage}{0.49\linewidth}
        \centering{\includegraphics[width=0.99\linewidth]{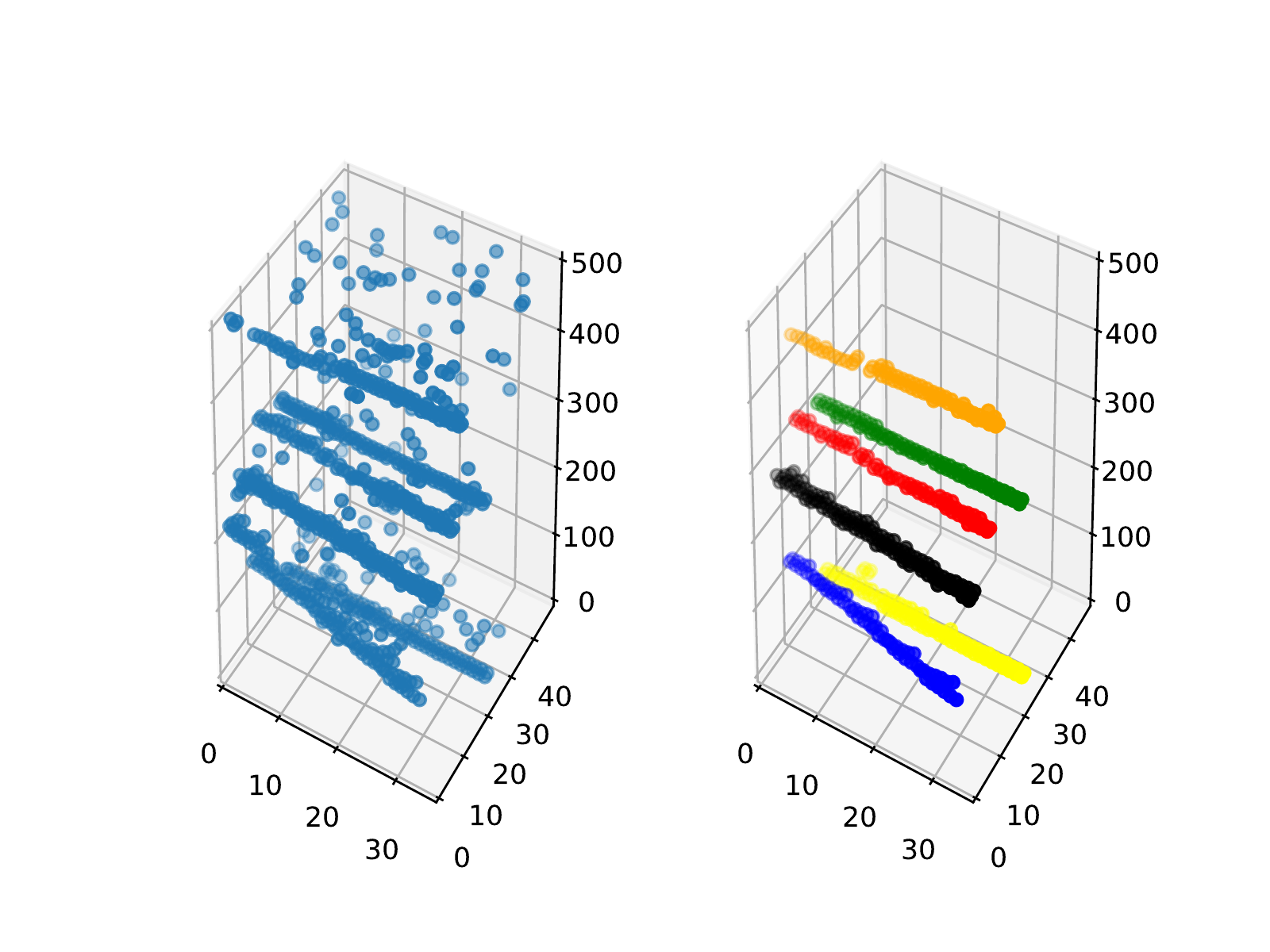} \\ a) }
    \end{minipage}
    \begin{minipage}{0.49\linewidth}
        \centering{\includegraphics[width=0.99\linewidth]{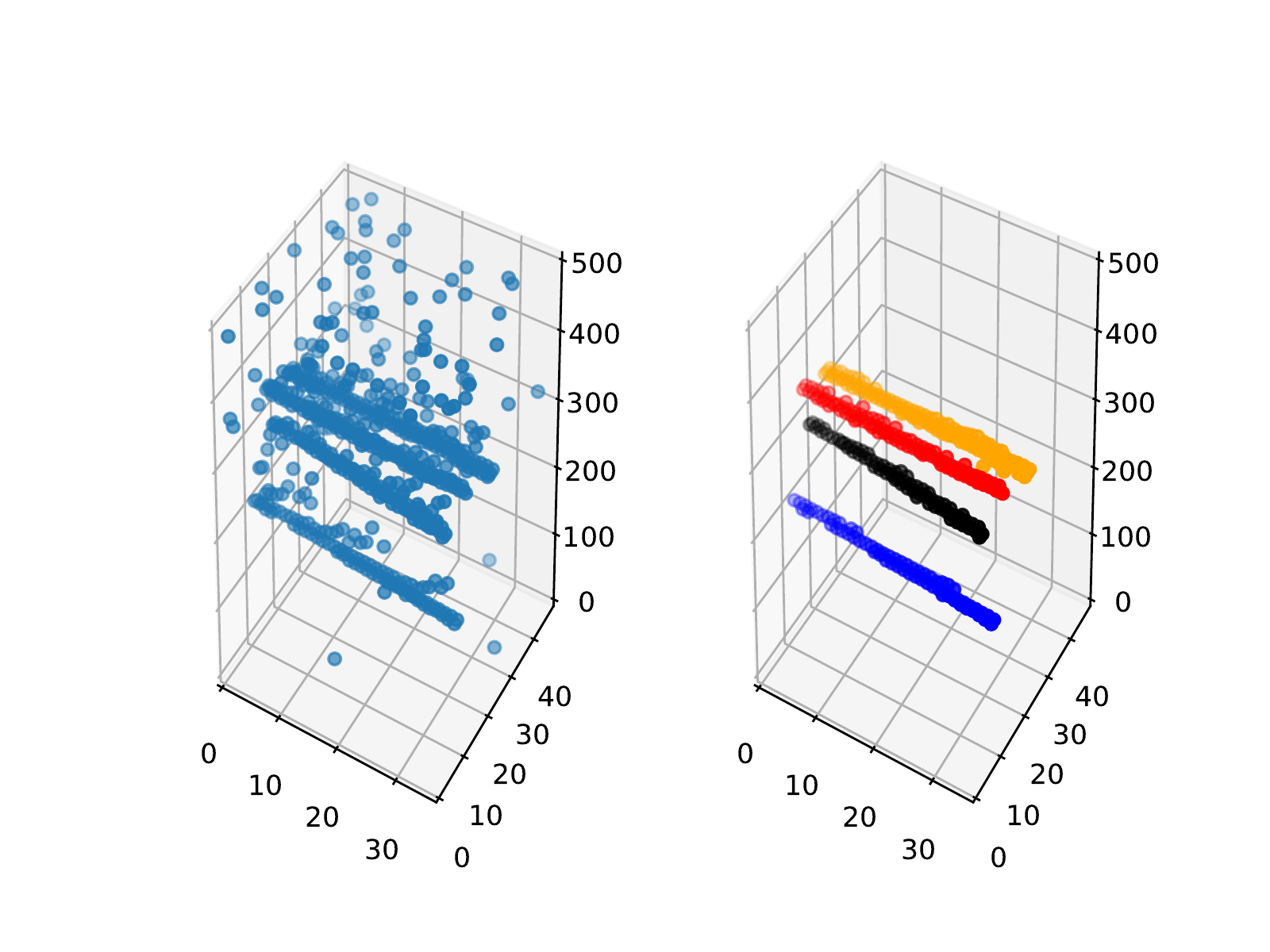} \\ b) }
    \end{minipage}
    \caption{Views of two events a) and b) before (left) and after (right) the selection.}
    \label{fig:3d_sel}
    \end{center}
\end{figure}

Reconstruction methods to select tracks are still under development. Since the selection of the tracks for the analysis is a critical factor the present results are still preliminary and therefore all the values provided in this section offer at most a lower bound of the future potential of resistive MM. The selection of tracks has been made using DBSCAN algorithm. It allows to separate and select tracks in events with high density of triggered pads, as shown in Figure ~\ref{fig:3d_sel}, with low noise and with high acceptance allowing to perform analysis with clean and large subsets of selected data. A intuitive view of the selection is also offered in Figure ~\ref{fig:sel} where the outcome of the selection is shown projected in the read-out plane.


\subsubsection{dE/dx}
\begin{figure}[hbtp]
    \begin{center}
    \includegraphics[width=1.\linewidth]{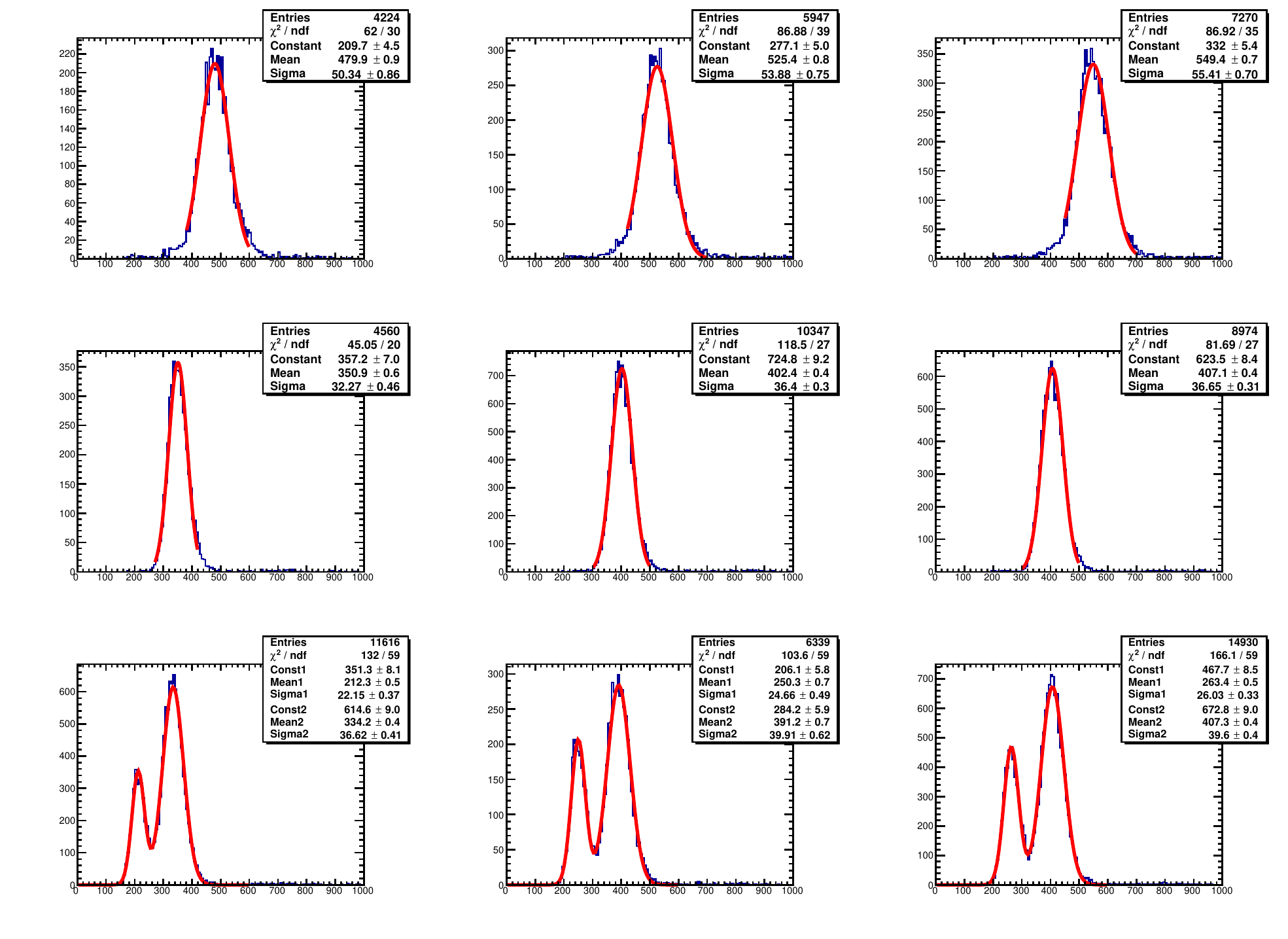}
    \hspace{0.16\linewidth} a) \hspace{0.33\linewidth} b) \hspace{0.33\linewidth} c) 
    \caption{The prototype dE/dx measurements for different values of the drift distance: (a) 80 cm, (b) 30 cm, (c) 10 cm. The rows corresponds to the proton, electron and pion triggers respectively.}
    \label{fig:text_beam_dedx}
    \end{center}
\end{figure}
The preliminary studies of the dE/dx were performed with the collected data. The goal was to analyze the measured ionization loss by the different particles selected by the triggers and estimate the dE/dx resolution. For this purpose a simple selection was developed to define beam tracks. The challenge was to suppress cosmic events, source signal and multiple beam tracks. 
After applying the selection we extract a single beam track for the analysis. The dE/dx study was done with the truncated mean method, widely used for TPCs~\cite{allison1980relativistic}:
\begin{itemize}
    \item pads in each column were grouped together into clusters
    \item the charge in the clusters was sorted in increasing order
    \item the 30 \% clusters with the highest charge were rejected
\end{itemize}
Thus we obtained nearly Gaussian distributions that describes the energy loss of the charged particles in the TPC prototype. The dE/dx distribution for various particle samples separated by the trigger are presented on the figure~\ref{fig:text_beam_dedx}.
The comparison with the current T2K TPCs dE/dx resolution is reasonable as we should be sure that the particle identification will not be worse with the new detectors. Taking into account the 2 times bigger size of the T2K TPCs (2 Micromegas with 36 pads each) we can conclude that the measured energy resolution is nearly the same value as we observed in the ongoing experiment.
\subsubsection{Point resolution}
\par A point resolution study has been done in order to estimate the potential improvement of using resistive Micromegas, see section~\ref{sec:MM}. In this analysis the approach in section 6.1.1 from Janssen (2008) has been followed. It is important to remark that the results are still preliminary since clusters with 4 or more pads have not been considered in the analysis. In addition the beam test data was collected with MM0 instead of the final version MM1, to be mounted in the final HA-TPCs. 
\begin{figure}[hbtp]
        \centering{\includegraphics[width=0.8\linewidth]{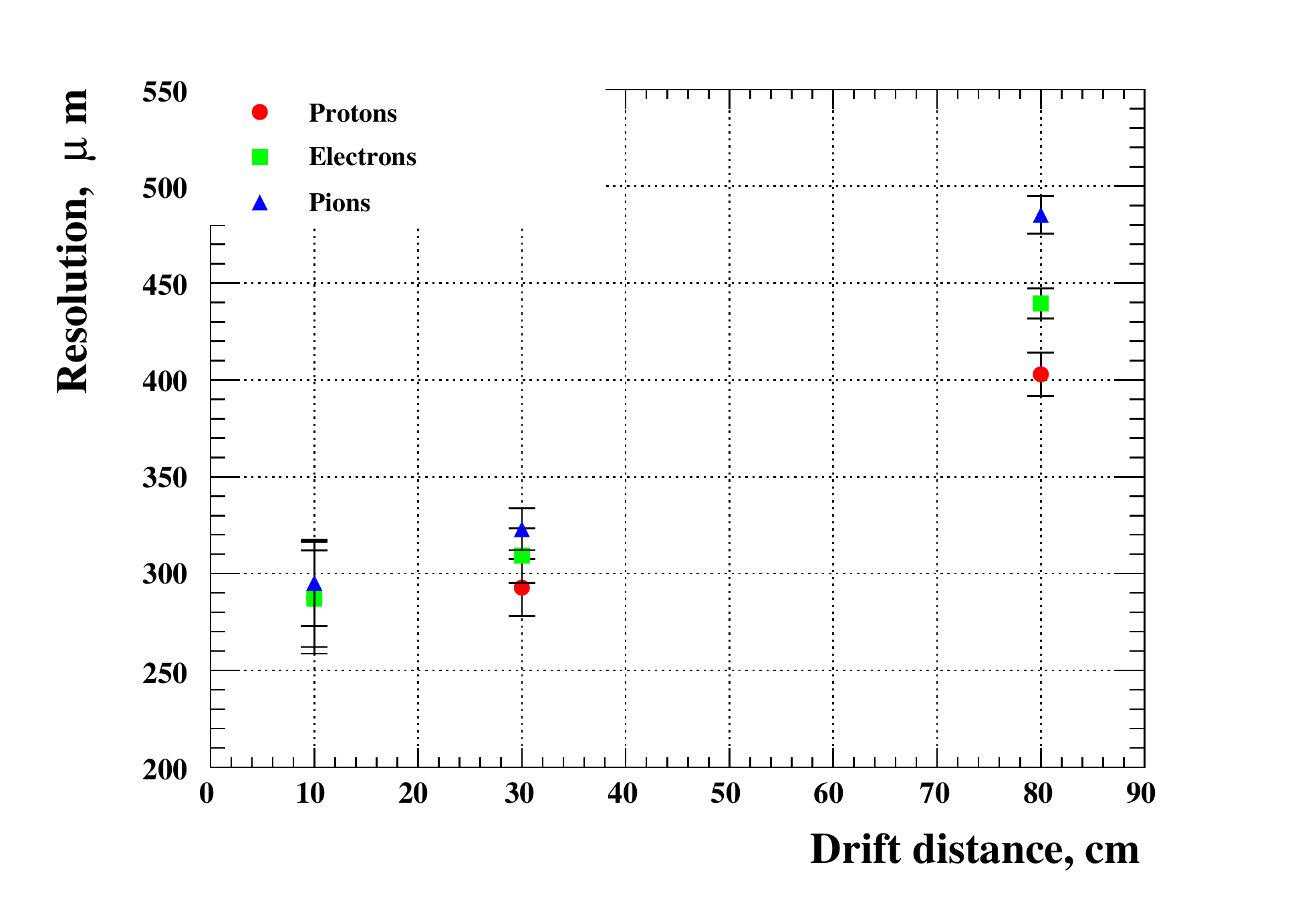} \\ a) }
    \caption{Space point resolution of MM0 as a function of the drift distance for different particle types of momentum 0.8 GeV/c.}
    \label{fig:point_res}
\end{figure}
\par Using a resistive MM has the advantage to increase the average number of hit pads per event due to the charge spreading in the read-out plane. Accordingly, the study of the Pad Response Function (PRF) may help improving the resolution if it is used to correct the input data. Assuming as true information the extrapolation of a linear fit at the cluster position and as reconstructed information the mean of the charge distribution in the same cluster it is possible to obtain the PRF information. 
\par
\begin{figure}[hbtp]
\centering{\includegraphics[width=0.99\linewidth]{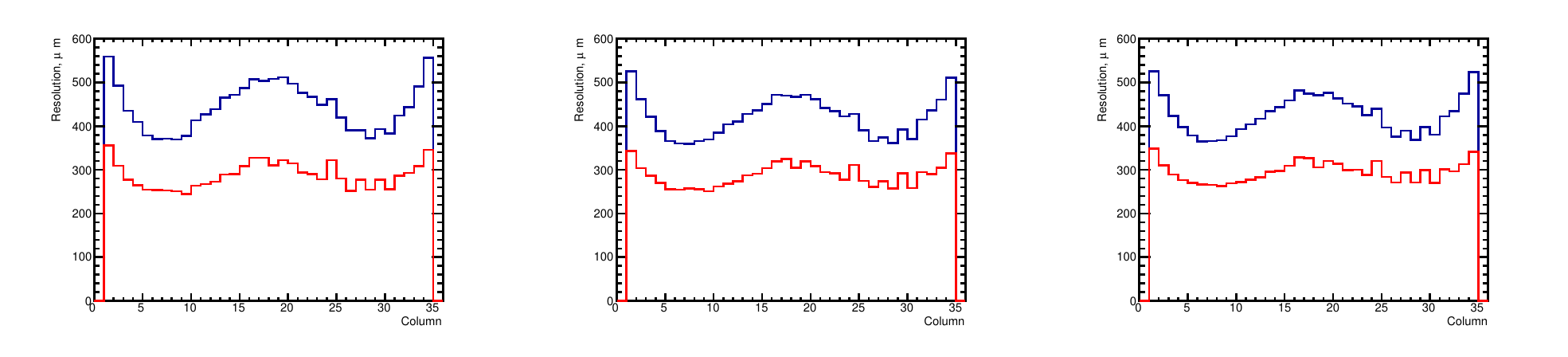}} 
    \caption{Spatial resolution of each MM column using 0.8 GeV/c momentum beam at 10cm drift distance for: Pions (left), Electrons (center) and Protons tracks (right). The black line shows the resolution using a simple charge barycenter method, the red line the resolution using the PRF method.}
    \label{fig:res_column}
\end{figure}

 To fully understand the behaviour of MM0 the spatial resolution has been computed for each one of the columns of the read-out plane (Fig. ~\ref{fig:res_column}). 
\par It is worth noting that even if the point resolution studies are still under development and can be fairly considered as preliminaries the values in Figure ~\ref{fig:point_res} are a factor of two better than the ones achieved by the current ND280's Micromegas~\cite{Abe:2014usb} showing the potential benefit of resistive Micromegas.
\subsubsection{Gas amplification studies}
\begin{figure}[hbtp]
    \begin{center}
    \begin{minipage}{0.49\linewidth}
        \centering{\includegraphics[width=0.95\linewidth]{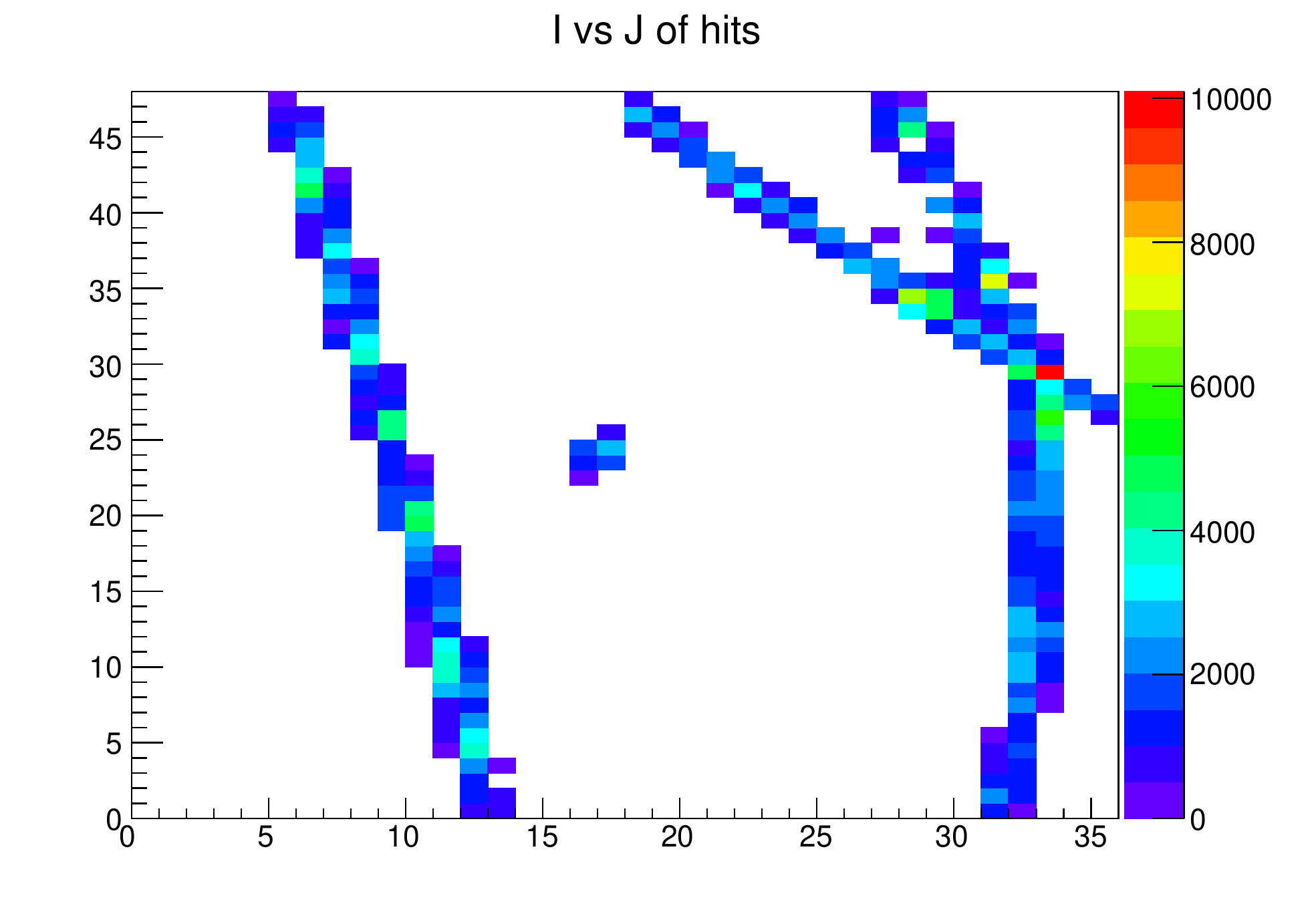} \\ a) }
    \end{minipage}
    \begin{minipage}{0.49\linewidth}
        \centering{\includegraphics[width=0.95\linewidth]{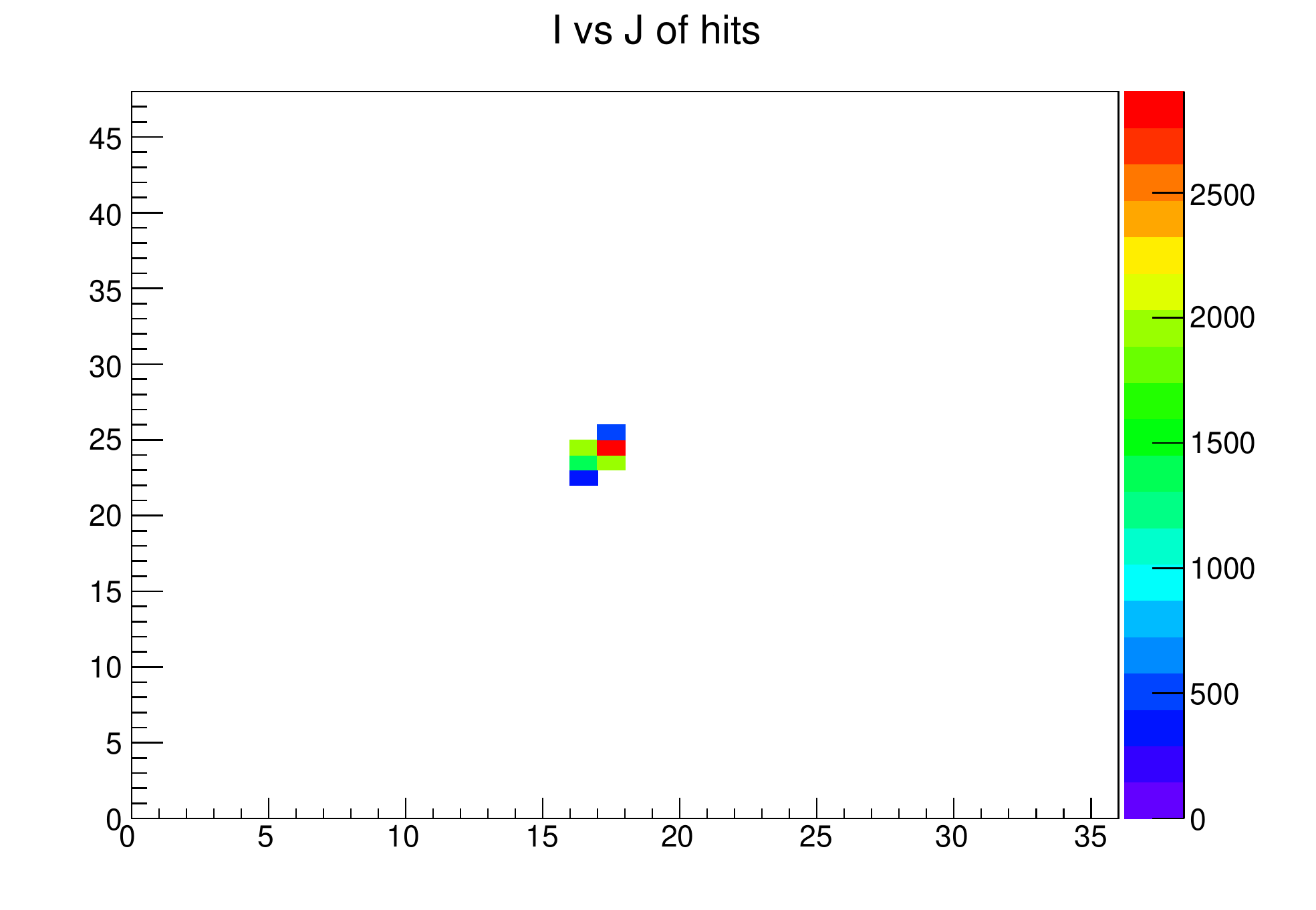} \\ b) }
    \end{minipage}
    \\
    \begin{minipage}{0.49\linewidth}
        \centering{\includegraphics[width=0.95\linewidth]{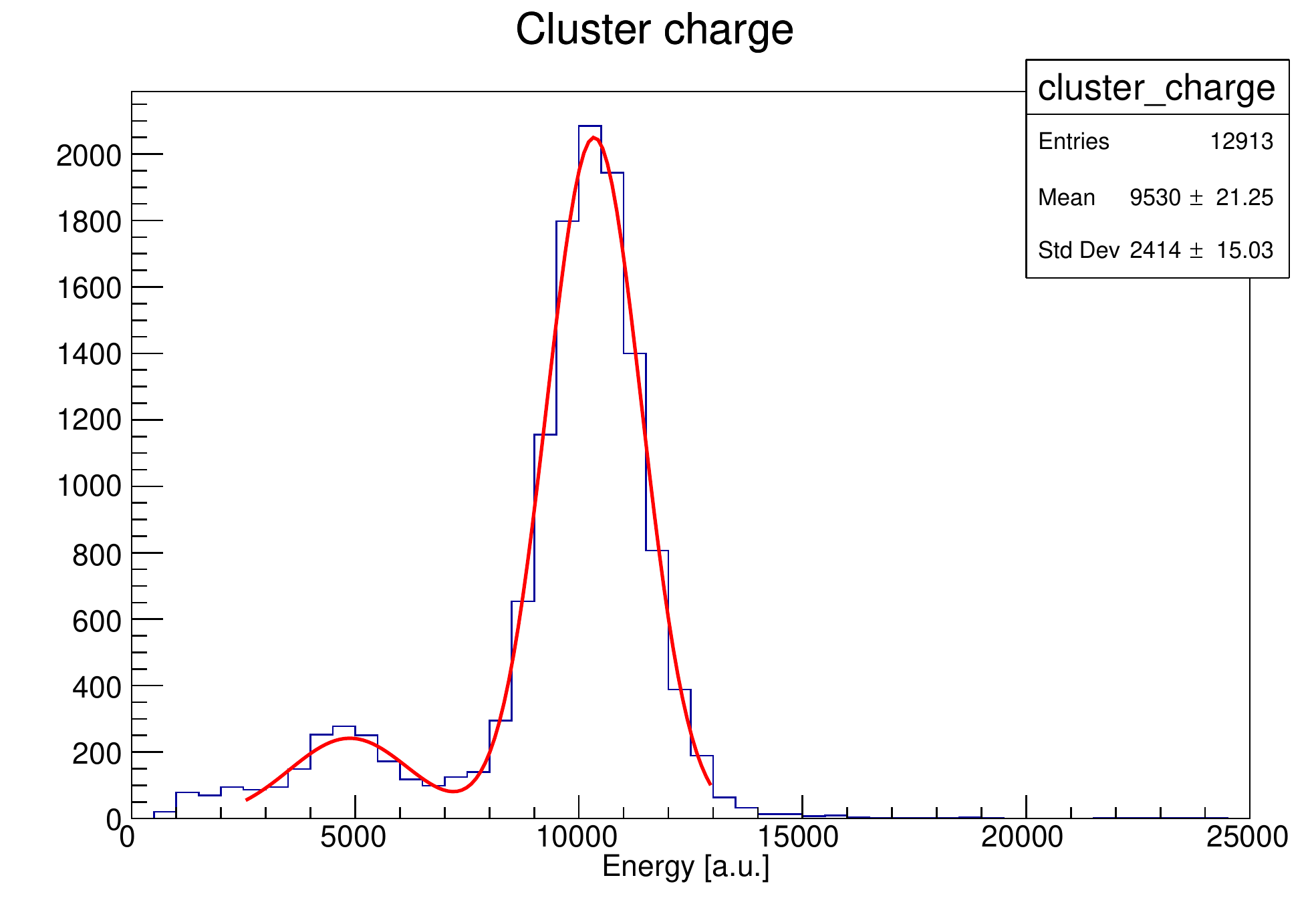} \\ c) }
    \end{minipage}
    \begin{minipage}{0.49\linewidth}
        \centering{\includegraphics[width=0.95\linewidth]{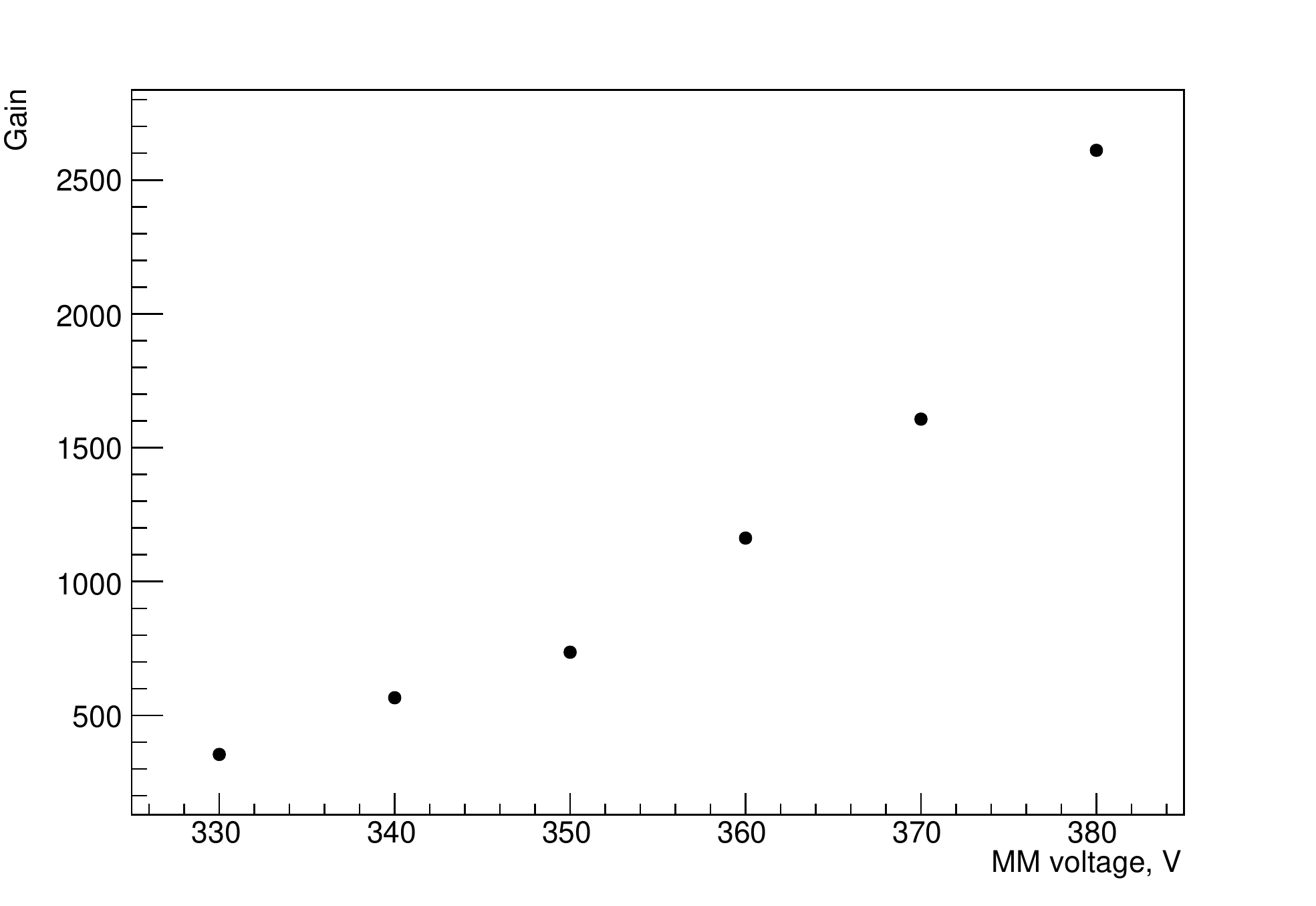} \\ d) }
    \end{minipage}
    \caption{a) View of one raw event b) Selected subsample from the raw event in \textit{a)} used for the analysis c) Iron source energy spectrum in arbitrary units d) Computed gain of the Micromegas from the observed $^{55}$Fe energy spectrum at different HV.}
    \label{fig:iron_sel}
    \end{center}
\end{figure}

An $^{55}$Fe radioactive source with a typical emission line of 6KeV was placed in the cathode of the HARP TPC chamber during data collection. A dedicated selection to look for isolated clusters was developed to select signal charge coming from argon ionized by one $^{55}$Fe X-ray photon. The results of such selection are exemplified in Figure ~\ref{fig:iron_sel} 

%% file: ToF.tex
\chapter{Time-of-Flight Detector}\label{Ch:TOF}

The Time-of-flight (TOF) system aims at precisely measuring the crossing time of charged particles in ND280. Combined with a timing measurement in the Super-FGD, this allows the determination of their direction to separate neutrino interactions in the target from backgrounds originating in the areas surrounding the detector. A time resolution better than 500~ps is required for an unambiguous determination of the flight direction of charged particles. An additional goal is to improve the particle identification, which will benefit from an even better time resolution (100-200~ps). In particular, discrimination between muons and electrons as well as protons and positrons in the energy ranges 0.1$-$0.3~GeV and 1$-$2~GeV respectively cannot be achieved with ionisation energy loss alone. Additionally, the fact that the TOF encloses the Super-FGD and TPC detectors makes it convenient for triggering on cosmic muons. Moreover, the TOF can assure a precision timing reference calibration of the Super-FGD. 

The concept of cast scintillator bars read out on both ends by arrays of large-area silicon photomultipliers (SiPMs also known as MPPC) described below allows for compact and economic design with a time resolution around 150~ps over the whole ND280 detector angular coverage.   

\section{Conceptual Design}
\label{TOFconcept}

In recent years, large-area silicon-photomultiplier MPPC sensors have appeared on the market at relatively low cost. Such modern devices offer several advantages over traditional PMTs: magnetic field tolerance, a much smaller volume and footprint allowing a compact design for bars without light guides, and an increased sensitivity to the part of the light spectrum (towards the green) which is least attenuated inside the bar, thus allowing to use longer bars with moderate degradation in the number of photon. 

Large-area MPPC applied directly to cast plastic scintillator bars on both ends to combine a time resolution of about 150~ps with a bar length of 2.3~m and a compact and robust design~\cite{Betancourt2017}. Design choices for the bar material and dimensions, MPPC type and arrangement, readout electronics, and mechanics, are summarised below.  

\section{Scintillator bar dimensions and material}

\begin{figure}
\centering
\includegraphics[width=0.495\textwidth]{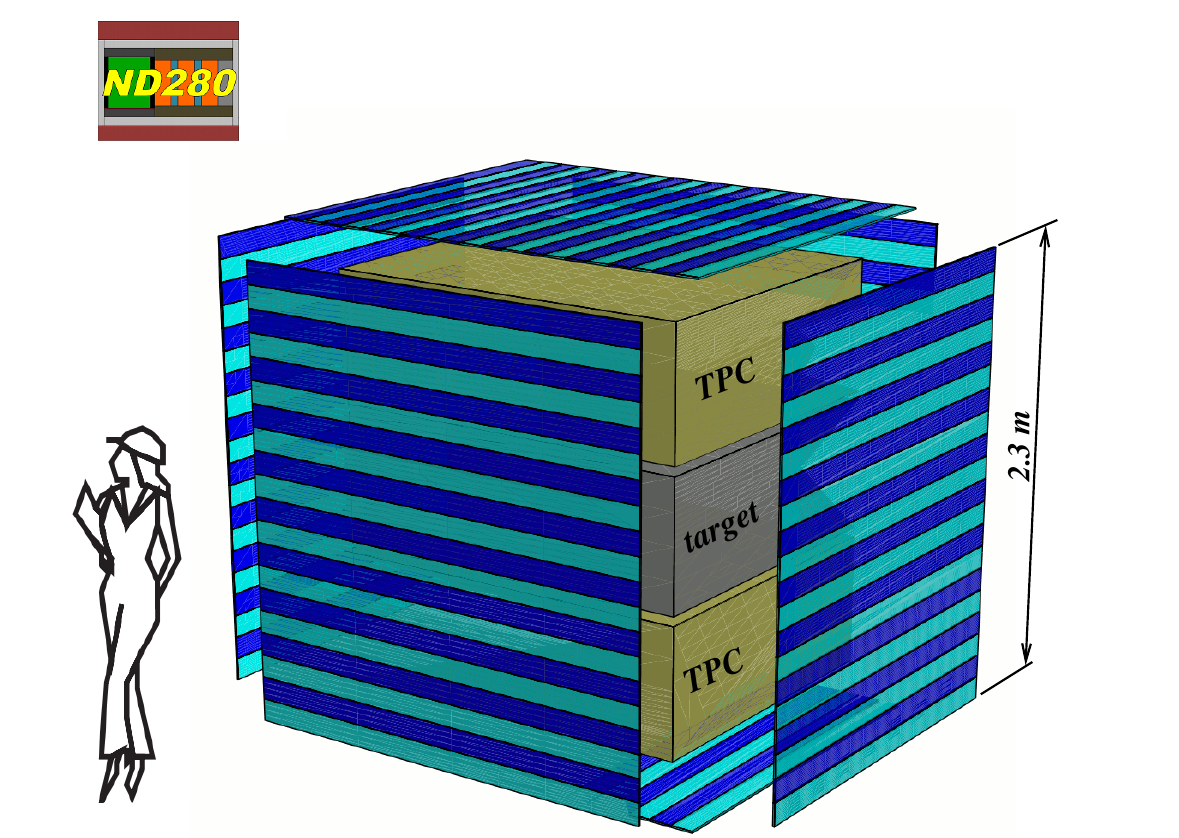}
\includegraphics[width=0.495\textwidth]{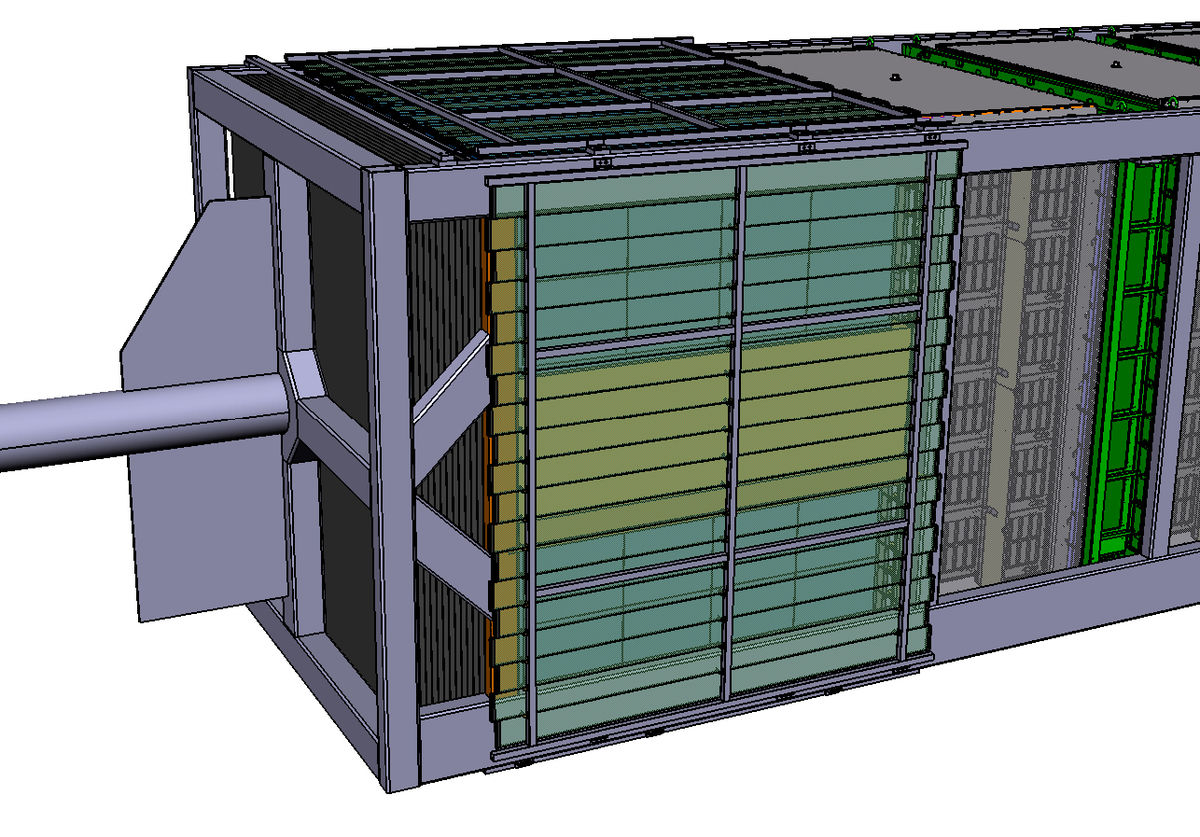}
\caption{
  Left: Schematic layout of the TOF detector planes surrounding the target and TPCs. Right: technical drawing showing how the planes are attached to the exterior of the ND280 basket and supported by aluminium structures.  
  }
\label{TOF_layout}
\end{figure}

Six TOF planes ensure full enclosure of the ND280 target and horizontal TPCs, as shown in Fig.~\ref{TOF_layout}. The bars running along the beam have a length of 2.0~m to cover the full length of the target, while the bars perpendicular to the beam (upstream and downstream planes) have a length corresponding to the basket width of 2.3~m. A bar thickness of 1~cm allows a good rigidity while being well adapted to light collection with 6$\times$6~mm$^2$ MPPC, and a breadth of 12~cm is chosen to accommodate for a arrays of either 8 or 16 MPPC, as detailed in Section~\ref{TOFsensors}. The bar dimension is thus 200$\times$1$\times$12~cm$^3$ or 230$\times$1$\times$12~cm$^3$. The planes oriented along the beam (2.0~m long bars) are to be fixed to the outside of the basket with a staggered arrangement, while the upstream and downstream planes (2.3~m long bars) are aligned in a plane to accommodate for the limited space.

\begin{figure}
\centering
\includegraphics[width=0.52\textwidth]{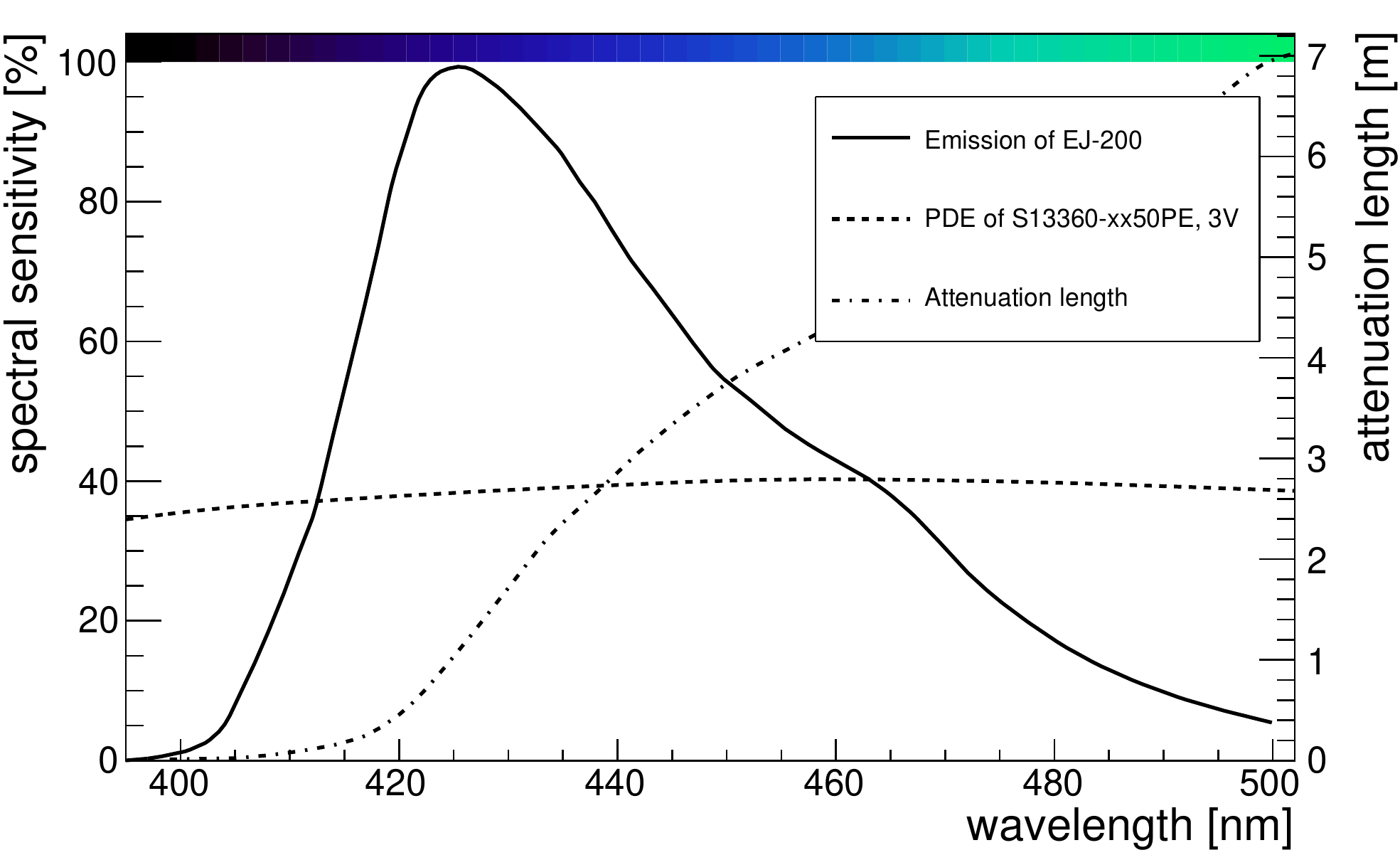}
\hfill
\includegraphics[width=0.47\textwidth]{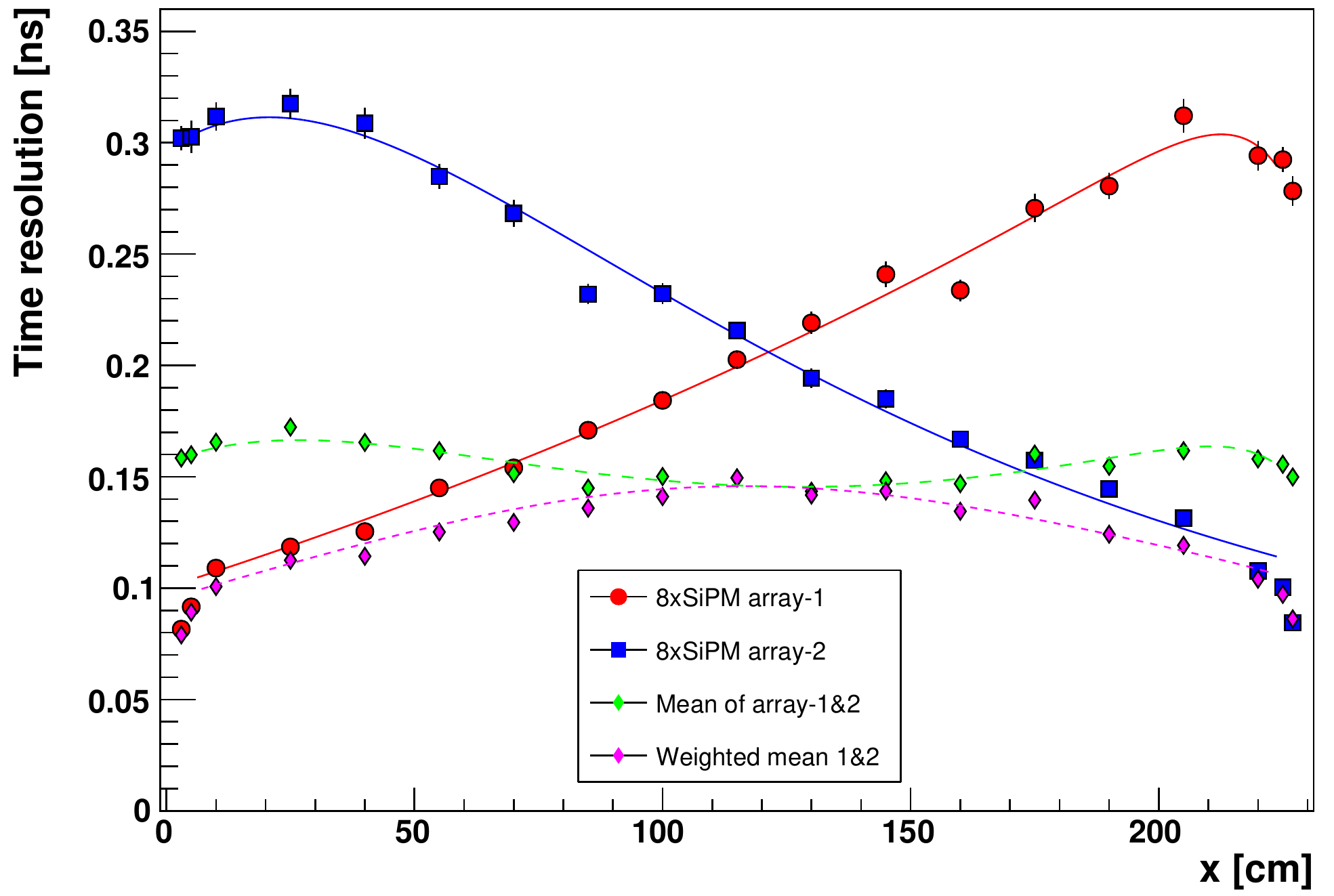}
\caption{
  Left: emission spectrum, photo-detection efficiency and attenuation length as a function of light wavelength for the  EJ-200 plastic scintillator. 
  Right: measured time resolution with an array of 8 6$\times$6~mm$^2$ MPPC at both ends of a 2.3~m long, 12~cm wide and 1~cm thick bar (same dimensions as the ND280 TOF detector) as a function of the beam impact position along the bar.
  }
\label{TOF_spectrum}
\end{figure}

The choice of plastic is driven by the need to achieve precision timing by detecting as many photons as possible for interactions occurring all along a $\sim 2$~m bar. EJ-200 provides an optimal combination of a high light output, suitable optical attenuation length (average 4~m, see Fig.~\ref{TOF_spectrum}, left), and fast timing (rise time of 0.9~ns and decay time 2.1~ns). Its emission spectrum resides in the near-UV region of the visible spectrum. The fact that it has a component that extends towards the green, at longer wavelength than eg EJ-204 and EJ-420, is a very useful property to achieve a better photon yield in long bars due to a combination of two effects: the MPPC photon detection efficiency is typically higher at longer wavelengths as compared to PMTs, and the attenuation length also increases. 

\section{Light sensor and coupling}
\label{TOFsensors}

The principal requirement for precision time measurements is a short rise time of the signal. A large MPPC capacitance increases the rise time and width of the signal, thus worsens the time resolution. In this regard, a large monolithic sensor or many smaller sensors with common cathodes and anodes~\cite{Bonesini2014} are naturally limited in area. A reduction of the capacitance can be achieved by connecting MPPC in series which decreases the rise time of the leading edge but also deteriorates the signal-to-noise ratio \cite{Cattaneo2014,Cervi2017}. Instead, a parallel connection of sensors with an independent readout and amplification to isolate the sensor capacitances from each other is the option chosen here. The signals are then summed up at the end. This scheme is described in more detail in Section~\ref{TOFelectronics} and demonstrated a time resolution around 80~ps all along the bar in the case of an array of 8 6$\times$6~mm$^2$ sensors S13360-6050PE coupled to a 1.5~m long and 6~cm wide bar as described in detail in Ref.~\cite{Betancourt2017}. 

\begin{figure}
\centering
\includegraphics[width=0.95\textwidth]{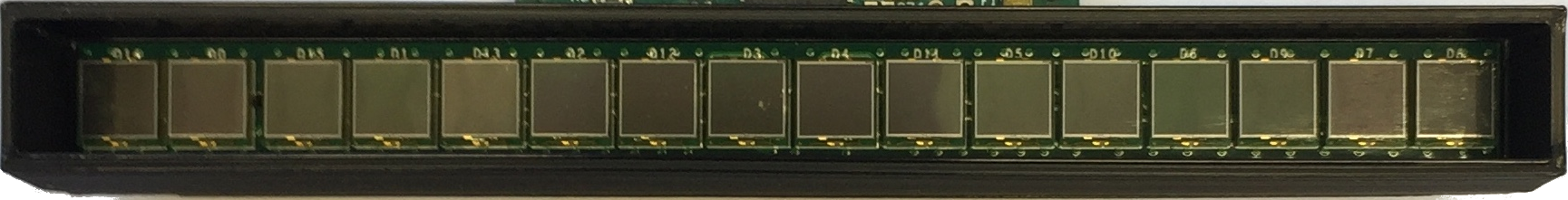}
\includegraphics[width=0.95\textwidth]{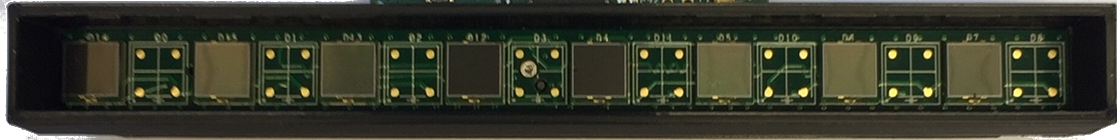}\caption{
  Arrangement for the large-area MPPC sensors at each end of the bar for the cast scintillator TOF, with 16 MPPC connected in a PCB with pairs connected in parallel (top picture) and then applied directly to the bar surface. The connection allows for the option of having only 8 MPPC for each PCB (bottom picture) and reducing the cost of MPPC by a factor 2 at the cost of a loss of photons which leads to a poorer time resolution. 
  }
\label{TOF_sensor}
\end{figure}

For bar dimensions relevant to the ND280 TOF detector (2.3~m long and 12~cm wide) with coupling to a similar MPPC array, the measured time resolution is of the order of 150~ps, as shown in Fig.~\ref{TOF_spectrum} (right). This meets the design requirements of the TOF detector. The sensor boards, shown in Fig.~\ref{TOF_sensor}, are designed such as to have the possibility to have 16 sensors with 8 pairs of sensors connected in parallel with individual sensors within each pair also connected in parallel, to allow the possibility of a time resolution around 100~ps if needed.

\section{Readout electronics}
\label{TOFelectronics}

The signal readout scheme described in the previous section can be implemented either as a discrete circuit \cite{Aguilar2016} or as an ASIC. The former offers a cheaper solution which can be implemented in University electronics workshops, while latter has the advantages of compactness and possibility of remote configuration. Both options are considered for the TOF detector, depending on the amount of available funding. 


\begin{figure}
\centering
\includegraphics[width=0.4\textwidth]{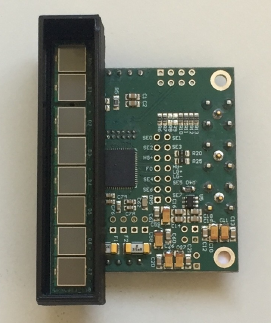}
\hfill
\includegraphics[width=0.45\textwidth]{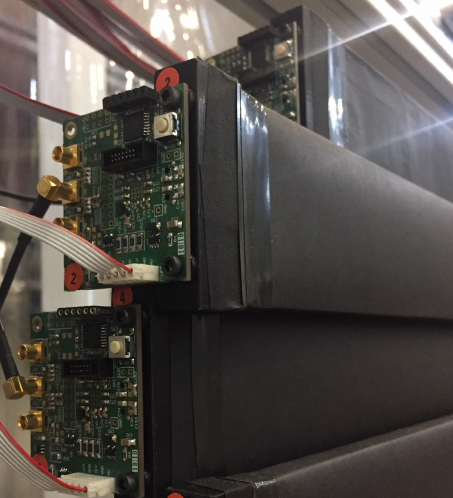}
\caption{
  Left: Array of 8 6$\times$6 mm$^2$ MPPC S13360-6050PE connected to the eMUSIC miniboard used for signal readout and summation for the 22-bar TOF detector prototype. Right: picture of the front-end electronics while taking data with the prototype.
  }
\label{fig:TOF_electronics}
\end{figure}

A 50~mm $\times$ 45~mm general-purpose electronic board based on the MUSIC chip~\cite{Gomez2016}, called eMUSIC miniboard, was employed for the readout of a 22-bar prototype detector array (see Section~\ref{sec:TOFprototype}). Pictures of the MPPC arrays and the eMUSIC miniboard are shown in Fig.~\ref{fig:TOF_electronics}. The board can be connected to 8 MPPC outputs through a high-density connector, and the outputs of the chip can be monitored via MCX connectors. It provides 8 individual analogue and discriminated outputs and two summation channels in the differential and single-ended mode for further digitisation. System settings as well as calibration parameters can be determined beforehand and thus the board control is reduced to a simple micro-controller which can be programmed once before the detector is installed. The board thus contains only two ASICs: the MUSIC and the micro-controller. The rest of the PCB is dedicated to the power regulator and various connectors such as low and bias voltage connectors, SPI connector, and analogue output connector. Omitting the readout of individual MPPC and using of only one (negative) signal polarity can reduce the board size by a factor of 2.


\section{Layout and mechanics}
\label{TOFmechanics}

\begin{figure}
\centering
\includegraphics[width=0.54\textwidth]{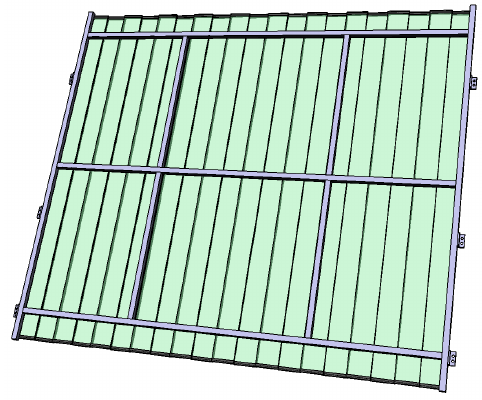}
\hfill
\includegraphics[width=0.4\textwidth]{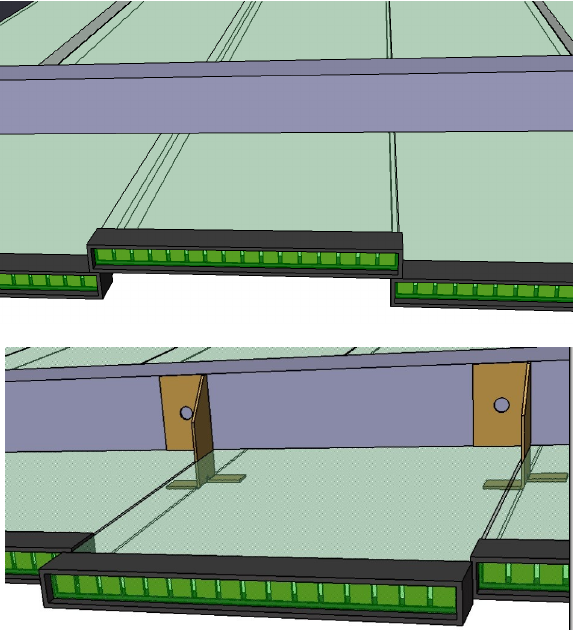}

\caption{
  Representation of a TOF plane with a 30~mm thick holding frame (left). The staggered design with overlap between bars (top right) has no dead space and is chosen for the top, bottom, left and right TOF planes to be placed outside the basket. The planar design (bottom right), to be used of the upstream and downstream planes inside the basket, features a 20~mm shift between adjacent bars to accommodate for the sensor covers and reduce the gap between bars to 1~mm (below 1\% of the active area). 
  }
\label{TOF_mechanics}
\end{figure}

The TOF detector arrays on the top, bottom, left and right will be placed outside the basket with 2.0~m bars staggered with a few mm overlap between them, as shown in Fig.~\ref{TOF_mechanics} (top, right). The staggering eliminates dead space between bars and allows for a flexible total array width by tuning the overlap. For these arrays, the design also includes a 30~mm thick aluminium frame fixed to the exterior of the ND280 basket to hold the bars and reduce sagging due to the gravitational pull and possible earthquakes. An example of such a frame is shown in Fig.~\ref{TOF_mechanics} (left). Calculations yield a maximum deflection of 9~mm which can be further reduced if needed by adding an aluminium plate to the structure. The bottom array is made of left and right parts with a hole in the middle for services but is otherwise similar in design. The total thickness of one detector plane, including the holding frame, is 4.5~mm, which leaves a comfortable margin of several cm separation from the EM calorimeter surrounding the arrays when the magnets are closed. 

For the upstream and downstream detector arrays, space is limited due to their placement inside the basket. Accordingly, they are designed as a single plane without staggering, with bars placed next to each other with a 1~mm gap between them and a 20~mm shift in length to accommodate for the protecting caps covering the sensors and the hooks inserted between bars to hold them, as shown in Fig.~\ref{TOF_mechanics} (bottom, right). These arrays will be inserted inside the basket and attached to existing structures. 

\section{Assembly, calibration and integration}

All TOF arrays will be assembled, tested and calibrated at CERN before being dismounted shipped to Tokai in separate boxes. The calibration is to be performed using cosmic muons crossing the whole array vertically such as the event shown in Fig~\ref{TOF_proto}, with a rate of approximately one event per minute. This allows to synchronise the timing signals to within 10~ps. 

At the ND280 site, the upstream and downstream TOF arrays need to be installed first by assembling individual bars directly into the ND280 basket before the installation of the target and TPC prevent the access. The top, left and right arrays will be assembled on the surface and then craned and attached in one piece to the exterior of the basket. The left and right arrays, once installed, prevent access to the target and TPC and therefore need to be easily removable. Finally, the bottom array, which will be assembled on the surface in two pieces, will be installed last, because it is located below the place where the cables of all subdetectors are gathered and channelled to the exit path below the basket. 

\section{DAQ}

The major design criterion for the data acquisition (DAQ) system is an internal time resolution which has to be much better than the expected resolution of the scintillator counter. For the TOF detector we use a system based on a SAMPIC ASIC~\cite{SAMPIC2015}. SAMPIC is a 16-channel chip implementing a novel type of digitizing electronics which performs both the function of a TDC and of a waveform
  sampler based on a switched capacitor array (SCA). The use of an analogue memory which is added in parallel with a delay line allows for analogue signal sampling at a very high rate. In addition, having the waveform recorded, one can extract various kinds of information such as baseline, amplitude, charge and time. The circular buffer of SAMPIC contains 64 cells which makes possible to cover a 20~ns window at the sampling frequency of 3.2 GS/s. It is enough to cover the rising edge of a signal (typically 3$-$6~ns) which is used for the digital Constant Fraction Discrimination analysis. In addition to the TDC, the ASIC contains an on-chip ADC which
  digitizes the waveform. Each channel of SAMPIC integrates a discriminator which can trigger itself independently of other channels. This is an important feature for a neutrino experiment such as T2K in which the incoming particle is not detected.

A 256-channel DAQ module will be assembled in LAL/Orsay. It will include four 64-channel SAMPIC new boards, one controller board and the backplane. The size of the module will be $10\times18\times25$~cm$^3$.
It will be placed at the bottom of the basket, right below the HA-TPC.

An accurate synchronisation between the timing signals of the active target and the TOF detector is mandatory. Indeed, the time-of-flight of a track identified in the TPC is measured as the difference between the time measured by the matched signal in the target and the time measured by the matched signal in the surrounding TOF bar. This requires a common start/stop and a common clock, and thus a unified solution for the DAQ electronics.

\section{Prototype Results}
\label{sec:TOFprototype}

\begin{figure}
\centering
\includegraphics[width=0.7\columnwidth]{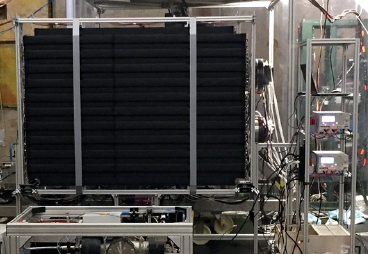}
\vspace{0.5cm}

\includegraphics[width=0.495\textwidth]{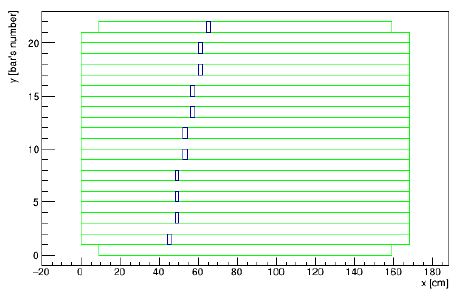}
\hfill
\includegraphics[width=0.495\textwidth]{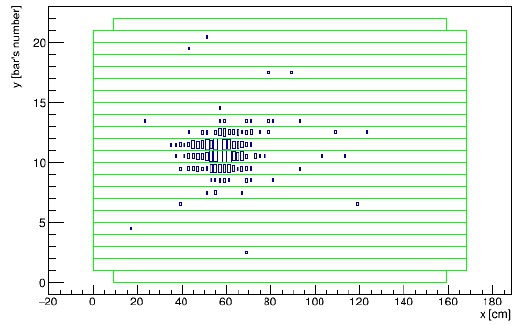}
\caption{
  Top: picture of the TOF detector prototype comprising 22 bars, placed in front of a high-pressure TPC prototype and exposed to test beams at the CERN PS. Bottom: event displays showing reconstructed particle positions using the difference in time between the two ends of each bar, for a cosmic event (left) and the beam profile without blocks (right). 
  }
\label{TOF_proto}
\end{figure}

In Summer 2018, a prototype array of 22 1.68~m long bars staggered with 5~mm overlap between them (shown in Fig.~\ref{TOF_proto}) was successfully operated at CERN PS test beams, providing time-of-flight information to a high-pressure TPC prototype.
This 44-channel prototype is very similar to an ND280 TOF detector plane in size and complexity, demonstrating practical solutions for the power distribution and heat dissipation, the signal readout and DAQ with a 64-channel SAMPIC module, the time synchronisation between bars, and the integration with other sub-detectors. 
Hamamatsu S13360-6050PE (area 6x6 mm$^2$ , pixel pitch 50 $\mu$m)
MPPC have been used for the light detection.
Displays showing reconstructed spatial distributions of hits for a cosmic event and for beam-induced events are also shown in Fig.~\ref{TOF_proto}. The prototype exhibits a timing resolution around 90~ps (similar to that of a single bar~\cite{Betancourt2017}) over its whole 2.1~m$ ^2$ active area.  

%% file: Integration.tex
\chapter{System Integration}

\section{Facility}

\subsection{Neutrino Assembly Building} 
The Neutrino Assembly (NA) building is located next to the experimental hall at the J-PARC site. It covers an area of 359 m$^2$. The T2K ND control room (45 m$^2$) can be found in this building. For the assembly and testing of the subdetectors 253 m$^2$ are available and the height in this area is 10 m. For storage of material and tools 61 m$^2$ are for disposal. Two movable cranes are available: a 5 crane and a 1 ton crane. Both are more than sufficient for the weights of the subdetectors. A clean room tent of 3.8$\times$5.3 m$^2$ and a height of 4.7 m is also available already. The assembly area will be shared between the different subdetectors. The exact needs for every subdetector is currently under discussion.

\subsection{Neutrino Monitor Building} 
\begin{figure}
\begin{center}
\includegraphics[width=0.5\textwidth]{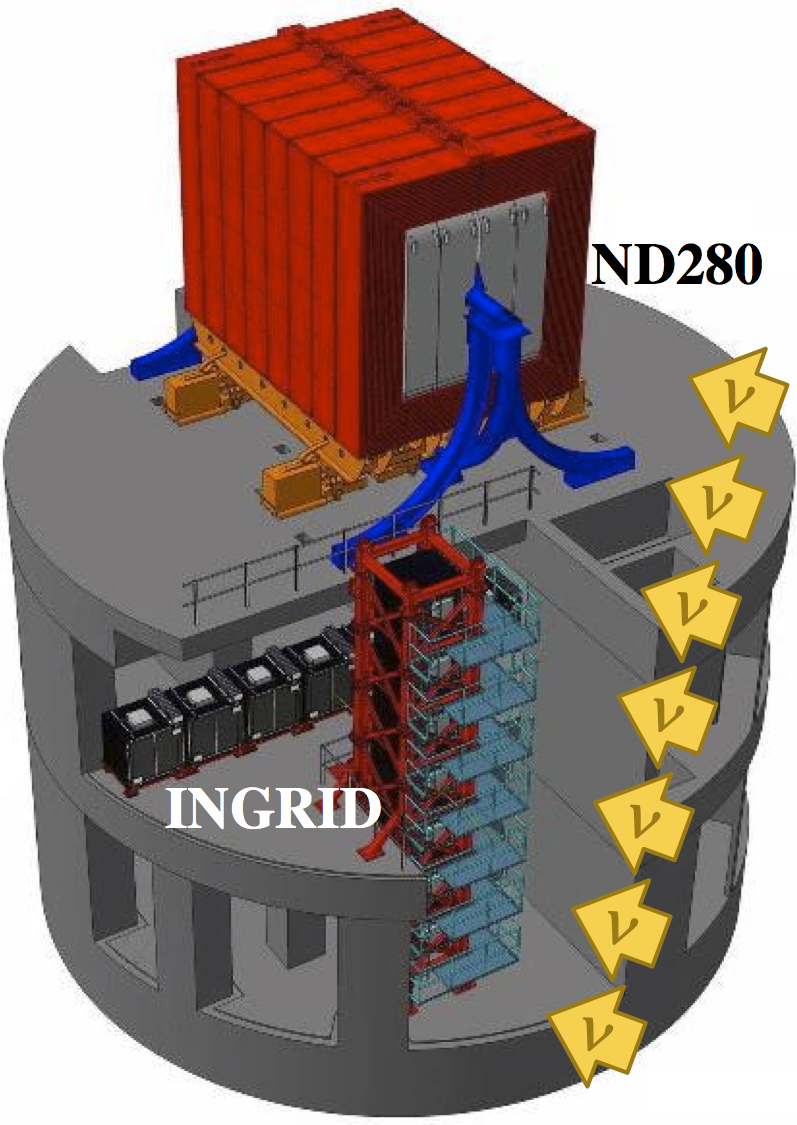}
\end{center}
\caption{3D sketch of the underground floors of the NM building. The B1 floor contains the ND280 complex, while the INGRID detector is situated in the SS and B2 floors.}
\label{fig:Integr_NM_det}
\end{figure}

The Neutrino Monitor (NM) building is located at about 280m from the target point. 
The NM building has been designed as follows, which is drawn in Fig.~\ref{fig:Integr_NM_det}.
It has a pit with a diameter of 17.5m and a depth of about 34m, which incorporates
both the on-axis detector (INGRID) and off-axis detectors. The B1 floor, which is about at a depth of 23m,
is for the off-axis detector. The off-axis detector is nearly located on the line between the target point
and the SK position. The SS (service stage) floor, which is about 29 m deep, is for the horizontal part of the on-axis detector. 
The B2 floor, which is about 34 m deep, is for the deepest part of the vertical on-axis
detector. The current nominal off-axis angle is 2.5 degrees and the on-axis beam line passes at about
0.8m above the SS floor. This facility design is adequate for off-axis angles of 2.0 - 2.5
degrees.
The hut with a size of about 21 m x 37 m covers the pit, and has a 10ton crane. The hut is a
little bit shifted to the north with respect to the pit center in order to use the north area in the hut for
the unloading of detector components and for the detector preparation (loading area). The effective
height of the crane is 4m and its dead space is about 3m from the north and south walls and 2m from
the east and west walls. The hut has an entrance shutter 5m wide and 3.9m high. There are a 6-people
elevator and stairs. Some area in the hut at the ground floor is used for the electricity preparation and
the cooling water preparation. \\
A 3D sketch of the underground infrastructure can be found in Fig.~\ref{fig:Integr_NM_det}.\\

\section{Surface Activities} 

\subsection{SuperFGD}
\begin{figure}[!htb]
  \begin{center}
    \includegraphics*[width=0.9\textwidth]{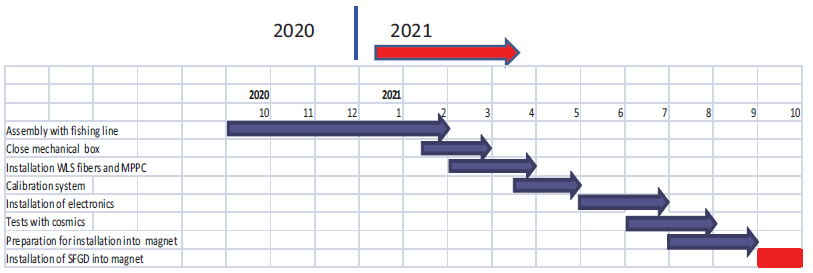}
  \end{center}
\caption{Preliminary plan for the assembly of SuperFGD at J-PARC. }
\label{fig:sfgd_plan}
\end{figure}

The SuperFGD will have about $2\times 10^6$ plastic scintillator cubes. All cubes will be produced in Russia. Mass production will begin in January 2019 and should be finished in January 2021. INR will provide the permanent control of the quality of scintillator cubes. Dimensions of cubes and position of holes inside cubes will be measured and controlled.  The light yield will be also measured using cosmics.  About 2-3\% of produced cubes will be tested at INR before shipping for 2 years of the mass production. Communications with the chemical company will be carried out constantly.   2-3 deliveries of plastics cubes to Japan are foreseen. A storage space for 3 boxes with cubes will be needed in the NA buildings. The assembly procedure is expected to start in October 2020. The special stand should be made for assembly. The discussion and the design of such a stand will start in the beginning of  2019. The preliminary plan for the assembly at J-PARC is shown in Fig.~\ref{fig:sfgd_plan}.
According to this plan, the detector  should  be ready for installation of electronics by June 2021. Tests of the assembled detector with cosmics and preparation for installation into the ND280 magnet will take about 3 months. The detailed plans for design, tests, production of electronics, calibration system, optical interface, mechanics will be defined and fixed in early 2019.   

\subsection{TPC}

The two HA-TPCs will be shipped separately from CERN to J-PARC. After arrival, the front end electronics will be remounted and the TPC will be tested for gas tightness, electrical connectivity etc. For these operations, an area of approximately 5 $\times$ 5 m$^2$ will be needed in the NA building. We plan to reuse the cleaner area with a plastic tent already used in the assembly of the existing TPCs in 2009. The support structure will be the same or similar to the one already used at CERN. These operations will last 2-3 weeks per TPC before the final installation in the basket.

\subsection{TOF}
The 6 TOF panels consisting each of 20 scintillator bars will be assembled first at CERN. All panels will be tested and characterized with cosmics before the panels will be dismantled again and shipped to J-PARC. To simplify the assembly at J-PARC the position of all pieces, especially the scintillator bars, will be documented. The motivation for this approach is to minimize the volume of the pieces to be shipped. To assemble the TOF panels which have a size of about 2.3 x 2.3 m$^2$ an assembly area of about 4 x 4 m$^2$ will be needed. There are 2 options under consideration where the assembly could take place: at the NA building together with the SuperFGD and TPC or on the surface level of the NA building. Considering that the panels will be tested already at CERN, a modest time for assembly and testing at J-PARC is foreseen.

\section{Overall Design}

\section{Basket Modifications}
\begin{figure}[ht!]
    \includegraphics*[width=15cm]{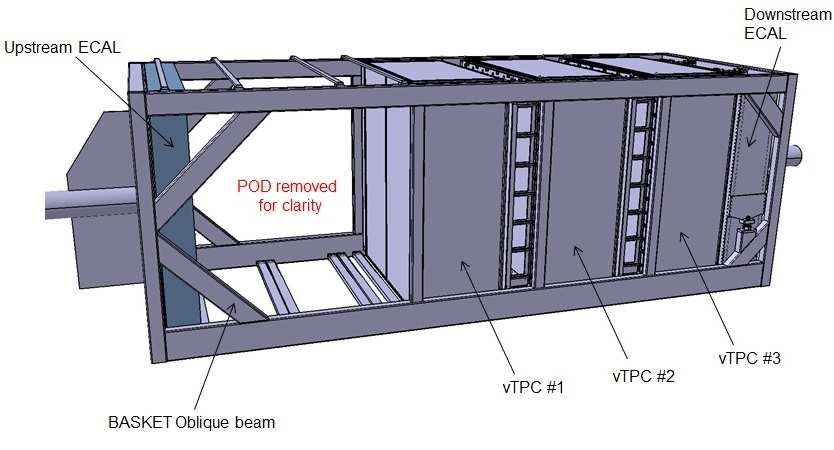}
\caption{The current Basket design. The two upstream oblique beams have to be modified to clear the access for the Upgrade detectors.}
\label{fig:TDR3}
\end{figure}

\begin{figure}[h!]
   \includegraphics*[width=13cm]{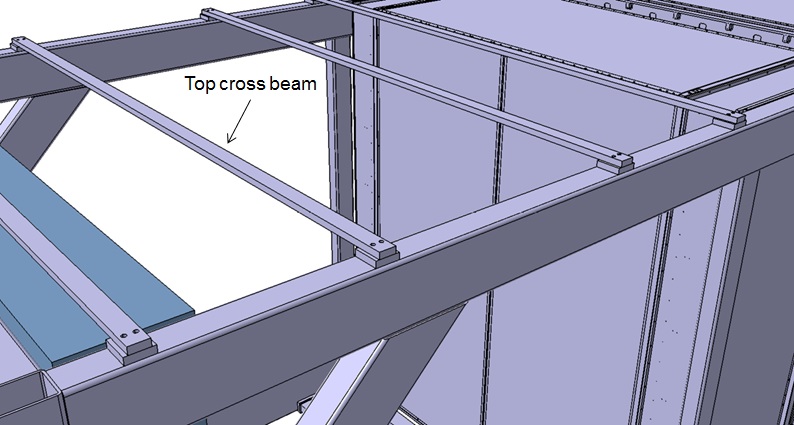}
\caption{Basket top cross beams that will be unscrewed from the basket structure.}
\label{fig:TDR4}
\end{figure}

\begin{figure}[ht!]
   \includegraphics*[width=10cm]{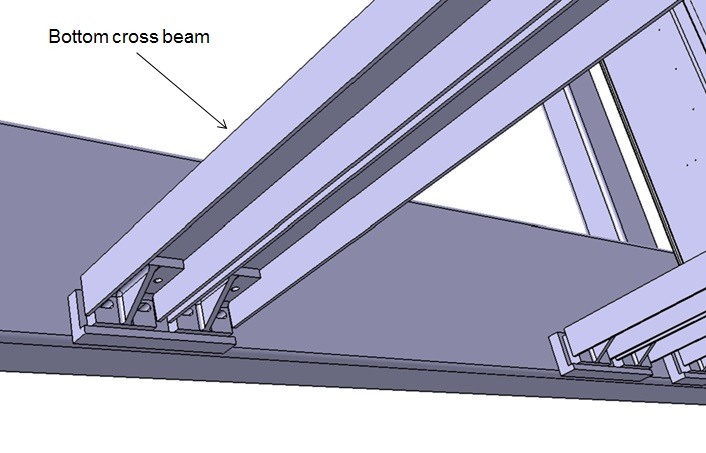}
\caption{Basket bottom cross beams that will be unscrewed from the basket structure if validated by FEA.}
\label{fig:TDR5}
\end{figure}

\begin{figure}[ht!]
   \includegraphics*[width=10cm]{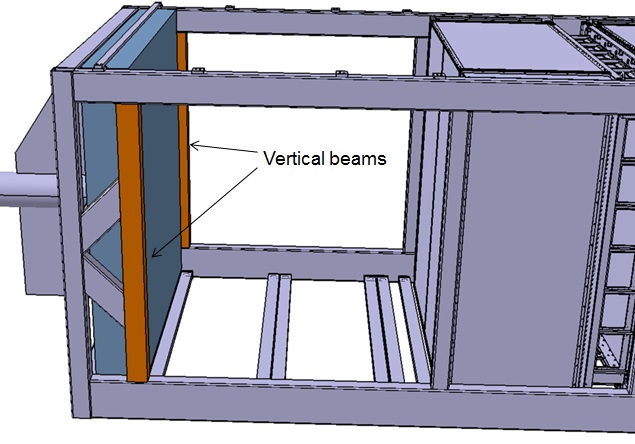}
\caption{Overview of a possible modification to the basket: vertical beams in orange and partial cuts on the oblique beams.}
\label{fig:TDR6}
\end{figure}

The Upgrade of T2K will require to remove the POD detector, hosted between the Upstream P0D ECAL and the first vertical TPC, labelled vTPC in figure~ \ref{fig:TDR3}. 
The first envisaged modification is to be done on the two oblique beams welded on each side of the Upstream ECAL (in blue). This will allow for an easy access for the electronics of both the SFGD and the two HA-TPC. Keeping them as they are might strongly impact the maintenance of the new detectors (front end electronics for SFGD, and Micromegas removal for HA-TPCs). The other modifications will be performed on the top and bottom cross beams (at P0D location). On the top side, three out of four cross beams will be removed to permit the fixation of the upperTOF (uTOF), while on the bottom side a Finite Element Analysis (FEA) will be done to validate a complete or partial removal of the beams (bolted to the inner side of the basket). This is to limit the amount of steel material in front of the bottom TOF (bTOF).  
On figure~ \ref{fig:TDR4} the top cross beams can be easily detached from the Basket structure (2x2 screws per beam). They will be replaced by a frame in aluminum casting the support of the uTOF. Hence, there will be no issue related to the stability of the whole Basket.
The figure~\ref{fig:TDR5} shows the cross beams from below. The fixation is achieved by means of small brackets between the U shaped beams in stainless steel, and the resting brackets which are welded onto the inner side of the Basket. As for the top part, only the welded blocs will remain for a potential re-use (new detector supports).
The main modifications applied to the oblique beams will require to cut the beams by means of circular saw or equivalent. It has to be limited as much as possible due to the metallic dust that is created by the sawing. The figure~ \ref{fig:TDR6} shows a possible design for modification with a partial cut of the oblique beams to clear the needed access to the new detectors. The vertical beams (orange color) link the top and bottom oblique beams to keep the overall stability of the Basket when loaded by the new detectors. A dedicated FEA is mandatory to validate the design of the new configuration respect to the specifications at the J-PARC site (gravity sag, earthquake). However, it has to be considered that the weight of the new detectors will be much lower than the current one of the POD itself.

\section{Sub-Detector Support Structure} 

All sub-detectors must be stably supported in the basket during long-term operation. In addition, support structure should be designed to maximize detector envelopes and to allow the insertion of sub-detectors during detector installation. We plan to adopt similar bracket design as current ND280, which has been successfully supported a few tons of several detectors for about 10 years. 

\begin{figure}[htbp]
  \begin{center}
    \includegraphics*[width=8 cm]{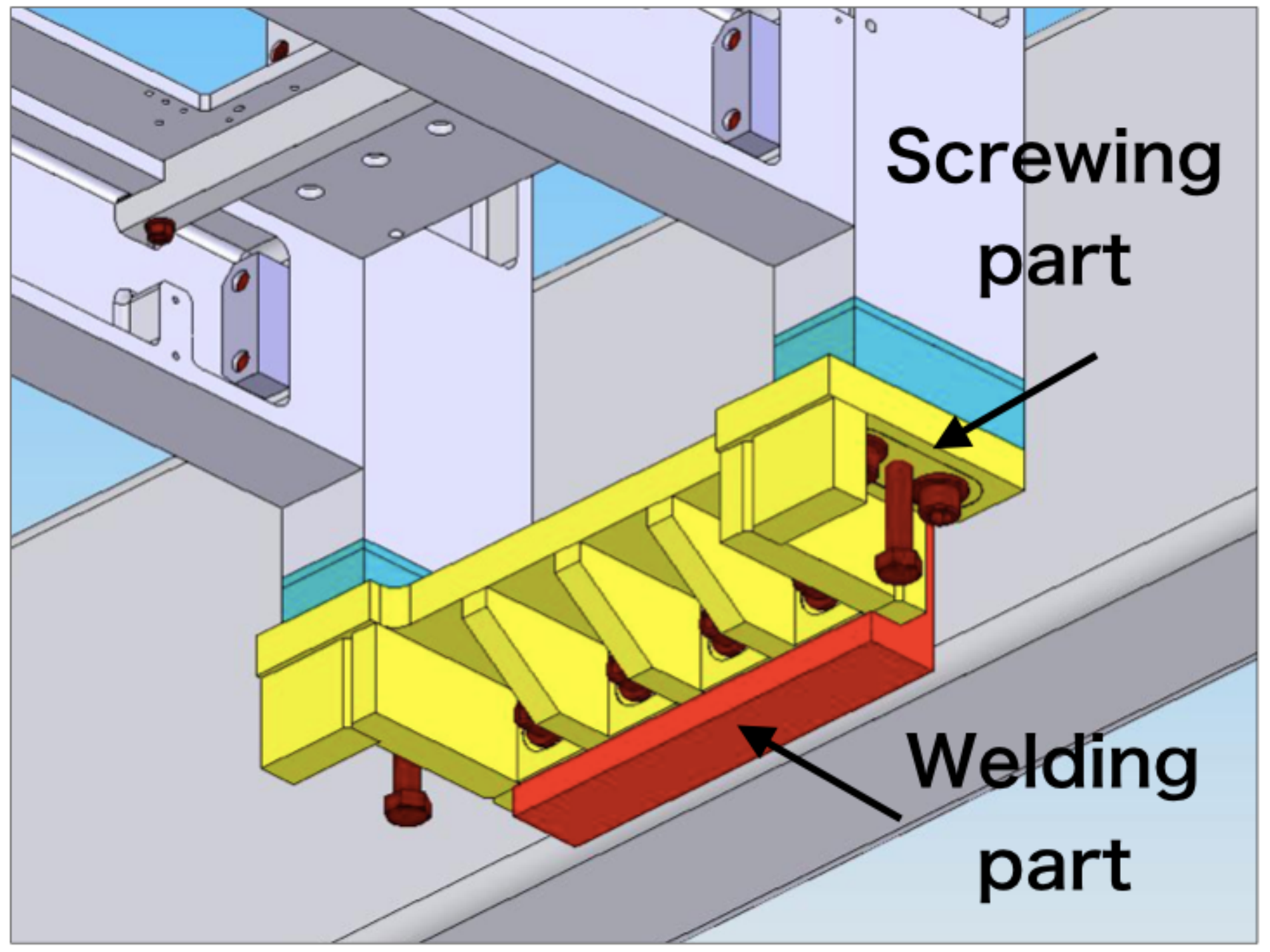}
  \end{center}
\caption{Bracket design of current FGD. It consists of welding part (red) and screwing part (yellow). Screwing part can be dismounted during detector installation inserting from the basket top.}
\label{fig:bracket}
\end{figure}

Bracket design of current FGD is shown in Figure~\ref{fig:bracket}. The bracket is made of stainless steel (SUS304). It consists of two parts, called welding part and screwing part. The welding part is welded on vertical pillar after the basket modification. The screwing part can be then screwed by bolts during installation work. We will place brackets at four corners of each detector for two High-Angle TPCs and SuperFGD. Expected load of those detectors are about 0.34 and 2 tons, respectively. We would support SuperFGD electronics with an independent brackets as it is assembled after the detector insertion to the basket. 

The basket top is open to allow the insertion of sub-detectors during installation work. We plan to install sub-detectors stacking those one-by-one. The removable screwing part of the bracket allow us to avoid overlap with a detector and a bracket maximizing available detector volumes. This simple design does not require many works to manufacture and construct. Shim plate can be used to provide level surface of the detector. We plan to perform FEA study with expected loads and design of modified basket. 

\subsection{Sub-Detector Envelopes} 
\begin{figure}[ht!]
\begin{center}
\includegraphics*[width=10cm]{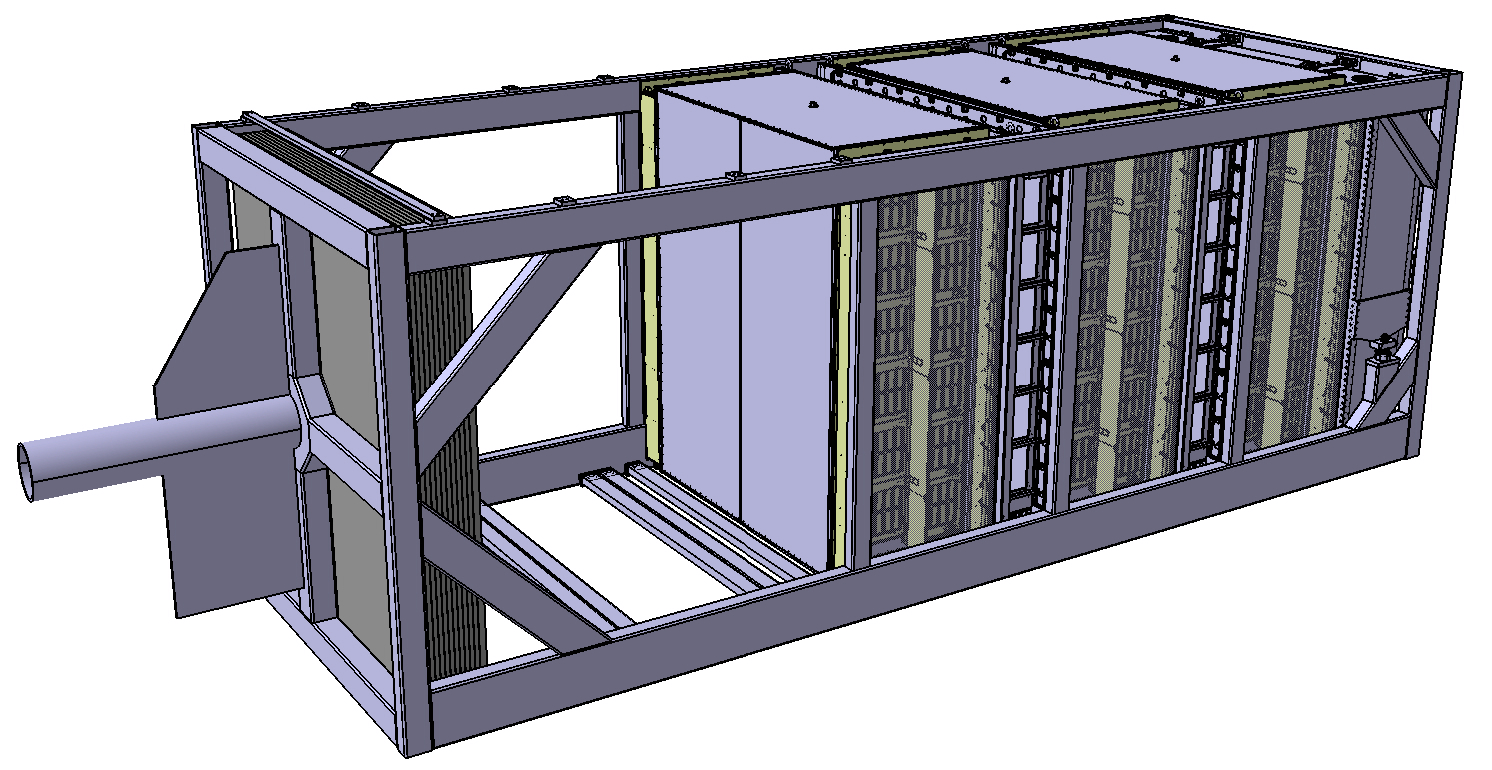}
\end{center}
\caption{Basket CAD model with POD removed.}
\label{fig:Basket_TDR7}
\end{figure}

\begin{figure}[th!]
\begin{center}
\includegraphics*[width=10cm]{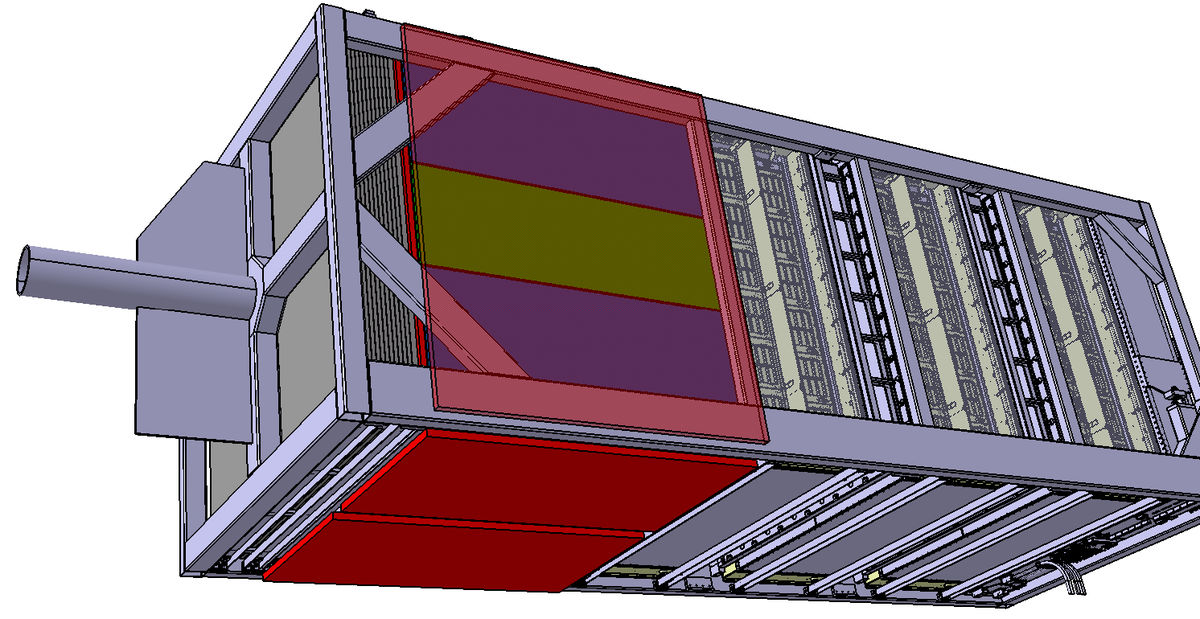}
\end{center}
\caption{Overview of the envelopes of new detectors.}
\label{fig:Basket_TDR9}
\end{figure}

Based on the CAD assembly model of the current T2K detector (Figure~\ref{fig:Basket_TDR7}), and including the current Basket structure, it has been proposed to build a new CAD model (Figure~\ref{fig:Basket_TDR9}) with the volume representing each new detector: 1 sFGD, 2 HA-TPCs, 6 TOF's. The volume includes the active parts of the detectors, the mechanical frames or structures and their electronics, excluding the services going to the control room (racks). The aim of this model is create a tool to keep track of all the modifications on the sizes and the potential impacts on the nearby detectors. This model has been checked by a few laser measurements in Tokai so that the key dimensions are now validated and the model reliable enough to proceed with the next steps (basket modifications, detailed CAD models of detectors).

\begin{figure}[ht!]
\begin{center}
\includegraphics*[width=13cm]{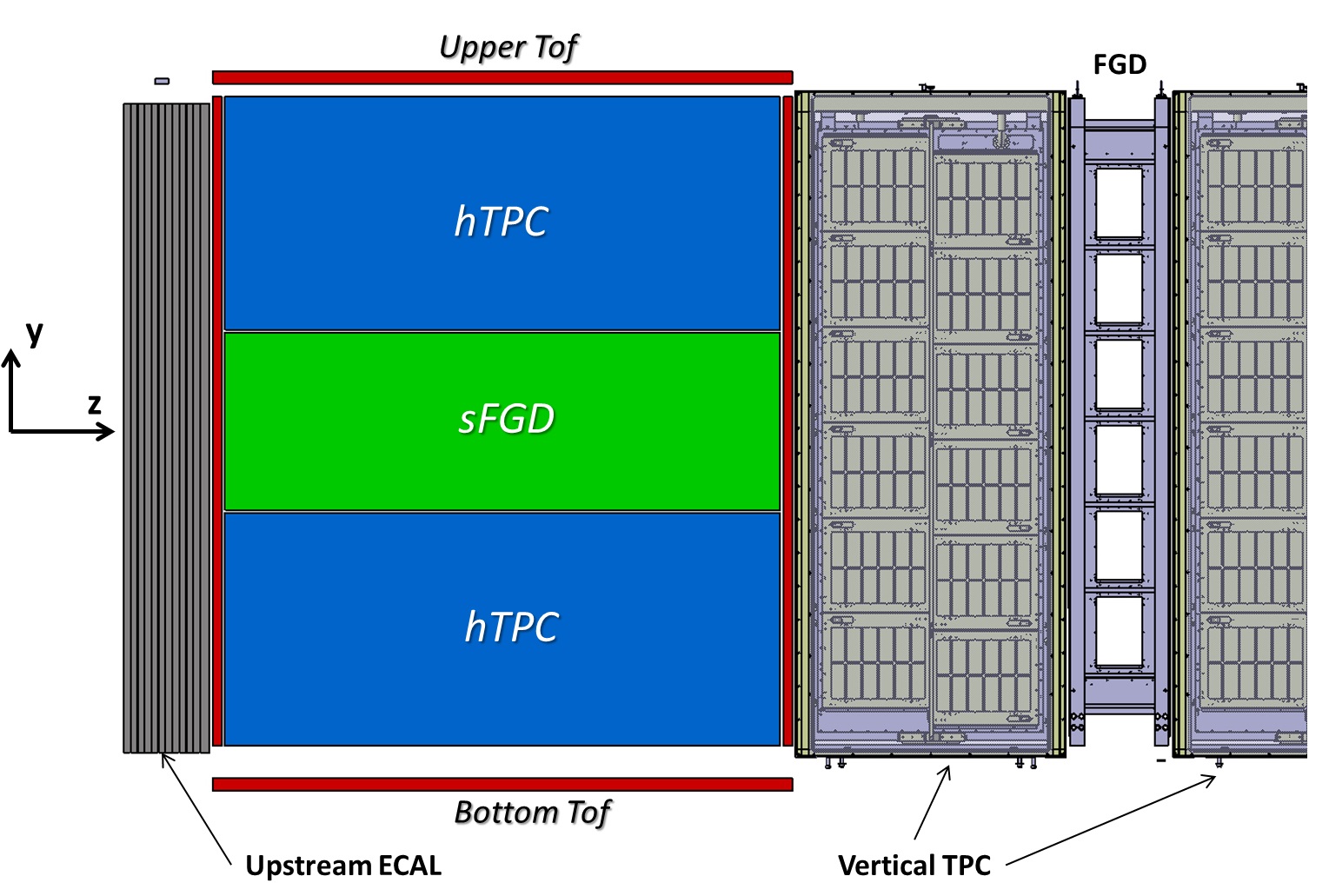}
\end{center}
\caption{Overview of the envelopes of new detectors}
\label{fig:TDRenvelope}
\end{figure}

The model in (Figure~\ref{fig:TDRenvelope}) shows the current detectors (Upstream ECAL, and Vertical TPC) with the insertion of the envelopes, that is the maximum volume that can be used by each detector. The rule is to keep 10mm minimum clearance between detectors in all axes. This is to prevent any clash or interference when detectors are integrated into the basket, and also to account for gravity sag (and possibly side loads for earthquake scenario). Note that the bigger gap between TOF and both HA-TPCs can be used for inner servicing. 

\begin{figure}[th!]
\begin{center}
\includegraphics*[width=13cm]{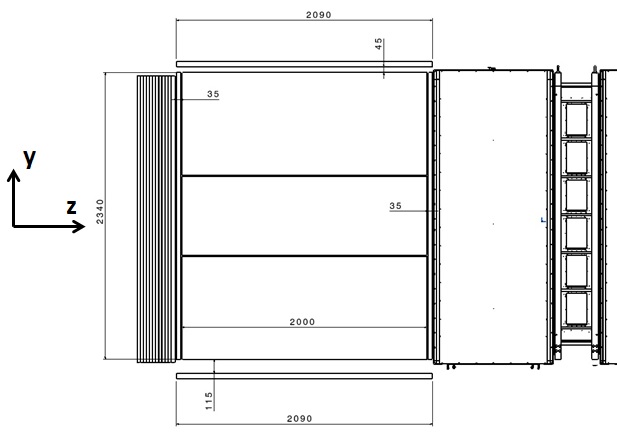}
\end{center}
\caption{Envelope dimensions in the y-z plane.}
\label{fig:TDRenvelope2}
\end{figure}

\begin{figure}[th!]
\begin{center}
\includegraphics*[width=10cm]{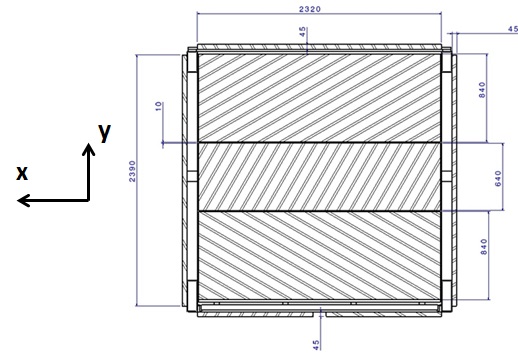}
\end{center}
\caption{Envelope dimensions in the x-y plane (cross section).}
\label{fig:TDRenvelope3}
\end{figure}

The (Figure~\ref{fig:TDRenvelope2}) shows the status of the envelope dimensions in the y-z plane, while in (Figure~\ref{fig:TDRenvelope3}) the x-y plane is sketched. 

The opening that is shown on the bottom TOF is to allow the main servicing for the new detectors. 
\section{Insertion/Removal Procedure} 
\subsection{POD Removal} 

The ND280 Pizero Detector (P0D) was installed in the ND280 Off-axis
detector during the original T2K installation and was designed to measure
the rate of $\pi^0$ production on water.  It consists of four modules that
are mounted at the upstream end of the basket.  The weight of each module
is independently supported on a mounting frame that is attached to the
basket.  The mounting frames provide the lift points by which the modules
are transported.  Because each module is independently supported, they can
be independently removed.  The modules are prevented from moving laterally
within the basket by a system of clamps and shims which can be removed.

In addition to the P0D modules and supports installed in the basket,
lifting fixtures and transport carts were manufactured.  This equipment is
currently stored at J-PARC, and will be used as part of removing the P0D.
Based on lessons learned during the installation of the P0D, it is
anticipated that minor modifications to the lifting fixture will be
required.  The lifting fixture was constructed according to both U.S. and
Japanese safety standards, and was certified before use, however, since the
lifting fixture has not been used in several years, it is anticipated that
it will need to be reinspected and certified by the appropriate regulating
organizations.

The P0D was installed in the basket prior to the installation of the first
TPC module which is situated immediately downstream of the detector.  This
allowed the P0D to be installed in sequence, starting from the most
upstream module (the Upstream ECal), and then proceeding to the Upstream
Water Target, the Central Water Target, and, finally, the Central ECal.
After the installation of the Central ECal.  Since the P0D is often filled
with water, braces were installed to support the lateral force of the water
pressure.  These braces support the downstream end of the Central ECal, and
are now trapped between the Central ECal and the first TPC.

Installation of the T2K upgrade detectors requires that three of the
existing modules be removed, while the most upstream module (the Upstream
ECal) will be left to function as the Upstream ECal for the new
configuration.  The removal will be done without modifying the currently
installed TPC modules.  For this reason, the removal procedure must protect
the TPC modules and the Upstream ECal module from damage.  In addition, the
supports at the downstream end of the Central ECal will not be accessible
until the Central ECal has been removed.

During the design of the P0D detector, it was envisioned that one of the
modules may need to be removed for maintenance.  The design to remove a
module was never fully developed, but the mounting method includes the
necessary features to remove single modules.

The key to removing a single module is that space must be created between
it and the neighboring modules.  The modules are installed with
approximately 1~cm of clearance between adjacent modules.  This space is
filled using shims that can be removed.  To remove a module, all of the
shims will be removed, and modules will be moved to maximize the space
around one of the modules.  The module supports rest on the basket, and are
fixed in place by bolts.  These bolts can be removed, allowing the modules
to slide.

The proposed plan to remove the P0D is to first move the Upstream ECal, and
Upstream Water Target as far upstream as possible.  This will free up
about 2~cm of clearance on either side of the Central Water Target.  This
space will be sufficient to allow a lifting fixture to be installed, and
for the Central Water Target to be removed.  After the removal of the
Central Water Target, the Upstream Water Target will be moved to create
clearance between it and the Upstream ECal so that the lifting fixture can
be installed.  Finally the procedure can be repeated for the Central ECal.
After the Central ECal is removed, the downstream support braces will be
accessible and can be removed.  

\subsubsection{Lifting fixtures}

During installation, several lessons were learned, and it is clear that the
lifting fixture will require minor modifications.  The most important issue
that was identified is that there was an unanticipated interaction between
the fiber lifting straps that carried the weight of the modules and the
motion of the crane.  This caused the modules to oscillate up and down in
resonant motion as they were lowered by the crane.  While the lifting
straps were chosen with appropriate safety factors, the modules should be
further stabilized during the removal, and the straps should be replaced
with a rigid material.  At the same time, the straps used during
installation required significant clearance between the modules to be
removed.  This can be mitigated by replacing the fiber lifting straps with
a metal bar which can be mounted directly on to the module support frame.

\section{Cable and Service Distribution} 

\begin{figure}[th!]
\begin{center}
\includegraphics*[width=13cm]{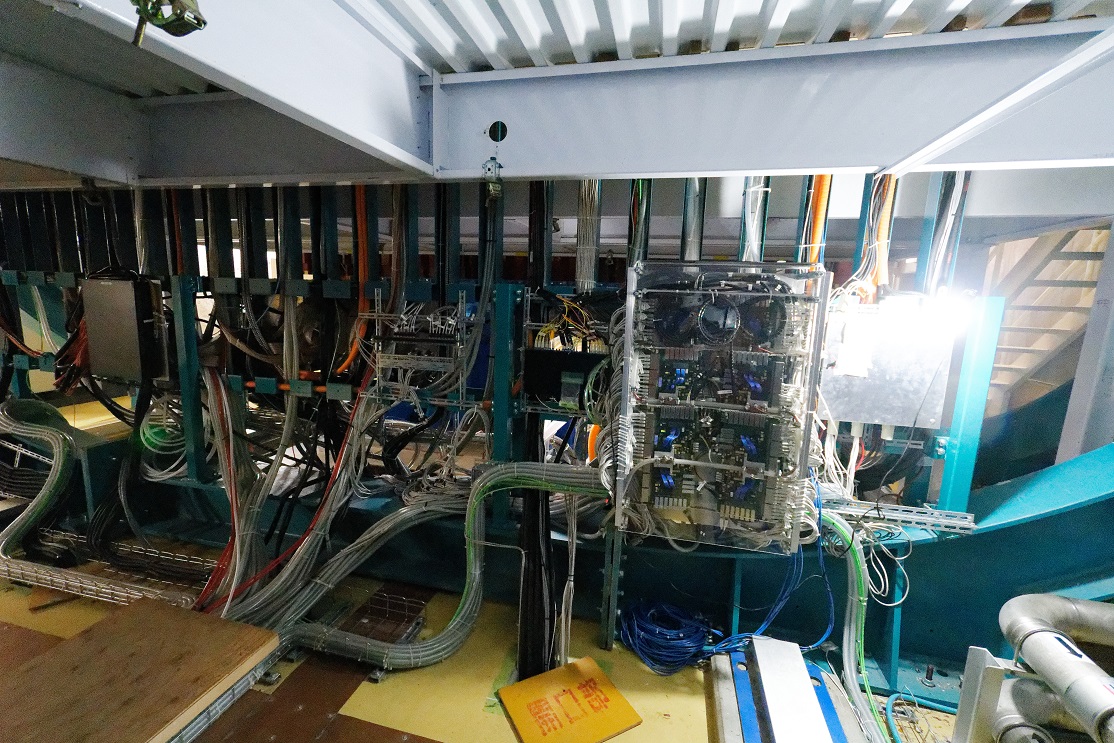}
\end{center}
\caption{Cable trays to bring cables from outside the magnet to the basket. 16 of the 24 cable trays are used for the current tracker system and therefore only 8 cable trays will be available for the new subdetectors.}
\label{fig:IntegrCableTrays}
\end{figure}

The cable and service distribution is an important task within the integration project considering the limited space available. Especially the number of cables and services which have to be brought from outside the magnet to the basket have to be minimized since only 8 cable trays, each of 8.5$\times$9.6 cm$^2$ large, are available for this.
Fig.~\ref{fig:IntegrCableTrays} shows the cable trays. For all subdetectors it is estimated that around 380 cables (SuperFGD: 58, TPC: 82, TOF: 240) will be necessary. Further effort will be undertaken to reduce especially the number of cables needed for the TOF by providing a distributor for the MPPC bias voltage inside the basket. 
Additional 240 signal cables for the TOF will have to be guided from the 120 scintillator bars of the TOF to the electronics module installed inside the basket. Also 18 water cooling pipes will be installed to transport the electrical power released in the basket to the outside. \\
Within the magnet the cables and services will be guided below the bottom TOF to the sides and from there up to the different subdetectors. In order not to interfere with the access to the SuperFGD and TPC electronics the cables and services will be installed on the sides of the upstream ECAL/TOF and the downstream TOF and outside of the basket. Since the barrel ECAL in this region is thinner, about 24 cm around the basket are available for this. In this way the amount of material in form of cables and services between the inner subdetectors is minimized.








%% file: Physics.tex
\chapter{Physics Motivations and expected performances}
\section{Physics achievements of ND280 and requirements for the upgrade}
In a long-baseline experiment the role of the Near Detector is to predict the unoscillated and oscillated spectra at the Far Detector reducing as much as possible uncertainties related to the neutrino fluxes and neutrino cross-sections. 
 
 
In the case of T2K it had been decided to build a magnetized off-axis near detector, ND280. Thanks to a set of three Time Projection Chambers (TPCs) surrounding two Fine Grained Detectors (FGDs), ND280 has excellent capabilities of measuring the momentum and the charge of the charged particles entering one of the TPCs. This allows to distinguish, for example, between negatively charged leptons produced by neutrino interactions and positively charged leptons produced by anti-neutrinos. This capability is particularly important when data are collected in anti-neutrino mode, when $\sim30\%$ of the interactions in ND280 are induced by neutrinos.

In addition ND280 has excellent particle identification capabilities, thanks to the presence of three TPCs and the surrounding electromagnetic calorimeter, it is possible to distinguishing between muons and electrons, selecting a clean sample of charged current $\nu_e$ interactions.

ND280 has been used for all the T2K oscillation analyses and it has been able to reduce the uncertainties due to neutrino fluxes and cross-sections from $\sim15\%$ to $\sim4\%$ as shown in Tab.~\ref{tab:nd280syst}.\footnote{The Super-Kamiokande--related systematics uncertainties will not be treated into this document but the collaboration is actively working to reduce them to the 1\% level. It should be noted that in the current treatment SK systematics are obtained from atmospheric neutrinos control samples and hence affected by flux and cross-sections uncertainties.}

\begin{table}[htb]
\centering
\caption{Effect of 1$\sigma$ variation of the systematic uncertainties on the predicted event rates at Super-Kamiokande of the $\nu$-mode samples.} 
\begin{tabular}{l c c  c} \hline \hline
Source of uncertainty& $\nu_e$ CCQE-like & $\nu_{\mu}$& $\nu_{e}$ CC$ 1\pi^+$ \\ 
 &$\delta N/N$&$\delta N/N$&$\delta N/N$\\ \hline

Flux &3.7\%& 3.6\%&3.6\% \\ 
(w/ ND280 constraint) & &  & \\ \hline
Cross section &5.1\%& 4.0\%&4.9\% \\ 
(w/ ND280 constraint) & &  & \\ \hline
Flux+cross-section&&& \\
(w/o ND280 constraint)&11.3\% &10.8\%&16.4\%  \\
(w/ ND280 constraint)&4.2\% &2.9\%& 5.0\% \\ \hline
FSI+SI+PN at SK& 2.5\%& 1.5\%&10.5\% \\ \hline
SK detector & 2.4\%& 3.9\%&9.3\% \\ \hline
All & && \\
(w/o ND280 constraint)& 12.7\%& 12.0\%&21.9\% \\ 
(w/ ND280 constraint)& 5.5\%& 5.1\%&14.8\% \\ \hline \hline
\end{tabular}\label{tab:nd280syst}
\end{table}

\begin{figure}[ht!]
  \includegraphics[width=0.45\linewidth]{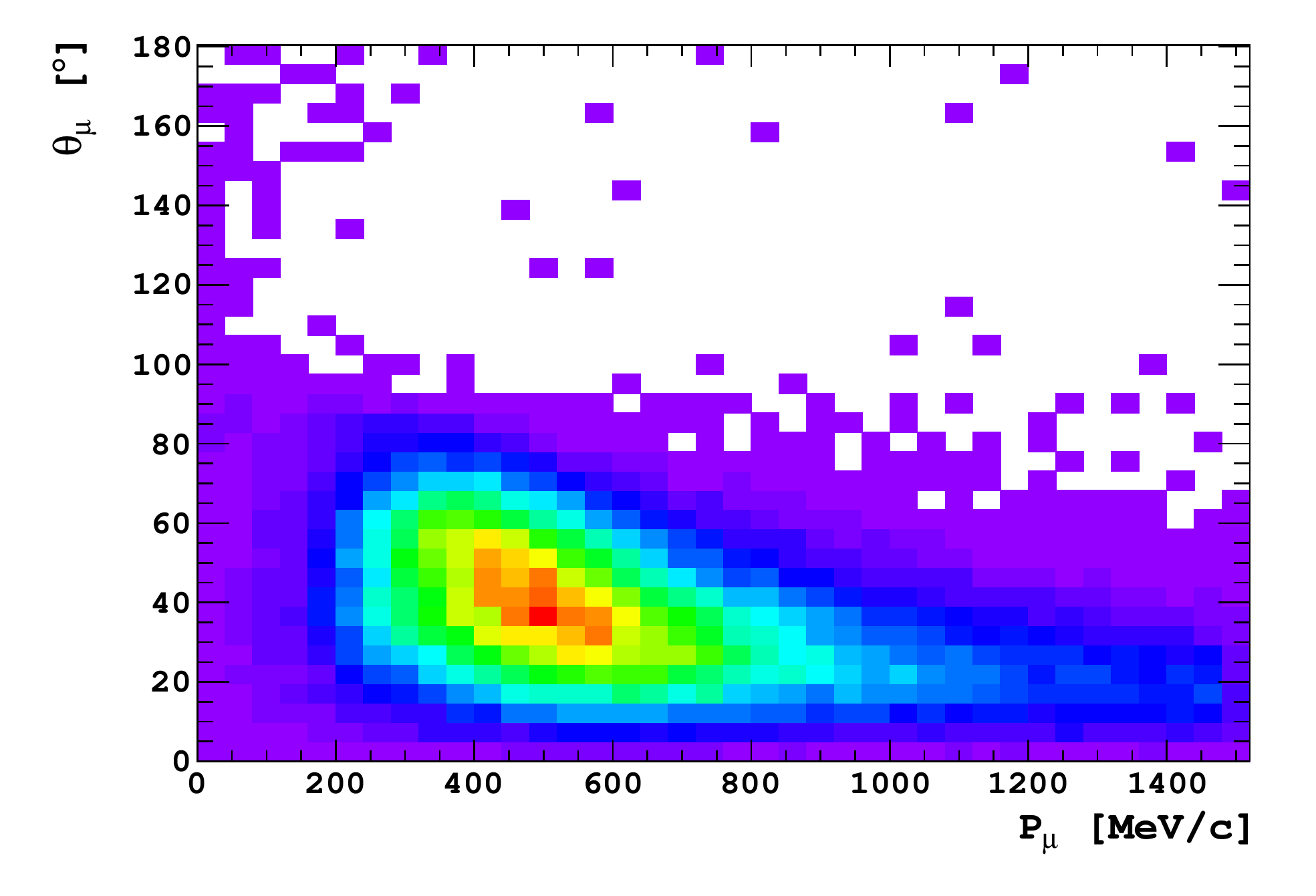}
  \includegraphics[width=0.45\linewidth]{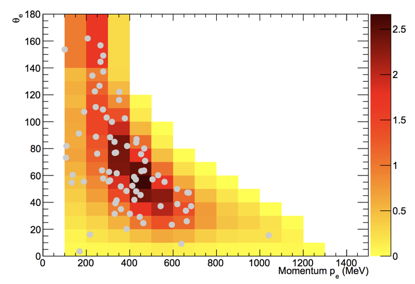}
\caption{Reconstructed momentum and angle for muons selected at ND280 (left) and electrons selected at SK (right). }
\label{fig:nd280skcomparison}
\end{figure}

The main limitation that has been identified in the current ND280 design is that, in order to precisely determine the properties of the leptons emitted in neutrino interactions, they have to be reconstructed in one of the TPCs. As a consequence the efficiency in the forward region is excellent but it drops considerably for scattering angles with respect to the beam direction larger than $\sim50$ degrees.
 At Super-Kamiokande, instead, given the $4\pi$ symmetry of the detector, the efficiency is flat with respect to the beam direction. The different acceptance between ND280 and SK is clearly shown in Fig.~\ref{fig:nd280skcomparison}.
 
In the extrapolation from the expected spectra extracted
using forward going tracks at ND280 to the ones at SK, cross-section models are needed to describe the dependency on the momentum transferred $Q^2$ or on momentum and angle of the lepton.  

\begin{figure}[ht!]
 \centering
  \includegraphics[width=0.6\linewidth]{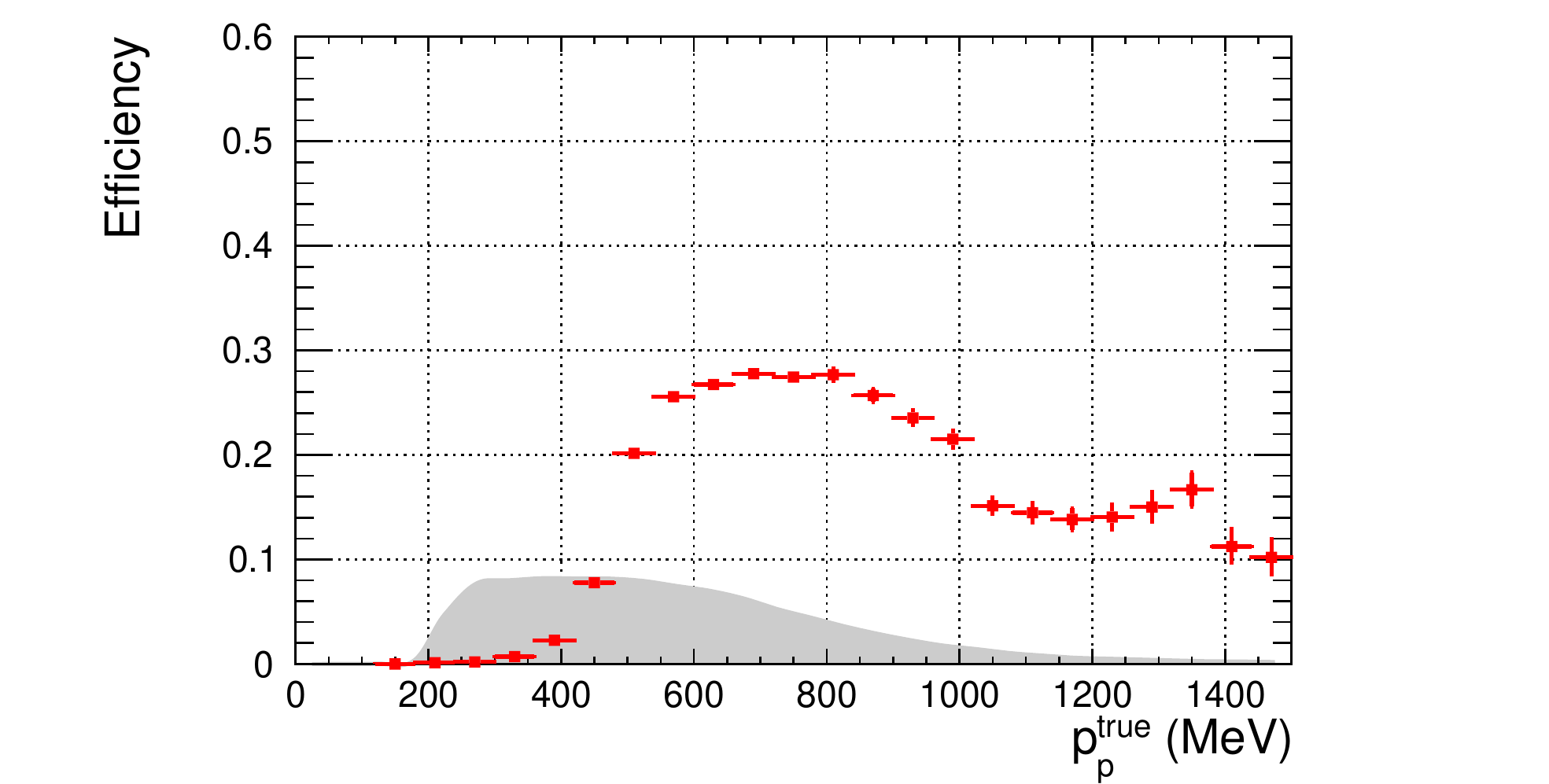}
\caption{Proton reconstruction efficiency in ND280. The grey histogram corresponds to the spectrum of generated protons according to NEUT MC 
}
\label{fig:nd280current_protons}
\end{figure}

In addition, tracks not entering the TPCs can only be reconstructed in two dimensions with the FGD. This implies limited tracking efficiency and a  relatively high momentum threshold, especially for protons. As an example, the protons reconstruction efficiency in ND280 is shown in Fig.~\ref{fig:nd280current_protons}. As it will be explained in Sect~\ref{sec:stvsuperfgd}, the reconstruction of low momentum pions and protons is fundamental in order to investigate nuclear effects in neutrino interactions.

\begin{figure}[ht!]
  \includegraphics[width=0.45\linewidth]{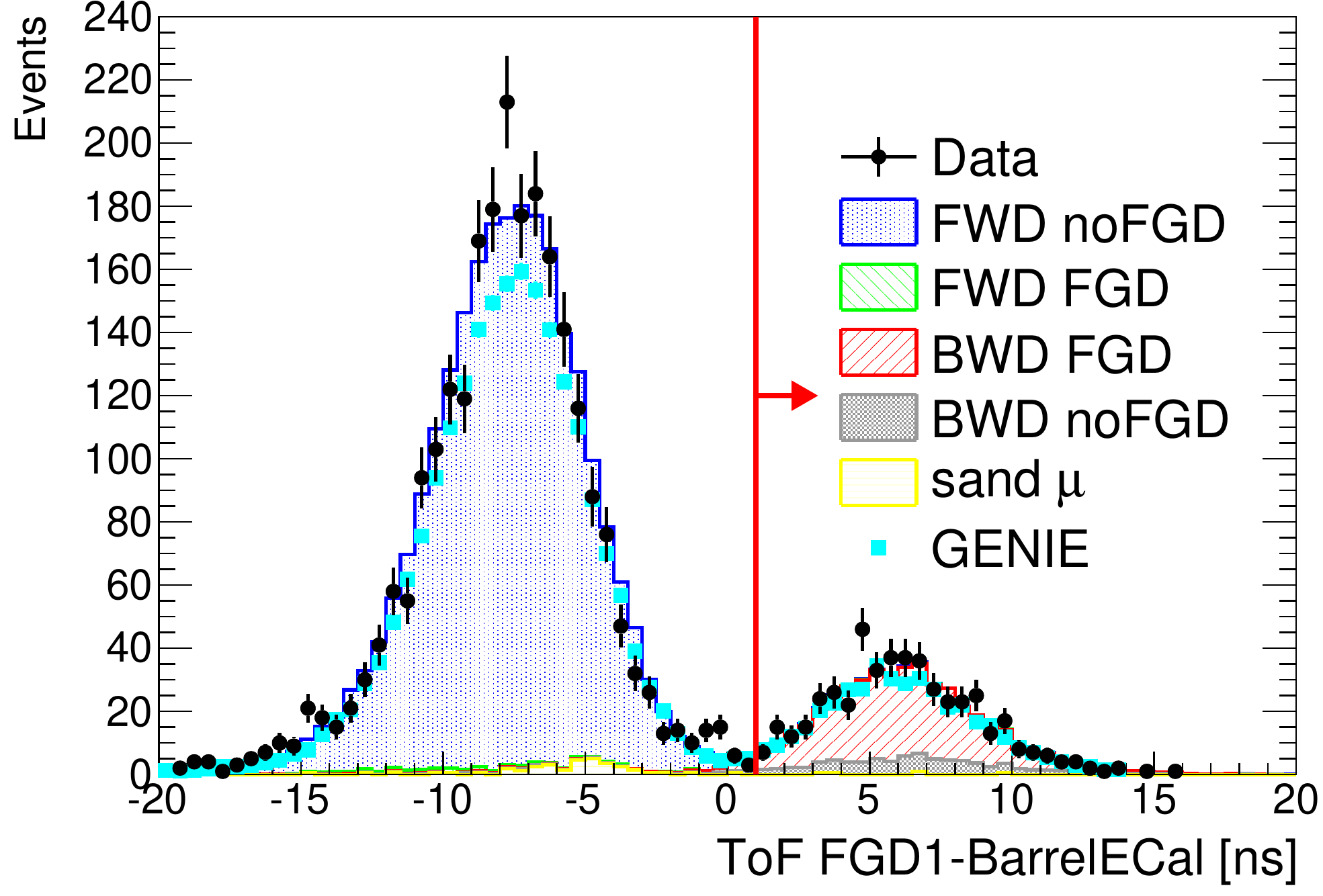}
  \includegraphics[width=0.45\linewidth]{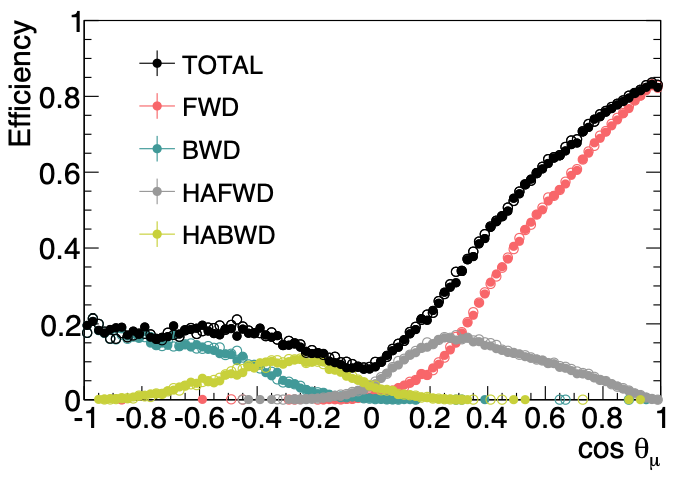}
\caption{ToF between FGD1-BarrelECal for tracks crossing
BarrelECal-TPC1-FGD1 (left) and efficiency as a function of muon polar angle for the ND280 selection with improved angular acceptance (right).  }
\label{fig:nd280_backward}
\end{figure}

The possibility of improving the angular acceptance of the current ND280 design has been investigated by the collaboration~\cite{Abe:2018uhf}. Leptons emitted with large polar angles can be reconstructed in the ECal or in the upstream TPC and can be distinguished from forward going tracks emitted by neutrino interactions upstream, thanks to the time of flight between two scintillator detectors, for example the P0D and the FGD1. As shown in Fig.~\ref{fig:nd280_backward}, the direction of the track can be easily determined, since the time difference is of the order of 6~ns, but the efficiency is relatively small ($\le$20\%), due to the requirement of having a track reconstructed in two high density detectors.

\begin{figure}[ht!]
 \centering
  \includegraphics[width=0.6\linewidth]{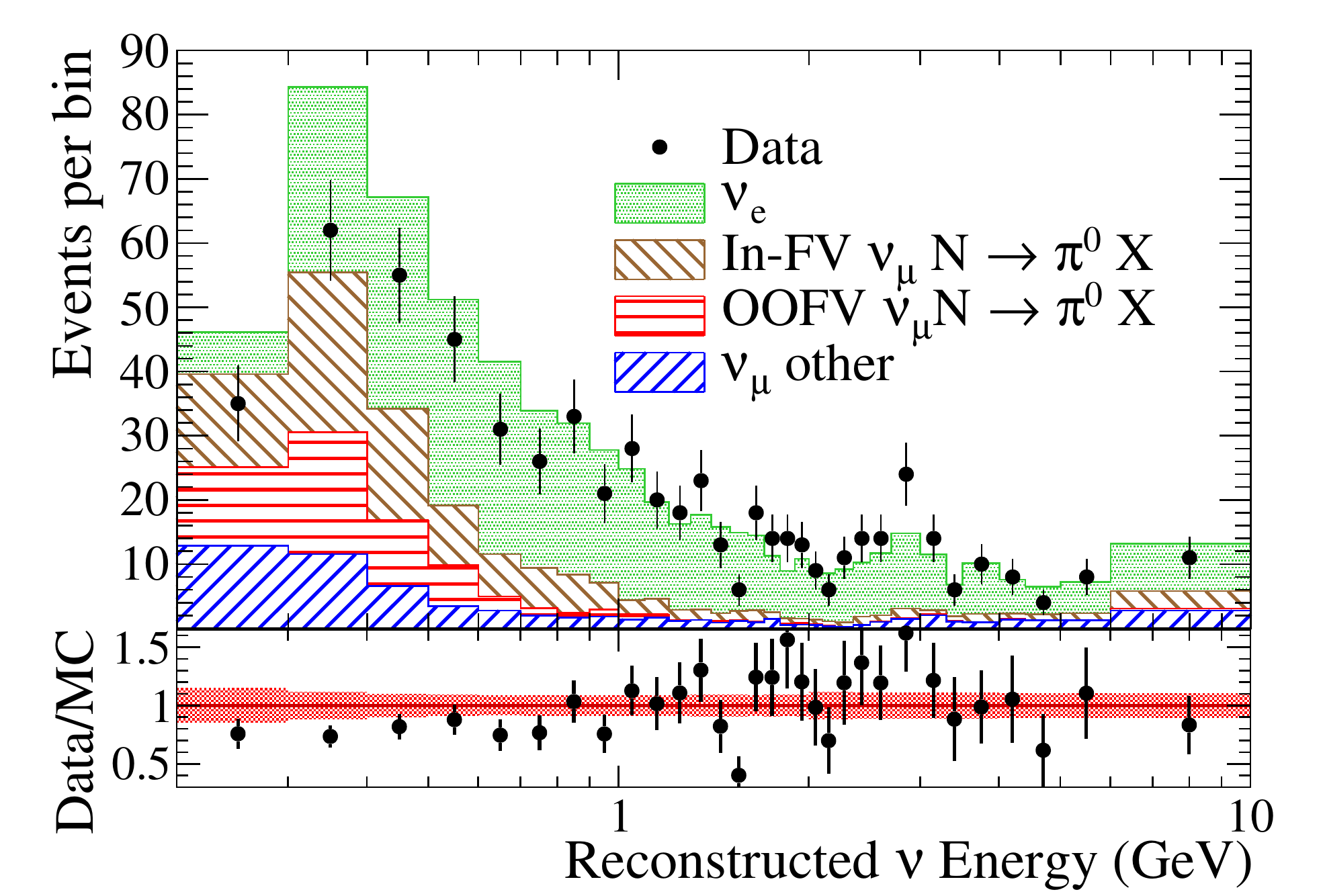}
\caption{Reconstructed neutrino energy distributions of the $\nu_e$ sample in the ND280 tracker. }
\label{fig:nd280current_nue}
\end{figure}

Another limitation of the Near Detector is its poor efficiency in selecting electron neutrinos with energies below 1 $GeV/c^2$, related both to limited efficiency for tracks at high angles, and to a large contamination due to converted gammas (see Fig.~\ref{fig:nd280current_nue}. The small number of $\nu_e$ selected at ND280 prevent the use of this sample in the oscillation analyses and the method used in T2K to constrain flux and cross-section systematic uncertainties solely relies on the selection of muon neutrinos at ND280 to constrain uncertainties for both $\nu_{\mu}$ and $\nu_e$ at SK. 
An additional uncertainty of 3\% due to possible cross-section model differences between $\nu_{\mu}$ and $\nu_e$ is then included in the oscillation analysis and has a non-negligible effect on the final systematic error budget. ND280 has already measured $\nu_e$ interactions in the Tracker and in the P0D and with the current statistics and detector ability it is able to constraint the $\nu_e/\nu_{\mu}$ cross-section difference at the 10\% level~\cite{Abe:2014usb,Abe:2015mxf}.

In summary ND280 proved very useful to select clean sample of $\nu_{\mu}$ and $\bar{\nu}_{\mu}$ interactions thanks to the presence of the TPCs and of the magnetic field. This allowed to reduce the flux and cross-section uncertainties at the level of 5\% that is more than enough for the oscillation analyses with the statistics collected by T2K so far. For the phase II of the experiments, when more statistics will be collected, an upgraded version of ND280 will be necessary. The goal of the upgrade will be to have a more efficient detector in selecting high angle and low momentum particles, as well as a larger sample of $\nu_e$ interactions.


Following the considerations above, the requirements for the upgraded near detector are:
\begin{itemize}
\item Full polar angle acceptance for  muons produced in Charged Current neutrino interactions with similar performance in term of momentum resolution, dE/dx, charge measurement as the current ND280.
\item Fiducial mass of few tons (each of the two present ND280 targets, the FGDs, has a fiducial mass of approximately one ton). 
\item High tracking efficiency  for low energy pions and protons contained inside the active target detector, in order to determine the event topology, with proton-pion identification. 
\item High efficient Time-Of-Flight detector, to reconstruct the direction (backward versus forward or inward versus outward) of all the tracks crossing the TPCs. If possible the TOF detector should also contribute to the particle identification.  
\end{itemize}

\begin{figure} [h!!]
\centering
\includegraphics[width=0.8\linewidth]{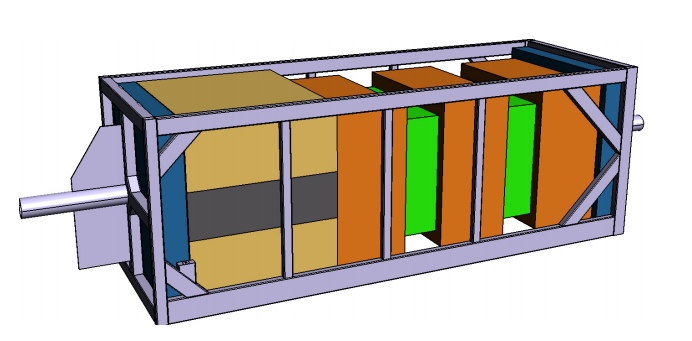}
\caption{
CAD 3D Model of the ND280 upgrade detector. In the upstream part (on the left in the
drawing) two High-Angle TPCs (brown) with the scintillator detector Super-FGD (gray) in the middle
will be installed. In the downstream part, the tracker system composed by three TPCs (orange) and
the two FGDs (green) will remain unchanged. }
\label{fig:nd280upgrade}
\end{figure}

These requirements lead to the design presented in this TDR, schematically shown in Fig.~\ref{fig:nd280upgrade}. It modifies the current ND280 configuration only in the upstream part and retains all other detectors except the P0D central part. 
Proceeding along the neutrino beam direction, after the Upstream ECal P0D  (lead scintillator sandwich, 4.9 $X_0$), we introduce a sandwich of a high granularity Scintillator Detector (SuperFGD) of approximately 2 ton, with two High-Angle TPCs (referred to as High Angle TPC, HA-TPC, in the following), one above and one below.
This central block of detectors is surrounded by a thin layer of TOF detectors, mounted in front of the large angle P0D ECAL. This geometry has been used for the simulations and the studies of the detector performances presented in this chapter.

\section{Monte Carlo simulation}

The performance of the ND280 Upgrade detector is evaluated with simulations. Neutrino interactions in the detector are simulated with GENIE~\cite{Andreopoulos:2009rq}, while the detector geometry and the particle trajectory in the detector are simulated by GEANT4~\cite{Agostinelli:2002hh}. 

The official T2K flux simulation for both, neutrino and antineutrino modes, is used as input to GENIE. 

In order to compare the upgraded detector with the current one, also the current ND280 geometry was simulated using the same framework.

\subsection{Simulated detector geometry}
\label{sec:detgeo}
\subsubsection{TPC}
\label{sec:dettpc}
The HA-TPCs are defined in GEANT4 simulation as rectangular volumes of gas contained in a hollow box made of a multilayered composite material. The size of one of the HA-TPCs is 2140x740x1780mm$^3$ and is splitted in two separated regions by placing a G10 volume with a thickness of 13.2mm  in-between of the two halves. The G10 volume accounts for the expected non-sensitive region associated to the cathode and represents its contribution to the material budget. Both halves of the drift volume are the sensitive regions of the HA-TPCs.
The gas volume is defined to be 95$\%$ Ar, 3$\%$ CF$_4$ and 2$\%$ C$_4$H$_{10}$ with 1.738 mg/cm$^3$ density as in the current forward ND280 TPCs. The field cage is simulated by surrounding the drift volume with seven consecutive layers as defined in Table ~\ref{tab:hatpc_fc_layers} from inside, layer 1, to outside, layer 7, to mimic the  design for the field cage prototype in section \ref{sec:tpc_fc}.

\begin{figure}[hbtp]
    \begin{center}
    \begin{minipage}{0.49\linewidth}
        \centering{\includegraphics[width=0.99\linewidth]{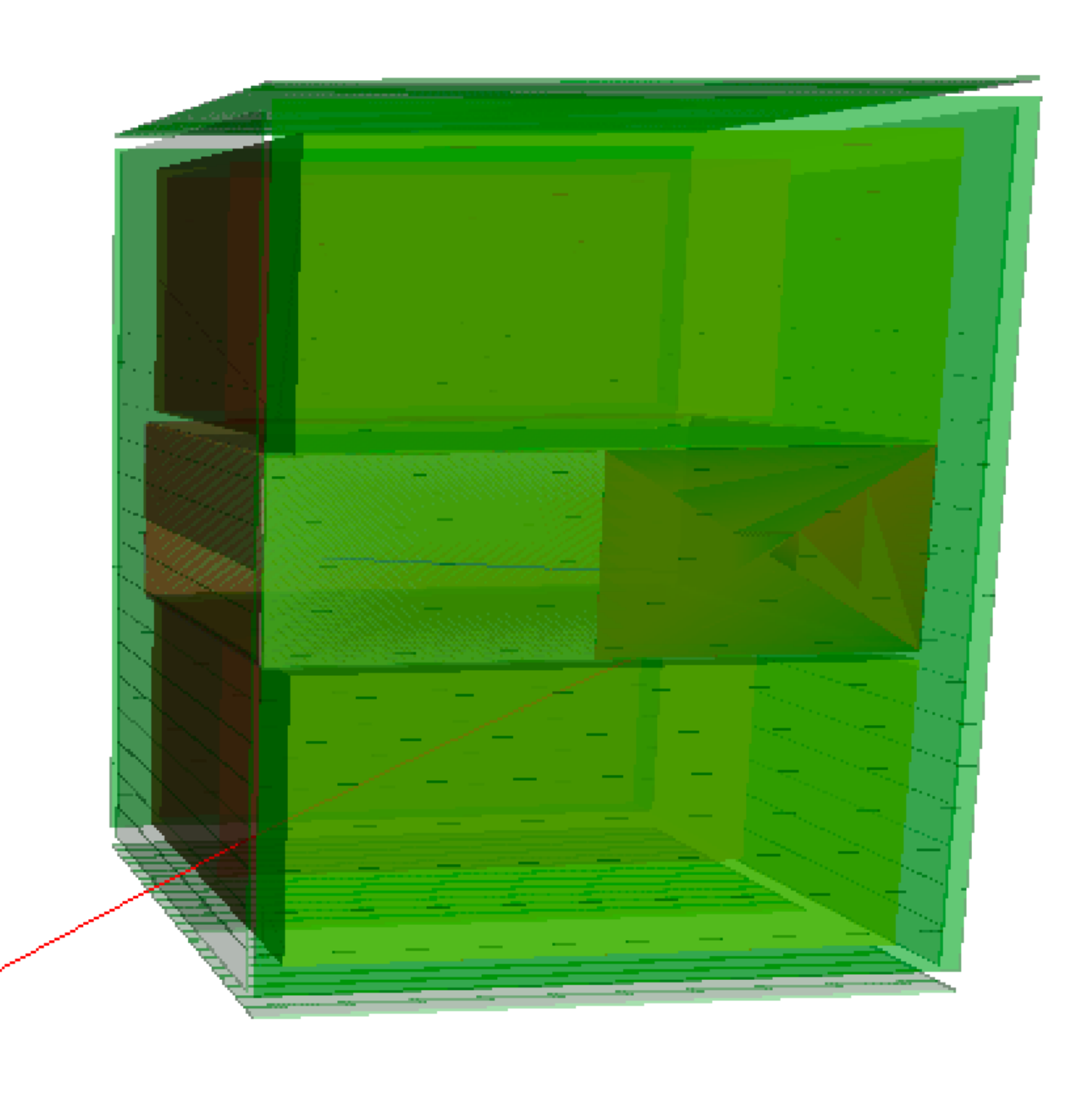}  a)}
    \end{minipage}
    \begin{minipage}{0.42\linewidth}
        \centering{\includegraphics[width=0.99\linewidth]{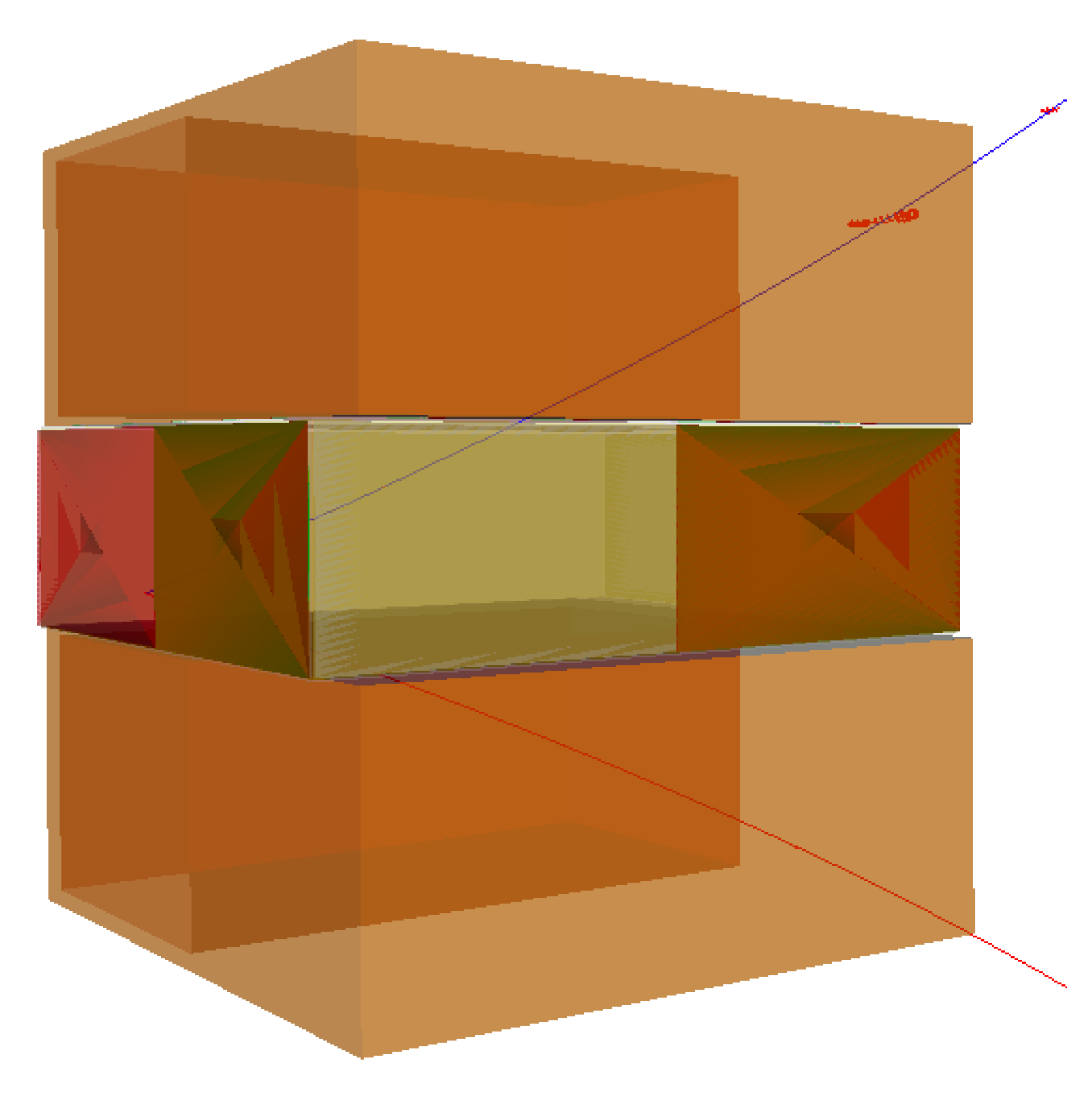}  b)}
    \end{minipage}
    \caption{View of two different simulated events from neutrino interactions in SFGD producing crossing tracks in HA-TPCs. The image in a) shows the detectors surrounded by the TOF panels while the image in b) only keeps HA-TPCs and SFGD.}
    \label{fig:Sim_Geom}
    \end{center}
\end{figure}

\begin{figure}[hbtp]
    \begin{center}
        \centering{\includegraphics[width=0.99\linewidth]{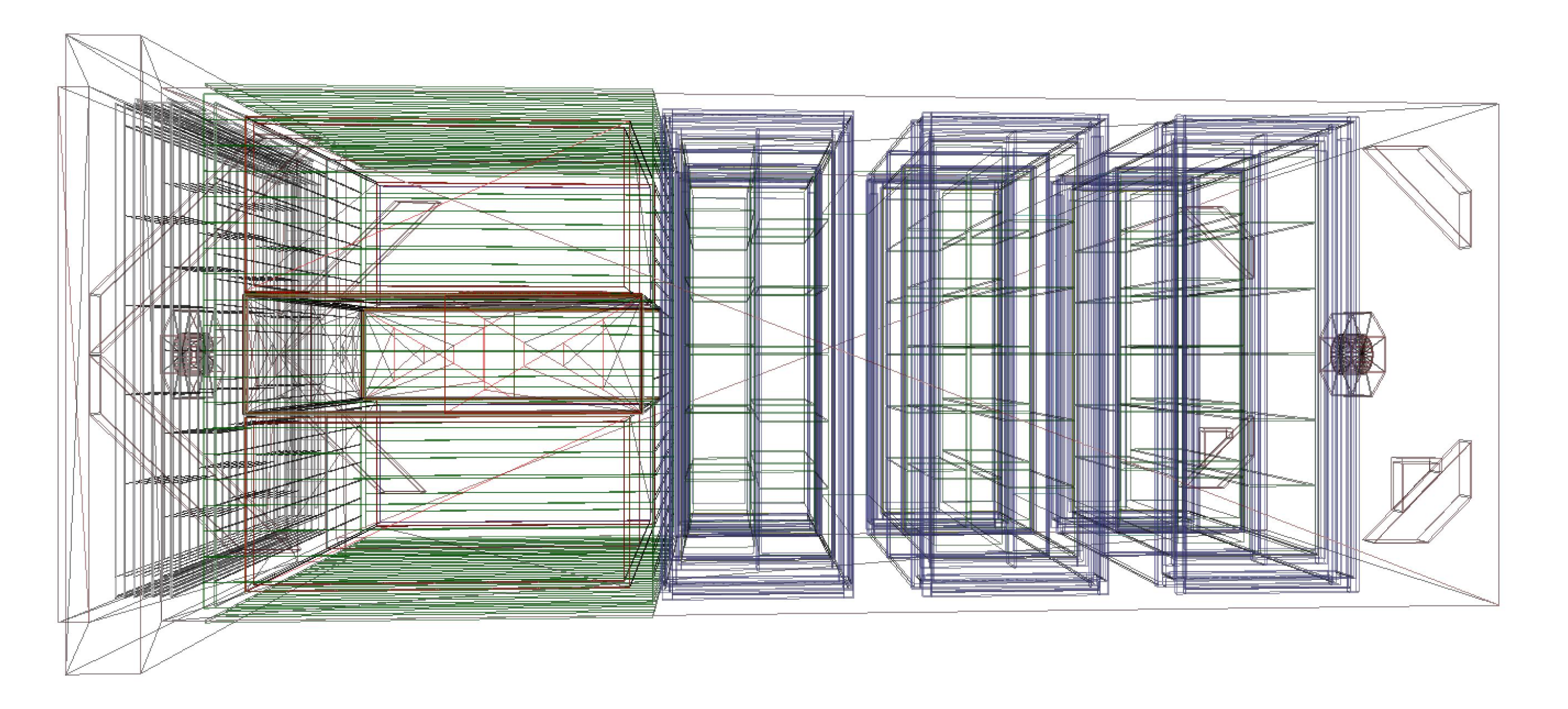}}
    \caption{GEANT4 rendering of the Monte Carlo simulation.}
    \label{fig:Sim_Basket}
    \end{center}
\end{figure}

\begin{table}[hbtp]
    \centering
    \begin{tabular}{c|ccccccc}
        \hline\hline
        Layer & Material  & Composition & $\rho$ [g/cm$^3$] & Thickness [mm] & $X_0$[cm] & $X/X_0$ [\%]  \\
        \hline
       1 & Kapton & C$_{22}$H$_{10}$O$_{5}$N$_{2}$ & 1.42 & 0.05 & 258.7 & 0.02 \\
       2 & Copper & 100$\%$ Cu & 8.94 & 0.10 & 14.4 & 0.70 \\
       3 & Kevlar & C$_{14}$H$_{10}$O$_{2}$N$_{2}$ & 1.44 &2.00 & 286.7 & 0.70 \\	
       4 & Honeycomb  & C$_{14}$H$_{10}$O$_{2}$N$_{2}$ & 0.03 &25.00 & 14237.6 & 0.18 \\
       5 & Kevlar & C$_{14}$H$_{10}$O$_{2}$N$_{2}$ & 1.44 & 2.00 & 286.7 &0.70 \\
       6 & Kapton & C$_{22}$H$_{10}$O$_{5}$N$_{2}$ & 1.42 &0.15 & 258.7 & 0.05 \\
       7 & Copper & 100\% Cu & 8.94 & 0.10 & 8.9 & 0.70 \\
        \hline\hline
    \end{tabular}
    \caption{Estimation of the HA-TPCs field cage material budget and their composition implemented in the GEANT4 simulation. The expected material budget is evaluated by radiation length of each component.}
    \label{tab:hatpc_fc_layers}
\end{table}
\par The two HA-TPCs are placed in a sandwich structure together with SFGD, one on top of it and the other below of it. The HA-TPCs are aligned in the center of the SFGD target with a clearance of 10mm between them. An illustrative image of the upgrade geometry is offered in Figure~\ref{fig:Sim_Geom} where two typical events used for the simulation studies are depicted together with a full view of the basket as it is in the simulation.

\subsubsection{SuperFGD }
\label{sec:detsfgd}
SuperFGD geometry is simulated by creating replicas of scintillator cubes in GEANT4. 
Each cube has the fixed dimensions of $10\times10\times10$~mm$^3$ and is made of plastic scintillator ($C_8 H_8$), covered by a reflector surface.
Three fiber holes with the radius of 0.75~mm are located 3~mm from the outer edges of the SuperFGD cube.
The wavelength shifting fiber (WLS) material with radius of 0.5~mm is placed inside the fiber holes. 
The plastic scintillator material in the SuperFGD cubes is set as sensitive materials in the GEANT4 simulation, 
i.e. the energy deposited by the charged particle outside this material is not detected. 


\par The SuperFGD box, MPPC interface, and the MPPC-electronics cables are implemented in the GEANT4 simulation as non-sensitive materials to predict the effect of the material budget. The SuperFGD detector is contained in the hollow box which consists of 16~mm AIREX R82 foam sandwiched by 2~mm carbon fiber reinforced polymer (CFRP) skins. Three of the outermost surfaces of the SuperFGD box in each axis are covered by the MPPC interface materials. The materials from the four-layer printed circuit board (PCB) are currently simulated with copper and G10. Cables to provide connections between MPPCs and electronics are placed on top of the PCBs. 
Table~\ref{tab:sfgd_materialbudget} shows the preliminary list of the material budget in radiation length $X_0$ and the composition currently implemented in the GEANT4 simulation. The simulation will be updated accordingly as the research and development of the SuperFGD integration progresses. With the current approximation, the total material budget of the SuperFGD box and PCB is estimated to be about 3.58\% radiation lenght, and the material budget from the microcoaxial cable is expected to vary between 0.11\% to 0.68\% radiation lenght due to the piling-up of the cables from the center to the outer layers of the SuperFGD. 

\begin{table}[hbtp]
    \begin{tabular}{c|cccc}
        \hline\hline
        Material  & Composition & Total thickness [mm] & $X_0$ [cm] & $X/X_0$ [\%]  \\
        \hline
       AIREX R82 & C$_{37}$H$_{24}$O$_{6}$N$_{2}$ &16 & 689.2& 0.23 \\
   CFRP skin  & 69\% CF, 31\% epoxy &4 & 27.6 & 1.45 \\
       G10 & 57\% glass, 43\% epoxy & 1.4 &19.4 & 0.72 \\	
       Copper & (Cu in the PCB) & 0.17& 1.4 & 1.18 \\
      Cables & 100\% Al & 0.101-0.606 & 8.9& 0.11-0.68 \\
        \hline\hline
    \end{tabular}
    \caption{Preliminary estimate of the SuperFGD material budget and their composition implemented in the GEANT4 simulation. The expected material budget is evaluated by radiation length of each component.}
    \label{tab:sfgd_materialbudget}
\end{table}

\subsubsection{Time of Flight detectors}
\label{sec:dettof}
TOF counters located on each side of ND280 Upgrade detector are simulated as layers of plastic scintillator that surround the tracker (SuperFGD and HA-TPCs) on each of the six sides.
The TOF counter size differs by each pairs located in front-back, left-right, and top-bottom with respect to the neutrino beam perspective. 

\subsection{Simulated detectors performances}
\label{sec:performances}
The GEANT4 simulation used for the studies described in this chapter. Simulated information for all the produced tracks are smeared based on the expected performances of the TPCs, the SuperFGD and the TOF detector. 
The SuperFGD detector response is parametrized as described in sec.~\ref{sec:superfgdresponse}.

The reconstruction and the Particle Identification performances of FGD1, FGD2 and Electromagnetic calorimeter are also parametrized in the simulation, based on the ND280 official results.

\subsubsection{TPC detector response} 
\label{sec:tpcresponse}
The TPC detector response, for the forward and the High Angle TPCs, is simulated according to the performances observed in the existing ND280 TPCs~\cite{Abgrall:2010hi}.

\begin{itemize}
\item A charged track is assumed to be reconstructed in a TPC if its true length projected on the readout plane is larger than $20$\,cm, the same requirement used in the ND280 reconstruction.
\item The curvature of the track, and hence its charge, is assumed to be reconstructed with 100\% efficiency. The measured charge misidentification of 1\% at 1 GeV/c is neglected in the simulation. 
\item The energy loss by unit length in the TPC is smeared from its true value according to the expected deposited energy resolution. A resolution of 8\% for MIPs crossing the entire TPC is assumed, based on the performances of the existing TPCs.

\end{itemize}

\subsubsection{SuperFGD detector response } 
\label{sec:superfgdresponse}
The SuperFGD detector performance is estimated with GEANT4 simulation. 
The Birks equation~\cite{birks1951scintillations} is applied to estimate the fraction of the deposited energy that was emitted as scintillation light. From the measurements in the FGD we know the numbers of photons that are emitted and collected in the fiber per MeV of scintillation energy (156 $\gamma$/MeV).
The light attenuation in the fibres is taken into account as an exponential law with attenuation constants based on the measurements in the FGD, since exactly the same fiber type will be used in the SuperFGD~\cite{Amaudruz:2012agx}. 
Finally the MPPC photo detection efficiency (PDE) is taken into account in order to evaluate the number of detected photo-electrons. 

\begin{figure}[hbtp]
    \begin{center}
    \begin{minipage}{0.49\linewidth}
		\centering{\includegraphics[width=0.9\linewidth]{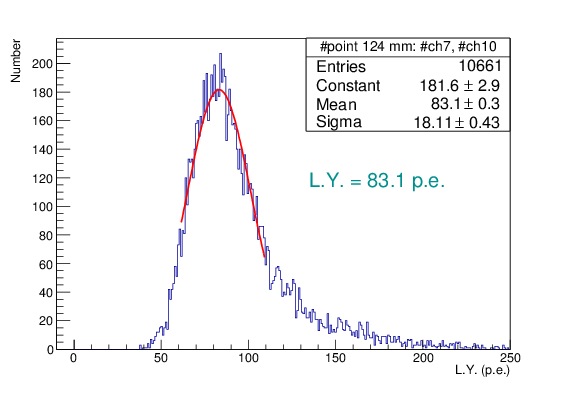}}
	\end{minipage}
	\begin{minipage}{0.49\linewidth}
		\centering{\includegraphics[width=0.9\linewidth]{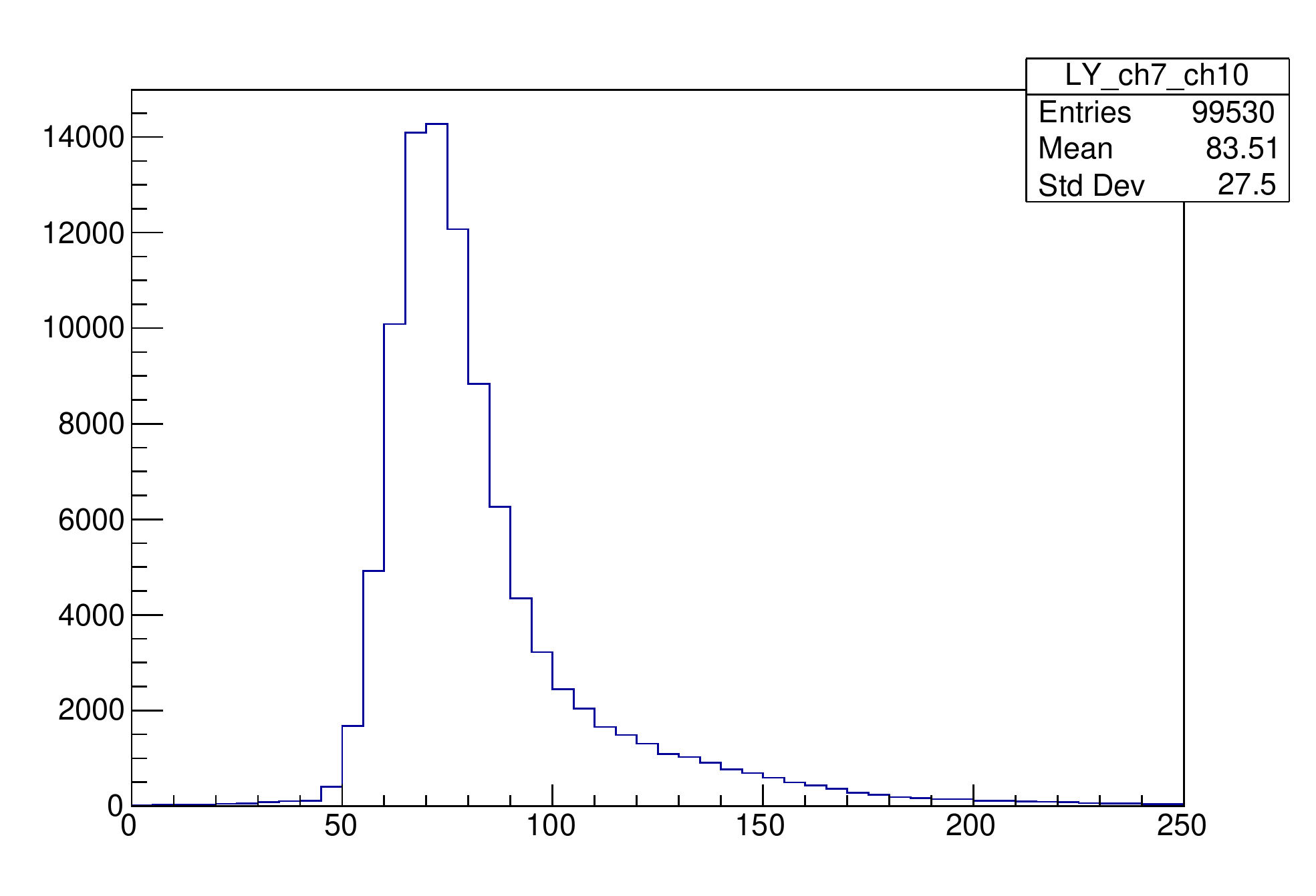}}
	\end{minipage}
    \caption{The comparison between October 2017 beam test results (left) and corresponding MC simulation (right). In both cases the light yield as number of photoelectrons is obtained from the sum of the same two simulated channels of the prototype.}
    \label{fig::sfgd_beam_test_comparison}
    \end{center}
\end{figure}

This method includes several empirical constants. The fiber attenuation and MPPC PDE are  well measured, while the number of photons emitted and collected in the fiber per MeV of the scintillation energy severely depends on the detector geometry. This constant need to be properly calibrated. For this purpose we use results of the October 2017 beam test~\cite{Mineev:2018ekk}. The comparison of the measured prototype light yield with the results of the simulation is presented in figure~\ref{fig::sfgd_beam_test_comparison}: we observe a good agreement between the simulation predictions and the beam test results.

This simulation have been used to model the performance of the SuperFGD in reconstructing and identifying charged particles as a function of their momentum and angle. 

\subsubsection{TOF detector response } 
\label{sec:tofresponse}
The TOF detector is used to reconstruct the direction of the tracks produced in neutrino interactions in SuperFGD and FGD, in order to improve the reconstruction efficiency of backward and high-angle tracks.

The true time of the hits in the ToF detector are smeared based on the performances observed in the ToF test beam (Chapter\ref{Ch:TOF}). Such resolution allows to improve the identification of ~1 GeV/c protons versus positrons. 


\section{Impact of ND280 upgrade on the T2K oscillation analyses}
In this section we will describe the expected impact of the upgraded ND280 detector on the constraint of the systematic uncertainties that affect the measurement of neutrino oscillations at T2K. 
For these studies we will limit to describe what will be done better thanks to the upgrade.
For this reason we will not  exploit the additional capabilities that the upgrade will provide, such as for example the possibility of reconstructing low momentum muons 
or of distinguishing electrons from gammas that will be described in section~\ref{sec:egmammaSFGD}. 
The analysis strategy is the same as used in T2K.

In the current T2K analyses samples of $\nu_{\mu}$ and $\bar{\nu}_{\mu}$ charged-Current interactions selected in the Tracker (FGD+TPC) are used to constraint flux and cross-section uncertainties.  The selection requires that the muon produced in the neutrino interaction, is reconstructed into one TPC, typically the one downstream the FGD. Stopped muons in the FGD or muons directly entering the ECal are not used for the T2K oscillation analyses yet. 
However a selection that exploits ECal has been developed by the T2K collaboration and used in $\nu_{\mu}$ cross-section analysis \cite{Abe:2018uhf}, so it will be used for the results of this chapter as well.
The selected inclusive sample is then divided into different sub-samples according to the number of reconstructed pions (0, 1, more than 1) in the final state.

In order to evaluate the impact of the upgrade we have simulated neutrinos and antineutrinos interactions in the FGDs and in the SuperFGD with GENIE and tracked the emitted particles with GEANT4, modelling the detector response as described in Sect.~\ref{sec:performances}. 

We then computed the expected efficiencies in selecting muons and pions in the FGDs and in the SuperFGD, according to the performances described in Sect.~\ref{sec:performances}. Then the sensitivity of both the current and upgraded versions of Nd280 were investigated with the fitter used in the T2K oscillation analysis to constraint flux and cross-section systematics uncertainties.

\subsection{Muon neutrino selection} 
\label{sec:muoneff}

A selection of $\nu_{\mu}$ and $\bar{\nu}_{\mu}$ charged-Current interactions has been developed in order to evaluate the performance of the new detector design with respect to the current design. 



For each neutrino (antineutrino) interaction, the most energetic negative (positive) track is selected as the muon candidate. The event is then retained if the muon candidate cross one of the TPCs active volumes for more than 20 cm and if it is identified as a muon according to the PID algorithms. High angle tracks are also added if the muon candidate enter ECal and is identified as a muon there.
Once the muon candidate is selected, we search for pions emitted in the interactions. Mimicking the ND280 algorithms, pions are reconstructed if they enter the TPC or if they are stopped in one of the scintillating detector with a track length longer than 20 cm. More details on the SuperFGD performances in reconstructing pions and protons will be given in Sect.~\ref{sec:superfgdonly}.

\begin{figure}[hbtp]
\begin{subfigure}[b]{.5\linewidth}
\centering
\includegraphics[width=\linewidth]{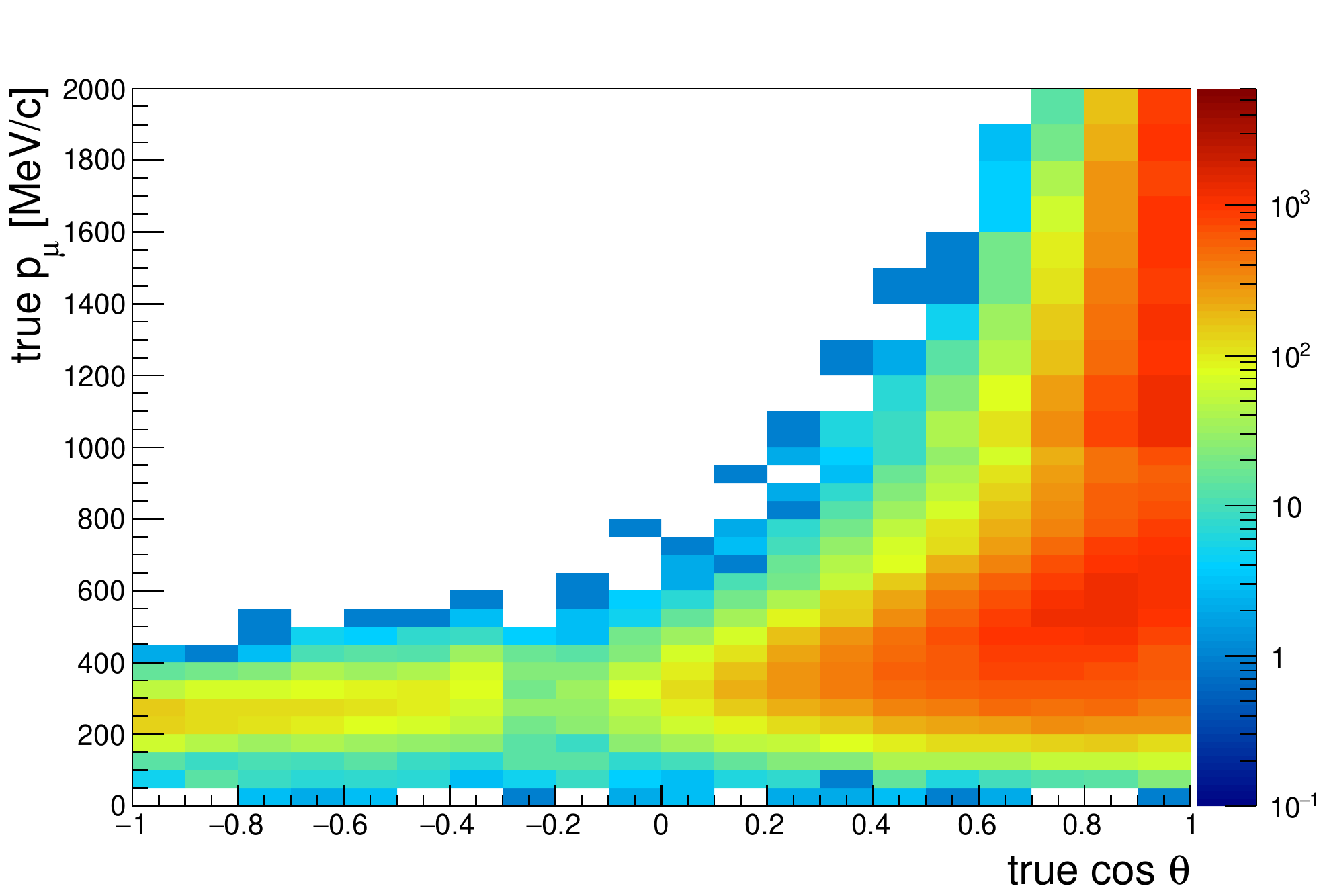}
\caption{Current ND280}
\end{subfigure}
\begin{subfigure}[b]{.5\linewidth}
\centering
\includegraphics[width=\linewidth]{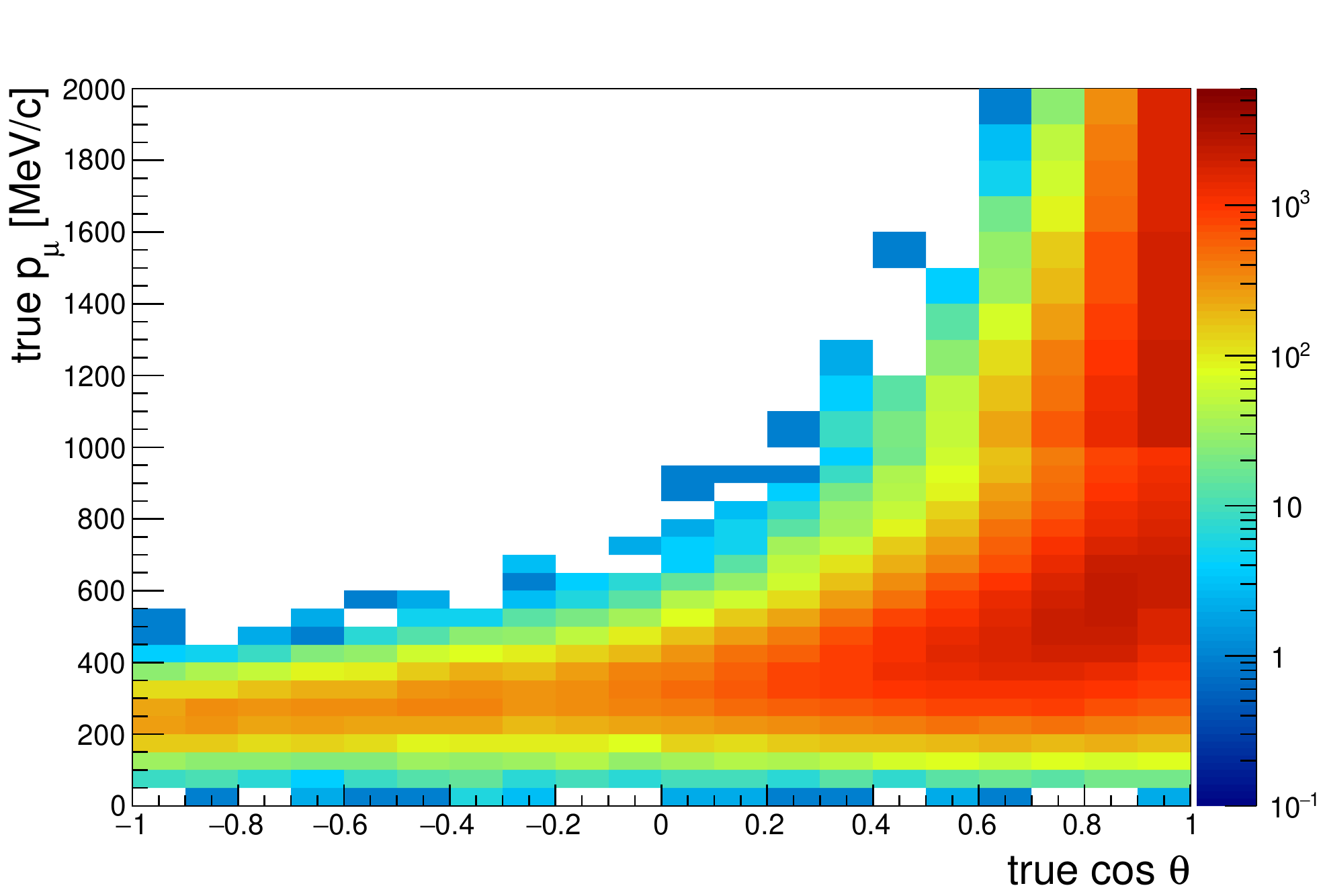}
\caption{Upgraded ND280}
\end{subfigure}
\caption{Distribution of selected $\nu_{\mu}$ Charged-Current events in the two configurations, in neutrino mode, as a function of true muon momentum and polar angle.}
\label{fig:DetPerf:dis2D_numu}
\end{figure}

\begin{figure}[hbtp]
\centering
\includegraphics[width=.49\linewidth]{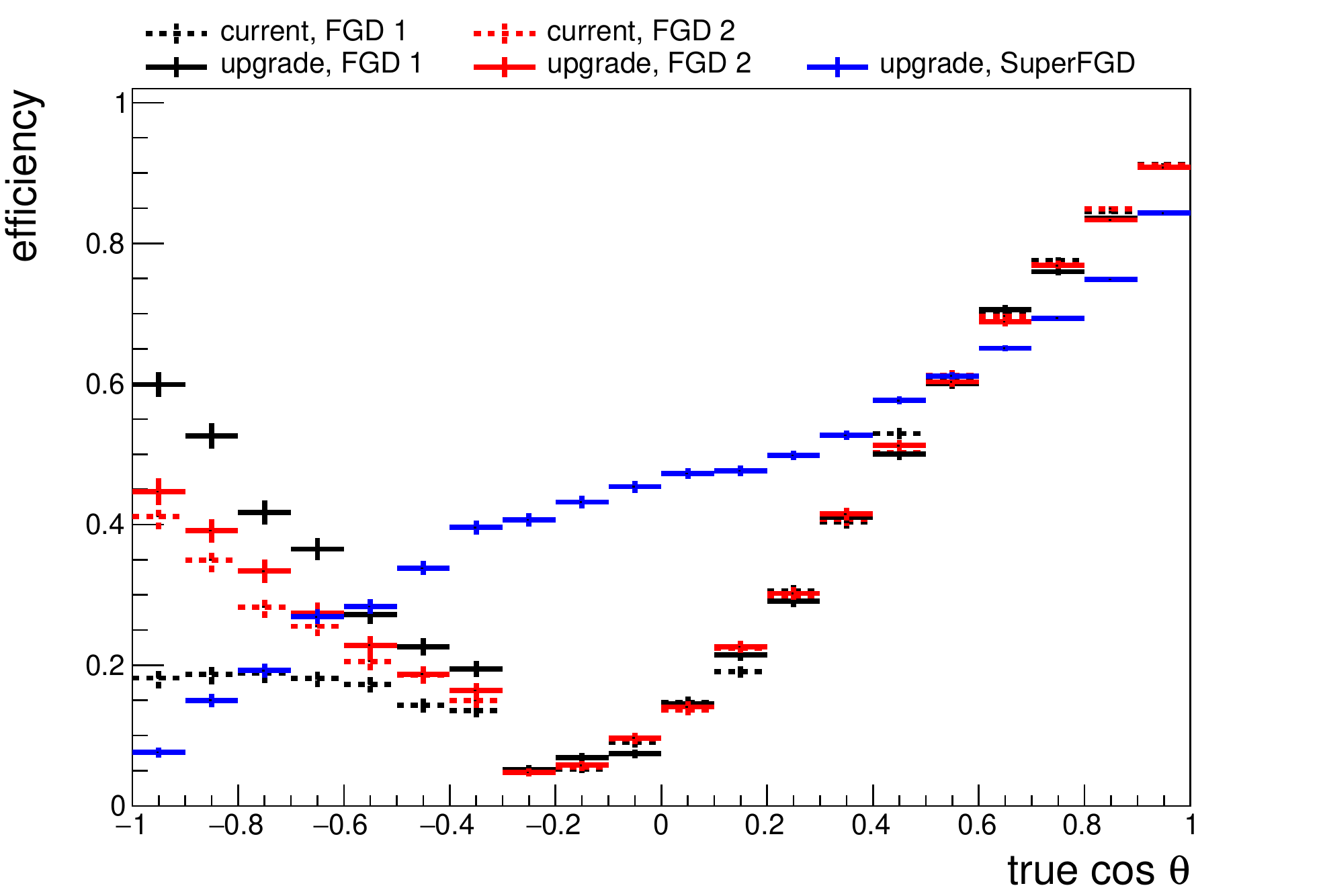}
\includegraphics[width=.49\linewidth]{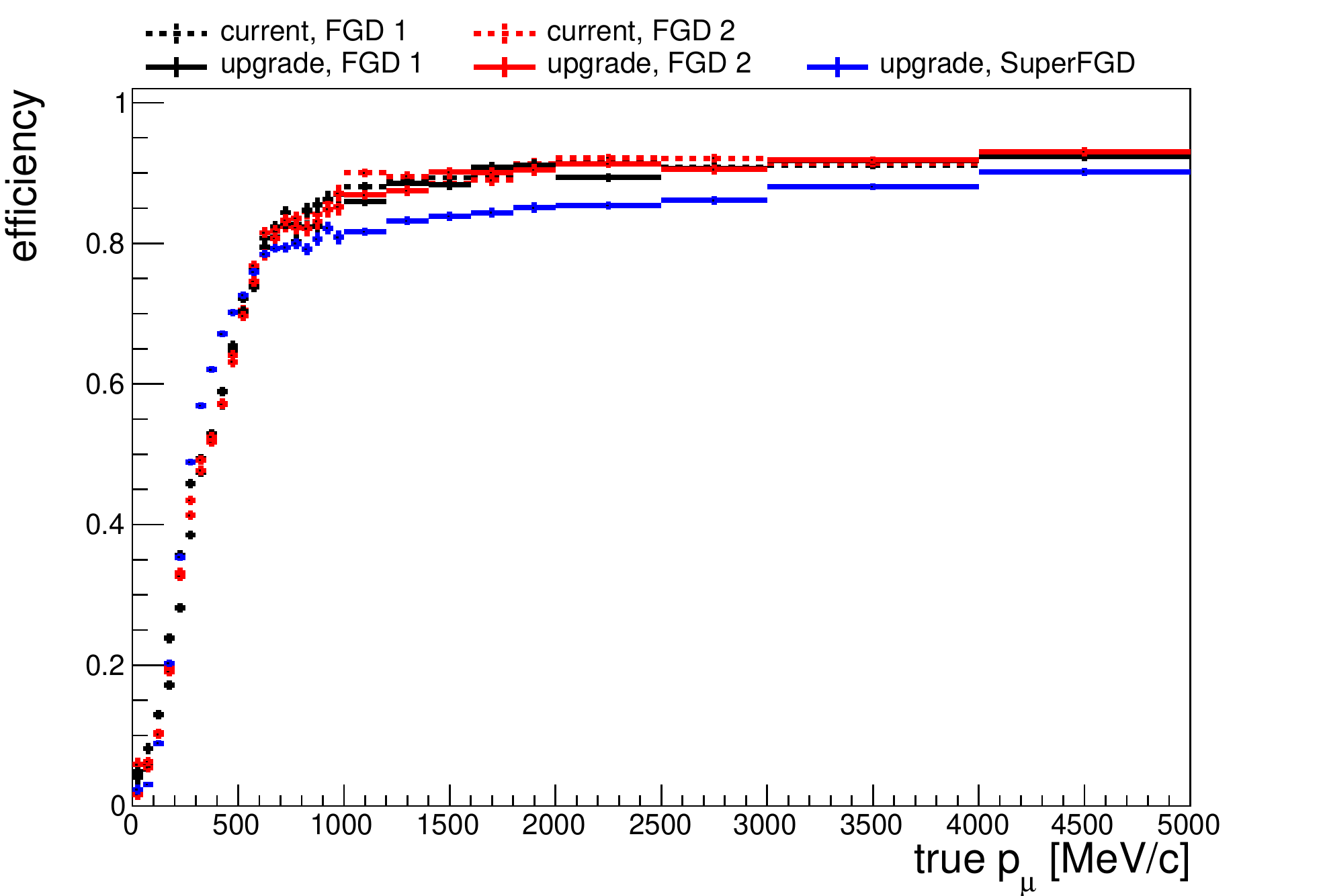}
\caption{$\nu_{\mu}$ Charged-Current event selection efficiency as a function of the true muon polar angle (left) and momentum (right), for both the current ND280 (dashed lines) and the upgrade configuration (solid lines), in neutrino mode. The different curves correspond to neutrino interactions in either FGD 1 (black), FGD 2 (red) or Super-FGD (blue).}
\label{fig:DetPerf:eff_numu}
\end{figure}

Figure~\ref{fig:DetPerf:dis2D_numu} shows the distribution of the muon true momentum versus polar angle for selected events, while Fig.~\ref{fig:DetPerf:eff_numu} presents the selection efficiency for $\nu_{\mu}$ Charged-Current (CC) inclusive events in neutrino mode. The upgraded configuration clearly improves the angular acceptance of the detector both for high-angle muons thanks to the new HA-TPCs and backward thanks to the ToF detector box.

The numbers of expected events in each beam mode and in each configuration are shown in Tab.~\ref{Tab:DetPerf:num_evts}. The larger target mass and the improved performances of the upgraded configuration allows providing about twice the number of selected events with respect to the current configuration. The purity for neutrino mode selection is shown in Tab.~\ref{Table:nd280ratetopology}. A clear improvement in the purity of the CC1$\pi$ sample is observed, mostly thanks to the better performances of the SuperFGD in tracking low momenta contained particles.  

\bgroup
\def\arraystretch{2}%
\begin{table}[htbp]
\centering
\begin{tabular}{c|cc}
\hline
\hline
Selection & Current-like & Upgrade-like \\
\hline
$\nu_{\mu}$ ($\nu$ beam) & $100632$ & $199605$ \\
\hline
$\bar\nu_{\mu}$ ($\bar\nu$ beam) & $32671$ & $60763$ \\
\hline
$\nu_{\mu}$ ($\bar\nu$ beam) & $16537$ & $29593$ \\
\hline
\hline
\end{tabular}
\caption{Predicted total number of selected events for each detector configuration and beam mode, for an exposure of $10^{21}$ protons-on-target. The Out-of-Fiducial-Volume background is not included and the wrong-sign component is included only in the $\bar\nu$ beam as it corresponds to a large fraction of the events there.}
\label{Tab:DetPerf:num_evts}
\end{table}
\egroup

\bgroup
\def\arraystretch{2}%
\begin{table}[htbp]
\centering
\begin{tabular}{cc|c|ccc}
\hline
\hline
\multirow{2}{*}{} & & $\#$ of events & & Purity ($\%$) & \\
\multirow{2}{*}{} & & (/$10^{21}$ POT) & CC$0\pi$ & CC$1\pi$ & CC Other \\
\hline
\multirow{2}{*}{current} & FGD 1 & $50507$ & $72.5\%$ & $64.0\%$ & $68.2\%$ \\
& FGD 2 & $50125$ & $71.5\%$ & $62.3\%$ & $63.8\%$ \\
\hline
\multirow{3}{*}{upgrade} & FGD 1 & $52655$ & $72.9\%$ & $64.1\%$ & $64.7\%$ \\
& FGD 2 & $51460$ & $71.6\%$ & $62.9\%$ & $63.3\%$ \\
& SuperFGD & $95490$ & $72.5\%$ & $70.3\%$ & $72.7\%$ \\
\hline
\hline
\end{tabular}
\caption{Predicted total number of selected $\nu_{\mu}$-CC events in neutrino enhanced mode
	for both ND280 upgrade-like and current-like configurations in each available neutrino target detector.
	Also the purity for each event topology is shown. 
	The prediction corresponds to $1\times 10^{21}$ POT.  
	The out-of-FV and the wrong-sign backgrounds are not included because the are expected to give an almost negligible effect.
	}\label{Table:nd280ratetopology}
\end{table}
\egroup



As already mentioned, the selection described in this section requires to reconstruct the muon track in one of the TPCs surrounding the SuperFGD or the FGDs. Additional statistics and sensitivity could be gained by selecting CC-$\nu_{\mu}$ interactions 
with a muon stopping in the SuperFGD.
Such sample will be affected by a worst purity ($\sim$80\%), 
due to the contamination of Neutral Currents with a proton or a pion misidentified as a muon,  
but we expect to add  
10-15\% 
$\nu_{\mu}$-CC events, 
with an efficiency that is flat with respect to the muon direction. 

This sample would be particularly interesting because it will contain mostly low momentum muons, and is expected to be particularly powerful to constrain the nuclear recoil models, like 2p2h as it will be shown in Sect.~\ref{sec:superfgdonly}. 

\subsection{Impact on T2K systematic uncertainties} 
\label{sec:performance_fit}

Sensitivity studies were performed to estimate the impact of an upgrade of 
ND280 onto the oscillation analysis of T2K. 
The goals of the study were the following: 

\begin{itemize}
\item evaluate how much we can improve the constraints on the flux and cross-section models thanks to the upgrade;
\item estimate the power to discriminate between different cross-section models. 
\end{itemize}

The Near Detector fitter that is used to constraint the flux and cross-section uncertainties in the T2K oscillation analyses is described in details in~\cite{Abe:2017vif}. It maximizes a binned likelihood ratio as a function of the neutrino flux, cross section and detector 
systematic parameters, all constrained with penalty terms reflecting our prior knowledge of neutrino fluxes, cross-sections and detector systematic uncertainties.

We developed a tool that is functionally identical to the fitter used in T2K oscillation analyses and was adapted to fit simulated samples of neutrino and antineutrino interactions in the simulated "current" and "upgrade" configurations. The Monte-Carlo (MC) events were reweighted according to the efficiencies and purities obtained from detector simulations in order to select the samples of interactions that were given as input to the Near Detector fitter. Despite a more detailed simulation of the upgrade detector configuration, described in sections~\ref{sec:detgeo}-\ref{sec:performances}, very similar detector performances to the ones reported in Ref.~\cite{Blondel:2299599} have been observed so the input to the present study were not updated with respect to the previous study.
 
For flux and cross-section uncertainties, the same model as the one used for T2K oscillation analyses was given as input to the fitter. In addition, a set of uncorrelated systematic parameters 
was used to describe the detector systematics as a function of the muon true angle and momentum,  for both the ND280 current and upgrade configurations fits with the values shown in Tab.~\ref{Table:banff_detsys}.

The main difference between the two configurations is that in the ND280 current configuration the high angle region is covered only by 
the ECAL detector, where the detector systematic uncertainties are larger than 30\%~\cite{Abe:2018uhf}. Since in the ND280 upgrade configuration the high angle region is mostly covered by TPCs, we expect the detector systematic uncertainties to be about 2.5\% above 0.5 GeV/c,  assuming the same performance of the vertical TPCs currently used in ND280.  



\begin{table}[!tbp]
	\centering
	\caption{Detector systematic uncertainties parametrized as a function of the muon true momentum and
	angle with respect to the Z direction.}
	\begin{tabular}{c|c|c|c}
		\hline
		\hline
		Detector configuration		& Momentum / $\cos \theta$  	& $0 < p < 0.5 \text{ GeV/c}$ 	& $p > 0.5  \text{ GeV/c}$  \\
		\hline
								& $ -1 < \cos \theta < -0.6$  	& 20\%					& 20\%  \\ 
		FGD1, FGD2				& $ -0.6 < \cos \theta < 0$  	& 50\% 					& 60\% \\ 
								& $ 0 < \cos \theta < +0.6 $  	& 30\%  					& 50\%  \\ 
								& $ +0.6 < \cos \theta < +1 $  	& 9\%	 				& 2.5\%  \\ 
		\hline
								& $ -1 < \cos \theta < -0.6$  	& 9\%					& 2.5\%  \\ 
		SD				& $ -0.6 < \cos \theta < 0$  	& 9\% 					& 2.5\% \\ 
								& $ 0 < \cos \theta < +0.6 $  	& 9\%  					& 2.5\%  \\ 
								& $ +0.6 < \cos \theta < +1 $  	& 9\%	 				& 2.5\%  \\ 

		\hline
		\hline
	\end{tabular}
	\label{Table:banff_detsys}
\end{table}

The impact of the different detectors on the neutrino flux and cross-section constraints is evaluated
by performing a fit of the Asimov data set, 
the most probable data set, corresponding to the MC expectation.
The simulated beam exposure, for both configurations, correspond to 
$8\times 10^{21}$ POT, about a third of the expected total data collected at the end of T2K-II~\footnote{We could not simulate a larger exposure because the official T2K MC production was used for this study and the available statistics is limited}.  
%
%
The sensitivity was obtained for both the ND280 upgrade and current ND280 configurations.
The post-fit errors of the most significant systematic parameters are shown
in Table~\ref{Table:banff_asimov}.
On average the error on the systematic parameters is reduced by about 30\% in the upgrade configuration. A larger reduction is observed for FSI parameters since they are more sensitive to low momentum pions. 


\begin{table}[htbp]
	\centering
	\caption{Sensitivity to  flux and cross-section parameters of interest for the 
	current ND280 and the upgrade configuration.}
	\begin{tabular}{c|c|c}
		\hline
		\hline
		Parameter 					& Current ND280 (\%) 	& Upgrade ND280 (\%) \\
		\hline
		SK flux normalisation  			& 3.1 			& 2.4 \\ 
		($ 0.6 < \text{E}_{\nu}  < 0.7$ GeV) 		&				& 	\\			
		MA$_{\text{QE}}$ (GeV/c$^2$) 			& 2.6 			& 1.8 \\ 
		$\nu_{\mu}$ 2p2h normalisation	& 9.5 			& 5.9 \\ 
		2p2h shape on Carbon 			& 15.6 			& 9.4 \\ 
		MA$_{\text{RES}}$ (GeV/c$^2$) 			& 1.8 			& 1.2 \\ 
		Final State Interaction ($\pi$ absorption)		& 6.5		& 3.4 \\
%
%
%
%

		\hline
		\hline
	\end{tabular}
	\label{Table:banff_asimov}
\end{table}

\begin{figure}[htbp]
	\begin{center}
		\includegraphics[width=0.45\linewidth]{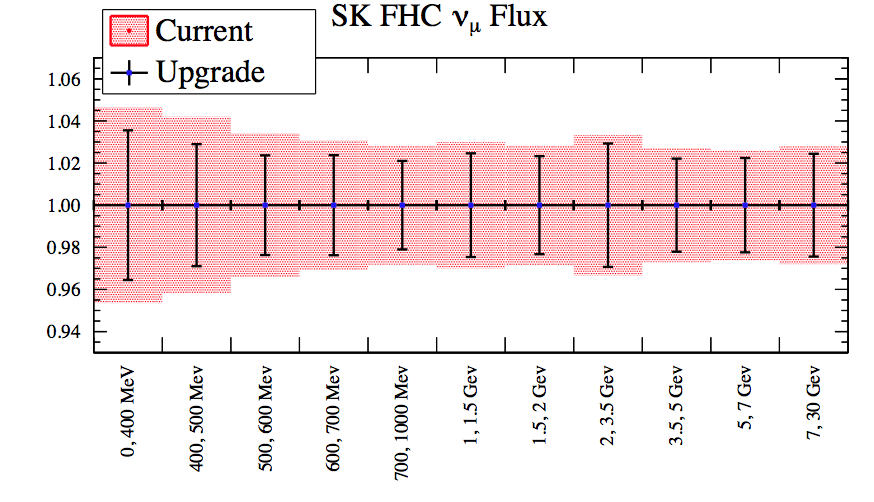} 
		\includegraphics[width=0.45\linewidth]{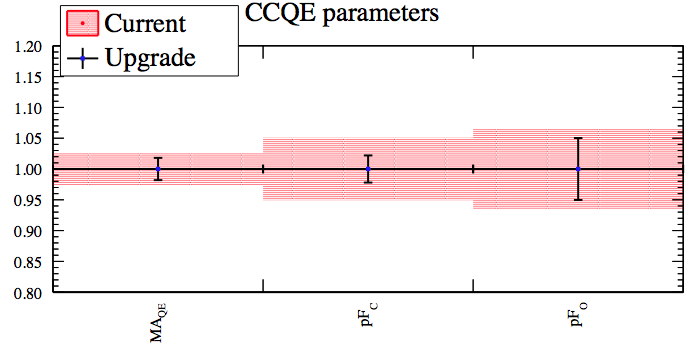} 
		\includegraphics[width=0.45\linewidth]{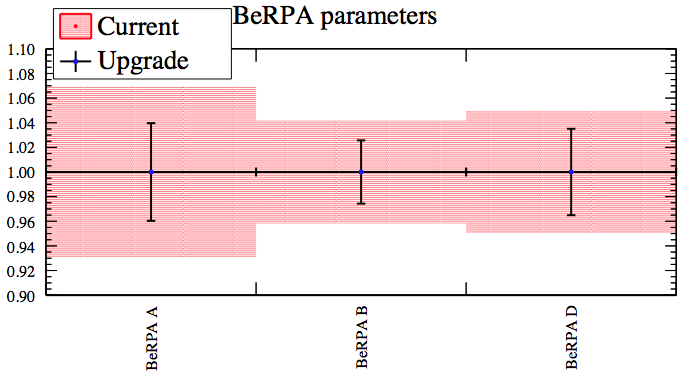} 
		\includegraphics[width=0.45\linewidth]{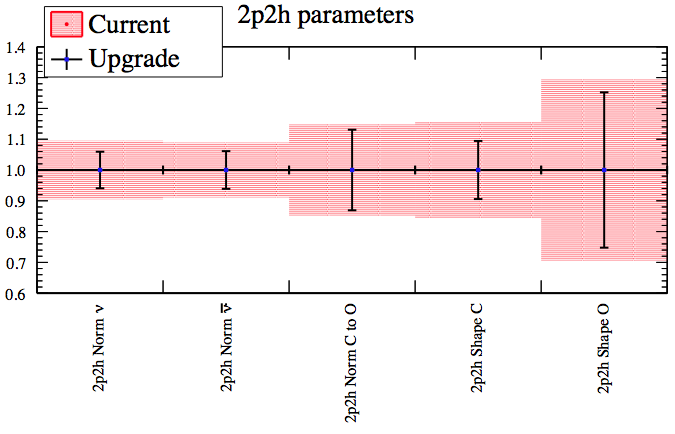} 
		\caption{\label{rpa_banff_constraint} 
		The post-fit errors on the main systematic parameters
		are shown for both the ND280 upgrade (blue dots) 
		and the current ND280 configuration (red bars).  
		These sets of parameters comprise
		the far detector $\nu_{\mu}$ flux (top left), 
		the CCQE cross-section (top right),
		the Random Phase Approximation (bottom left) and
		2p2h parameters (bottom right).
		}
	\end{center}
\end{figure}

In Fig.~\ref{rpa_banff_constraint} the main post-fit systematic errors are shown.
The ND280 upgrade-like configuration can provide overall smaller 
systematic uncertainties to the neutrino oscillation measurement. 

The uncertainty on the total number of events selected at the T2K far detector, Super-Kamiokande (SK), 
was evaluated using the best-fit ND280 covariance matrix obtained by the Asimov data set fit.
The neutrino cross-section parameters that cannot be constrained by the ND280 detector, 
like $\sigma_{\nu_e} / \sigma_{\nu_{\mu}}$ ratio and the NC parameters,
are not propagated from ND280 to SK analysis.
While the absolute value of the uncertainty depends on the simulated exposure, the relative difference between current and upgrade does not depend on the exposure and a reduction of the uncertainty on the total number of events at SK is reduced by 20-30\%.


\begin{table}[htbp]
	\centering
	\caption{Sensitivity to flux and cross-section parameters constrained by the near detector for ND280 upgrade with $8\times 10^{21}$~POT.}
	\begin{tabular}{c|ccc}
	\hline
	\hline
	Source of uncertainty 	&  	 $\nu_e$ CCQE-like & $\nu_{\mu}$ 	& $\nu_e~ CC1\pi^{+}$ \\ 
	 & $\delta N / N$ (\%) 	&$\delta N / N$ (\%) & $\delta N / N$ (\%)  \\
	\hline
Flux + cross-section 		   &   &  &    \\
	(constrained by ND280)	&  1.8 & 1.9 & 1.4  \\  
	\hline
	\hline
	\end{tabular}
	\label{Table:sk_err}
\end{table}

These studies provide an indication of the sensitivity of an upgraded detector configuration
but are limited to the specific neutrino cross-section model that is used for the oscillation analysis.
We know that the current model is not necessarily the correct and complete parametrization of the neutrino interactions for the full phase space.
While this model is adequate for the current T2K analyses, and potential biases on the extraction of the oscillation parameters are carefully studied by the T2K collaboration, with the use of simulated data (see \cite{Abe:2017vif} for details), its limitations could be an issue when the systematic uncertainties will become as large as the statistical ones. 

In order to provide useful information on the importance of improving the ND280 angular acceptance, 
complementary studies were performed: 
assuming Nature behaves differently from the cross section model used for the neutrino events prediction,
the bias on the neutrino flux and cross section systematic parameters was evaluated.
It is expected that a more sensitive detector configuration will provide larger biases in the best-fit parameters as well as a poorer goodness of fit (g.o.f.) if the wrong model is used in the fitter.
Several alternative models were tested instead of the nominal prediction and 
it was found that, thanks to the largely improved angular acceptance and 
the increased target mass, the ND280 upgrade configuration was able to reject
the alternative model with a better significance than the current ND280 configuration.
As an example, we changed the Random Phase Approximation (RPA) parameters, that describe the behavior of the CC0$\pi$ cross-section as a function of the transferred momentum, $\text{Q}^2$: the RPA parameters were set at $+1\sigma$ with respect to the prior systematic uncertainties. When the current configuration is used, the $\Delta\chi^2$ between the nominal data set and the one obtained with modified RPA parameters is 38.3. When the upgrade configuration is used, the  $\Delta\chi^2$ is 79.9, showing the greater potential of the upgrade in distinguishing the two cases.

\section{Super-FGD stand--alone  performances} 
\label{sec:superfgdonly}



The SuperFGD has been conceived to have optimal track reconstruction capabilities and identification performances for particles produced in neutrino interactions. A preliminary quantification of such capabilities will be presented in this section by using the simulation described in Section~\ref{sec:superfgdresponse} and focusing on tracks that stop in the SuperFGD volume.
For such study we assume perfect pattern recognition and apply the following track reconstruction criteria:
\begin{itemize}
\item
more than two MPPC hits in at least 2 views (XY, XZ, YZ), similarly to what is done in the ND280 FGD reconstruction;
\item
no MPPC hits in the outermost cubes, to assure that the particle stops in the SuperFGD;
\item
tracks must be separated between them: for each track pairs, the endpoint of the shorter track should be separated by at least one SuperFGD cube (1~cm) from the longer track.
\end{itemize}

\subsection{SuperFGD reconstruction efficiency}
\begin{figure}[hbtp]
\centering
\includegraphics[width=\linewidth]{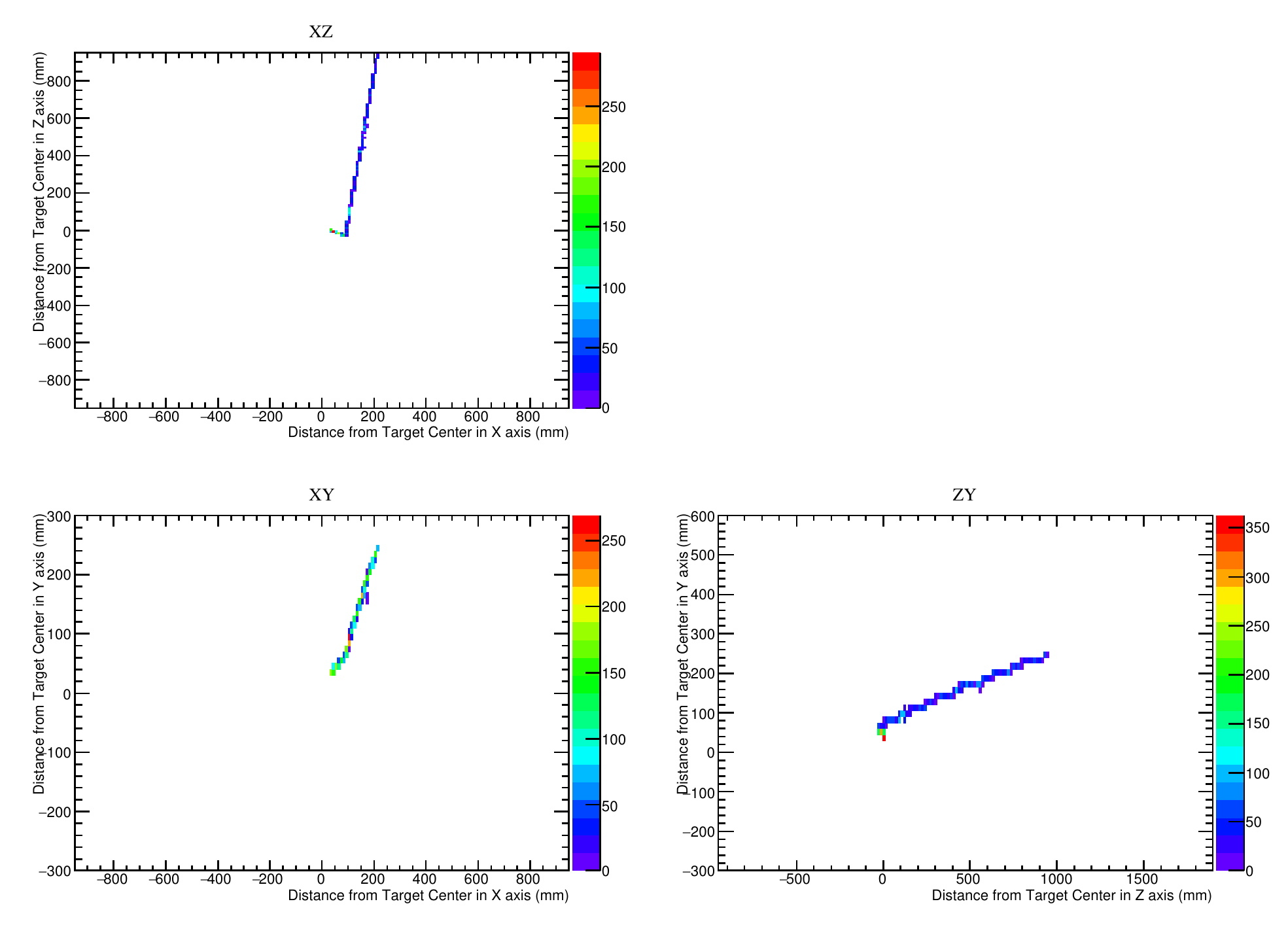}
\caption{The SuperFGD event display of a $\nu_{\mu}$ CCQE interaction generated with GENIE. The muon and the low momentum protons are visible.}
\label{fig:TrkReco:numuccqe}
\end{figure}

\begin{figure}
\centering
\includegraphics[width=0.49\linewidth]{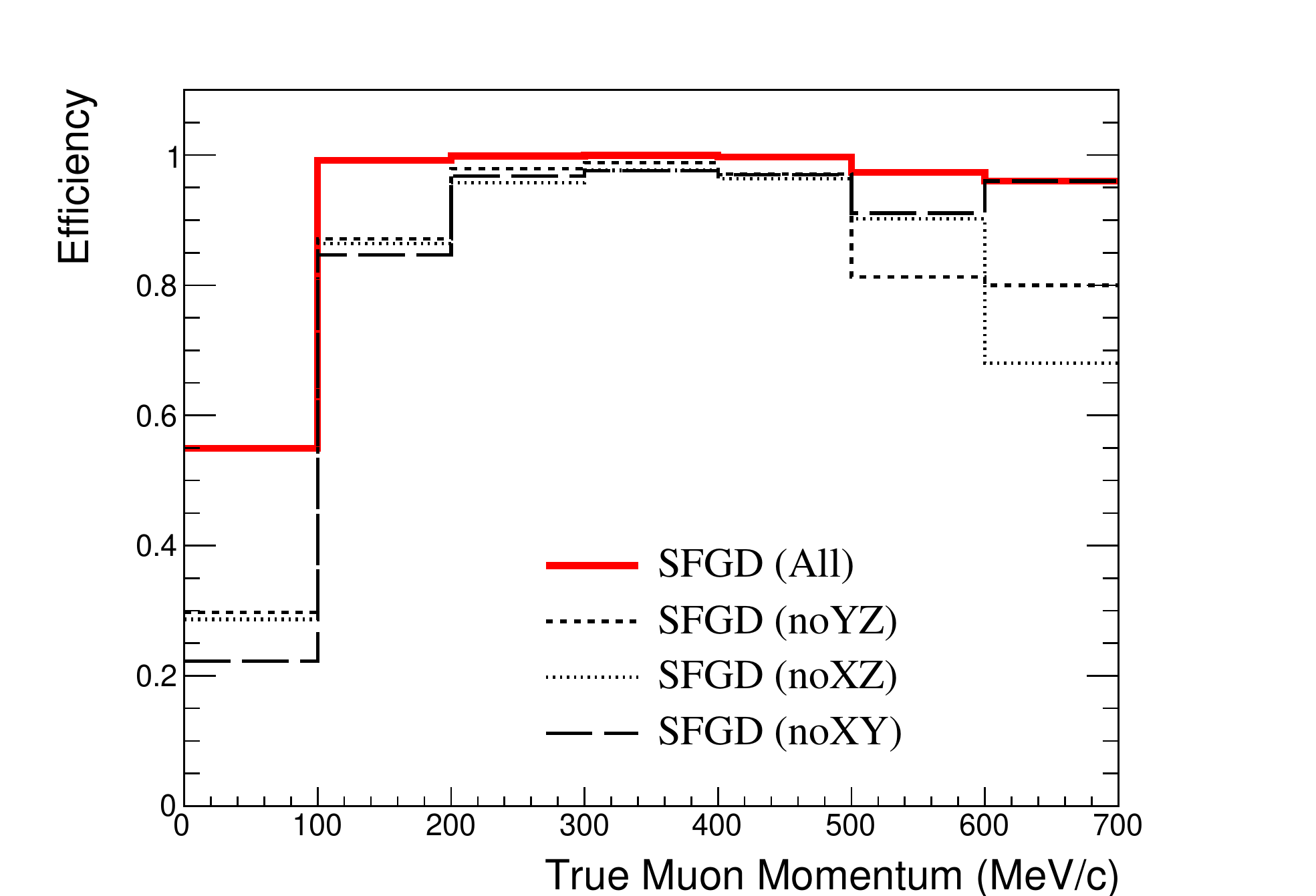}
\includegraphics[width=0.49\linewidth]{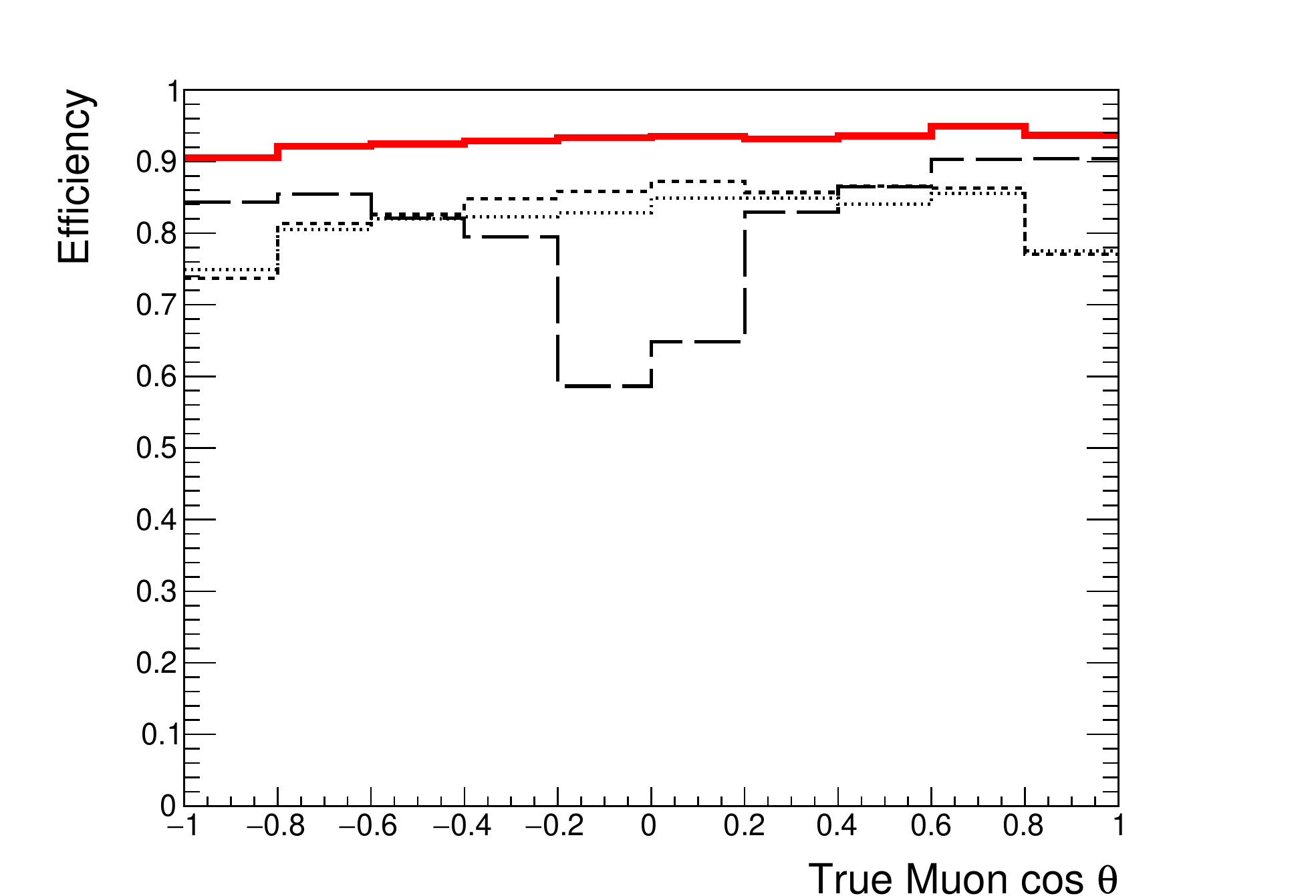}
\includegraphics[width=0.49\linewidth]{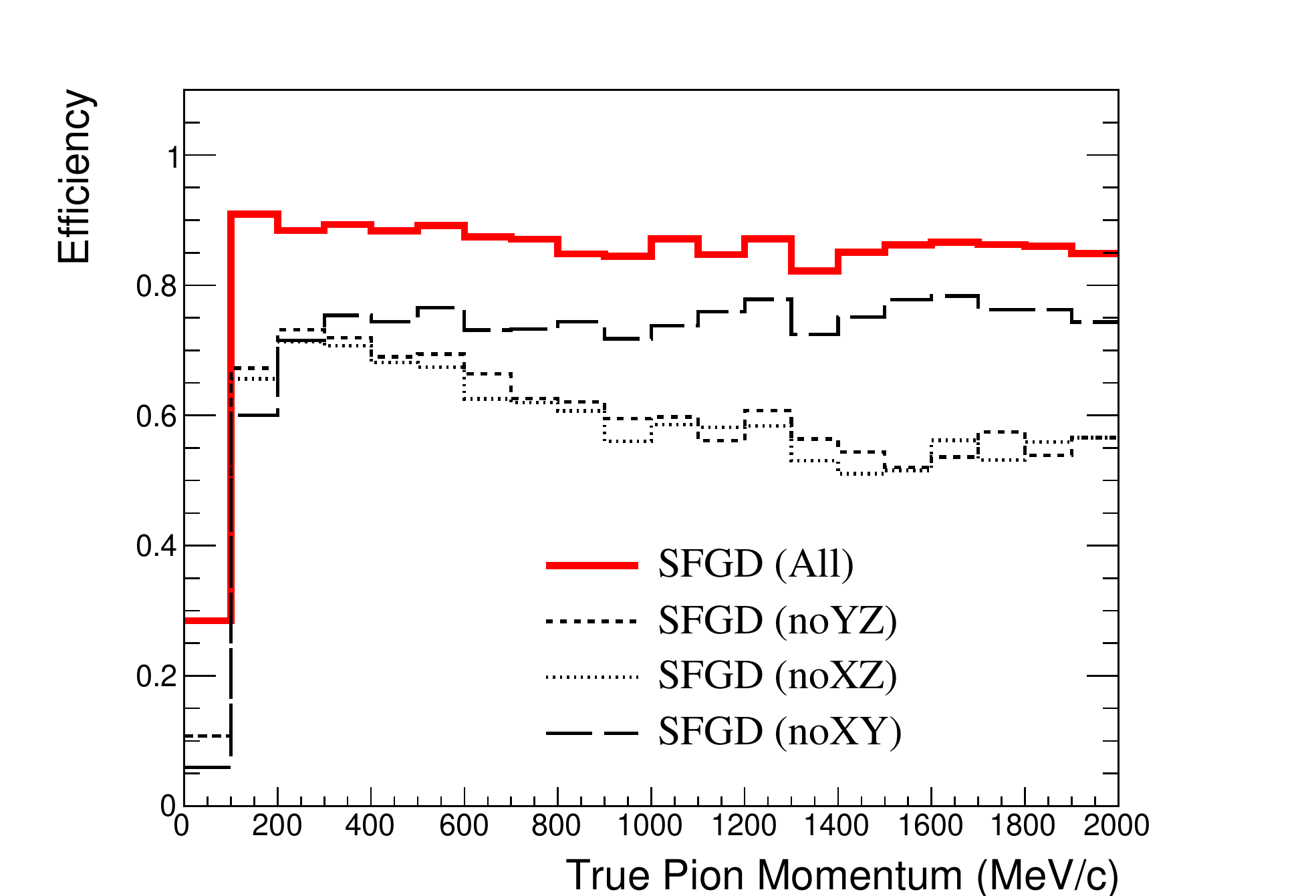}
\includegraphics[width=0.49\linewidth]{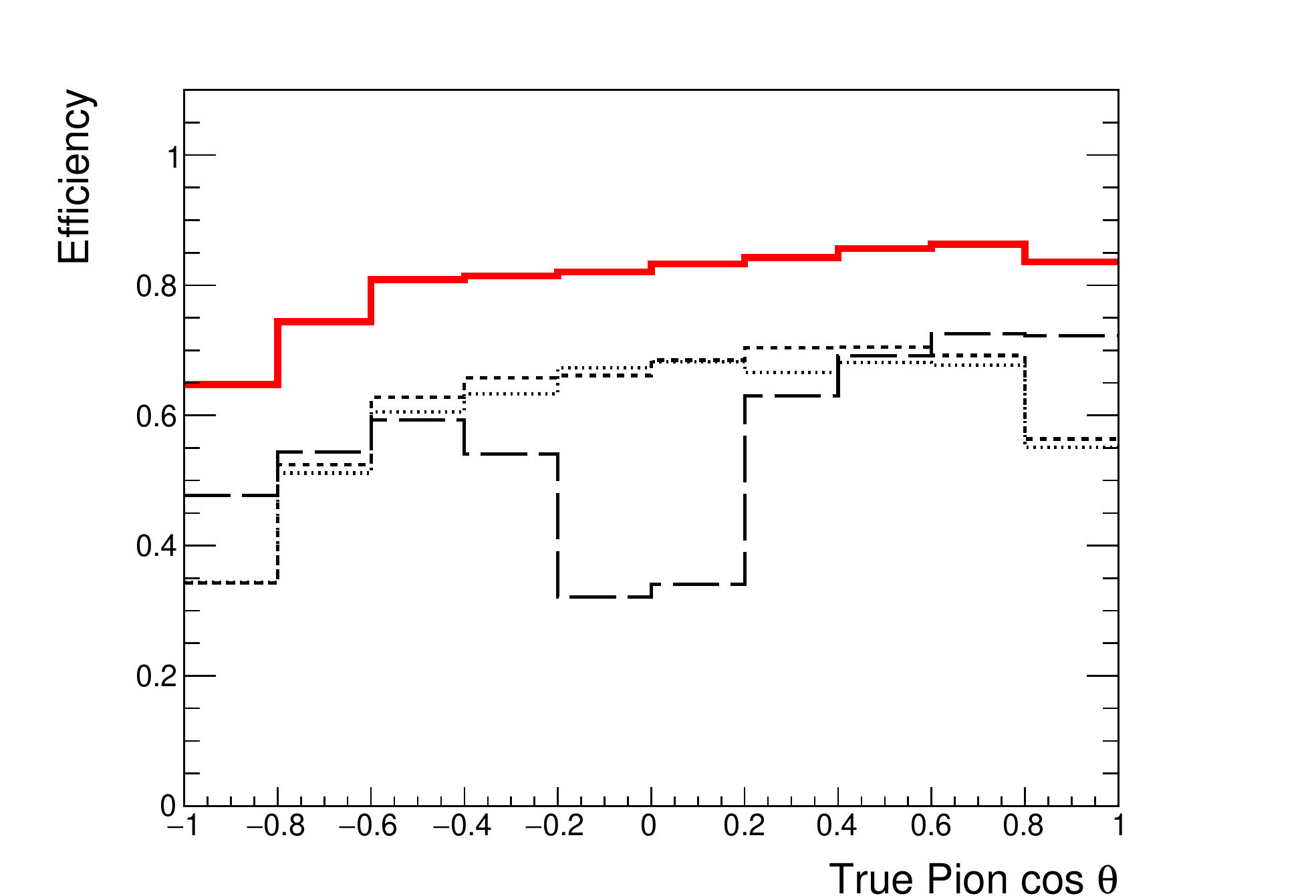}
\includegraphics[width=0.49\linewidth]{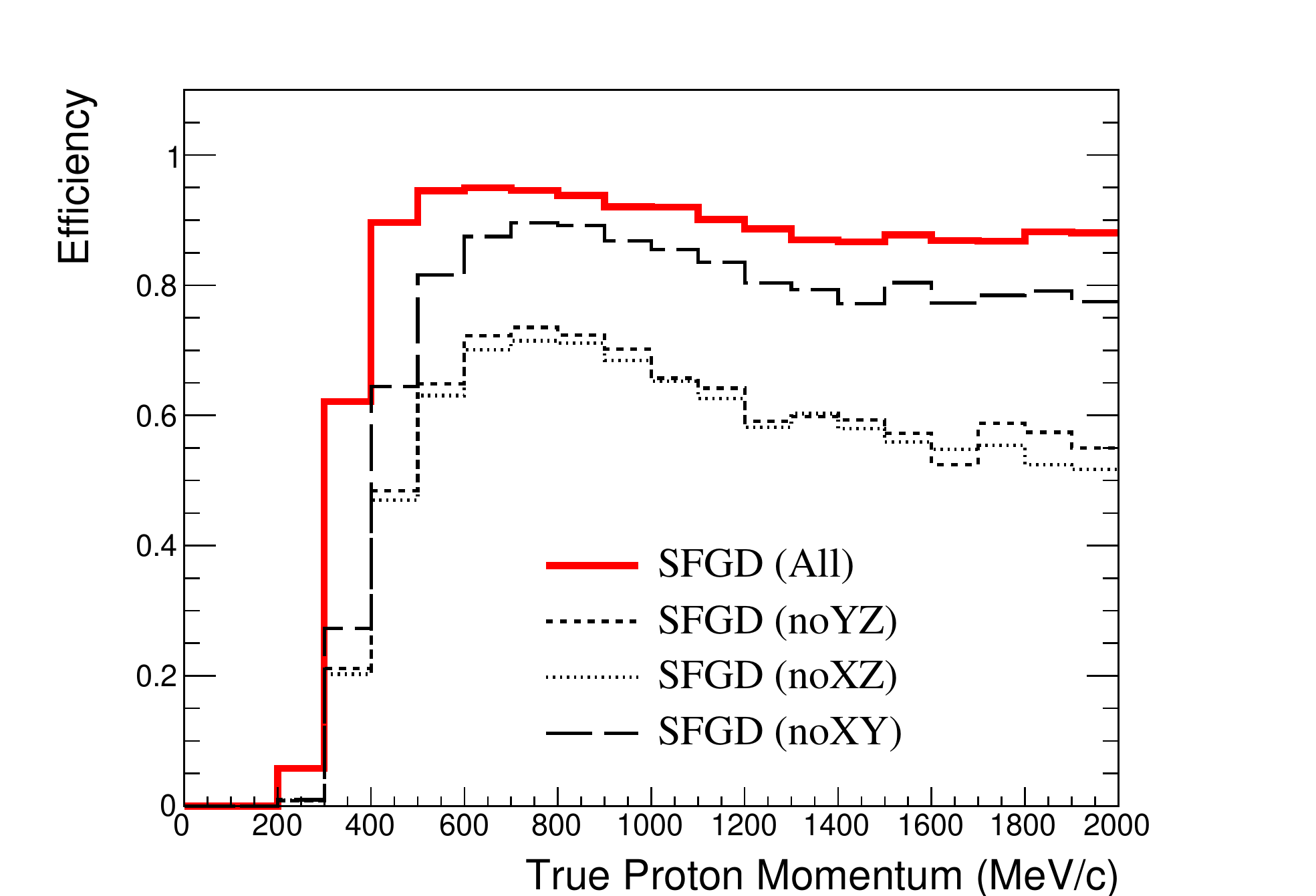}
\includegraphics[width=0.49\linewidth]{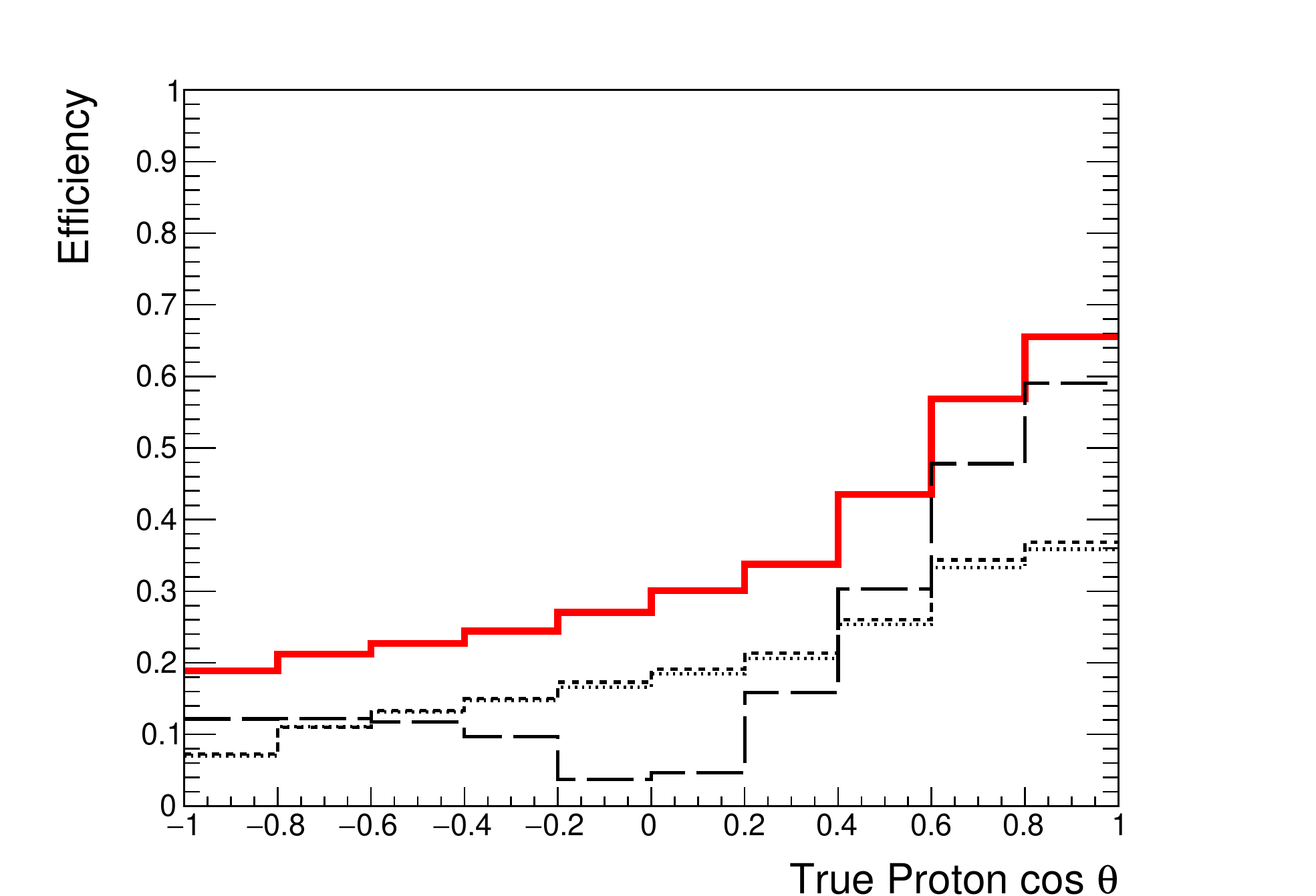}
\caption{Track reconstruction efficiencies for muons (top), pions (middle) and protons (bottom) in SuperFGD with three readout views or with only two readout views.}
\label{fig:TrkReco}
\end{figure}

The track reconstruction efficiency in SuperFGD is evaluated for muons, pions and protons simulated with GENIE with T2K $\nu_\mu$ flux. Figure~\ref{fig:TrkReco:numuccqe} shows a typical event display obtained for a $\nu_{\mu}$ charged current quasi-elastic (CCQE) interaction. Figure~\ref{fig:TrkReco} shows the reconstruction efficiency for muons, pions, and protons as a function of the momentum and the angle of the particle with respect to the neutrino beam direction. In order to emphasize the importance of the 3D reconstruction, the efficiencies are compared with those expected for the same detector but exploiting only 2 views (alternatively XY, XZ, and YZ).  

While muon tracks can be reconstructed with an efficiency higher than 90\% for all the angles in SuperFGD with all three views, the efficiency is about 20\% lower for forward and backward going tracks in SuperFGD without YZ and XZ view. SuperFGD without XY view has track reconstruction efficiency comparable to the detector with three views for the forward tracks. However, it loses approximately 30\% efficiency for the track angle perpendicular to the beam direction. The detector with three views has also a lower momentum threshold. SuperFGD with three views can reconstruct protons down to approximately 300~MeV/c, while SuperFGD with two views has a threshold approximately at 500~MeV/c.


\subsection{SuperFGD momentum resolution}
The momentum resolution of SuperFGD is estimated with particle-gun muons produced upstream of the detector with uniform momentum distribution up to 600~MeV. The track lenght is used as an estimator of the momentum applying the smearing matrix obtained from the simulation. The resolution is of the order of 3\%. 

\subsection{SuperFGD particle identification}
The SuperFGD particle identification performances for protons, pions and muons are evaluated using the ratio between the number of MPPC photoelectrons and the track lenght. The misidentification probabilities are evaluated for a given working point, defined by the intersection of the probability distribution functions of such particle identification parameter between different particle species: a stopping muon/pion is misidentified as proton in 8\% of the cases, while a proton is misidentified as muon/pion in 17\%/22\% of the cases. This performance is found to be similar to the FGD detector one, simulated in the same framework. In addition, thanks to the 3D tracking capability of SuperFGD combined with high granularity, even better performances can be achieved exploiting a more precise parametrization of the measured dE/dx as a function of the track length.

\subsection{SuperFGD electron/$\gamma$ separation}
\label{sec:egmammaSFGD}
\begin{figure}[hbtp]
\centering
\includegraphics[width=0.8\linewidth]{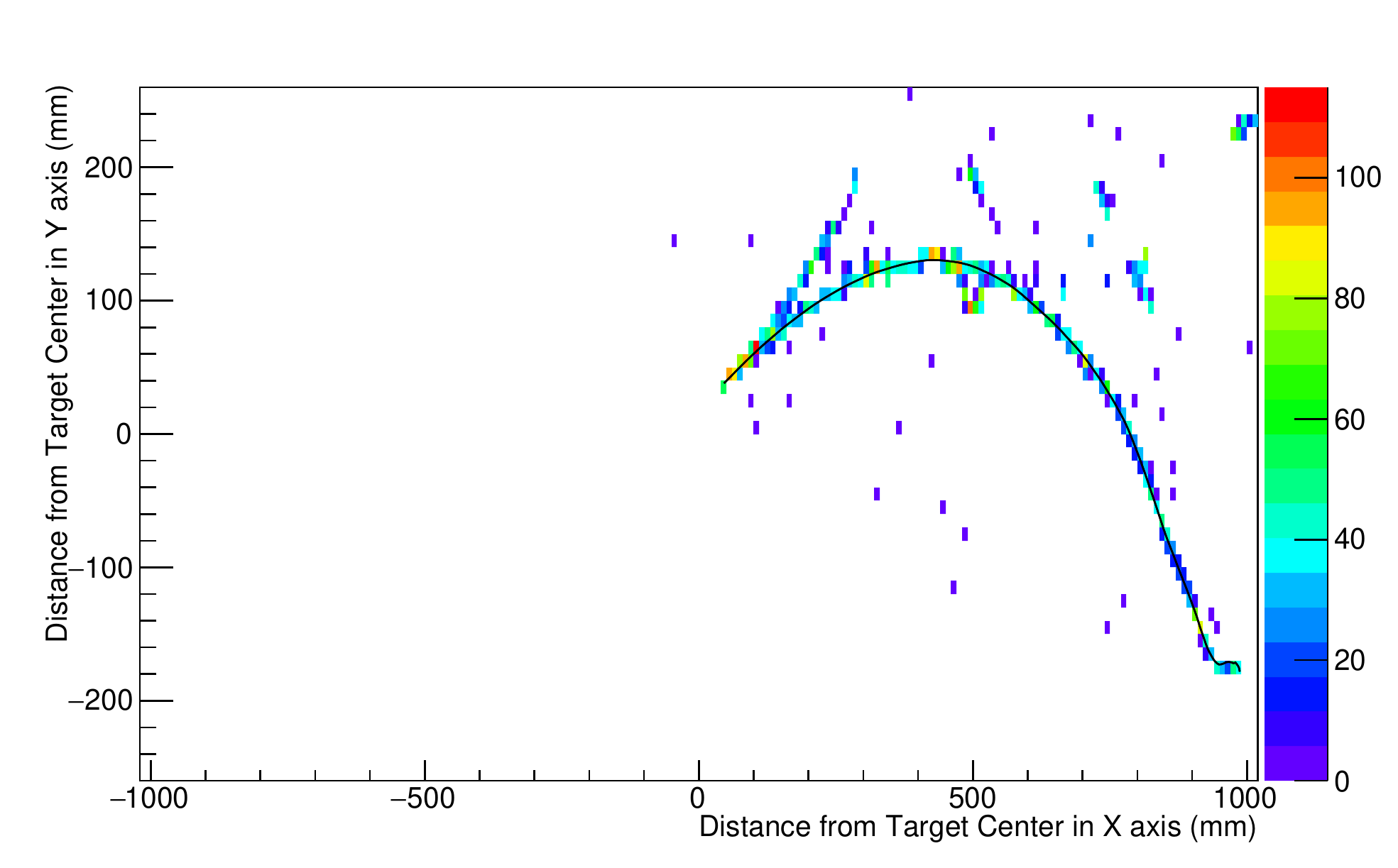}
\caption{The SuperFGD event display of a $\gamma$ particle-gun event. The number of photo-electron obtained from the MPPCs in XY view is shown in this event display. The black line shows the true electron trajectory.}
\label{fig:nue:gamma_display}
\end{figure}

The high granularity of SuperFGD provides an additional avenue to electron/$\gamma$ identification. The production of $\gamma$ from $\pi^0$ decays in neutral current $\nu_\mu$ interactions, followed by $\gamma \rightarrow e^{-}e^{+}$ conversion, is the dominant background to the $\nu_e$ selection in the current ND280. It is indeed difficult to reject electron-positron pair tracks if the low momentum positron stops in the target before the two tracks can be reconstructed. On the other hand, for such events with low positron momentum, twice larger ionization is expected in the upstream part of the electron track, with respect to single electron tracks from $\nu_e$ interactions. Such feature is shown in Figure~\ref{fig:nue:gamma_display} and can be exploited in SuperFGD for electron/$\gamma$ identification.

The electron track is divided in two upstream/downstream segments with maximal light yield ratio. The light yield of the two track segments in each of the three view is considered. A view is rejected if its optical fiber is parallel to the particle direction or the ratio of light yield between the two segments is abnormally high or if it is the view with smallest number of MPPC hits. Finally, the light yield ratio closest to 2 (the value expected for $\gamma$ events) between the remaining views, is used as particle identification parameter.
The expected distribution of such parameter for $\nu_e$ interactions events is affected by energy deposits around the interaction vertex. Such vertex activity is highly dependent on nuclear effects (like nucleus de-excitation and production of untracked low momentum pions and protons) which are not well known in the framework of the available neutrino-nucleus interaction models and simulations. A more sophisticated study is needed to assess the impact of such uncertainties on the proposed particle identification algorithm, which is not yet used in the ND280 analysis. In principle $\nu_\mu$ interactions could be used as control sample to validate the $\nu_e$ selection efficiency. In order to minimize the dependence to such issue for the simplified study discussed here, the first MPPC hit of the track, corresponding to the cube where the neutrino interaction happens, is not included in the evaluation of the light yield of the upstream electron track segment.

\begin{figure}[hbtp]
\centering
\includegraphics[width=0.8\linewidth]{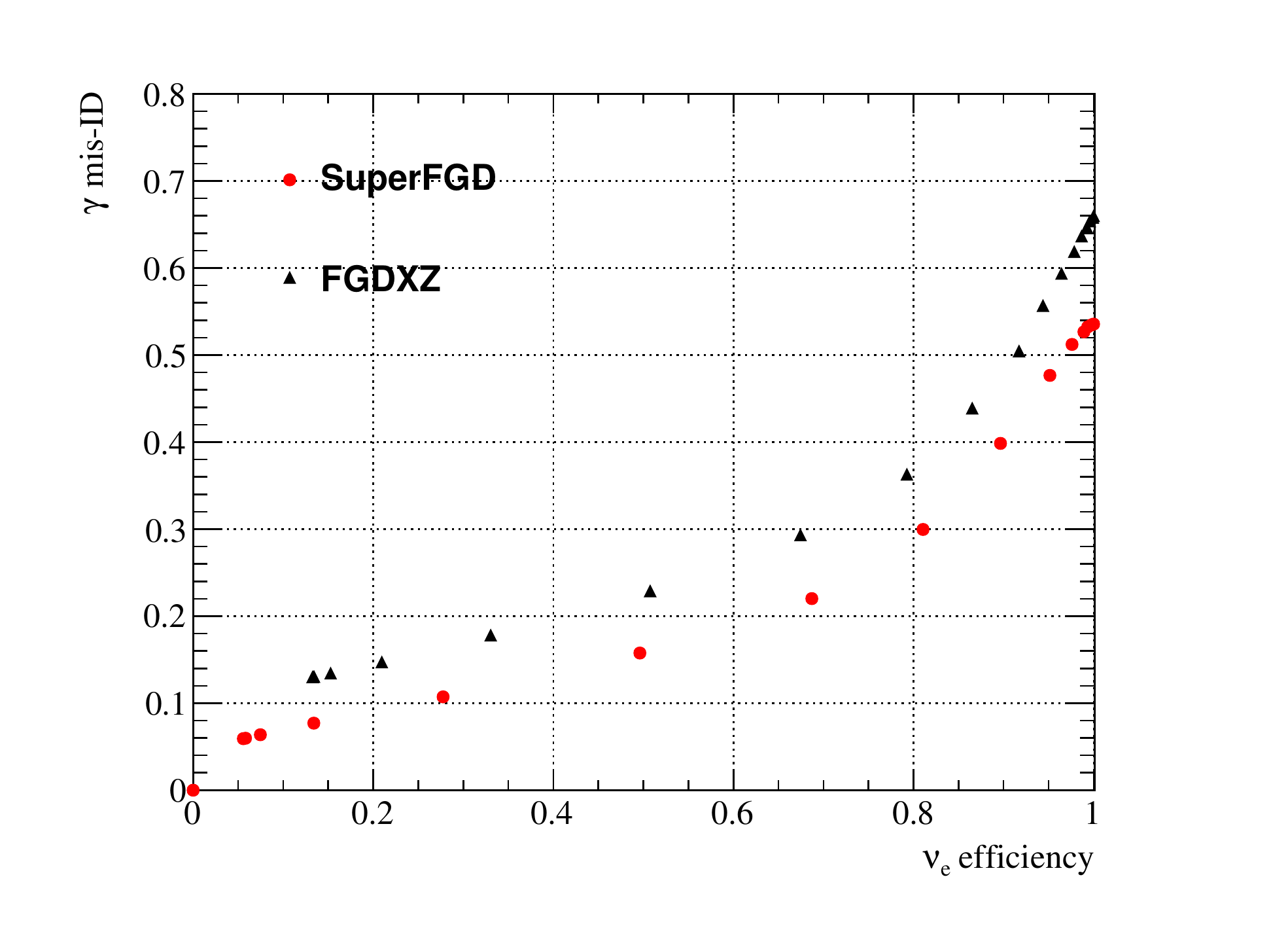}
\caption{$\gamma$ mis-identification probability and $\nu_e$ selection efficiency requiring one-track events and different cuts on the particle identification parameter for SuperFGD and FGD with XZ views.}
\label{fig:nue:egamma_pid}
\end{figure}

The electron/$\gamma$ separation study is performed with charged current quasi-elastic (CCQE) events generated with GENIE with T2K $\nu_e$ flux and $\gamma$ particle gun events generated at the center of the detector and with uniform angular and momentum distribution up to 1~GeV.  Figure~\ref{fig:nue:egamma_pid} shows the $\gamma$ mis-identification probability as a function of the $\nu_e$ efficiency requesting one track events and applying different cuts on the particle identification parameter. The superior performances in electron/$\gamma$ separation for a three-views detector are clearly visible: considering the same selection efficiency as in the current ND280 analysis, i.e. about 30\%, the $\gamma$ mis-identification probability in SuperFGD is about half with respect to a two-views detector. 

Such performances in distinguishing electrons from gamma, coupled with the large target mass of the SuperFGD, and its better efficiency in selecting particles emitted at all angles, will allow to select a clean sample of  $\nu_e$ interactions with energies below 1 GeV. A full analysis of $\nu_e$ interactions requires the development of additional reconstruction algorithms in the SuperFGD, in order to select electrons that can shower before entering the TPCs. Moreover a simulation of the entire detector is needed in order to estimate the amount of $\gamma$ background from out of the SuperFGD. Such detailed studies will be developed after the completion of this TDR.

\section{Probing nuclear effects with the SuperFGD} 
\label{sec:stvsuperfgd}

\newcommand{\pt}{p_\textrm{T}}
\newcommand{\dalphat}{\delta\alpha_\textrm{T}}
\newcommand{\dphit}{\delta\phi_\textrm{T}}
\newcommand{\dpt}{\delta \pt}

The phase of neutrino flavour oscillations depends on the distance between neutrino production and detection as well as the neutrino energy. In long-baseline neutrino oscillation experiments, this distance is fixed and well known, but the true neutrino energy needs to be reconstructed for each event. In order to reconstruct the neutrino energy from outgoing particle kinematics, assumptions must be made about the nature of the interaction. For the `kinematic' method used by T2K~\cite{Abe:2017vif} the neutrino energy is reconstructed using the kinematics of a selected outgoing lepton, assuming that the neutrino scatters off a stationary target nucleon and that interaction was quasi-elastic. However, in reality the initial state nucleon is bound within a complex nuclear environment and a variety of so-called `nuclear effects' obfuscate any attempt to reconstruct the incoming neutrino energy. Although current neutrino-nucleus interaction simulations provide some modelling of these effects, the associated uncertainties are already the dominant systematic on current T2K measurements of oscillation parameters and will soon become the principle limitation if an improved understanding cannot be achieved. This section will demonstrate that the large acceptance and low tracking thresholds of the SuperFGD may be able to provide such an understanding. This will be shown mostly by using one particularly powerful probe of nuclear effects (transverse kinematic imbalance) but conclusions regarding the SuperFGDs potential to probe nuclear effects through different observables can be generalised. 

\subsection{Nuclear effects and transverse kinematic imbalance}

Nuclear effects can broadly be factorised into three categories:

\begin{itemize}
\item the initial state motion of nucleons inside a nucleus (Fermi motion);
\item nucleon correlation effects, which can sometimes lead to two nucleon, or `two particle two hole' (2p2h) final states;
\item final state re-interactions (FSI) of the struck nucleon inside the nuclear medium which can both alter the kinematics of the final state nucleon and stimulate nuclear absorption or emission (of other nucleons or pions) thereby altering the topology of the interaction.
\end{itemize}

\begin{figure}[ht!]
  \includegraphics[width=0.9\linewidth]{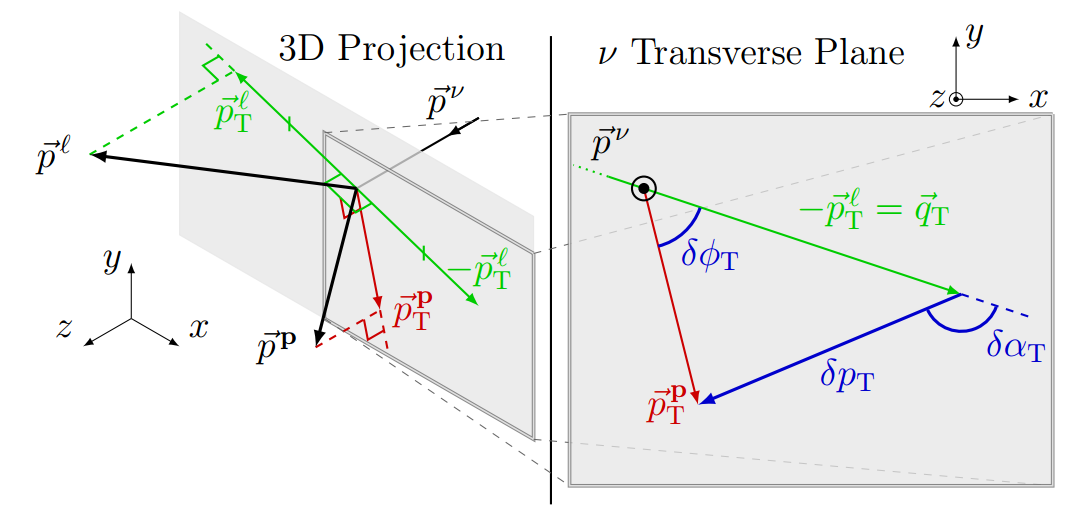}
\caption{Schematic view of the definition of the observables: $\dpt$, $\dalphat$ and $\dphit$. The left side shows an incoming neutrino interacting and producing a lepton ($\ell$) and a proton $\mathrm{p}$, whose momenta are projected onto the plane transverse to the neutrino ($\nu$). The right side then shows the momenta in this transverse plane and how the observables are formed from considering the imbalance within it.  Reproduced from~\protect\cite{Abe:2018pwo}. }
\label{stv}
\end{figure}

One particularly powerful tool to probe these nuclear effect is to utilise the kinematic imbalance between the final state lepton and hadrons in the plane transverse to the neutrino direction~\cite{Lu:2015tcr}. When measured for neutrino-nucleus interactions containing only the final state lepton and nucleons, these `transverse' observables are typically defined as:
\begin{align}
\dpt &= |\overrightarrow{p}_T^l + \overrightarrow{p}_T^p|, \label{eq:dpt}\\
\dalphat &= \arccos \frac{-\overrightarrow{p}_T^l \cdot \delta \overrightarrow{p}_T}{p_T^l \dpt}, \label{eq:dat}\\
\dphit &= \arccos \frac{-\overrightarrow{p}_T^l \cdot \overrightarrow{p}_T^p}{p_T^l p_T^p} \label{eq:dphit}.
\end{align}
where $p^p$ and $p^l$ are the (highest momentum) proton and lepton momenta, and the $T$ index is the projection of the vector on the plane transverse to the incoming neutrino direction. The observable definitions are also shown schematically in Fig.~\ref{stv}. In the absence of nuclear effects, $\dpt$ and $\dphit$ vanish, while $\dalphat$ is undefined. These observables have recently been measured by both the T2K~\cite{Abe:2018pwo} and MINERvA~\cite{Lu:2018stk} experiments.  \\

\subsection{Simulation}
\label{sec:stvSimulation}
In order to determine the SuperFGD's sensitivity to distinguish nuclear effects it is necessary to produce simulations with variations of nuclear models. To do this the NEUT 5.4.0. simulation~\cite{Hayato:2002sd} is used to generate muon-neutrino interactions with a hydrocarbon target. For the charged current quasi-elastic (CCQE) interactions that are most relevant to T2K, NEUT is capable of several different descriptions of the Fermi motion. It can simulate events according to: the Llewellyn-Smith formalism~\cite{llewellyn-smith} based on a relativistic Fermi gas (RFG) model of the Fermi motion; the spectral function (SF) from Ref.~\cite{Benhar:1995}; and a local Fermi gas (LFG) 1p1h model based on the work of Nieves et. al in Ref.~\cite{Nieves:2012}. In all of these the axial mass used for quasi-elastic processes ($M_A^{QE}$) is set to $\sim1.0$ GeV. 

Resonant pion production process is described by the Rein Sehgal model~\cite{rein-sehgal} with the axial mass $M_A^{RES}$ set to 1.21 GeV, whilst the simulation of 2p2h interactions is based on the model from Nieves et. al in Ref.~\cite{Nieves:2012}. The FSI, describing the transport of the hadrons produced in the primary neutrino interaction through the nucleus, are simulated using a semi-classical intranuclear cascade model.

NEUT is used to produce large ensembles of neutrino-nucleus interactions using each of the available descriptions of the Fermi motion and additionally with and without both 2p2h and FSI effects. To isolate a realistically measurable cross-section, `CCQE-like' interactions with only a single muon, at least one proton and no mesons are selected. For each of these interactions the transverse observables are calculated as in Eqns.~\ref{eq:dpt}-\ref{eq:dphit} and a differential cross-section is produced. 

Following the production of the cross-section from NEUT, the detector resolution is approximated by a conservative  4\% Gaussian smear to each component of the outgoing particles momentum vectors 
. The detectors acceptance is simulated as hard momentum thresholds: only protons with momenta between 300 MeV and 1 GeV, and muons with momenta above 50 MeV are considered. 
 A very approximate representative error in each bin of the observables was calculated by scaling the statistical error in the T2K publication~\cite{Abe:2018pwo} (statistical error taken from~\cite{dolanThesis}) by the square root of the ratio of the number of SuperFGD events\footnote{Although a naive statistical error could be calculated simply using the number of events expected in the SuperFGD, this does not take account of possible error inflation in a cross-section extraction procedure from unfolding.} expected with a 30\% integrated efficiency (this is very conservative, it is the same as was achieved using FGD1 in~\cite{Abe:2018pwo}) and $3\times 10^{20}$ P.O.T (the full expected statistics) in that bin, before adding an ad-hoc 5\% `systematic' error which is around the size of the combined detector and model systematics in the current T2K analysis~\cite{dolanThesis}. An effective flux systematic is not included since this is predominantly a normalisation systematic and the transverse observables offer most sensitivity to nuclear effects through their shape. 

\begin{figure}[ht!]
\begin{subfigure}{.5\textwidth}
  \centering
  \includegraphics[width=0.9\linewidth]{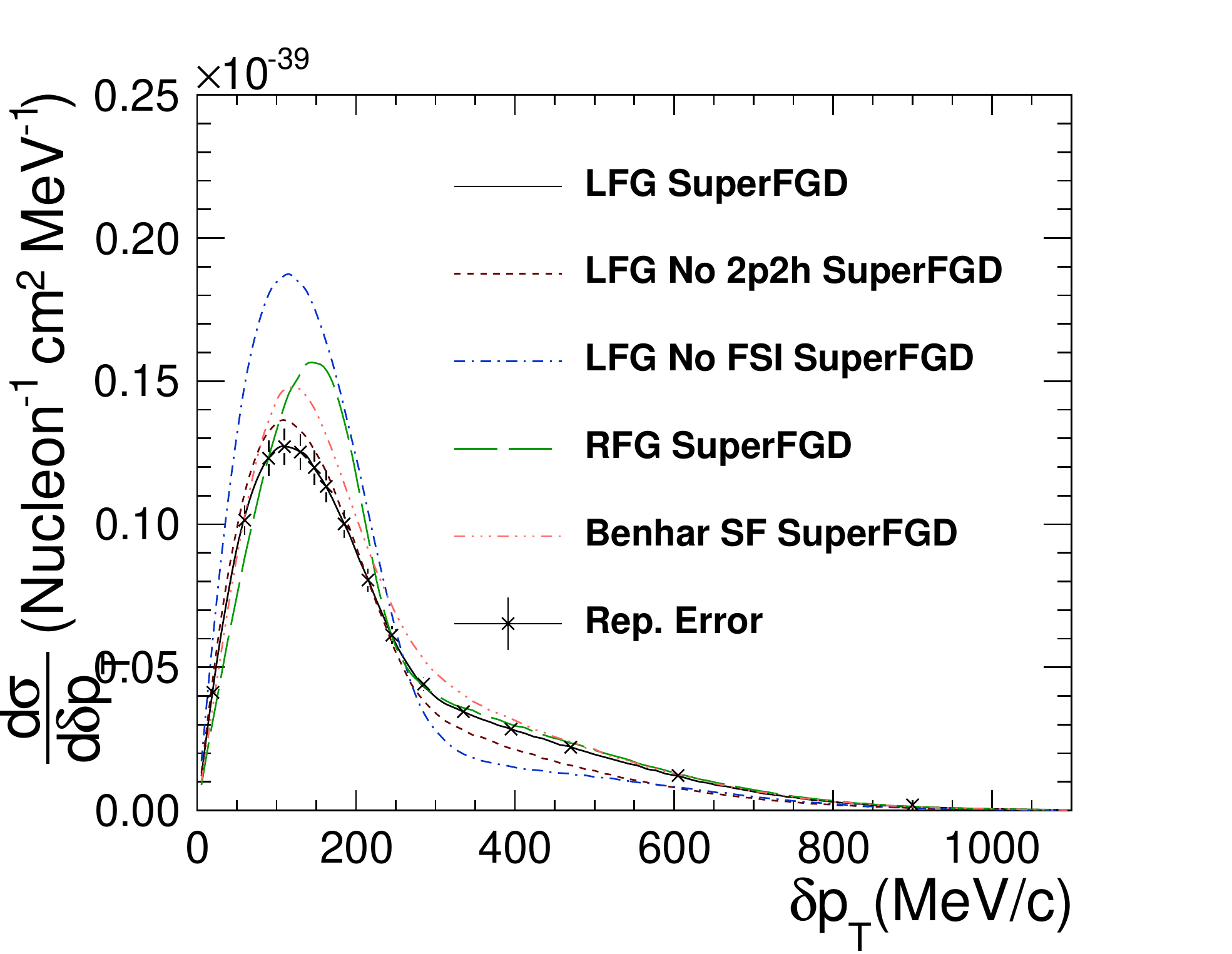}
\end{subfigure}%
\begin{subfigure}{.5\textwidth}
  \centering
  \includegraphics[width=0.9\linewidth]{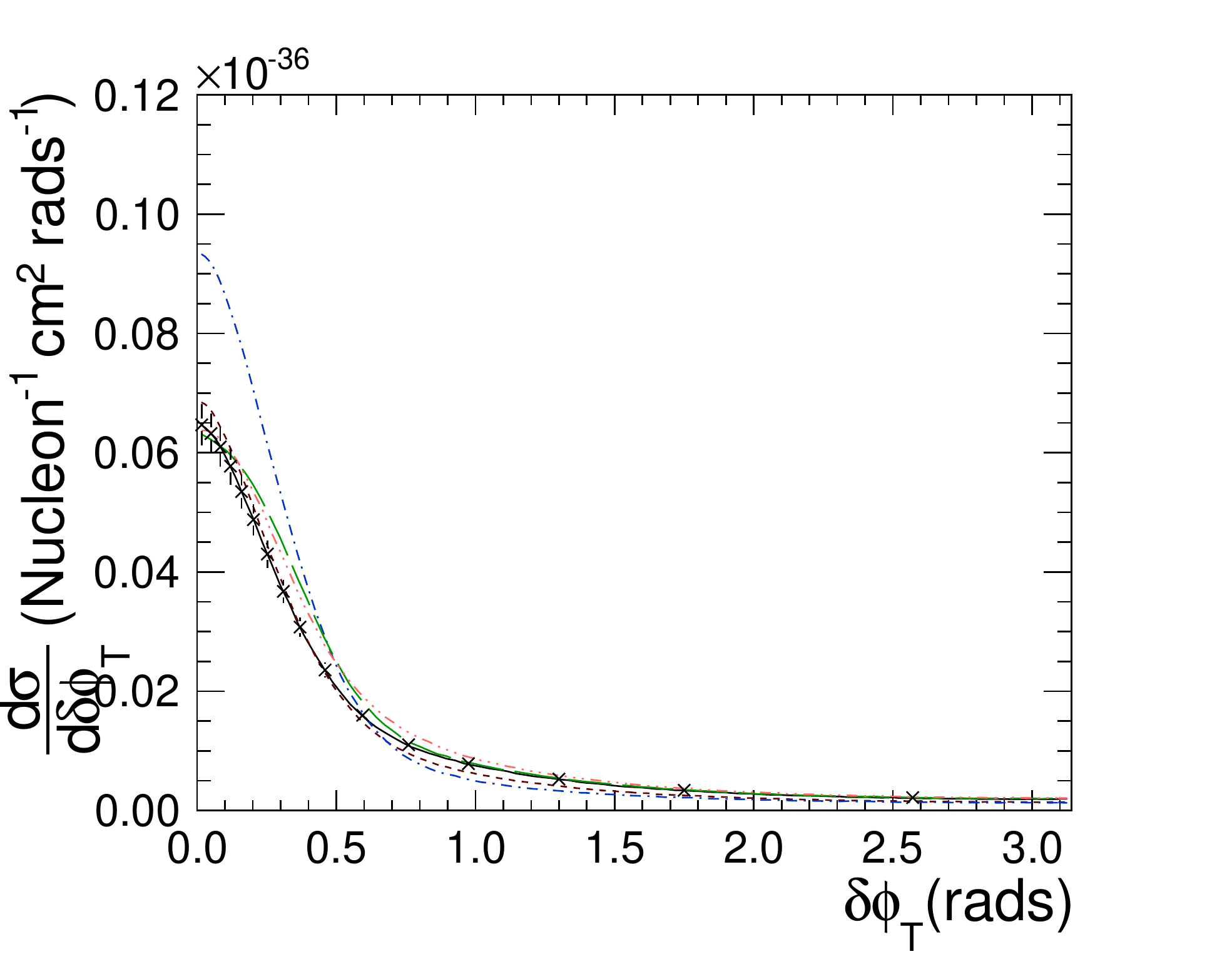}

\end{subfigure}\\
\begin{subfigure}{.5\textwidth}
  \centering
  \includegraphics[width=0.9\linewidth]{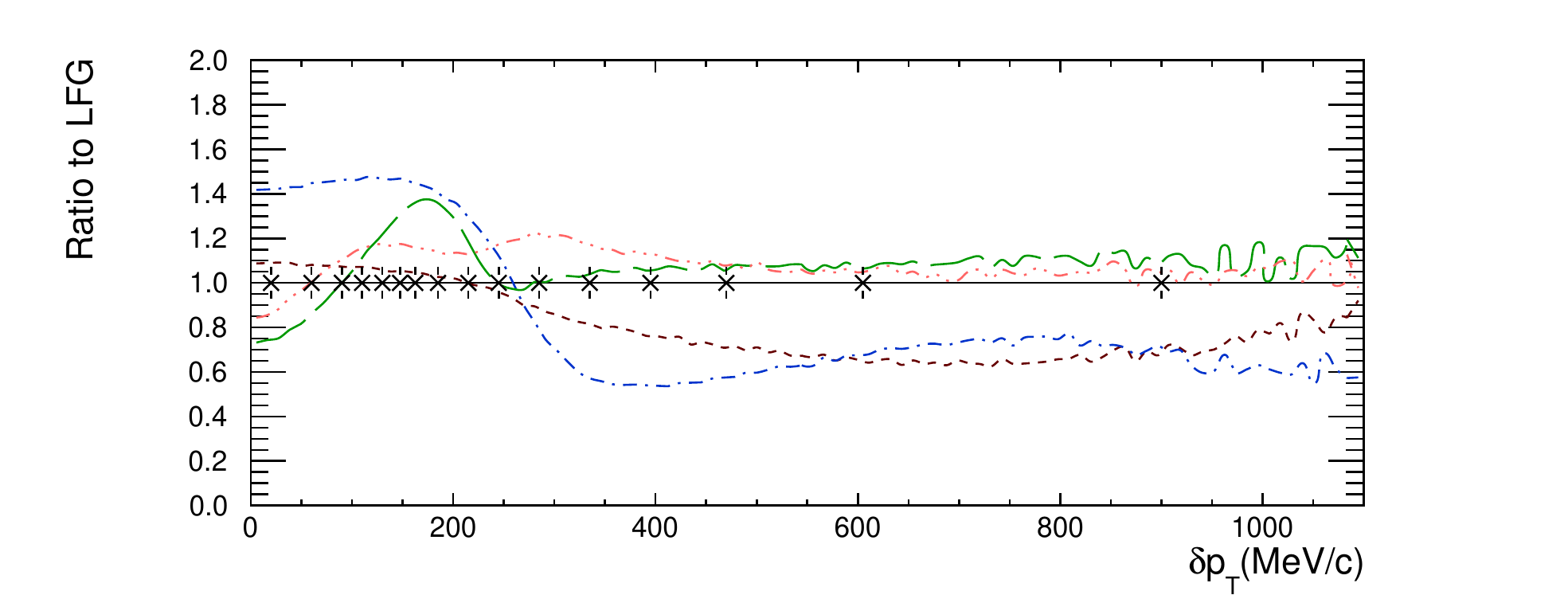}
    \caption{}
  \label{dptmodels}
\end{subfigure}
\begin{subfigure}{.5\textwidth}
  \centering
  \includegraphics[width=0.9\linewidth]{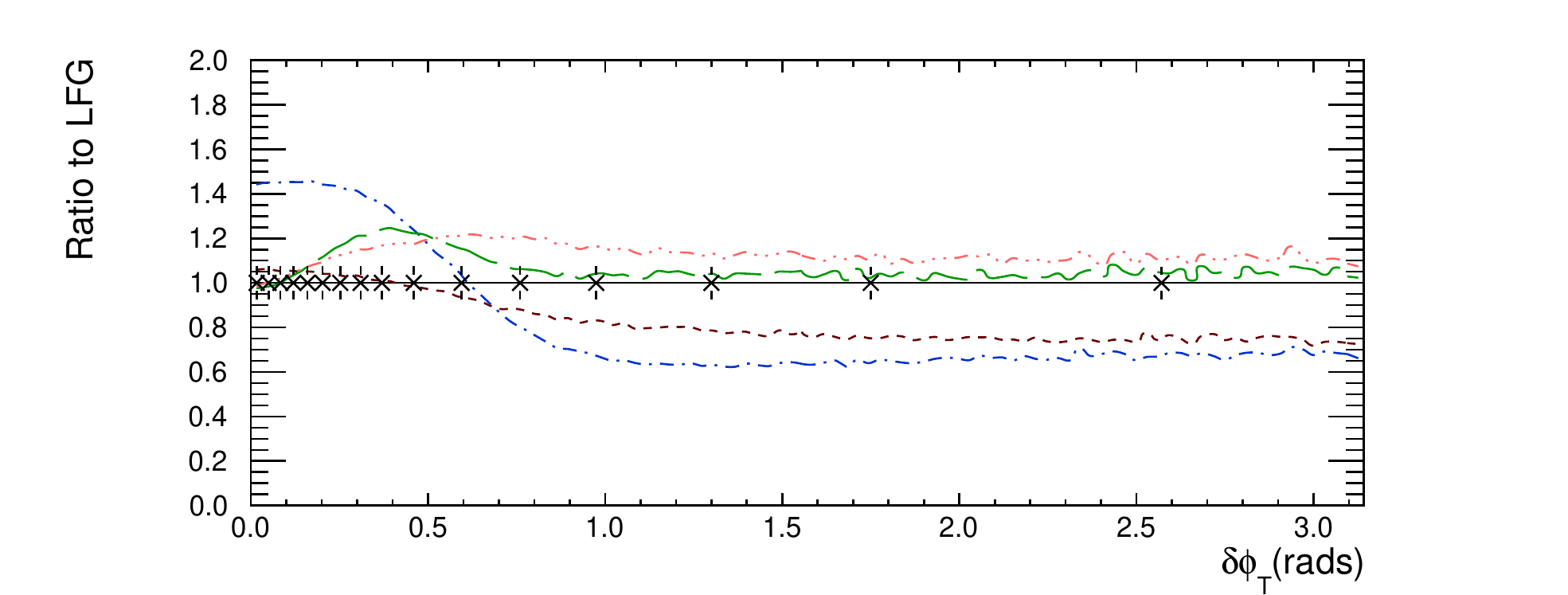}
  \caption{}
  \label{dphitmodels}
\end{subfigure}
\caption{The differential cross section of CCQE-like neutrino-hydrocarbon interactions in $\dpt$ (a) and $\dphit$ (b) for different nuclear models, smeared and acceptance corrected based on the expected SuperFGD performance. The LFG prediction shows an approximate error based on assumptions discussed in Sec.~\ref{sec:stvSimulation}. The lower figures present the same information as ratios to the LFG case.}
\label{model_discrimination}
\end{figure}

\begin{figure}[ht!]
\begin{subfigure}{.9\textwidth}
  \centering
  \begin{tikzpicture}
    \node(a){\includegraphics[width=0.9\textwidth]{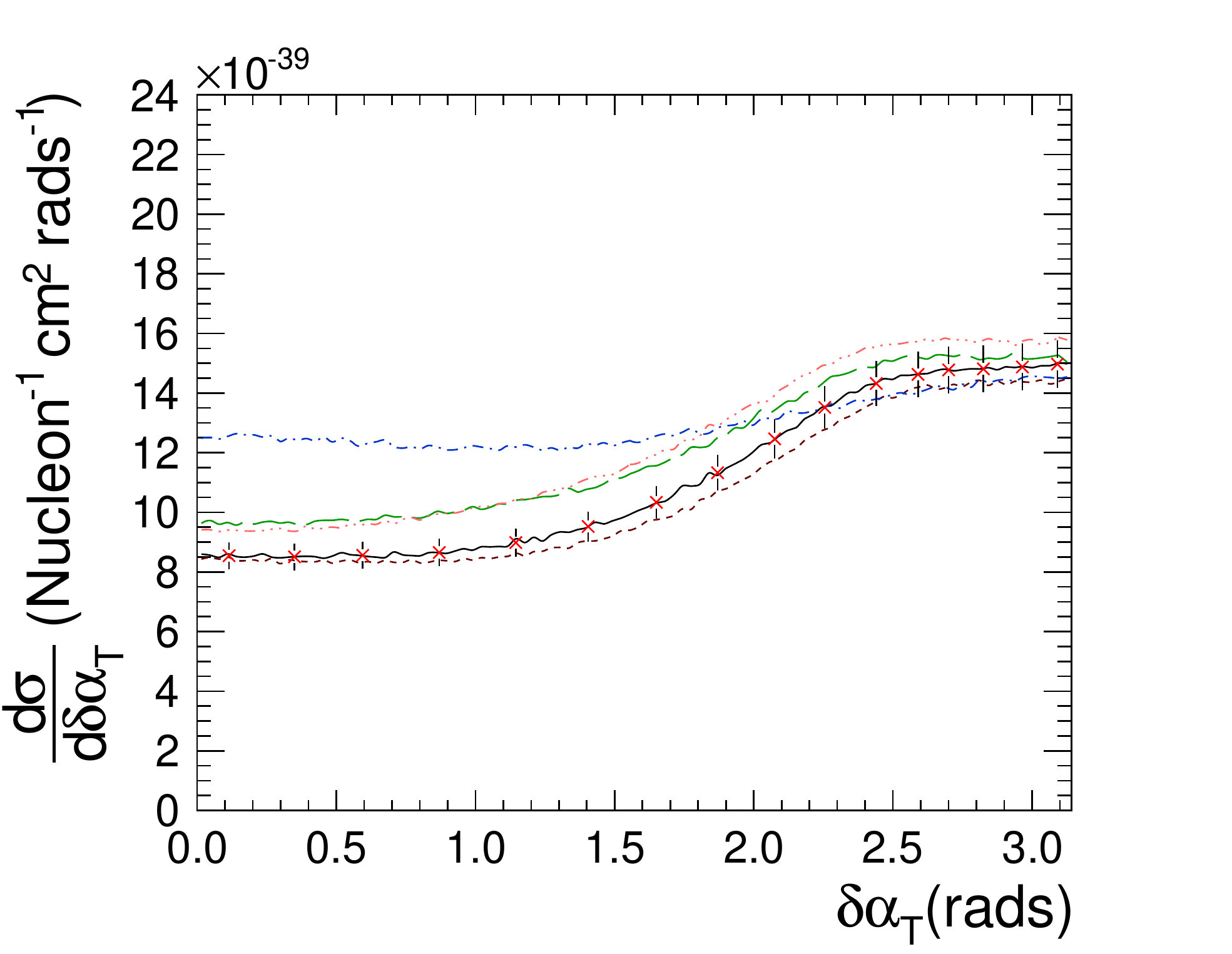}};
    \node at (a.south east)
    [
    anchor=center,
    xshift=-37mm,
    yshift=35mm
    ]
    {
        \includegraphics[width=0.22\textwidth]{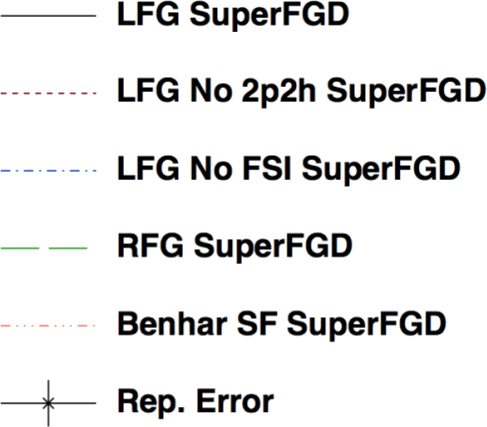}
    };
    \end{tikzpicture}
\end{subfigure}
\\
\begin{subfigure}{.9\textwidth}
  \centering
  \includegraphics[width=0.9\linewidth]{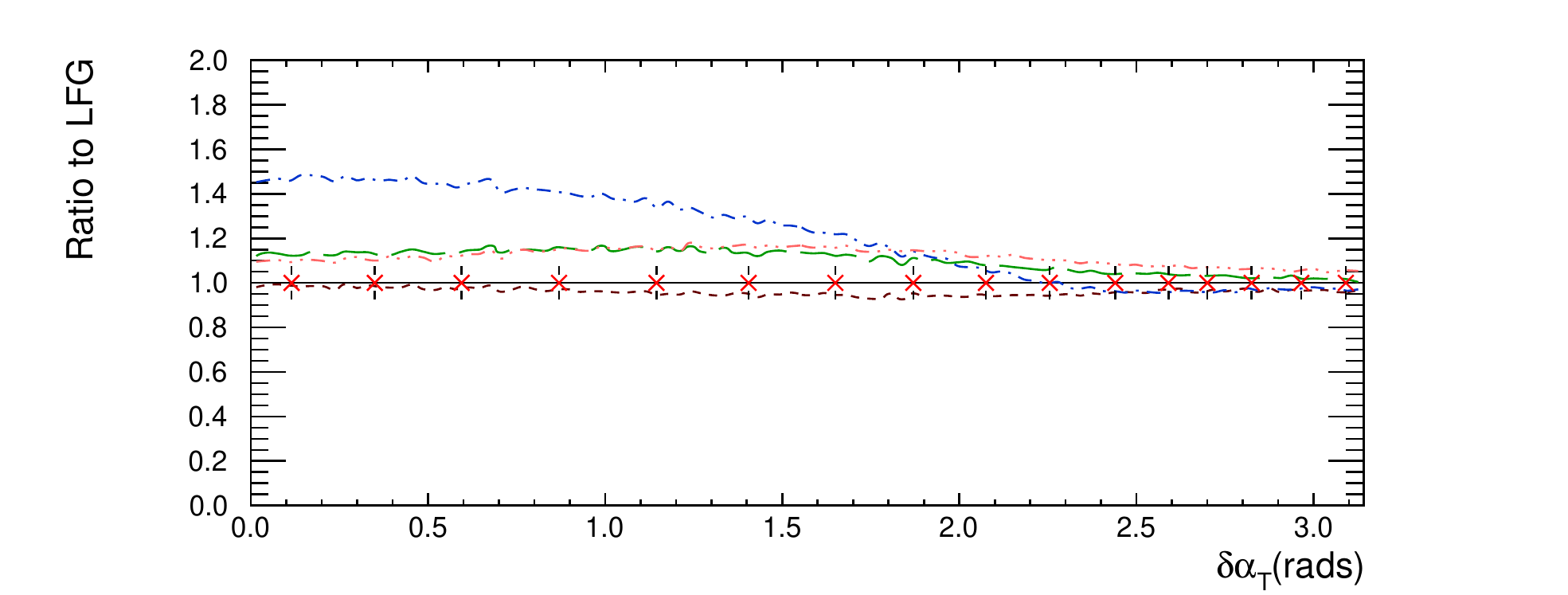}
\end{subfigure}
\caption{The differential cross section of CCQE-like neutrino-hydrocarbon interactions in $\dalphat$ for different nuclear models, smeared and acceptance corrected based on the expected SuperFGD performance. The LFG prediction shows an approximate error based on assumptions discussed in Sec.~\ref{sec:stvSimulation}. The lower figures present the same information as ratios to the LFG case.}
\label{dalphatmodels}
\end{figure}

Fig.~\ref{model_discrimination}~and~\ref{dalphatmodels} shows the resultant smeared and acceptance-corrected cross-sections for and the ratio of each model to NEUT's default model (LFG) for both $\dpt$ and $\dphit$. The `representative errors' are placed on the LFG model. Note that this simulation does not account for difficulties associated with background subtraction. However, it is expected that a CCQE-like selection should be of a high purity (the current ND280 selections achieve around 80\%) and that the major backgrounds (mostly associated with resonant pion production) are able to be well-constrained with control regions~\cite{Abe:2018pwo}. It is therefore not likely that the background subtraction will be pivotal in determining the SuperFGDs sensitivity to nuclear effects.

\subsection{Model discrimination potential}
\label{sec:stvModelDiscrim}

Compared to the precision offered by the current T2K analysis of transverse observables~\cite{Abe:2018pwo}, Fig.~\ref{model_discrimination} demonstrates that the additional statistics and kinematic acceptance offered by the SuperFGD will likely allow a much more sophisticated probe of nuclear effects. In particular, the bulk region of $\dpt$ shown in Fig.~\ref{dptmodels} shows a very clear separation between the RFG and SF/LFG models whilst the tail region shows a clear separation of no FSI and no 2p2h cases from the others. Since turning off FSI conserves the full phase-space CCQE-like cross section, this is partially distinguished from 2p2h from the large impact FSI has on the bulk of the distribution. 

However, there remains some degeneracy between the impact of 2p2h and FSI effects on both $\dpt$ and $\dphit$. Within the NEUT models, this degeneracy can be partially lifted by additionally considering $\dalphat$, shown in Fig.~\ref{dalphatmodels}, which is fairly insensitive in shape to nuclear model variations other than FSI. In the absence of FSI, the distribution of $\dalphat$ is expected to be flat~\cite{Lu:2015tcr}, whilst the presence of FSI will shift the distribution towards high values of $\dalphat$, corresponding to the deceleration of the final state protons. In the SuperFGD, three regimes of $\dalphat$ can therefore be distinguished:
\begin{itemize}
\item $0<\dalphat<\tfrac{\pi}{3}$: low FSI region;
\item $\tfrac{\pi}{3}<\dalphat<\tfrac{2\pi}{3}$: intermediate FSI region;
\item $\tfrac{2\pi}{3}<\dalphat<\pi$: high FSI region.
\end{itemize}

\begin{figure}[htbp!]
\begin{subfigure}{0.5\textwidth}
\centering
  \includegraphics[width=1\linewidth]{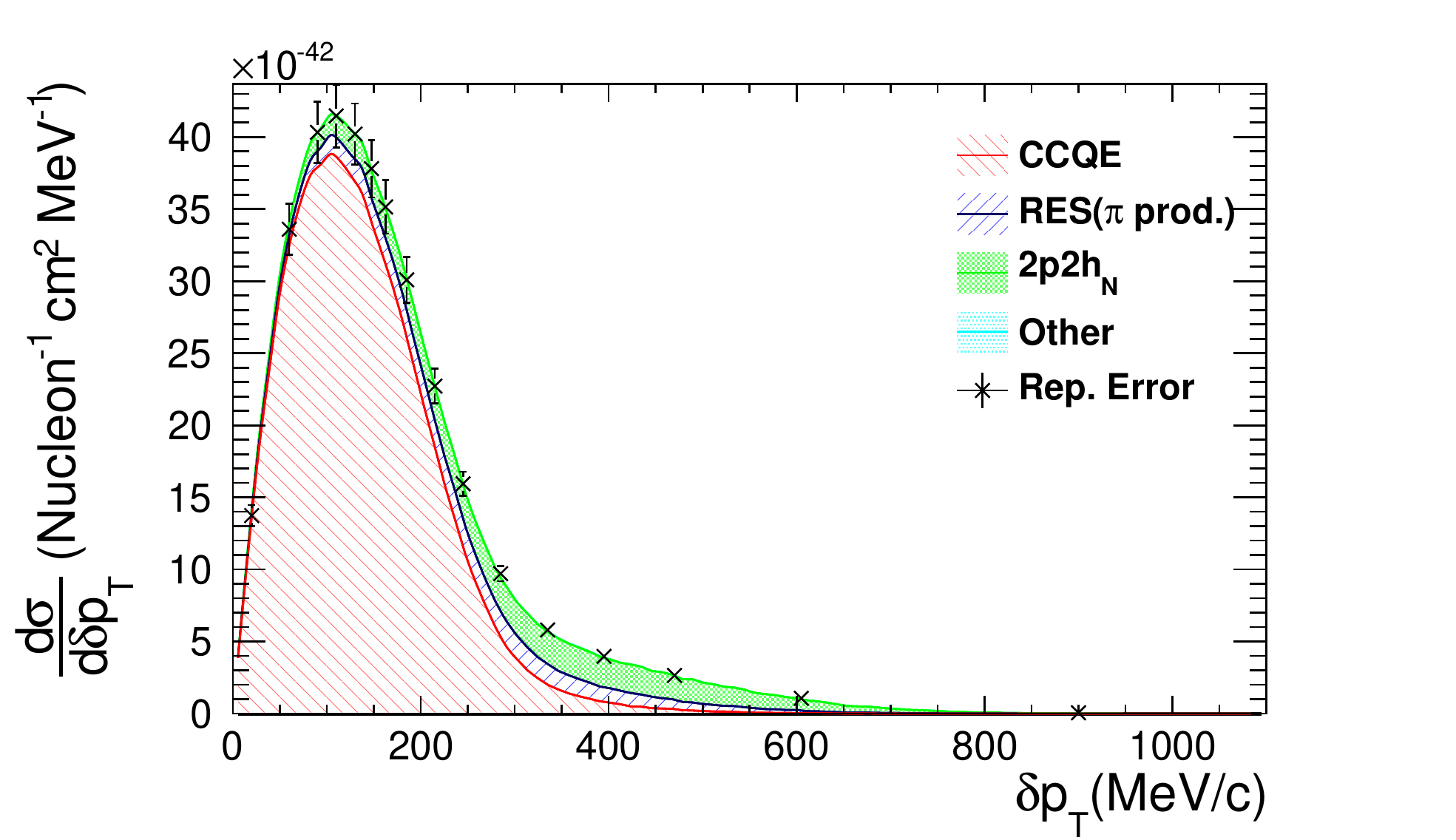}
  \caption{LFG - Low $\dalphat$}
  \label{dat_lfg_low}
\end{subfigure}
\begin{subfigure}{0.5\textwidth}
\centering
  \includegraphics[width=1\linewidth]{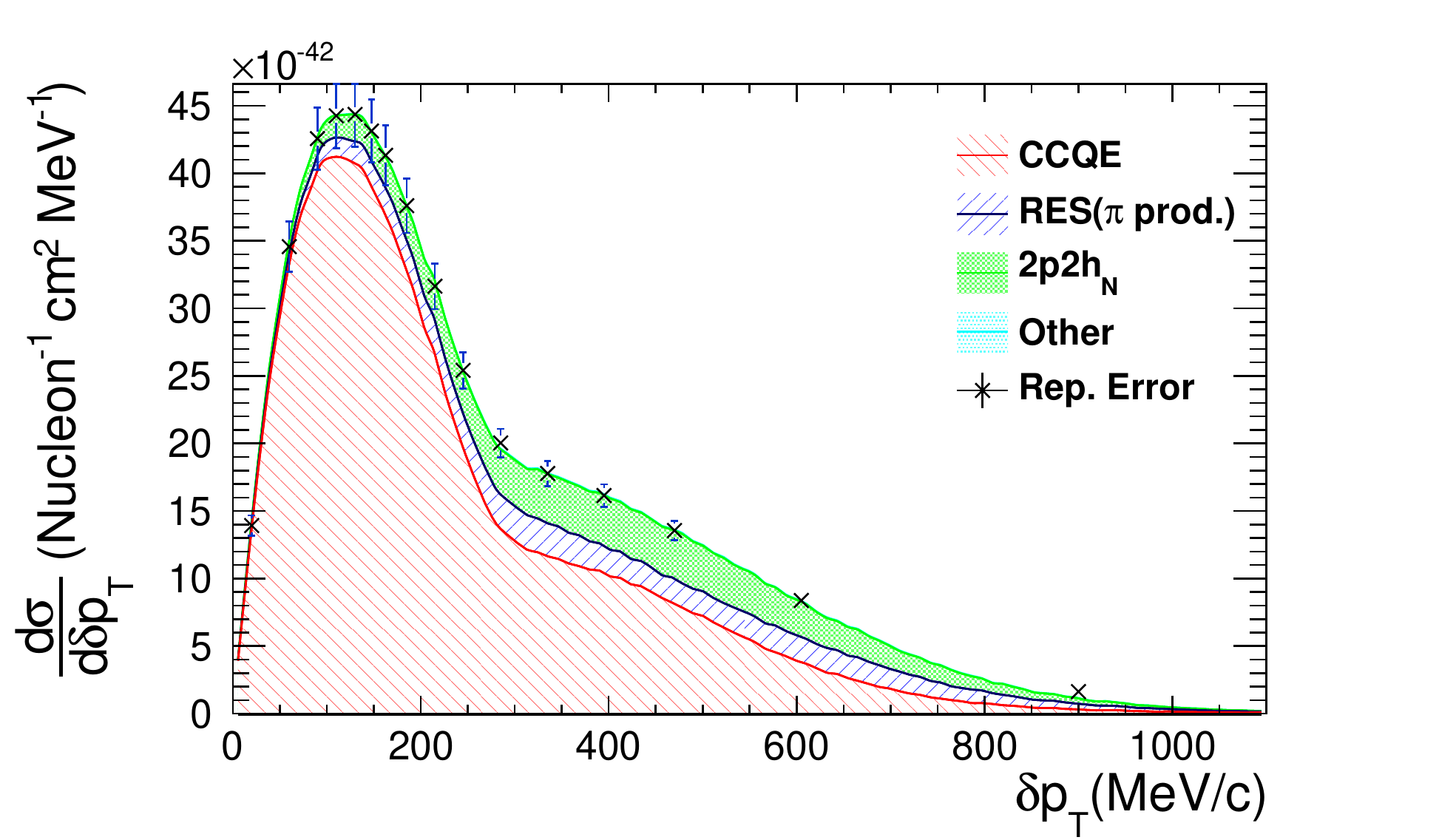}
   \caption{LFG - High $\dalphat$}
   \label{dat_lfg_high}
\end{subfigure}\\
\begin{subfigure}{0.5\textwidth}
\centering
  \includegraphics[width=1\linewidth]{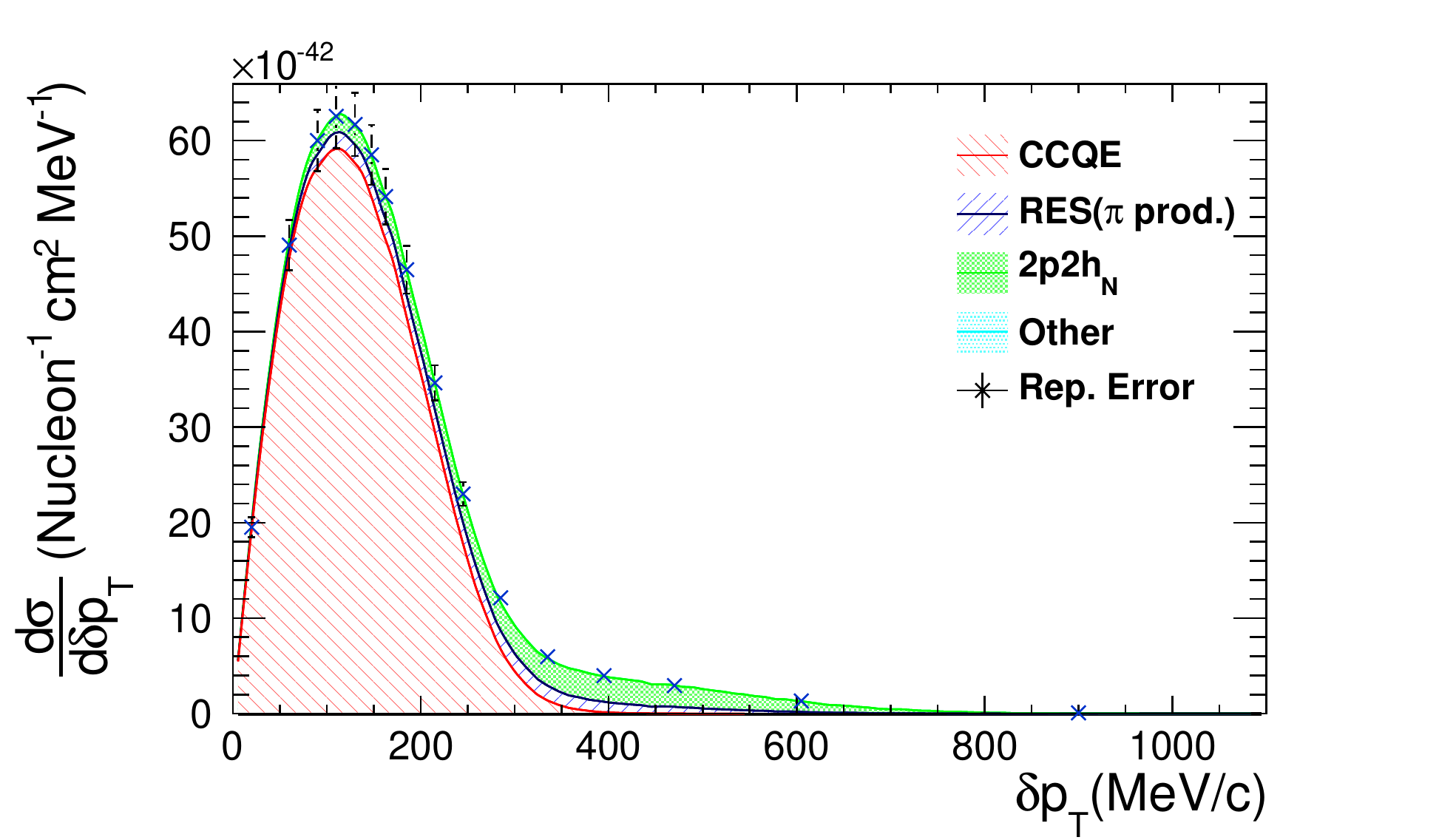}
   \caption{LFG no FSI - Low $\dalphat$}
   \label{dat_lfgnoFSI_low}
\end{subfigure}
\begin{subfigure}{0.5\textwidth}
\centering
  \includegraphics[width=1\linewidth]{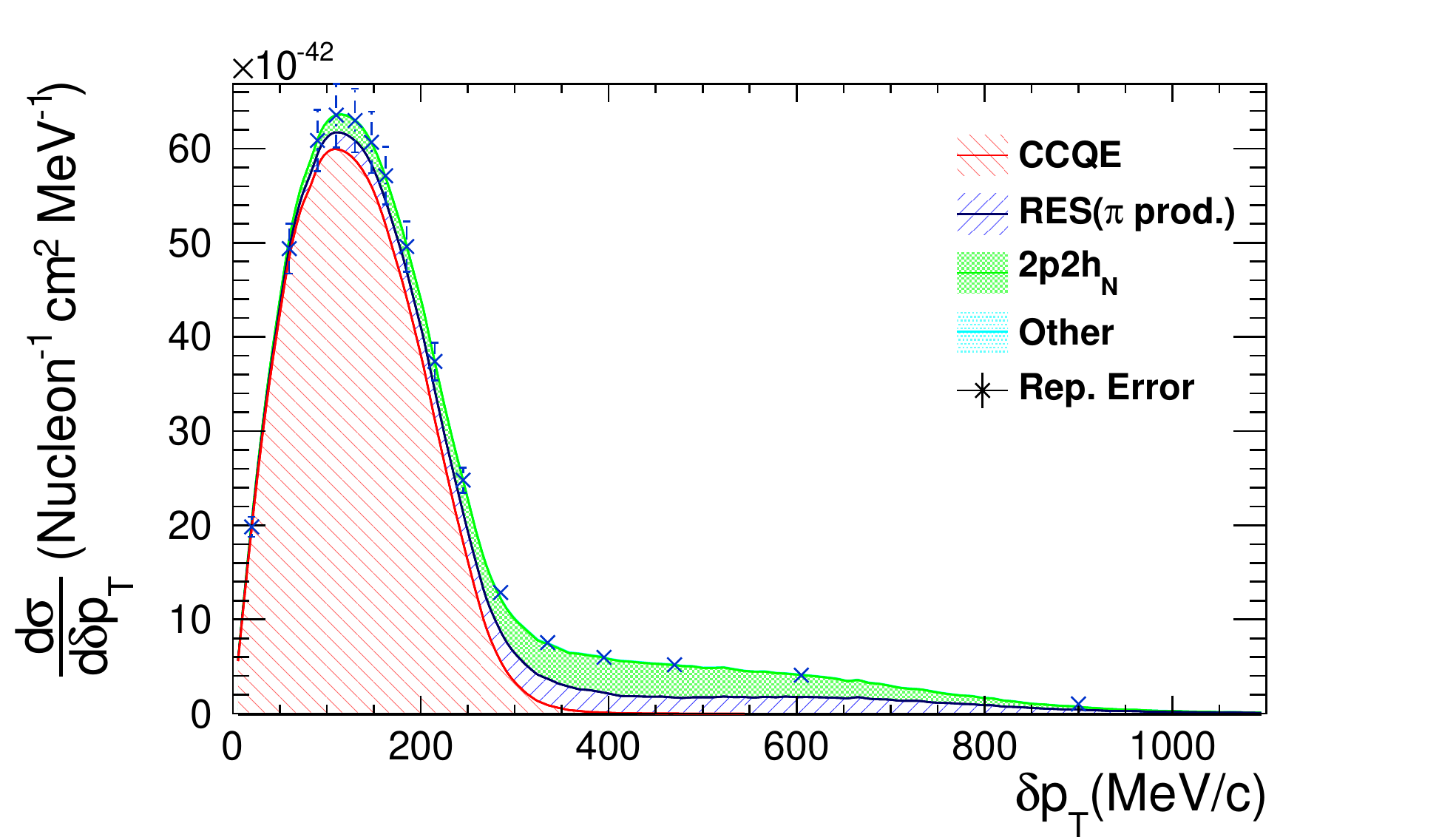}
   \caption{LFG -no FSI - High $\dalphat$}
    \label{dat_lfgnoFSI_high}
\end{subfigure}
\caption{$\dpt$ distributions broken down by interaction modes in different regions of $\dalphat$. Figures \ref{dat_lfgnoFSI_low} \& \ref{dat_lfgnoFSI_high} show the sample which has no FSI processes, whereas figures \ref{dat_lfg_low} \& \ref{dat_lfg_high} show a realistic LFG model.}
\label{dalphat_2p2h}
\end{figure}

In this way, working within a particular region of $\dalphat$ allows the selection of FSI strength such that the tail of $\dpt$ becomes either dominated by 2p2h effects (in the low FSI region) or FSI effects (in the high FSI region). This is demonstrated in Fig.~\ref{dalphat_2p2h}, which presents $\dpt$ cross-sections, separated by interaction mode, in the low and high FSI regions of $\dalphat$. Fig.~\ref{dalphat_2p2h} further demonstrates that selecting the low FSI region of $\dalphat$ it is almost equivalent in the LFG and LFG without FSI models, therefore demonstrating that this technique allows access to an otherwise unphysical scenario. The SuperFGD is expected to gather sufficient data to make such a combined measurement of $\dpt$ and $\dalphat$.

\subsection{Advantages over a two readout plane horizontal target}
\label{sec:FGDXY}

This section has so far demonstrated that the SuperFGD will likely be able to offer a high precision probe of the nuclear effects responsible for some of the dominant systematics in neutrino oscillation analyses. However, it is pertinent to consider whether the apparent greatly improved precision compared to the current T2K analysis (using FGD 1 as a target) stems from a simple statistics increase or whether the SuperFGDs improved acceptance is critical.

\begin{figure}[ht!]
\centering
  \includegraphics[width=0.9\linewidth]{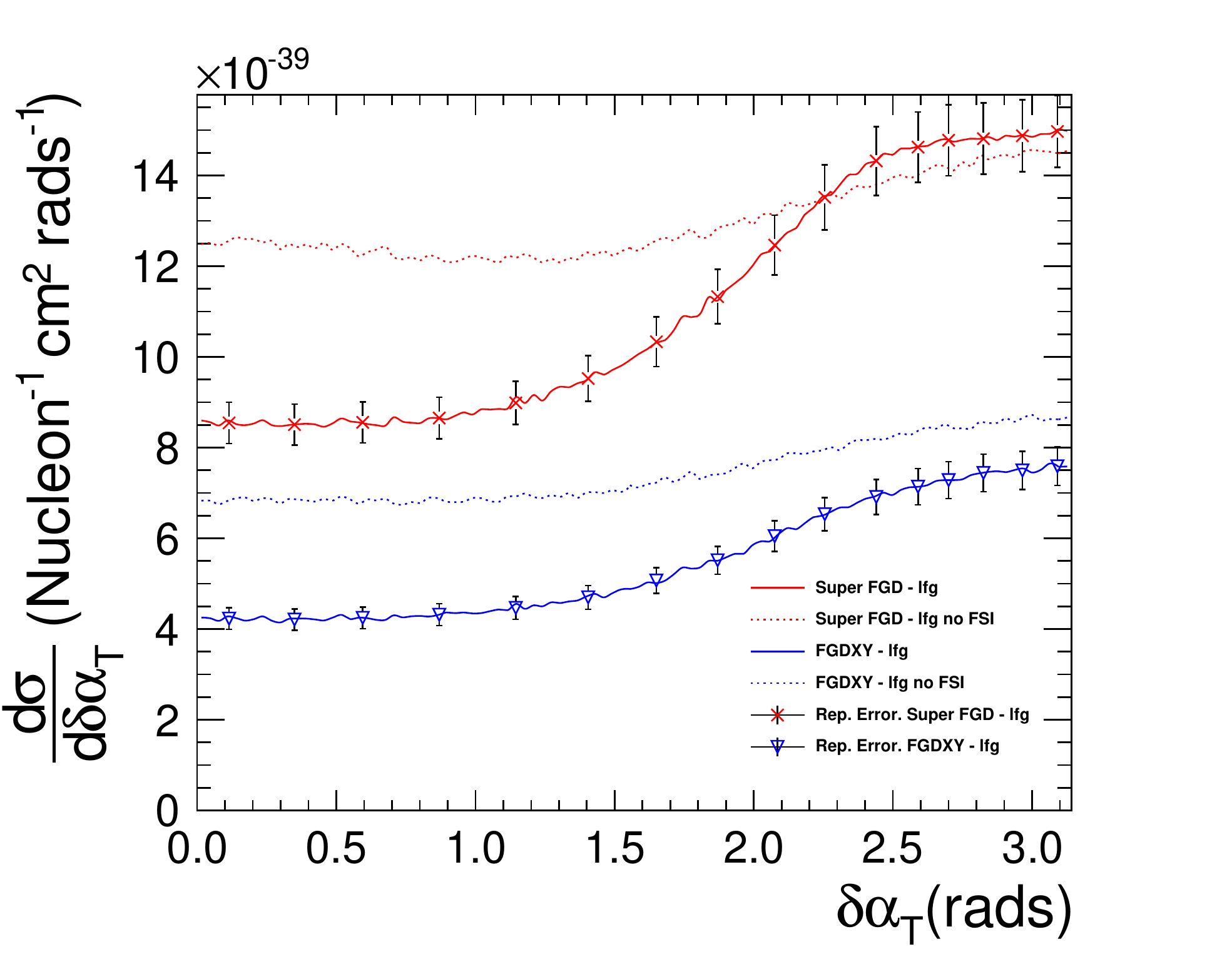}
  \caption{Comparison of the sensitivity to FSI effects through a measure of $\dalphat$ for the SuperFGD and an FGDXY. The y-axis reports the CCQE-like cross section within the phase space accessible by the relevant detector. Detector smearing and acceptance effects are applied as described in Sec.~\ref{sec:stvSimulation}.}
    \label{dalphat_superFGDvsFGDXY}
\end{figure}

To assess the impact of the SuperFGDs improved acceptance, the study of section~\ref{sec:stvModelDiscrim} is repeated but using the tracking thresholds of FGD~1 (taken from~\cite{Abe:2018pwo}). From this study, it was found that the largest advantage of the SuperFGD's improved acceptance is from its ability to measure $\dalphat$. This is demonstrated in Fig.~\ref{dalphat_superFGDvsFGDXY}, which shows a comparison of the expected measure of $\dalphat$ of an FGD-like detector (FGDXY) compared to the SuperFGD. The key advantage of the SuperFGD is in larger the relative shape differences between the FSI and no FSI case. For the SuperFGD the three FSI regions are clearly distinguishable, whilst they are not for the FGDXY. This extra sensitivity to FSI effects (and through this, the ability to separate FSI and 2p2h effects) exhibited by the SuperFGD stems mostly from its ability to identify low momentum protons, thereby demonstrating that the SuperFGD's unique design offers interesting advantages in probing the nuclear effects pertinent to neutrino oscillation analyses.

\subsection{Double transverse momentum imbalance}

\begin{figure}[ht]
\centering
  \includegraphics[width=0.9\linewidth]{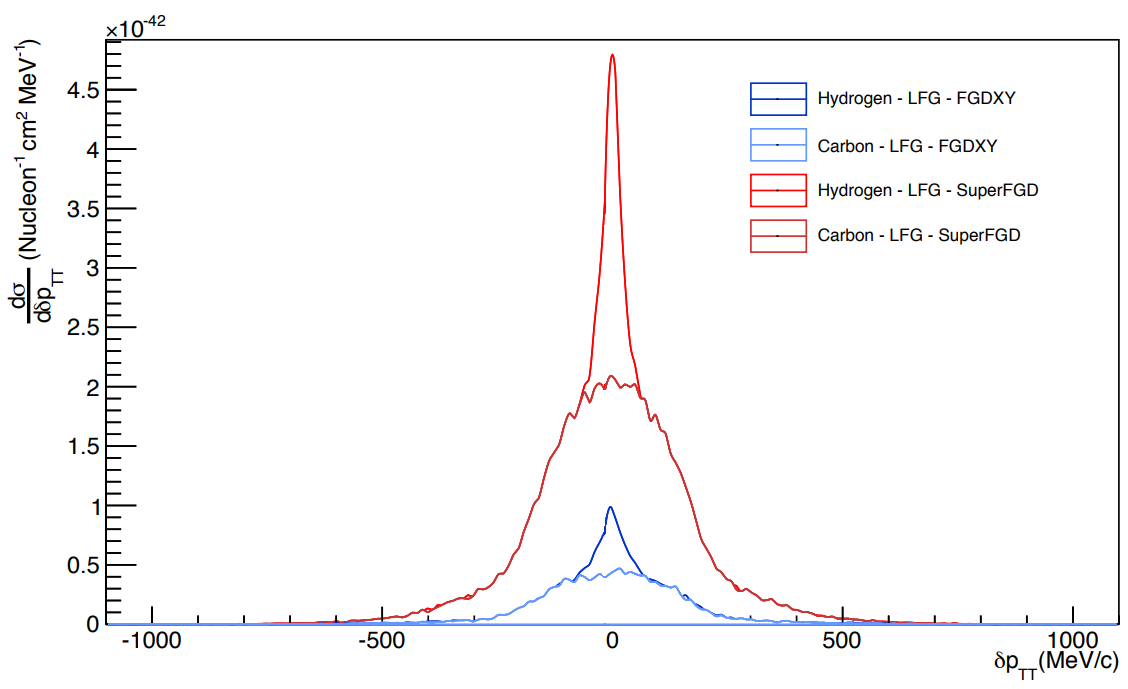}
  \caption{Comparison of the ability of a SuperFGD and an FGD XY to measure $\delta p_{TT}$. The y-axis reports the CCQE-like cross section within the phase space accessible by the relevant detector. Detector smearing and acceptance effects are applied as described in Sec.~\ref{sec:stvSimulation}. }
  \label{dptt}
\end{figure}

Another interesting method of providing a constraint on nuclear effects is to measure the `double transverse momentum imbalance', $\delta p_{TT}$, between the proton and the pion momentum, in neutrino interactions with at least one muon, one proton and one pion in the final state~\cite{double} on a composite target involving hydrogen, such as in the Carbon-Hydrogen scintillator of the SuperFGD.  For interactions on Hydrogen (which are therefore free of nuclear effects) $\delta p_{TT}=0$, whilst for non-hydrogen interactions the distribution is broadened from nuclear effects. By measuring $\delta p_{TT}$ the SuperFGD may be able to separate interactions on Hydrogen and Carbon to offer a direct factorisation of the neutrino free-nucleon interaction and nuclear effects. Using the same simulations as described in Sec.~\ref{sec:stvSimulation}, the ability for the SuperFGD to measure $\delta p_{TT}$ is compared to what could be achieved with an FDGXY (simulated as in Sec.~\ref{sec:FGDXY}, with pions treated the same as muons) is shown in Fig.~\ref{dptt}. This demonstrates that the SuperFGD's lower tracking thresholds are able to access a much wider phase space than those of an FGDXY and that the Hydrogen peak therefore becomes much more prominent. However, further study is required to determine whether an analysis with the SuperFGD could offer a reliable subtraction of the Carbon background. 

\section{Neutron detection in the SuperFGD}  
\label{sec:superfgdneutrons}

The large size and high granularity of the SuperFGD can also be used to tag and reconstruct neutrons produced in  anti-neutrino interactions. 
The possibility of measuring the neutron kinematics can contribute to improve the knowledge of nuclear effects, for instance studying 2particles-2holes events with neutron-proton or neutron-neutron final states.

Neutrons emitted in antineutrino interactions will, in some cases, break a nucleus, producing protons with energies of few tenths of MeV that can be detected in the SuperFGD. The measurement of the time delay between the antineutrino interaction and the detection of the neutron--induced proton, can provide the information about the nucleon energy. 

An analysis was performed to estimate the neutron detection efficiency and the energy resolution. 
Neutron
particle guns
were uniformly thrown at the center of the SuperFGD. 
Neutron--induced protons were selected if they were produced away from the 3x3x3 cubes around the neutron production point, in order not to be affected by vertex activity in real neutrino interactions. 

\begin{figure}[hbtp]
  \centering{\includegraphics[width=0.6\linewidth]{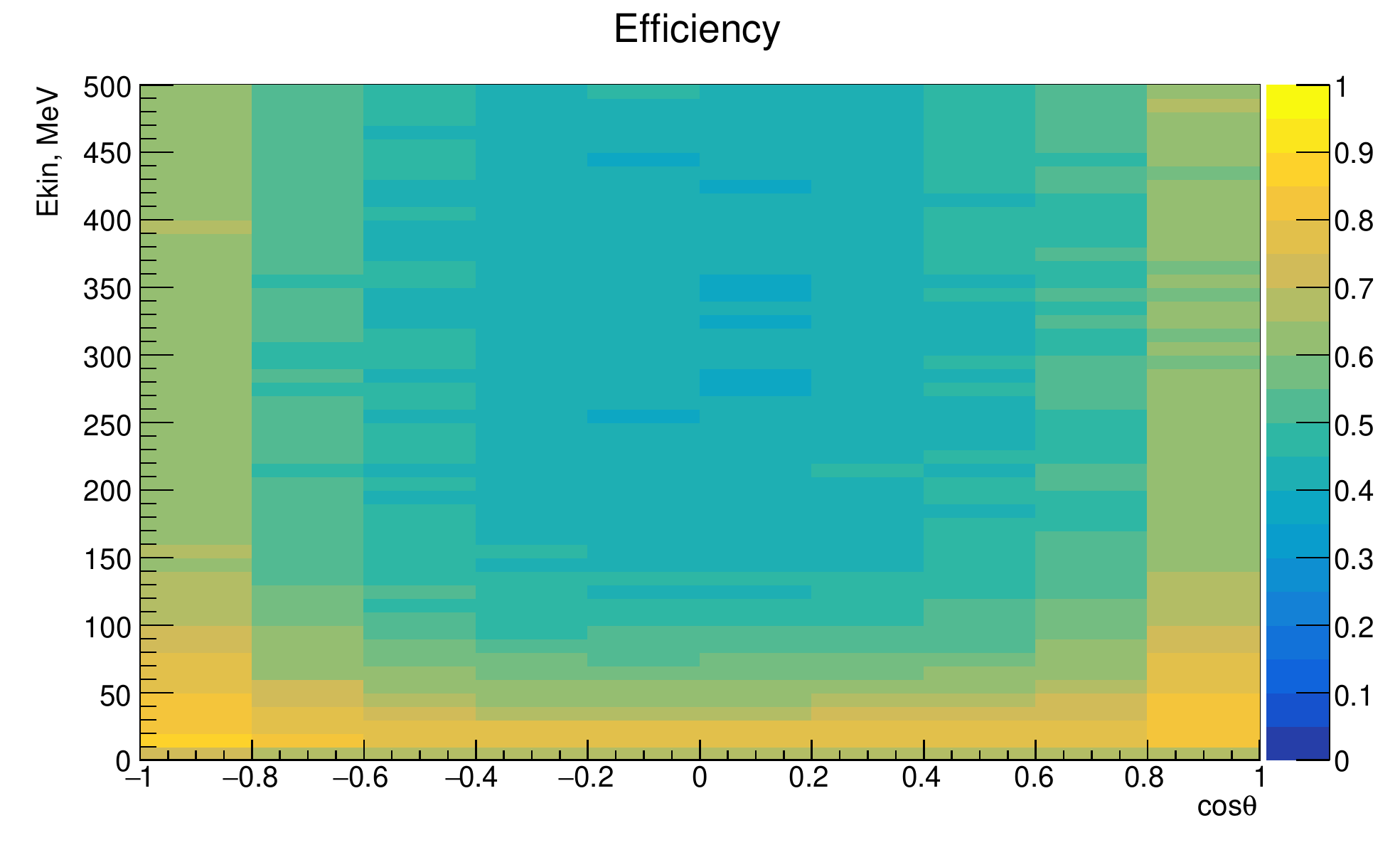}}
  \caption{Neutron detection efficiency in the SuperFGD from the particle gun study.}
  \label{fig::neutron_eff}
\end{figure}

The first hit in time is used to define the time of the neutron interaction. In order to simulate the detector response, the measured time is smeared based on the expected time resolution for a MIP in a single cube, for instance $1.5 ns/\sqrt{3}\approx 0.9$ ns assuming a perfect efficiency for all the three WLS fiber in the cube.
This approach could be conservative since a recoiled proton could produce more scintillation light than a MIP particle.
Furthermore the energy resolution would be improved by $\sqrt{N}$ if the proton produces scintillation light in N cubes. 
The expected efficiency for such selection is presented in figure~\ref{fig::neutron_eff}. 

\begin{figure}[hbtp]
    \begin{center}
    \includegraphics[width=0.8\linewidth]{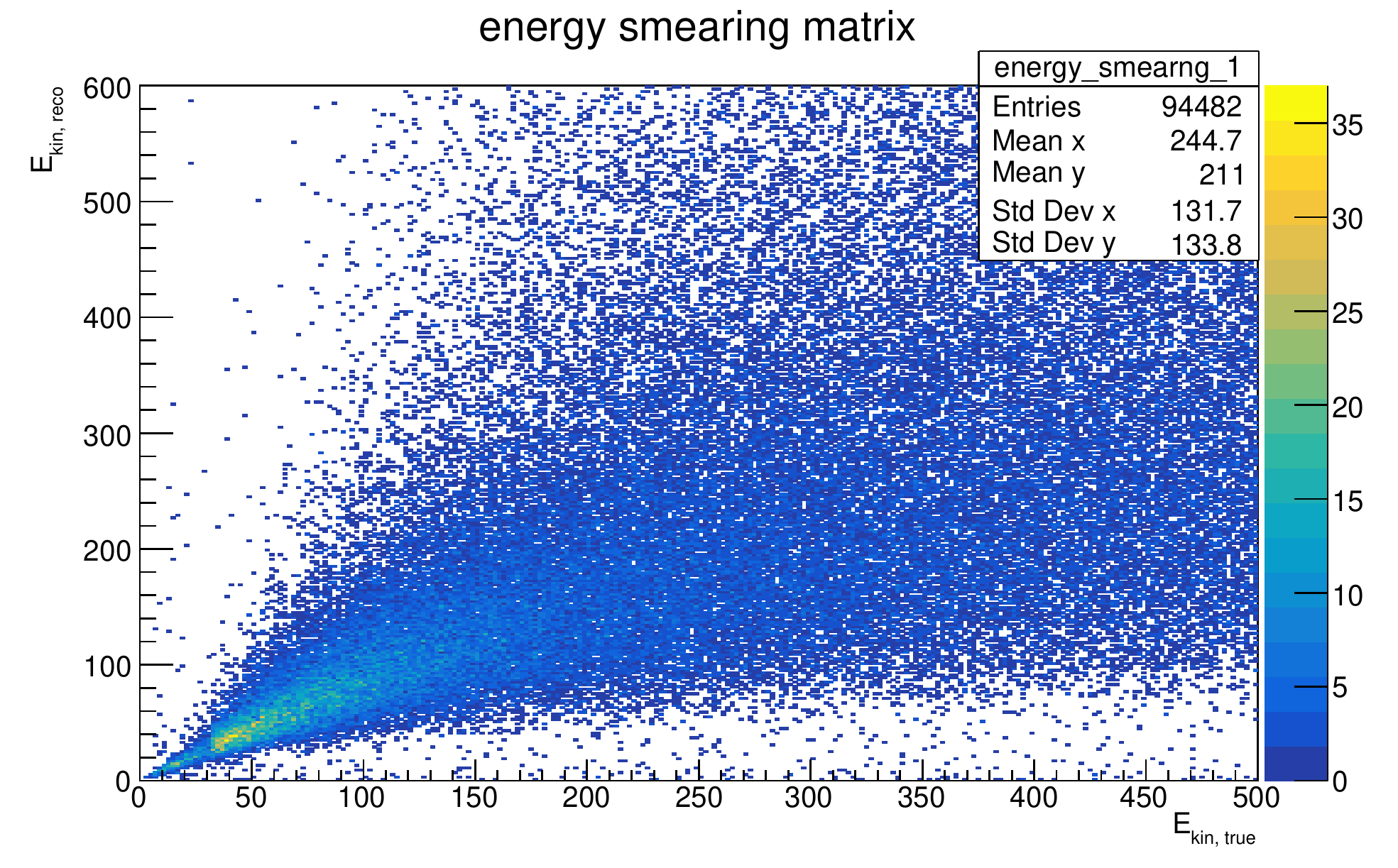}\\
    \includegraphics[width=0.46\linewidth]{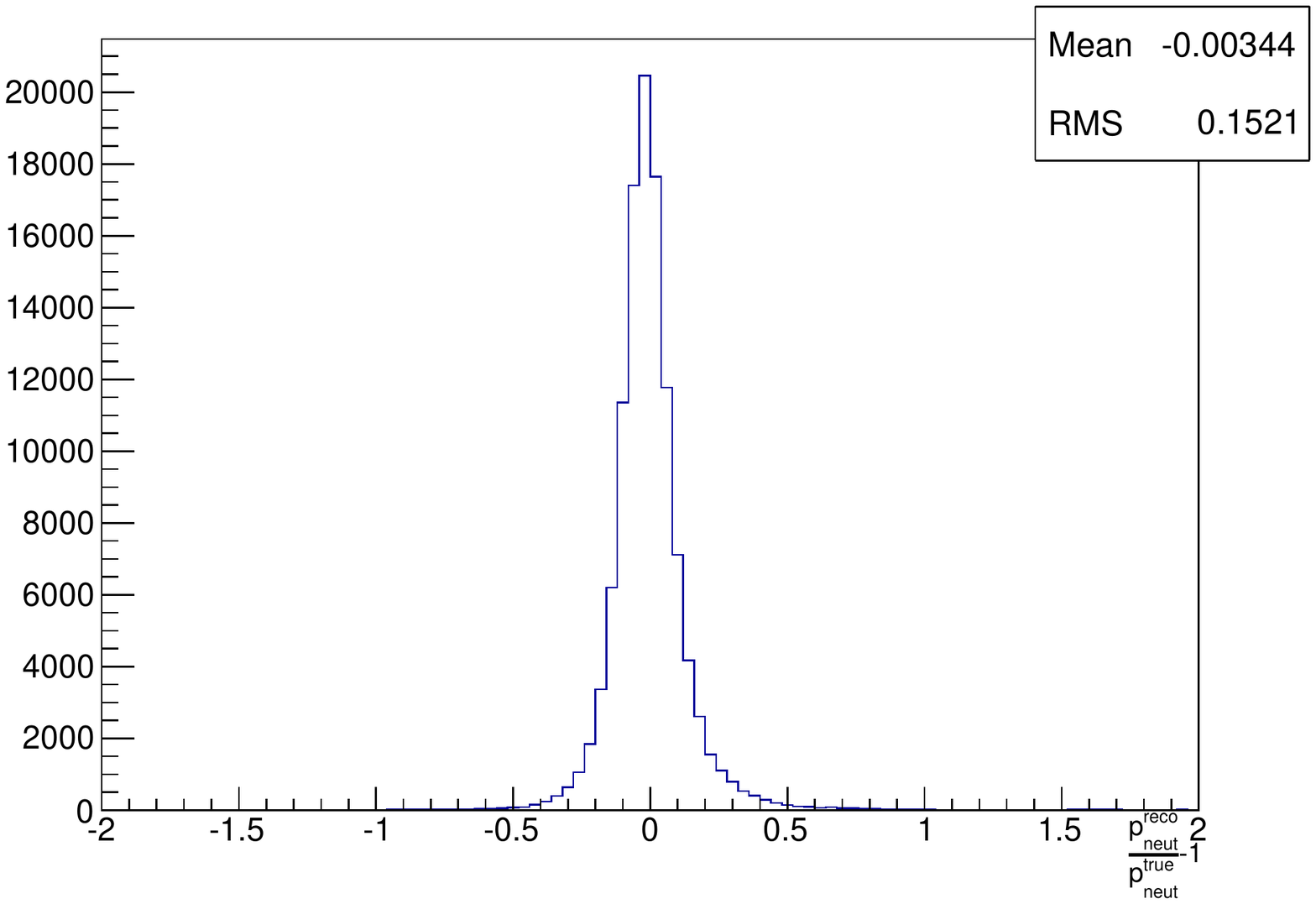}
    \includegraphics[width=0.49\linewidth]{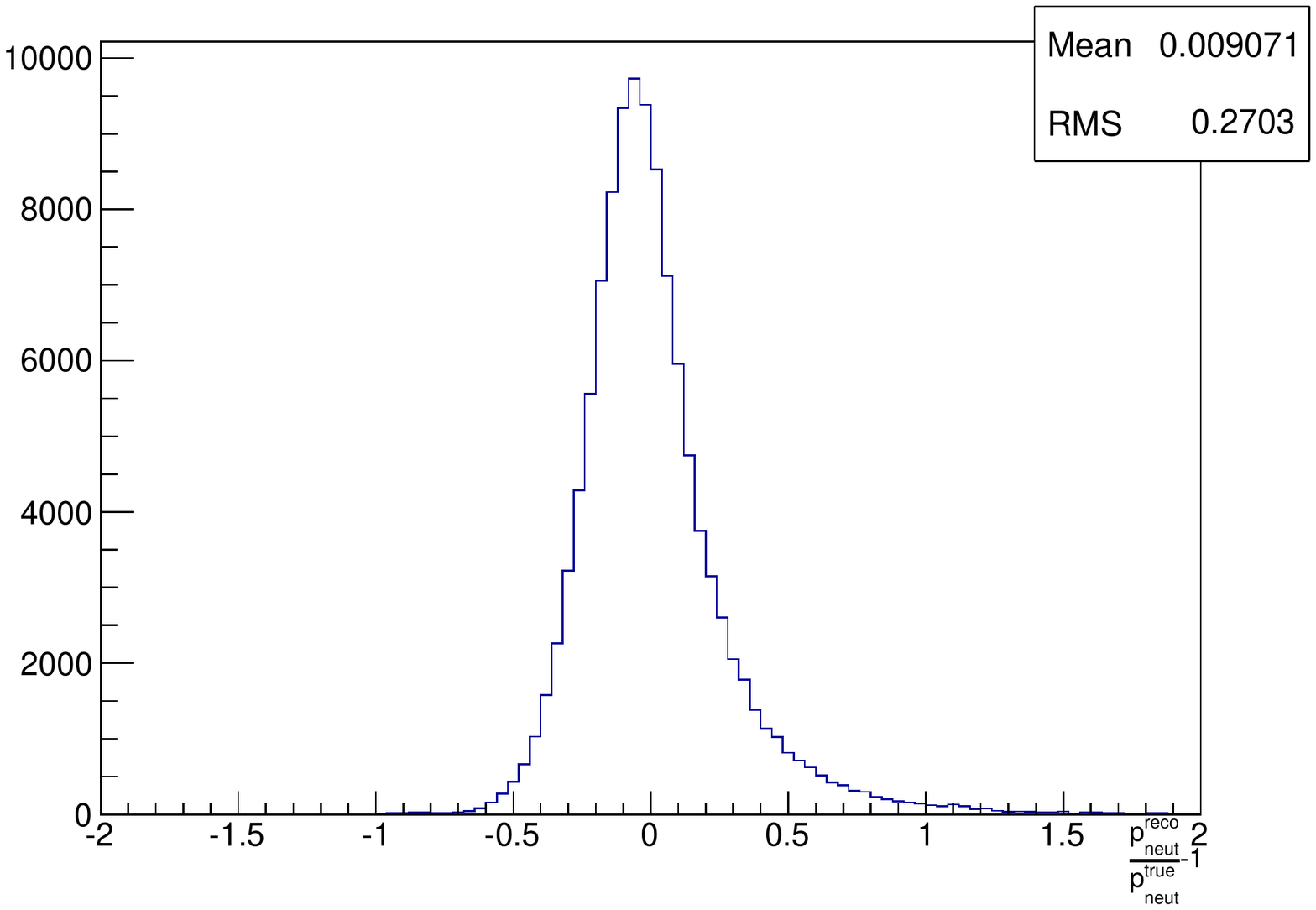}
    \caption{In the top figure the expected  energy smearing matrix (true vs reconstructed kinetic) for neutrons traveling longer than 40 cm. 
    The energy resolution for typical T2K neutrons is shown for time resolution of $0.6/ \sqrt{3}$~ns (bottom left) and $1.5/ \sqrt{3}$~ns (bottom right).}
    \label{fig::neutron_05}
    \end{center}
\end{figure}

Although the detection efficiency is good, 
for small traveling distances the time resolution cannot be good enough to precisely measure the neutron kinetic energy.
A sufficient accuracy on the neutron energy can be achieved for neutron--induced protons occurring far from the production point. 
The expected kinetic energy smearing matrix as well as the momentum resolution for neutrons traveling more than 40 cm and detected in SuperFGD are shown in figure~\ref{fig::neutron_05}.
The momentum resolution is shown for both an expected time resolution of a MIP particle
and an improved time resolution (e.g. higher scintillation light and/or proton traveling through more cubes).
It becomes clear that, depending on the neutron interaction topology, a quite precise measurement of the neutron momentum, between about 15\% and 27\%, is potentially achievable.
Additional improvements could be obtained by requiring a longer neutron flight path with the drawback of a reduction in statistics.

This study clearly shows the capability of SuperFGD in detecting neutrons with high efficiency and potentially of measuring their kinetic energy by time-of-flight.
While preliminary studies show a quite good separation between neutrons and photon produced by nucleus de-excitation, studies are ongoing to evaluate the background due to neutrons producted by neutrino interactions outside the SuperFGD fiducial volume.


\section{Prospects for the T2K oscillation analysis}
In order to estimate the impact of the ND280 upgrade over the future neutrino oscillation measurements of T2K, a reasonable estimate on the extrapolation at high statistics of the present uncertainties, quoted for instance in Ref.~\cite{Abe:2018wpn}, can be done. 

\begin{figure}[hbtp]
    \begin{center}
    \includegraphics[width=0.45\linewidth]{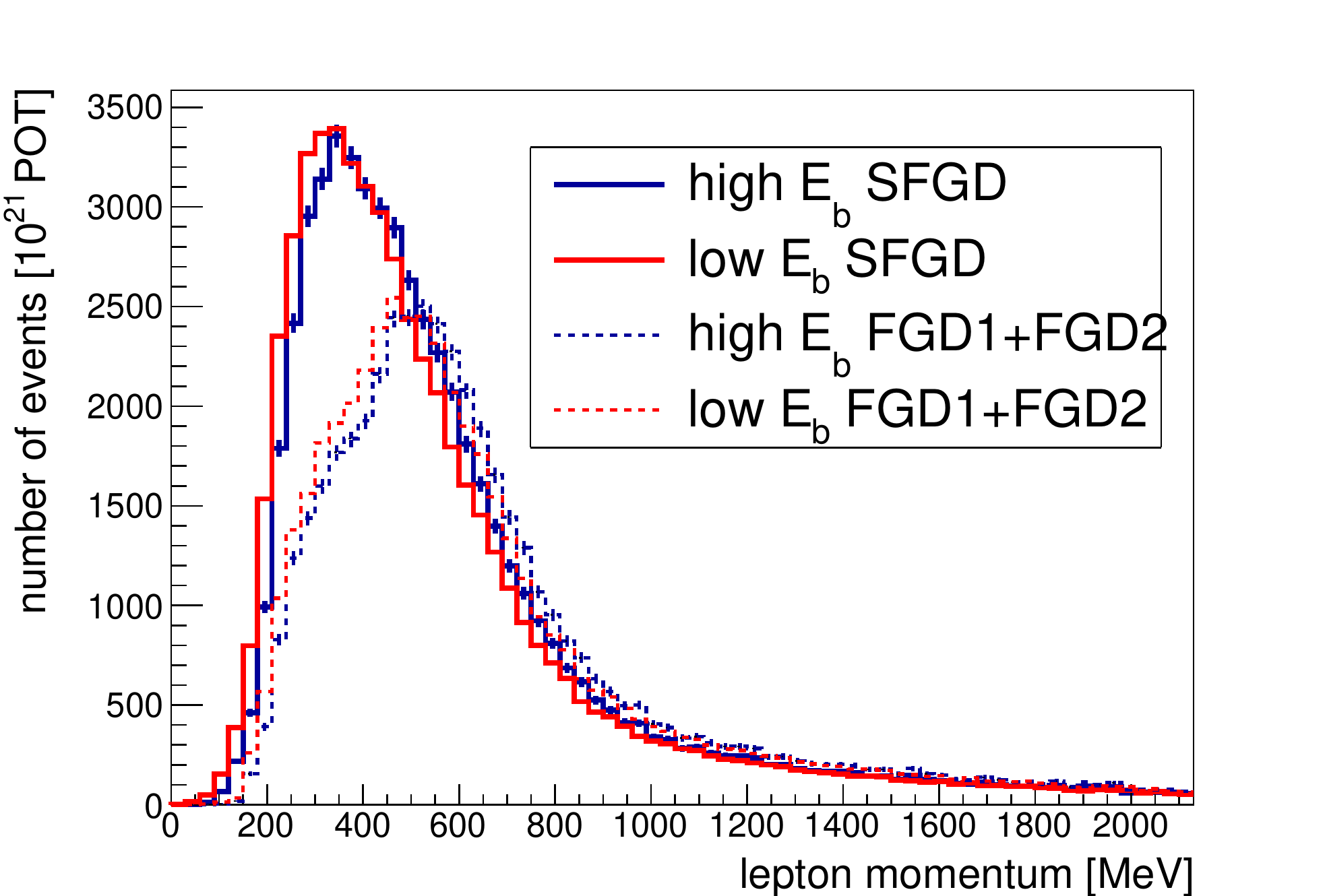}
        \includegraphics[width=0.45\linewidth]{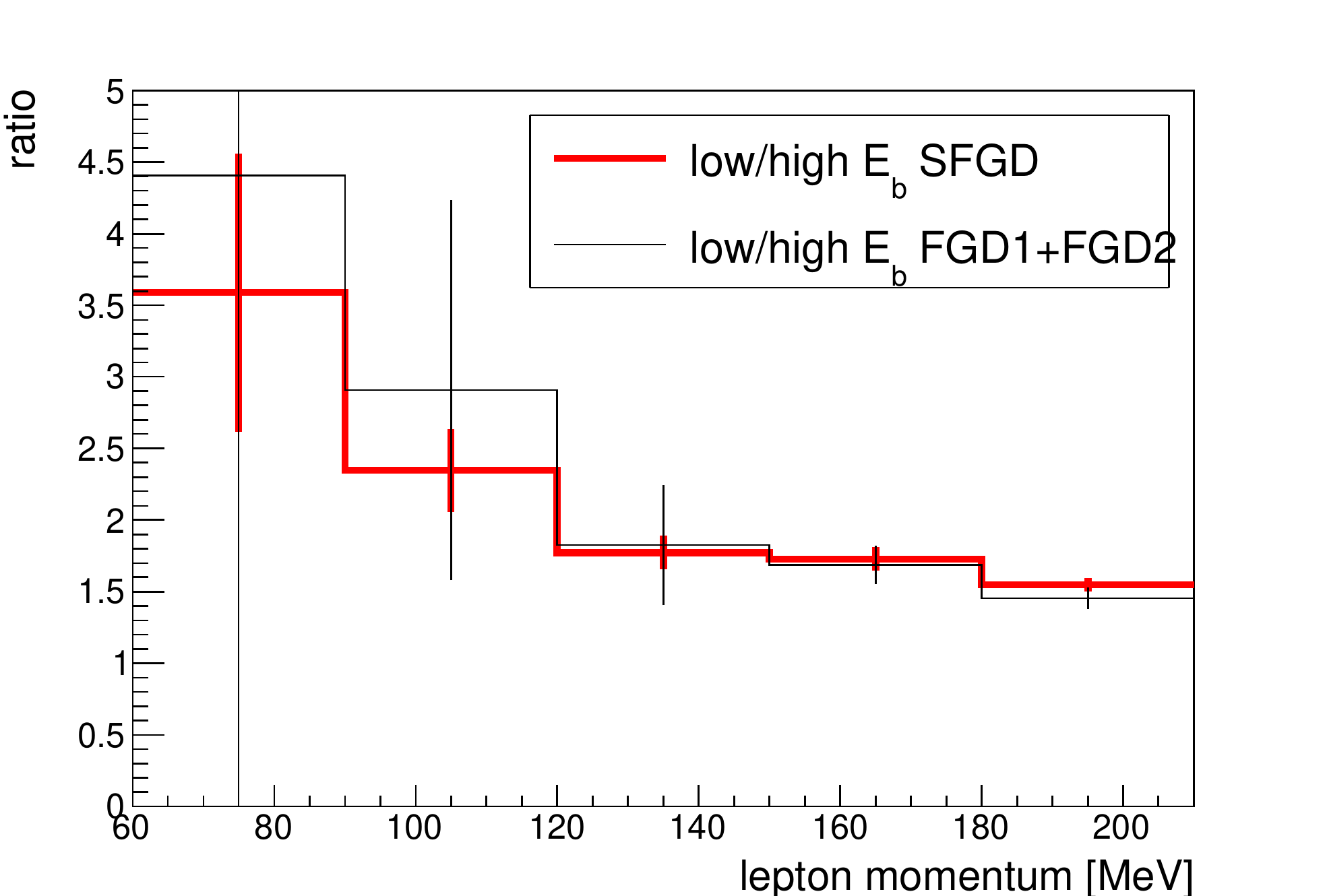}
    \caption{Left: spectrum of muon momentum with different binding energy values (16 MeV and 43 MeV, as evaluated in Ref.~\cite{Abe:2018wpn}) for CCQE events selected in FGD1 and FGD2 with ND280 detector and in SuperFGD with ND280 upgrade detector. Right: ratio of the spectra with different binding energy. Statistical errors only.}
    \label{fig:EbEffect}
    \end{center}
\end{figure}

The ND280 unconstrained cross-section uncertainties, notably on the number of $\nu_e$ events, are dominated in Ref.~\cite{Abe:2018wpn} by the effect of the binding energy. This is the energy needed to extract a nucleon from the nucleus in a neutrino-nucleus interaction. This quantity has been actually measured with good precision in electron scattering data
and it is expected to be the same in electron-nucleus and neutrino-nucleus interactions (see, for instance, ~\cite{Bodek:2018lmc}). Unfortunately the electron scattering constraints could not be included in the neutrino interaction model used in Ref.~\cite{Abe:2018wpn} due to time constraints. Moreover, in that analysis the binding energy uncertainty was not constrained by the ND280 data.
A new model~\cite{Nieves:2011pp}, with more careful treatment of binding energy and other nuclear effects, is already implemented in the new version of T2K Monte Carlo and ready to be deployed. The framework to exploit the ND280 data to constrain the binding energy is also being developed. Such constraints will be highly improved by the ND280 upgrade thanks to the new high statistics sample with low muon momentum selected in the SuperFGD standalone. Figure~\ref{fig:EbEffect} shows the muon momentum spectrum for events reconstructed and selected following neutrino interactions in the FGDs and the SuperFGD: the power of the latter is clearly visible enabling larger statistics, notably in the low momentum region where the effect of binding energy is particularly relevant. The SuperFGD sample, complemented by the mentioned improvements in the model, will allow to constrain the binding energy uncertainty well below the other systematic uncertainties.

In T2K-II the dominant ND280 unconstrained cross-section uncertainty will be due to the difference between $\nu_e$ and $\nu_\mu$ interactions. Such uncertainty is estimated to be 3\% and an intense work is needed on the theory of neutrino-nucleus interactions in order to reduce this. The impact of secondary class current and radiative corrections~\cite{Day:2012gb}, which depends on the mass of the outgoing lepton, should be calculated and included in the models. A more precise measurement of nuclear effects in $\nu_\mu$ interactions will also help in reducing the $\nu_e/\nu_\mu$ uncertainty, as shown for instance in~\cite{Nieves:2017lij}. Such theoretical improvements will be complemented by improved constraints from the ND280 upgrade, notably thanks to the improved electron/$\gamma$ separation in the SuperFGD discussed in Sec.~\ref{sec:egmammaSFGD}. Nevertheless, due to the purity of the T2K $\nu_\mu$ beam, the $\nu_e$ sample will have too low statistics to reach a precision of few \%, thus the $\nu_e/\nu_\mu$ uncertainty will be fully driven by theoretical considerations.

The systematic uncertainties related to the SuperKamiokande detector are estimated using a sample of atmospheric neutrinos and are today limited by the available statistics. In particular, larger statistics of atmospheric neutrinos will allow to refine the evaluation of systematic uncertainties in smaller bin of the kinematic variables. No official estimate is available yet, but we can reasonably evaluate these uncertainties at high statistics to be well below the present 2\%. 

Finally the uncertainties due to Final State Interactions (FSI), Secondary Interactions (SI) and Photo-Nuclear (PN) effects are today conservatively quoted to be around 2-3\%. The PN effects corresponds to the emission of low-energy photons from the excited nucleus following the neutrino interactions. This process may induce some NC background events to be misidentified as 1-Ring $\nu_e$ events but this effect is sub-dominant with respect to FSI and SI uncertainties. In the present T2K analysis the constraints on the FSI and SI from the near detector are not propagated to the far detector, because of lack of information on the correlation between Carbon and Oxygen uncertainties. This problem has been recently studied in Ref.~\cite{PinzonGuerra:2018rju} where a detailed fit to all the pion-nucleus scattering data, including different target materials, is performed and an improved FSI uncertainty is obtained. Moreover recent improvements in the Monte Carlo allowed to describe FSI and SI in the same model and thus improving the constraints on such effects, fully exploiting their correlation. Such developments will be complemented by a high statistics sample of neutrino interactions on Oxygen in the WAGASCI detector~\cite{Ovsiannikova:2017uzn}. Exploiting these new developments and the expected results on Oxygen, a residual FSI, SI and PN uncertainty of the order of 1\% can be considered for T2K-II. 

\begin{figure}[hbtp]
    \begin{center}
    \includegraphics[width=0.8\linewidth]{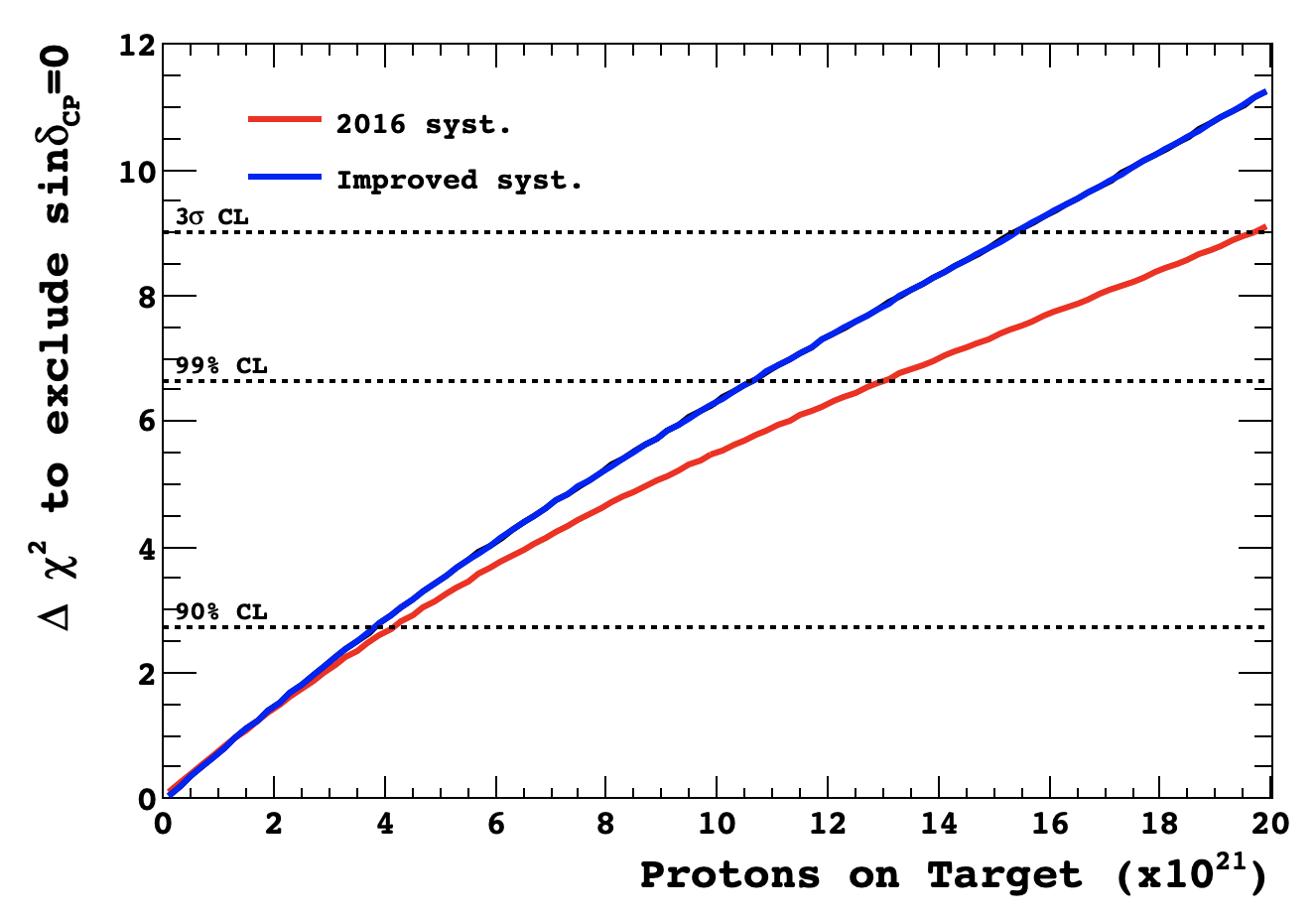}
    \caption{Sensitivity to Charge-Parity violation as a function of POT. The systematic uncertainties corresponding roughly to Ref.~\cite{Abe:2017vif} are compared to the case of 4\% systematic uncertainties on all the SuperKamiokande samples, as can be conservatively estimated in T2K-II using the constraints from ND280 upgrade (see Tab.~\ref{Table:oa_err}).}
    \label{fig:sensitivity}
    \end{center}
\end{figure}

In summary, on the basis of these estimates, the uncertainties on the number of events at SuperKamiokande can be extrapolated for T2K-II to be of the order of 1\% (3\%) due to ND280 unconstrained $\nu_\mu$ ($\nu_e$) cross-sections, 1\% due to FSI, SI and PN effects and 1\% due to the SuperKamiokande detector. It is therefore crucial to strengthen the ND280 constraints on flux and cross-section uncertainties well below 2\%. The analysis in Sec.~\ref{sec:performance_fit}, exploiting only the samples with muons reconstructed in the horizontal and vertical TPCs, have shown that a relative improvement of 30\% on such constraints can be obtained thanks to the ND280 upgrade, enabling uncertainties below 2\% with $8\times 10^{21}$~POT. The summary of the expected systematic uncertainties is reported in Tab.~\ref{Table:oa_err}. The expected sensitivity on Charge-Parity violation search for this level of systematics is shown in Fig.~\ref{fig:sensitivity}.

\begin{table}[htbp]
	\centering
	\caption{Projected systematic uncertainties for the oscillation analysis in T2K-II. The constraints of ND280 upgrade are evaluated for $8\times 10^{21}$~POT. The total is evaluated considering the various sources of uncertainties to be uncorrelated.}
	\begin{tabular}{c|cc}
	\hline
	\hline
	Source of uncertainty 	&  	 $\nu_e$ CCQE-like & $\nu_{\mu}$  \\ 
	 &$\delta N / N$  & $\delta N / N$  \\
	\hline
    ND280 unconstrained cross-section & 3\% & 1\% \\
     Flux + cross-section (constrained by ND280 upgrade)
		&  1.8\% & 1.9\% \\  
	SuperKamiokande detector systematics & 1\% & 1\% \\
	Hadronic re-interactions & 1\% & 1\% \\
	\hline
	\hline
	Total & 3.8 & 2.6\\
	\hline
	\hline
	\end{tabular}
	\label{Table:oa_err}
\end{table}

A possible evidence of Charge-Parity violation at 3$\sigma$ level in the neutrino oscillation will certainly require an unprecedented control of the complex systematic uncertainties due to neutrino-nucleus interaction modeling.
The previous results of T2K along the years have shown that the modeling of neutrino-nucleus interactions is a delicate task and, each time the precision of ND280 constraints increases with the statistics, new area of such modeling have been explored and new challenges arise. 
The new samples of low momentum muons and protons in SuperFGD and the new sample of high angle muons in the HA-TPC will be a crucial input to meet these challenges and to allow a robust estimation of the systematic uncertainties.